\documentclass[6x9]{newmath}
\usepackage[utf8]{inputenc}
\usepackage{tikz,pgfplots}
\usepackage{tikz-3dplot-circleofsphere}
\usepackage[compat=0.6]{yquant}
\usetikzlibrary{decorations.markings,decorations.pathmorphing,arrows,arrows.meta,positioning,automata,calc,patterns,fadings}
\usepackage{amssymb}
\usepackage{hyperref}
\usepackage[type={CC},modifier={by-sa},version={4.0},]{doclicense}
\usepackage{gensymb}
\usepackage{makecell}
\usepackage{pifont}
\usepackage{xcolor}
\usepackage[utf8]{inputenc}
\usepackage{float}
\usepackage{graphicx}
\usepackage{etoolbox} 
\tikzset{>=latex}

\let\oldding\ding
\renewcommand{\ding}[2][1]{\scalebox{#1}{\oldding{#2}}}
\newcommand{\xmark}{\ding[2.2]{56}}
\newcommand{\xmarksmall}{\ding[1]{56}}
\definecolor{darkred}{RGB}{200,0,0}
\colorlet{lightred}{red!80!black}
\definecolor{newgreen}{RGB}{33,172,47}
\definecolor{newblue}{RGB}{43,174,222}
\colorlet{darkgreen}{green!50!black}
\colorlet{lightgreen}{green!80!black}

\colorlet{wall}{blue!30!black}
\colorlet{myblue}{blue!70!black}
\colorlet{myorange}{orange!90!black!80}
\colorlet{myred}{red!70!black}
\colorlet{mydarkred}{red!50!black}
\colorlet{mylightgreen}{green!60!black!70}
\colorlet{mygreen}{green!60!black}
\colorlet{myredgrey}{red!50!black!80}
\colorlet{myshadow}{blue!30!black!90}
\tikzstyle{wave}=[myblue,thick]

\def\tick#1#2{\draw[thick] (#1) ++ (#2:0.1) --++ (#2-180:0.2)}

\newcommand{\bra}[1]{
    \ensuremath{\left\langle #1 \right|}\xspace}

\newcommand{\ket}[1]{
    \ensuremath{\left|  #1 \right\rangle}\xspace}

\newcommand{\braket}[2]{\ensuremath{\langle #1|#2 \rangle}}
\newcommand{\ketbra}[2]{\ensuremath{|#1 \rangle \langle #2|}}

\newcommand{\rrr}{{\bf Van Meter:{ }}}
\newcommand{\mmm}{{\bf Hajdu\v{s}ek:{ }}}

\newboolean{ShowComments}
\setboolean{ShowComments}{true}  
\ifthenelse{\boolean{ShowComments}}%
	{
		\newcommand{\ColorComment}[3]{%
				{\colorbox{#1}{\color{white}   \textsf{\textbf{#2}}} \textcolor{#1}{#3}}}

	}%
	{
		\newcommand{\ColorComment}[3]{}

	}%

\definecolor{rdvcolor}{rgb}{0,0.5,0}
\definecolor{satohcolor}{rgb}{0.5,0,0.5}
\definecolor{michalcolor}{RGB}{255,127,80}
\definecolor{shotacolor}{rgb}{0,0,1}
\definecolor{cocoricolor}{RGB}{238, 130, 238}
\definecolor{naphanncolor}{RGB}{112, 51, 173}
\definecolor{touchcolor}{RGB}{45,0,134}
\definecolor{shigeyacolor}{RGB}{198,53,39}

\title{Quantum Communications}
\subtitle{Online Lecture Notes}
\edition{This is not an official edition, but is instead a temporary facsicle made available only for review purposes on \today. It is nearing completion and almost ready for human consumption. Feedback from \emph{friendly} reviewers always welcome! Also, the book is not being published by MIT Press, we are just building off of their \LaTeX template.}
\author{Michal Hajdu\v{s}ek and Rodney Van Meter}

\begin{document}
\halftitlepage


\titlepage

\begin{copyrightpage}
\doclicenseThis%

\end{copyrightpage}

\dedication{For my sisters, who make the world a better place\\
---rdv\\
\newline
\newline
For my wife and children, my three deterministic sources of bright light\\
---michal}



\tableofcontents
\listoffigures

\begin{contributors}[twocolumn]

\contrib 
Michal Hajdu\v{s}ek\\
Graduate School of Media and Governance, Keio University\\
Fujisawa, Japan

\contrib 
Rodney Van Meter\\
Faculty of Environment and Information Studies, Keio University\\
Fujisawa, Japan

\contrib
Yuka ``shori'' Kataoka\\
Graduate School of Media and Governance, Keio University\\
Fujisawa, Japan

\contrib
Achmad Husni Thamrin\\
Graduate School of Media and Governance, Keio University\\
Fujisawa, Japan

\contrib
Leo Watson\\
Q-Leap summer intern, 2021

\contrib
Samanvay Sharma\\
Graduate School of Media and Governance, Keio University\\
Fujisawa, Japan

\contrib 
A long list of students!\\
AQUA (Advancing Quantum Architecture) Kenkyuukai/Research Group\\
Keio University

\contrib
Linda Van Meter\\

\contrib
Sheila Rose\\

\contrib
Lanty Rose\\

\end{contributors}

\begin{preface}
Q-Leap is a large effort supported by the Japanese government, including major projects in quantum hardware, software, AI, and education or human resources development.  This book is a compilation of the contents of lessons on quantum communication from the Q-Leap Education project known as Quantum Academy of Science and Technology~\footnote{\url{https://qacademy.jp/en/}}. Quoting the website, ``The Quantum Academy of Science and Technology operates as a common core program of the Q-LEAP Program of the Ministry of Education, Culture, Sports, Science and Technology based on 'Development of the Quantum Academy of Science and Technology Standard Program'.''

The Quantum Academy aims to produce over thirty modules of similar size over a six-year period, with the work being done by the National Institute of Informatics (NII), University of Tokyo, Nagoya University, Kyushu University, and Keio University.  These thirty-plus modules will collectively comprise about one-quarter of an undergraduate degree, enough to build a major in quantum engineering for universities that choose to do so. Within this overall project, Keio's responsibility is quantum communications.  This book represents the first module delivered by AQUA, the research group of Prof. Rodney Van Meter at the Keio University Shonan Fujisawa Campus (SFC).  This module will be followed by ``From Classical to Quantum Light'' and ``Quantum Internet'' as the next two modules.  Each module will be available first as an online course, then later as a CC-BY-SA book.

This book itself, the videos and slides, and accompanying software demos are all licensed Creative Commons. You are not only \emph{allowed} to reuse these materials, you are \emph{encouraged} to do so, provided you credit the original authors, the AQUA team, and our institution, Keio University.

If you want to recompile the \LaTeX, the source is available on Github. You are free to reorder, remove or add chapters to suit your needs for your particular purposes.  More suggestions on how to use the material are in the next section.

\author{Michal and Rodney}
\date{sometime in 2023}
\end{preface}

\chapter*{Using This Book}

\section*{In a Degree Curriculum}

This book recapitulates and extends about eleven hours of online video~\footnote{\url{https://www.youtube.com/playlist?list=PLCTGenrx1-SOC-b98RCC1uEGI-Sc-N3C-} or find the corresponding playlist at \url{https://www.youtube.com/c/QuantumCommEdu}.}, organized as a set of fifteen lessons corresponding to the chapters of this book. Coupled with homework assignments, it could be twenty to forty hours of work for well-prepared students (and substantially more for not-as-well-prepared students who build their understanding along the way).  As such, it is intended to represent one credit toward graduation by Japanese standards, where an undergraduate degree is typically 124 credits over eight 15-week semesters. As typical courses are two credits, this content is expected to be paired with additional material in a Japanese curriculum to form a single for-credit course. In the United States or elsewhere where the typical course is a larger chunk of learning, this material may form a third or a quarter of a course.

In conjunction with the videos and this book, we recommend using the material in an active learning context, such as a \emph{flipped classroom} environment.

Additional details, such as the recommended mathematics background, are covered in Sec.~\ref{sec:mod-over}.

\section*{In a Tutorial}

It is also possible to take a subset of this material and use it in a half-day or full-day tutorial, with appropriate background preparation.  For example, at the Third Workshop for Quantum Repeaters and Networks, held in Chicago in August 2022, we asked attendees to prepare by watching the following subset of the videos plus part of a seminar given by Prof. Van Meter:
\begin{itemize}
\item QSI Seminar: Prof. Rodney Van Meter, Keio University, Engineering the Quantum Internet, 30/06/2020~\footnote{\url{https://www.youtube.com/watch?v=FZsEOhM4mIY}}
At minimum, watch from 9:45 to 27:30 on 
“Applications of a Quantum Internet”
\item 2 Quantum States
\begin{itemize}
\item 2-1 Qubits
\item 2-2 Unitary operations				
\item 2-3 Measurement
\item 2-4 Probabilities, expectation, variance
\item 2-5 Multiple qubits
\end{itemize}
\item 3 Pure and Mixed States
\begin{itemize}
\item 3-1 Noisy world
\item 3-2 Outer product
\item 3-3 Density matrices
\item 3-4 Pure vs mixed states
\item 3-5 Fidelity
\end{itemize}
\item 4 Entanglement
\begin{itemize}
\item 4-1 CHSH Game
\item 4-2 Entangled states
\item 4-3 Bell states
\item 4-4 SPDC
\item 4-5 Entanglement as a resource
\end{itemize}
\item 8 Teleportation
\begin{itemize}
\item 8-1 Introduction to teleportation
\item 8-2 Teleportation protocol
\item 8-3 No-cloning theorem and FTL communication
\end{itemize}
\item 12 Quantum Repeaters
\begin{itemize}
\item 12-1 The need for repeaters
\item 12-2 Making link-level entanglement
\item 12-3 Reaching for distance: Entanglement swapping
\item 12-4 Detecting errors: purification
\item 12-5 Making a network 
\end{itemize}
\end{itemize}

At the workshop, we conducted a four-hour session consisting primarily of hands-on work.  The hands-on work demonstrated the three key concepts of \emph{teleportation}, \emph{entanglement swapping}, and \emph{purification}. The Qiskit Jupyter notebook~\footnote{\url{https://github.com/sfc-aqua/wqrn-tutorial}} and QuISP (Quantum Internet Simulation Package)~\footnote{\url{https://github.com/sfc-aqua/quisp}} demos used in the hands-on session are available on Github. 

\section*{Reusing, Recompiling, Reorganizing and Contributing to the Material}

You are welcome to recompile this book to meet your own local needs, adding, removing or rearranging material as you please. For example, if you use this book in a graduate-level course or after students have a substantial background in quantum information, the introductory material in Chapters 2 and 3 may be unnecessary. Or, you may wish to combine chapters from this book with other material of your own to create a single volume of lecture notes for a course at your institution.

The Creative Commons wiki provides some additional explanation on the license~\footnote{\url{https://wiki.creativecommons.org/wiki/ShareAlike_interpretation}.}.  The source for the book is available on GitHub~\footnote{\url{https://github.com/sfc-aqua/Overview-of-Quantum-Communications-E}}. 

Contributions to the book itself are welcome.  The text can always be improved; it began life as direct transcripts of the videos, and as such does not always flow as smoothly in book form.  Contributions of exercises for students are especially welcome.  Additions that substantially modify the flow of material may be organized as supplementary material at the end of chapters or the book, in order to maintain the correspondence between the online videos and the text.  The preferred form of contributions is as \emph{pull requests} on Github.  Those whose contributions are accepted will have their names and affiliations added to the contributors page, at their discretion.

\part{Quantum Mechanics for Quantum Communication}
\begin{partintro}
\partintrotitle{Introduction to the first chapter block}
In this first block of chapters, we open by introducing the entire module (or book, in this form) in Chapter 1.  All readers will benefit from reading this chapter, in order to understand the purpose and flow of the book.

Chapters 2-4 are a very brief introduction to quantum information and quantum computational notation and operations.  If you are already familiar with Dirac's ket notation for multiple qubits, you can easily skip Chapter 2; likewise, if you are familiar with density matrices for mixed states and their fidelity you can skip Chapter 3. Chapter 4 may similarly feel like familiar territory, but we encourage you to at least check out 4.1 (CHSH Game) and 4.4 (SPDC), as they have material that is not included in all introductions to the field.
\end{partintro}

\chapter[Introduction]{Introduction}
\label{sec:1_Introduction}

\chaptermark{Introduction}

In this chapter, we will give you an overview of how communication has evolved over the many thousands of years of human culture and civilization, then we will tell you about sending signals between parties of a network.
We will tell you the differences and similarities between digital and analog signals.
We will introduce the fundamental building block of modern communication, the \textbf{\emph{bit}}\index{bit}.
We will move on to quantum communication and explain why you should care about it, what new capabilities that it brings to the table, and what the new challenges are that face us in designing quantum communication systems. 
We will move on to the disruptive nature of quantum technologies and quantum communication in particular.
We will conclude this chapter by giving you an overview of the entire module, what the prerequisites are, what you will learn, and the outcomes of the module.

\section{History of Communication}

Methods of communication advance in order to accommodate a growing society while new technological advances allow society to grow and expand.

\begin{figure}[ht]
    \centering
    \includegraphics[width=0.6\textwidth]{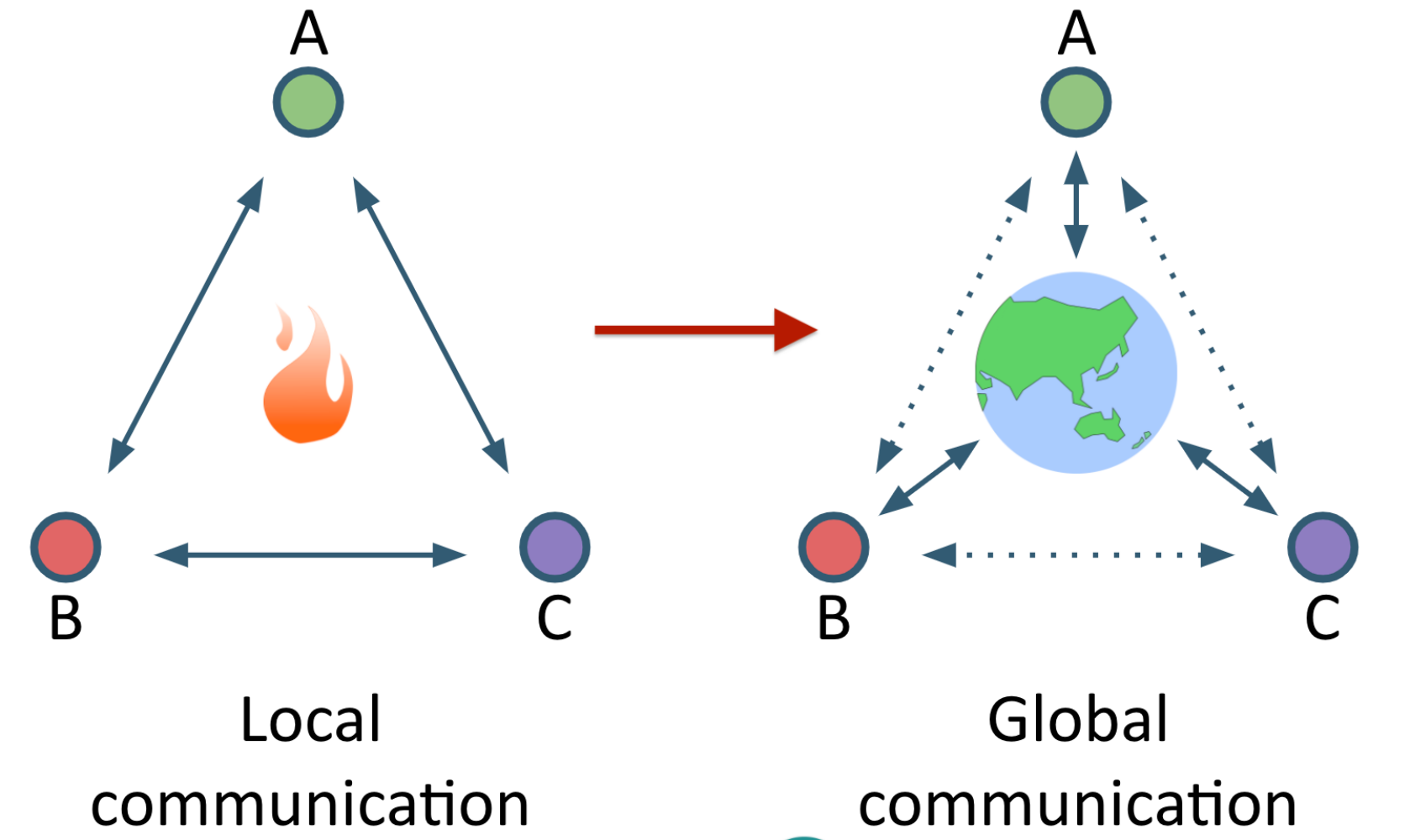}
    \caption[Evolution of communication.]{Evolution of communication. At the start, all communicating parties had to be within earshot of each other to exchange information quickly and effectively. Today, thanks to the Internet, we can still do the same without worrying about the parties' locations.}
    \label{fig:1-1_communication}
\end{figure}

Humans are social creatures by nature.
Effective communication has been crucial to our survival;
in particular, it has been very important to share information about potential new sources of food and danger.
We communicated by gathering around the fire and talking to each other as illustrated in Fig.~\ref{fig:1-1_communication}.
Over many thousands of years, the methods of communication have evolved to become more efficient and longer ranging until we have finally reached our modern age where nearly every device that we have in our possession, our phones, our TVs, our iPads, even our fridges are all connected to a massive internetwork, the Internet.
We can be separated by many thousands of kilometers, but when we communicate it almost feels as if we are all sharing the same fire like in the old times.
We have gone from local communication to truly global communication.

Let's see how such a massive transformation happened.
In the next couple of paragraphs, you are going to see some famous examples from history illustrating the many different ways in which we have communicated.
After the invention of paper (or papyrus, parchment or vellum), the most obvious way was to take your message, write it on a piece of paper and send the message directly.
A very famous example of this method was the Battle of Marathon in 490 BC, where the legend goes that a runner was sent from where the battle took place near the city of Marathon to Athens, a distance of about 40 kilometers.
But that is historically not true.
The Athenians were trying to gain support of the army from the city of Sparta, located around 225 kilometers away.
The Athenians sent a runner carrying their message to Sparta.
And that journey, believe it or not, took a little less than a day.

Another method of sending a message directly was using birds such as homing pigeons.
These pigeons were specially trained to always return to their homes or their coops. People used to put the pigeons in cages when traveling somewhere. Whenever they needed to send a message back home, they would take a little piece of parchment paper, write the message down, attach it to the pigeon's leg, and let the pigeon go. The pigeon would automatically fly back and deliver the message.
The range of these birds was around a very impressive 1600 kilometers.
The average speed was around 95 kilometers per hour.
The top speed of some very athletic pigeons was upwards of 160 kilometers per hour.

Our final example of sending messages directly is the Pony Express.
The Pony Express was a company set up in the mid-nineteenth century in the United States to connect the East and West Coasts.
If you were on the East Coast and you wanted to send a letter to somebody on the West Coast, you went to this company, and then they would send a horse and rider who would deliver your message.
The messenger would ride the horse to a relay station on the way, change to a fresh horse and continue with your message onwards to the next relay station.
Believe it or not, the entire journey of 4,000 kilometers took approximately 10 days.
In the context of communication speeds at the time, this was very, very fast.
Not only could you send letters, but you could also send some small parcels.
The Pony Express was the very early Amazon!
But despite the company's successful system, it only existed for one year; it was an early victim of the advance of technology, because it was created just as the method of communication was changing from direct transmission to electrical telegraphy.

But before we get to this revolution in communication technology, let's talk about these different ways of sending a message. A written letter can carry a fairly detailed and nuanced message.
However, sending a message directly in the form of a written letter is generally \textit{\textbf{slow}} and suffers from \textit{\textbf{reliability issues}}.
You may lose your pigeon, or your runner may become exhausted and just give up the task.
If this happens you have to find another pigeon or runner and resend the message, provided that you are even aware that the message has not been delivered yet.

An alternative to direct transmission was \textit{\textbf{optical telegraphy}}.
This is an old but ingenious method where the sender and the receiver share some pre-agreed signals.
The sender uses some optical means in order to generate these signals such that the receiver can see them.
A very good example is the Great Wall of China, which was designed and built to protect the northern border of the Chinese Empire.
Due to its vast length, it was crucial to devise a communication system that could quickly relay messages between the guard towers located along the Wall.
Whenever the enemy tried to attack the Wall, the nearest guard tower would light its signal fire, a pile of wood prepared in advance. This fire was then observed from the neighboring guard towers, which were approximately 2.5-5 kilometers apart.
Then they would light their fires.
Then their neighbors would see that and light their fires.
This is how the message spread across large portions of the Wall very quickly.
One drawback was that the \textit{\textbf{expressibility}} of this communication method was limited.
Only certain messages could be sent.
For example, the enemy is here or it's not, the fire is burning or it's not.
One could develop this method a little bit further by introducing a second fire at each guard house.
When one fire is burning, the enemy is there but in small numbers.
When two fires are burning, the enemy is there in large numbers.
Despite the limited expressiveness of this communication method, this system was very efficient and served well in protecting the Wall.

\begin{figure}[t]
    \centering
    \includegraphics[width=0.7\textwidth]{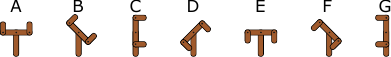}
    \caption[Napoleon's semaphore.]{Napoleon's semaphore encodings for the first seven letters of the alphabet.}
    \label{fig:1-1_napoleon}
\end{figure}

Our last example of optical telegraphy is known as \textit{\textbf{Napoleon's semaphore}}.
This system was a lot more expressive than the signalling system of the Great Wall.
It worked on the basis of crane arms which could be arranged at certain angles.
An operator controlled levers and pulleys in order to arrange the arms at various angles. Each configuration of the arms carried its own meaning.
The leftmost arm configuration in Fig.~\ref{fig:1-1_napoleon} corresponds to the letter ``A".
A network of semaphores, each located around 10 kilometers apart, spanned all of France.
It was possible to send a message from Paris all the way to Venice, a distance of 1100 kilometers, in as little as few hours.
The record time of sending a message between Paris and Strasbourg in the east of France, was an astounding 1 hour.
This would have been unthinkable using the direct communication discussed above.

As you can imagine, there were some drawbacks to this method, the biggest one being that you had to have \textit{\textbf{direct visual contact}} with your neighboring semaphore.
Therefore, the semaphore worked reliably only in good weather and during daytime.
In addition, even though you could express an arbitrary message using this method, it was \textit{\textbf{physically demanding}} to operate the crane arms.

\begin{figure}[t]
    \centering
    \includegraphics[width=0.8\textwidth]{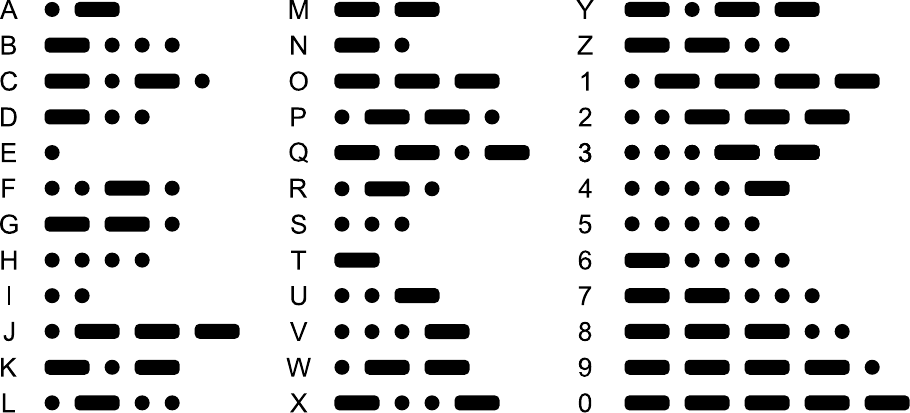}
    \caption{The Morse code.}
    \label{fig:1-1_morse}
\end{figure}

This brings us finally to the reason why the Pony Express operated only for a single year: the advent of \textit{\textbf{electrical telegraphy}} and the invention of the \textit{\textbf{Morse code}} that used electric signals to transmit messages.
It worked by encoding the letters of the alphabet into a series of ``dots'' and ``dashes'' as seen in Fig~\ref{fig:1-1_morse}.
An operator used a telegraph key to close an electric circuit to produce a signal of desired length.
Closing the electric circuit for a short time produced a ``dot'' while keeping the circuit closed for a longer time produced a ``dash''.
If the length of a ``dot'' is one unit, then a ``dash'' has length of 3 units.
Parts of the same letter are separated by 1 unit.
Different letters are separated by a space of 3 units while different words are separated by 7 units.
A skilled operator could transmit up to 30 words per minute, which were decoded by an operator at a distant telegraph station and passed onto the intended recipient.
This communication method allowed for messages to be delivered across greater lengths, spanning continents within minutes, making direct communication such as the Pony Express or optical telegraphy such as Napoleon's semaphore obsolete.

\begin{figure}[h]
    \centering
    \includegraphics[width=\textwidth]{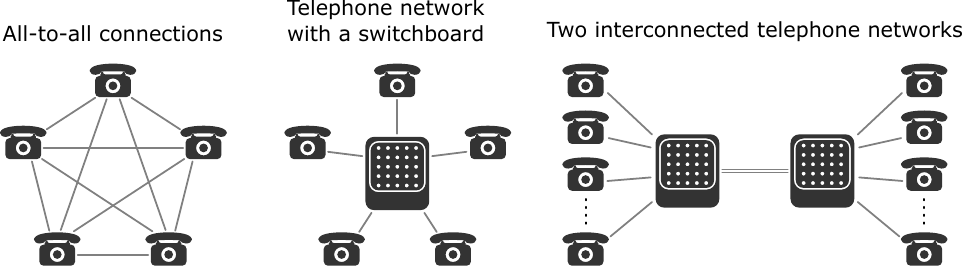}
    \caption[Telephone networks.]{Various topologies for telephone networks.}
    \label{fig:1-1_telephone}
\end{figure}

Eventually, the electrical telegraph gave way to the \textit{\textbf{telephone}}, which implemented the dream of communication between humans using the human voice.
Early telephone connections were simple point to point.
As the demand for communicating via the telephone grew, it became clear the \textit{\textbf{all-to-all}} approach would not work.
Imagine a network of $N$ telephones where all of them are directly connected to each other as seen in the left panel of Fig.~\ref{fig:1-1_telephone}.
Consider adding a new telephone to the network.
In order for the new user to be able to call any of the existing telephones on the network, we have to add $N$ physical connections to all of the existing telephones.
Adding yet another telephone would require another $N+1$ connections.
The all-to-all approach is intuitive but simply does not \textit{\textbf{scale}} with the size of the network.

The solution came with the introduction of a \textit{\textbf{switchboard}}.
In order to call anybody on the network, each unit had to be connected only to the switchboard.
You first called the switchboard, which would then connect you to the desired telephone unit.
With this approach, adding a new telephone to the network required adding a single connection to the switchboard, presenting a \textit{\textbf{constant scaling}}.
In other words, the effort of adding new users to the network did not increase with the size of the network.
Eventually, switchboards of different networks were interconnected together, allowing users from one network to call users on an entirely different network.

The Internet that our society has come to depend on so much works on the same principle.
It is not a single network but a \textit{\textbf{network of networks}}, allowing heterogeneous smaller networks to be interconnected.
Nowadays, a message sent from a laptop can be read by its recipient half-way around the world on their phone within seconds.
A broadcast stream of the FIFA World Cup Final can be enjoyed by millions around the world with minimal delay.
We can even start the air conditioning in some modern cars remotely without having to leave the comfort of our house.

\section{Analog to digital}

\begin{figure}[t]
    \centering
    \includegraphics[width=0.7\textwidth]{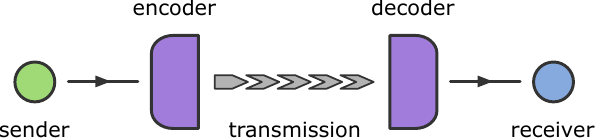}
    \caption[Abstraction of communication.]{Abstract representation of communication between a sender and a receiver.}
    \label{fig:1-2_communication}
\end{figure}

The methods of communication described in the previous section were all very different from each other.
However, at a fundamental level they all followed the same basic principle depicted in Fig.~\ref{fig:1-2_communication}.
A sender \textit{\textbf{encodes}} her message into a form suitable for transmission.
The encoded message is then sent to the \textit{\textbf{decoder}}, which transforms the message back to a more easily readable form and passes it to the receiver.
To make this abstraction a little more concrete, let's consider the electrical telegraph as an example.
The sender brings her message to an operator who knows how to use the Morse key.
The operator uses the key to produce a series of dots and dashes, represented by short and long electrical signals which are transmitted to a distant telegraph station.
There another operator receives these electrical signals, decodes them and reproduces the original message for the receiver.

The immediate question that arises is finding some good ways of encoding the message.
The first method that we consider is encoding the message using an \textit{\textbf{analog}} signal.
Analog signals can admit a continuum of values.
We perceive the world around us as analog.
The loudness of sounds varies continuously from a quiet whisper to a loud music concert.
The temperature rises and falls in a continuous way as measured by our thermometers.
The three primary colors can be mixed together to produce a continuous spectrum of colors.
Humans have evolved to process these types of analog signals; therefore, it makes sense to try using analog signals to encode our messages.
The key is that analog signals can take on values from a continuous interval, representing the range of some quantity.
For example, the voltage of an electric circuit can vary continuously depending on the pressure of sound waves in the microphone of an old telephone.
Early AM radio signals and old TV broadcasting used continuous sinusoidal waves that required simple technology to produce and decode.

\begin{figure}[t]
    \centering
    \includegraphics[width=\textwidth]{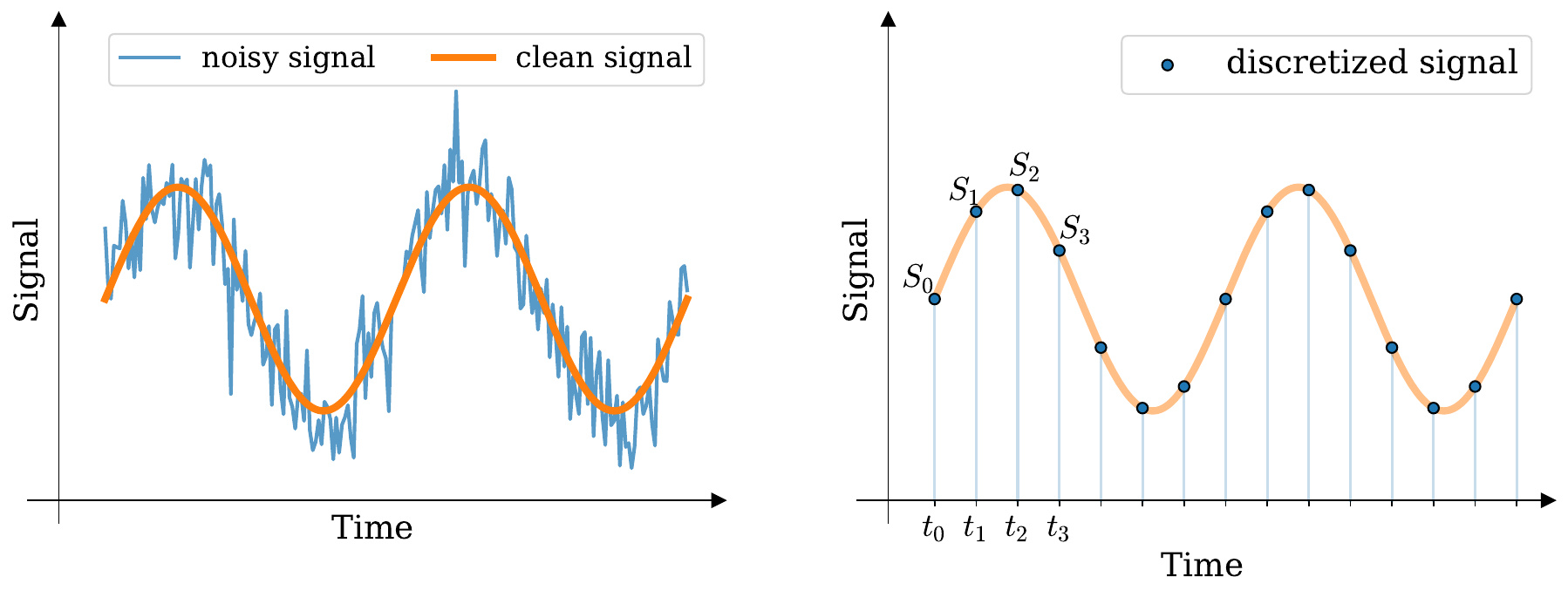}
    \caption[Continuous and discrete signals.]{Analog signal subjected to noise on the left. The right figure shows how the analog signal can be discretized by sampling at a fixed time interval.}
    \label{fig:1-2_signals}
\end{figure}

One problem with analog signals is that they are \textit{\textbf{susceptible to noise}}.
Even small changes to the signal can alter the meaning of the transmitted message.
Due to this sensitivity to noise, analog signals are also difficult to copy.
Every time an analog signal gets copied, it degrades in quality due to both the noise as well as the finite accuracy the signal read out.

Figure~\ref{fig:1-2_signals} shows a clean analog sinusoidal signal in orange.
The blue jagged line represents the original signal affected by noise.
The noisy signal may have a meaning similar to the original noiseless message or may be quite different.

An alternative to analog signals is \textit{\textbf{digital}} signals.
Digital signals are very different from analog signals; they only use a discrete set of values to represent the message and encode it.
We can take the original analog signal, and try to encode it digitally.
A way to do this to \textit{\textbf{sample}} the analog signal at discrete time steps $t_0$, $t_1$, $t_2$, $t_3$ and so on, as shown on the right side of Fig.~\ref{fig:1-2_signals}.
The encoded digital signal is now a set of discrete values,
\begin{equation}
    S_0, S_1, S_2, S_3, \ldots,
\end{equation}
each of some finite accuracy.

The accuracy with which we can reproduce this analog signal depends on how often we look at and measure the signal, which is known as the \textit{\textbf{sampling rate}}, and the precision of each sample, i.e. the number of discrete values that are used.
For slowly varying signals, we don't have to sample that often, while still doing a pretty good job of encoding the analog signal accurately.
But for analog signals that vary quickly, we should sample with higher frequency.

Digital signals are less affected by noise and are therefore easier to copy.
Other advantages include the ease of producing digital signals and the ease of processing with digital logic.

\section{Bits as building blocks}

\begin{figure}[t]
    \centering
    \includegraphics[width=0.8\textwidth]{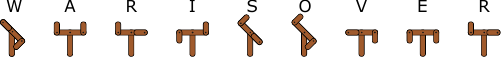}
    \caption[Message with Napoleon's semaphore.]{Encoding a short message using Napoleon's semaphore.}
    \label{fig:1-3_warisover}
\end{figure}

We have seen that digital signals can be more practical than analog signals when it comes to communication.
The question now is how do we represent these digital signals.
We saw an example of a digital system with Napoleon's semaphore.
Let's go back and consider it again.
In order to send a message, for example ``WAR IS OVER'', the arms have to be rearranged as shown in Fig.~\ref{fig:1-3_warisover}.

What does it take to change the state of the semaphore?
Changing the state of the semaphore from ``A'' to ``R'' does not require that much effort.
All we have to do is change the state of one of the side arms and fold it onto the main arm.
On the other had, going from ``W'' to ``A'' requires changing the state of the main arm as well as the two side arms.
This change requires both physical effort as well as longer time.

\begin{figure}[t]
    \centering
    \includegraphics[width=\textwidth]{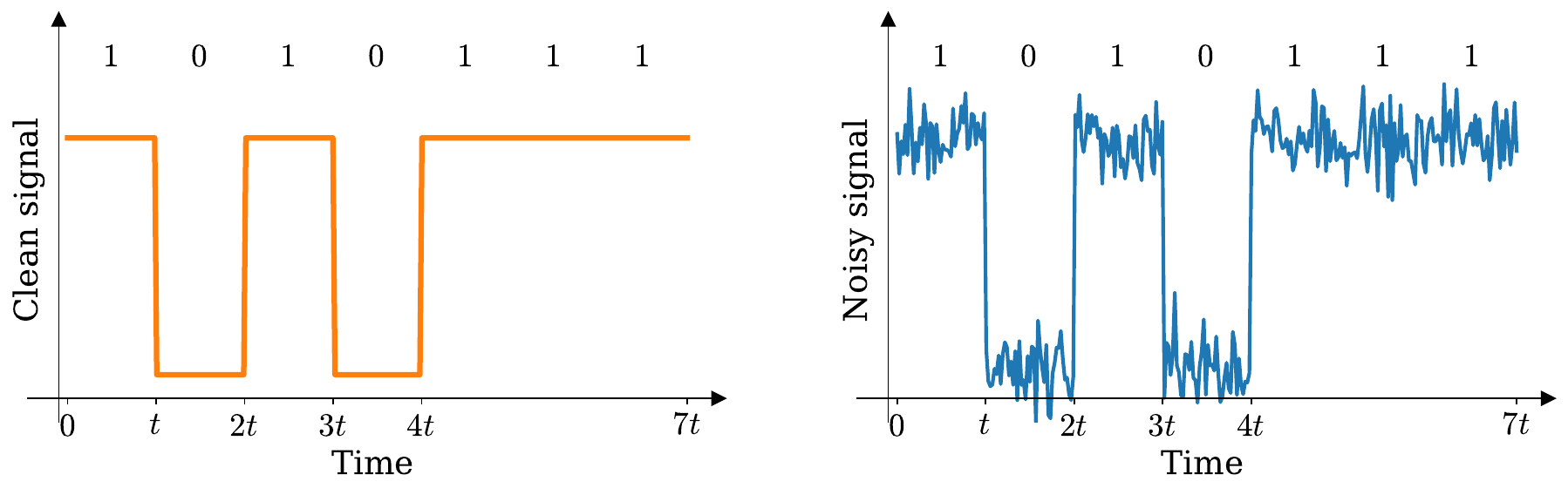}
    \caption[Morse code signal.]{Physical representation of the Morse code signal for the letter ``U''. The desired ideal signal is on the left, and the noisy signal is on the right. Even with noise, it is easy to recover the original message.}
    \label{fig:1-3_morse_signal}
\end{figure}

A good representation for digital signals should require as little effort and time as possible in order to change the state of the digital signal.
This suggests that having fewer internal states produces a better encoding for the message.
The Morse code is such an example, where we only have two internal states.
The letter ``U'' is encoded by two dots followed by a dash.
The physical signal that represents this encoding is shown in Fig. \ref{fig:1-3_morse_signal}.
The signal is switched on for a short time to represent a dot and for a long time to represent the dash.
It is the presence or absence of a signal that conveys the information.
When you change from \emph{no signal} to \emph{signal}, or when the signal \emph{doesn't} change, this carries a small amount of information.

This unit of information is known as a \textit{\textbf{bit}}, which is short for \textit{\textbf{binary digit}}.
In classical computation and in digital communication, the bit is the basic unit of information.
It conveys the message of something being true or false, the signal being on or off.
Usually we write the two states of a bit as 0 or 1.
The signal on the left of Fig.~\ref{fig:1-3_morse_signal} encoding the letter ``U'' in Morse code can be written using bits as 1010111, where zeros and ones denote the state of the bit at different times.
The right side of Fig.~\ref{fig:1-3_morse_signal} shows how the clean signal looks after being affected by noise.
The different states of the bit can still be easily distinguished from each other, hence the meaning of the message can be read out without any ambiguity.
Digital signals are generally more resilient to noise than analog signals.

We have said that a single bit can hold two different values.
How many different values can multiple bits, also called \textit{\textbf{bit strings}} hold?
The following table lists all possible bit string values for 1, 2, 3, and 4 bits.
\begin{table}[ht]
    \centering
    \begin{tabular}{c|c|c}
        Number of bits  & Possible bit strings & Total number \\
        \hline
        1 & 0, 1 & 2 \\
        2 & 00, 01, 10, 11 & 4 \\
        3 & 000, 001, 010, 011, 100, 101, 110, 111 & 8 \\
        4 & 0000, 0001, 0010, 0011, 0100, 0101, 0110, 0111 & 16 \\
        & 1000, 1001, 1010, 1011, 1100, 1101, 1110, 1111 & 
    \end{tabular}
\end{table}
It is clear that the total number of possible bit string values for $N$ bits is $2^N$.

Bits can be used to encode decimal numbers.
Let's examine how decimal notation works.
The decimal system uses ten numerals, 0-9.
In a decimal number, the rightmost digit represents ones, the next digit to the left represents tens, the next hundreds and so on.
For example, the number 1024 can be written in terms of powers of ten as follows,
\begin{equation}
    1024 = 1 \times 10^3 + 0 \times 10^2 + 2 \times 10^1 + 4 \times 10^0.
\end{equation}
This idea carries over to binary numbers where the digits can only take values of 0 or 1, and represent multiples of powers of two.
Let's see how we can write the binary number 1001 in decimal notation,
\begin{equation}
    1001 = 1 \times 2^3 + 0 \times 2^2 + 0 \times 2^1 + 1 \times 2^0 = 9.
\end{equation}

Bits are the building blocks of modern communication.
They are robust to noise, can be easily used to encode and decode information, and are easy for modern computers to process.
Given that bits admit only two values, we might easily consider them to be the most fundamental units of information.
In these lectures, we will learn that this is not quite true.
We will see that quantum communication relies on more fundamental units of information.
Classical bits are merely special cases of their quantum cousins.

\section{Quantum communication}

Information is physical (an aphorism coined by Rolf Landauer).
It is carried by physical systems. The laws of physics determine how we can process or communicate this information.
If we are only considering information processing in the context of classical mechanics, classical electromagnetism, and classical optics, then this will give us tools and ways of processing and communicating the information classically.
However, if we expand our toolbox to include also quantum mechanics and quantum optics, then we are also expanding the ways in which we can process and communicate this information.

The question that we should answer before we start learning about quantum communication is, why do we need quantum mechanics?
Why do we want to use quantum mechanics to process and communicate information? First, quantum mechanics is the fundamental theory of nature as we currently understand it.
It describes the microscopic world where classical mechanics does not apply.
It makes some stunning predictions, which, despite their counter-intuitiveness, have been tested very thoroughly over many decades.
So far, the theory has always been proven correct.
Furthermore, considering new laws of physics, and applying them to information processing and communication, often leads to new ways of processing and communicating this information.

These reasons in themselves are very fundamental, but there are also practical reasons.
Current computing technology is hitting a classical to quantum boundary.
Maybe you have heard of Moore's law, which despite its name is an observation rather than a physical law.
Moore's Law states that the number of transistors in an integrated circuit doubles about every two years.
It is astonishing that this observation has held for decades.
Chip manufacturers now can pack about ten orders of magnitude more transistors into the same area than when integrated circuits were first invented. The chips themselves are not getting any bigger, so in order for Moore's Law to hold, the transistors must get smaller and smaller.
In the beginning, the transistors were approximately 10 microns across.
In the 1990s, they moved to about 600 nanometers.
Now, we are at the level of single-digit nanometers.
The transistors have gotten so small that we need to worry about the effects that quantum mechanics has on the transistors' operation.

\begin{figure}[t]
    \centering
    \includegraphics[width=0.8\textwidth]{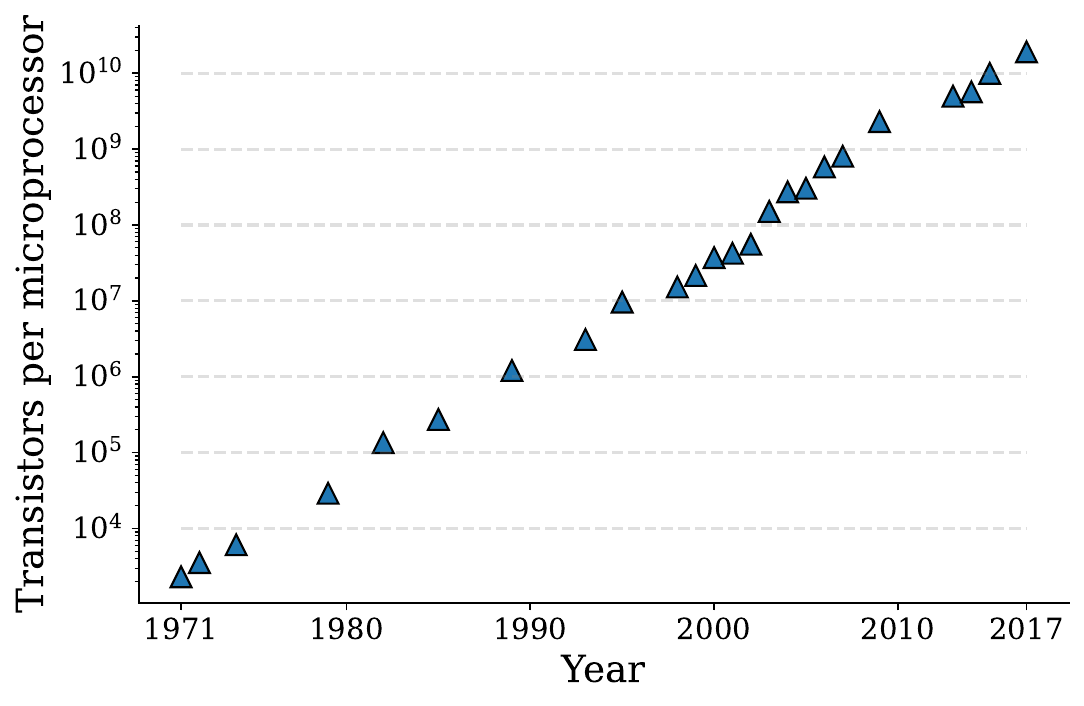}
    \caption[Moore's law.]{Number of transistors on a microprocessor has been doubling nearly every two years.}
    \label{fig:1-4_transistors}
\end{figure}

What are the two main ingredients that set quantum communication apart from its classical counterpart?
The first one is the \textit{\textbf{superposition principle}}.
This principle is not really anything new, as it is observed in the classical world as well.
We are all familiar with superposition of waves.
What is meant by superposition principle in quantum communication is the ability of quantum bits to be in a superposition of their usual classical states.
The possibility of a quantum bit to be both on and off at the same time is mind-bending but is at the heart of quantum communication.

Expanding the principle of superposition to multiple particles, we arrive at the concept of \textit{\textbf{entanglement}}.
Entanglement has no classical counterpart whatsoever.
It correlates distant quantum objects across large distances of space much more strongly than is possible using classical laws of nature.
The beauty of entanglement is that it allows for radically new ways to communicate and is used as a resource in quantum communication.
Distribution of entanglement is therefore one of the main jobs of quantum networks.

Quantum communication protocols are currently very difficult to implement.
Quantum systems are very delicate as they \textit{\textbf{decohere}} extremely rapidly, leading to loss of their quantum properties.
They go from being true and false at the same time, to being only true or only false.
Basically, they just become classical bits.

Quantum systems are difficult to build at the hardware level, but at the same time, it is also conceptually challenging to think about new ways of exploiting their quantum properties.
Designing new protocols for processing and communication in the quantum realm requires new tools.
It requires a completely new mental framework for how we approach problems, and how we solve problems.
This all seems very daunting but these challenges should be viewed as opportunities.
Quantum computation and quantum communication are vibrant and cross-disciplinary fields.
Engineers, physicists, mathematicians, and computer scientists are all working together on cracking difficult questions whose solutions will lead to incredible new quantum technologies.

\section{Security in the quantum age}

Quantum technologies carry the potential to impact a number of important areas.
Quantum simulation and computation are thought to bring about new methods in developing materials with novel and useful properties, as well as novel drugs.
Quantum metrology allows for measurements with unprecedented resolution and sensitivity.
Quantum machine learning aims to exploit the properties of quantum mechanics to enhance artificial intelligence.
Lastly, you may have heard that quantum computers will be able to crack some widely used encryption schemes deployed currently, which may sound like a doomsday scenario.
Since these lecture notes focus on quantum communication, we will briefly discuss what this claim means and show how quantum technologies can also enhance security.

\begin{figure}[t]
    \centering
    \includegraphics[width=0.5\textwidth]{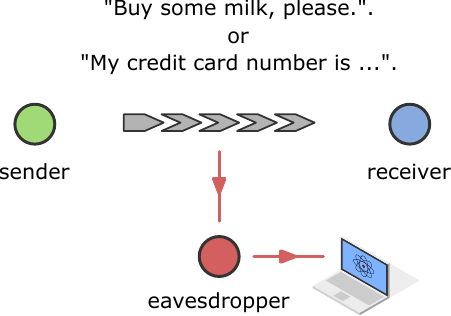}
    \caption[Security in the quantum age.]{An eavesdropper is trying to intercept and decrypt the message sent from the sender to the receiver.}
    \label{fig:1-5_security}
\end{figure}

Computer security experts sometimes refer to the "CIA triad"\index{CIA triad} of \emph{confidentiality}\index{confidentiality}, \emph{integrity}\index{integrity}, and \emph{availability}\index{availability}. \emph{Encryption}\index{encryption} refers to a set of mathematical techniques for protecting digital information, making it difficult to read or to modify undetected, often used to achieve one of the CIA goals. In this book, we will loosely refer to "security" to mean this use of encryption.

Let's consider a sender wishing to send a message to a receiver, as pictured in Fig.~\ref{fig:1-5_security}.
The message might be something mundane such as ``Buy some milk, please'' or might be something sensitive such as ``My credit card number is ...''.
Either way, messages sent over public channels are encrypted to preserve the privacy of the conversation between the sender and the receiver.
Let's consider a third party, an eavesdropper, who is trying to listen in on the conversation by intercepting the message.
The most commonly deployed urrent classical encryption techniques offer \textit{\textbf{computational security}}, meaning that cracking them is not impossible, merely very difficult and would require enormous classical resources.
Quantum algorithms exist that have the potential to break this encryption if they capture the message sequence starting from the initial exchange.
An eavesdropper with access to a quantum computer could in principle intercept the encrypted message, use their quantum computer to break the code and listen in on the conversation with impunity.

To overcome this problem, the communicating parties need to resort to using \textit{\textbf{quantum key distribution}} (QKD).
With the help of the superposition principle or entanglement, the sender and the receiver can discover the eavesdropper's attempts to intercept and tamper with their messages.
They can then simply not transmit their sensitive message.
This method of encryption offers \textit{\textbf{information-theoretic security}}.
Such security does not rely on the computational difficulty of certain mathematical problems and therefore is stronger than the computational security used currently (or at least offers a different attack surface).
For this reason, QKD is one of the primary applications of quantum networks.
We will explore the basics of QKD later in this book.

\section{Module Overview}
\label{sec:mod-over}

Before concluding the first chapter, let's have a brief look at the structure of this module.
The next three chapters deal with the basics of quantum mechanics relevant to quantum communication.
We will learn about quantum bits and how they are different from classical bits, consider how noise affects the state of quantum states and how we can describe this effect mathematically.
Finally, we will look at systems of multiple quantum bits, including entanglement and SPDC, a critical method of creating entanglement in optical systems.

Chapters 5-7 deal with the basics of optics.
Light is an excellent information carrier and we will learn how lasers produce light.
We will discuss interference, one of the fundamental phenomena in both classical as well as quantum physics.
We will conclude this block of chapters by learning about waveguides and how light is guided in a network.

Chapters 8-10 look at fundamental quantum communication protocols.
We will learn how teleportation can be used to transmit information without sending the physical system encoding this information as well as how quantum key distribution works.

In chapters 11-13, we will look at the basics of quantum repeaters, a quantum technology that makes long-distance quantum communication possible.
The last two chapters look at quantum repeater systems.

There are some basic prerequisites for this module such as linear algebra (meaning vector and matrix multiplication; eigenvectors and eigenvalues; also, we will use the tensor product which will be introduced in this book), discrete probability, and complex numbers (e.g., $i = \sqrt{-1}$ and Euler's equation, $e^{i\pi} + 1 = 0$).  The description of how lasers amplify light in Sec.~\ref{sec:5-4_lasers2} makes minimal use of derivatives from elementary calculus, but the core discussion is designed to be understandable without a background in calculus.  It's very helpful if you have some introduction to quantum computing; programming and classical computing; and computer networks.
Other than that there are no physics requirements.

If you don't have some of this background yet, there are a lot of online materials, particularly our MOOC (massively open online course), "Understanding Quantum Computers", which is targeted at learners at the high school level and requires very minimal math.
It is available in English, Japanese, Thai and Indonesian.
If you would like to learn a little bit more about some basic linear algebra you can have a look at the playlist on Professor Van Meter's YouTube channel. There are many other courses available. Other modules from the Quantum Academy of Science and Technology (supported by the Q-Leap Education office in Japan) cover some of this background material as well, and may be available to you.

\chapter[Quantum States]{Quantum States}
\label{sec:2_quantum_states}

\chaptermark{Quantum States}

In this chapter, we will learn about quantum states: how to write them down, what they represent, and how they differ from classical states.
Then, we will learn how to operate on and extract information from quantum states using unitary operations and measurements.
Lastly, we will discuss multiple quantum states and how to describe them. 
Most of this chapter will be familiar to those who have taken an introductory quantum computing course.

\section{Qubits}



First, let's discuss quantum bits, also known as \textbf{\emph{qubits}}\index{qubit}. We have seen in the previous chapter that information can be represented by classical bits. In the classical world, a classical bit can only be in two states: it can be in a state we label \emph{zero}, or in a state we label \emph{one}, nothing in between. In contrast, a quantum bit can be anything in between. It can be 100\% zero or 100\% one, but it can also be 50\% zero and 50\% one or 1\% zero and 99\% one. Such a state is called a \textbf{\emph{superposition}}\index{superposition} of zero and one. It's not that we don't know what the state is, it really is neither zero nor one but it's somewhere in between. (This notion of "in between" is not a completely accurate description, but it will do for now.  As we do the mathematics over the next several sections, your understanding will grow.)

\begin{figure}[H]
    \centering
    \begin{tikzpicture}
    \node[] at (0,0) {\Large $|\psi\rangle \; = \; \alpha \; |0\rangle \; + \; \beta \; |1\rangle$};
    
    \draw[-latex,red!60] (0.05,0.4) arc[start angle=160, end angle=130, radius=1.5];
    \draw[-latex,red!60] (1.8,0.4) arc[start angle=20, end angle=50, radius=1.5];
    \node[red!60] at (0.9,1.3) {basis states};

    \draw[-latex,blue!50] (-0.5,-0.4) arc[start angle=200, end angle=230, radius=1.5];
    \draw[-latex,blue!50] (1.3,-0.4) arc[start angle=340, end angle=310, radius=1.5];
    \node[blue!50] at (0.6,-1.3) {probability amplitudes};

    \draw[-latex,orange!80] (-1.9,-0.4) arc[start angle=340, end angle=310, radius=1.5];
    \node[orange!80] at (-2.6,-1.3) {``ket psi''};
    \end{tikzpicture}
    
    \caption[Dirac ket notation.]{Breakdown of the Dirac ket notation.}
    \label{fig:ket-notation}
\end{figure}

The most common notation for writing down the state of a qubit (or more than one qubit) is called the \textbf{\emph{Dirac notation}}\index{Dirac notation}, and it's extremely useful.
A general superposition of a qubit can be written as pictured in Fig.~\ref{fig:ket-notation}.
The funny angle bracket \ket{\cdot} is called a \textbf{\emph{ket}}\index{ket}. The symbol $\psi$ (Greek letter ``psi'') in a ket, which we will call ``state psi'' or ``ket psi'' interchangeably, is often used to describe a general state of a qubit. $\ket{0}$ and $\ket{1}$ are called the \textbf{\emph{basis states}}\index{basis states}, and they determine what our state is. This $\alpha$ and $\beta$ are \textbf{\emph{probability amplitudes}}\index{probability amplitudes} that tell us how much of the state is in zero and how much of the state is in one.  These probability amplitudes can be any complex numbers, provided that they satisfy the following normalization condition,
\begin{align}
    |\alpha|^2 + |\beta|^2 = 1.
    \label{eq:normalization-condition}
\end{align}
It should be read as ``mod alpha squared plus mod beta squared is equal to one''. This condition ensures that whatever measurements we do in the future on this state produce results with the correct probabilities.

\begin{figure}[H]
    \centering
    \def\r{4}
    \tdplotsetmaincoords{70}{115}
    \def\azimuthP{40}
    \def\polarP{60}
    \def\circleSize{88.5}
    
    \begin{tikzpicture}[tdplot_main_coords,scale=0.7,every node/.style={scale=0.7}]

        \tdplotsetcoord{O}{0}{0}{0}
        \tdplotsetcoord{P}{\r}{\azimuthP}{\polarP}
        \tdplotsetcoord{Q}{\r*sin(\azimuthP)}{90}{\polarP}

        \shade[tdplot_screen_coords,ball color=gray!40,opacity=0.5] (0,0,0) circle (\r);    
        \tdplotCsDrawLonCircle[tdplotCsBack/.style={black!50,dashed},thin]{\r}{\polarP-90}
        \tdplotCsDrawLatCircle[tdplotCsBack/.style={black!50,dashed},thin]{\r}{90-\azimuthP}
        \tdplotCsDrawLatCircle[tdplotCsBack/.style={black!50,dashed}]{\r}{0}

        \draw[-,blue!50,thick] (0:0.5*\r) arc (0:\polarP:0.5*\r);
        \node[] at ({0.4*\polarP}:0.27*\r) {\Large $\phi$};
        \tdplotsetthetaplanecoords{\polarP}
        \draw[tdplot_rotated_coords,-,blue!50,thick] (0:0.4*\r) arc (0:\azimuthP:0.4*\r);
        \node[tdplot_rotated_coords] at ({0.5*\azimuthP}:0.27*\r) {\Large $\theta$};

        \draw[-latex,thick] (O) -- (1.3*\r,0,0) node[pos=1.15] {\Large $x$};
        \draw[-latex,thick] (O) -- (0,1.3*\r,0) node[pos=1.07] {\Large $y$};
        \draw[-latex,thick] (O) -- (0,0,1.3*\r) node[pos=1.06] {\Large $z$};
        \draw[dashed,black!50] (O) -- (-\r,0,0);
        \draw[dashed,black!50] (O) -- (0,-\r,0);
        \draw[dashed,black!50] (O) -- (0,0,-\r);
    
        \draw[-latex,thick,blue!50] (O) -- (P);
        \draw[thick,dashed,blue!50] (O) -- (Q) -- (P);
    
        \tdplotCsDrawCircle[tdplotCsFill/.style=blue!50,thick]{\r}{\polarP}{\azimuthP}{\circleSize}
        \node[] at (4, 3.8, 4.7) {\Large $|\psi\rangle$};
    \end{tikzpicture}
    \caption[Bloch sphere.]{Bloch sphere visualization of a pure state $|\psi\rangle$.}
    \label{fig:bloch}
\end{figure}

Another very useful representation of quantum states is using the \textbf{\emph{Bloch sphere}}\index{Bloch sphere}, as shown in Figs.~\ref{fig:bloch} and \ref{fig:annotated-bloch}. This visual representation gives us a very intuitive way of thinking about quantum states. All the states are given as points on the surface of the sphere, parameterized by the angle $\theta$ and the angle $\phi$. Then the state $\ket{\psi}$ can be written in the following form,
\begin{equation}
|\psi\rangle=\cos \frac{\theta}{2}|0\rangle+e^{i \phi} \sin \frac{\theta}{2}|1\rangle
\end{equation}
where the probability amplitude for basis state 0 is given by $\cos(\theta/2)$ and the probability amplitude for basis state 1 is given by $\sin(\theta/2)$, multiplied by $e^{i \phi}$, known as the complex phase of the state. This phase does not affect the probability of finding a one when we measure the qubit, but it is critically important as part of the state and in quantum algorithms.

This \ket{\psi} is a general state, but let's look at some examples, as in Fig.~\ref{fig:annotated-bloch}. We have already encountered \ket{0} and \ket{1}, and they sit at the north and the south pole of the Bloch sphere, respectively. We also said that we can have an arbitrary superposition of zero and one. For example, we can have a state known as the \textbf{\emph{plus state}}\index{plus state}, written \ket{+}, which is an equal superposition of zero and one. The plus state appears on the equator of the Bloch sphere, at the point where the sphere's positive $X$ axis intersects the surface. We can have its friend the \textbf{\emph{minus state}}\index{minus state}, written \ket{-}, on the other side of the Bloch sphere. It also is an equal superposition, but this time it's on the negative side of the $X$ axis.  If you think about a rotation about the $Z$ along the equator, since $e^{i\pi} = -1$, it has the complex phase $\pi$.  We can also have two states on the $Y$ axis. One is called the ``plus i'' state, written \ket{i} or occasionally \ket{+i}, and the other is called ``minus i'', written \ket{-i}. You can see that again, both of these states are an equal superposition of zero and one, but this time the phase between zero and one is given by the complex number $i$ or the angle $\pi/2$ for \ket{i} and $-i$ or the angle $3\pi/2$ for \ket{-}.  Summarizing, these states are
\begin{equation}
    |\pm\rangle=\frac{1}{\sqrt{2}}(|0\rangle \pm|1\rangle), \qquad|\pm i\rangle=\frac{1}{\sqrt{2}}(|0\rangle \pm i|1\rangle).
\end{equation}

\begin{figure}[H]
    \centering
    \def\r{4}
    \def\circleSize{88.5}
    \tdplotsetmaincoords{70}{115}
    \begin{tikzpicture}[tdplot_main_coords,scale=0.7,every node/.style={scale=0.7}]

        \tdplotsetcoord{O}{0}{0}{0}

        \shade[tdplot_screen_coords,ball color=gray!40,opacity=0.5] (0,0,0) circle (\r);    
        \tdplotCsDrawLonCircle[tdplotCsBack/.style={black!50,dashed},thin]{\r}{0}
        \tdplotCsDrawLonCircle[tdplotCsBack/.style={black!50,dashed},thin]{\r}{90}
        \tdplotCsDrawLatCircle[tdplotCsBack/.style={black!50,dashed}]{\r}{0}

        \draw[thick] (O) -- (\r,0,0);
        \draw[thick] (O) -- (0,\r,0);
        \draw[thick] (O) -- (0,0,\r);
        \tdplotCsDrawCircle[tdplotCsFill/.style=blue!50,thick]{\r}{0}{0}{\circleSize}
        \tdplotCsDrawCircle[tdplotCsFill/.style=blue!50,tdplotCsBack/.style={-},thick]{\r}{0}{0}{-88.8}
        \tdplotCsDrawCircle[tdplotCsFill/.style=blue!50,thick]{\r}{0}{90}{\circleSize}
        \tdplotCsDrawCircle[tdplotCsFill/.style=blue!50,tdplotCsBack/.style={-},thick]{\r}{0}{90}{-\circleSize}
        \tdplotCsDrawCircle[tdplotCsFill/.style=blue!50,thick]{\r}{90}{90}{\circleSize}
        \tdplotCsDrawCircle[tdplotCsFill/.style=blue!50,tdplotCsBack/.style={-},thick]{\r}{90}{90}{-\circleSize}
        \draw[dashed,black!50] (O) -- (-\r,0,0);
        \draw[dashed,black!50] (O) -- (0,-\r,0);
        \draw[dashed,black!50] (O) -- (0,0,-\r);
        \draw[-latex,thick] (\r,0,0) -- (1.3*\r,0,0) node[pos=1.5] {\Large $x$};
        \draw[-latex,thick] (0,\r,0) -- (0,1.3*\r,0) node[pos=1.4] {\Large $y$};
        \draw[-latex,thick] (0,0,\r) -- (0,0,1.3*\r) node[pos=1.38] {\Large $z$};
    
        \node[] at (\r,0.5,-0.4) {\Large $|+\rangle$};
        \node[] at (-\r,0.5,0.4) {\Large $|-\rangle$};
        \node[] at (0,\r+0.4,-0.4) {\Large $|i\rangle$};
        \node[] at (0,-\r-0.7,0.2) {\Large $|-i\rangle$};
        \node[] at (0,0.35,\r+0.6) {\Large $|0\rangle$};
        \node[] at (0,0.35,-\r-0.6) {\Large $|1\rangle$};
    \end{tikzpicture}
    \caption{Bloch sphere with Pauli $X$, $Y$, and $Z$ basis states.}
    \label{fig:annotated-bloch}
\end{figure}

\section{Unitary Operations}
\label{sec:2-2_unitary_operations}

\begin{figure}[H]
    \centering
    \includegraphics[width=0.9\textwidth]{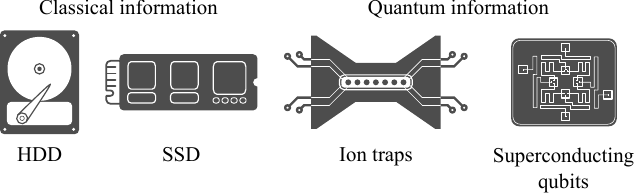}
    \caption[Physical information-processing systems.]{Examples of physical systems processing classical and quantum information.}
    \label{fig:physical-system}
\end{figure}

Let's see how we can manipulate quantum states, and therefore manipulate quantum information as well. We're going to do this with \emph{unitary operations}. One thing that you have to realize is that \emph{information is physical} (an aphorism coined by Rolf Landauer of IBM). This is because the information is represented by physical systems. Therefore, to change the information and process it, we have to interact with the physical systems that carry the information. In classical information, of course the primary processing technology is transistors manufactured using a photolithography process.  Computational states are represented using electrical charge, and information processing is done by using charge to control whether a switch is on or off, allowing charge to move from place to place within a computer chip.  Another very good example is HDDs (hard disk drives), where you read, write and manipulate information with very weak and precise magnetic fields. For an example of a younger technology, we can look at the "solid-state drive" where you do the same thing, but you achieve it by manipulating very weak electric currents.

In quantum information, on the other hand, we have physical systems such as \textbf{\emph{ion traps}}\index{ion traps}, from companies such as IonQ and Quantinuum. The atoms are suspended on magnetic fields and they represent individual quantum bits. To manipulate the states of these qubits, you apply some laser pulses.  One of the most prominent forms of qubits is \textbf{\emph{superconducting qubits}}\index{superconducting qubits} from companies such as IBM and Google, where microwave pulses manipulate the state of a quantum of electrical current. (Details of such processing hardware are beyond the scope of this module, but will be found in other modules in this series.) Some of these are represented in Fig.~\ref{fig:physical-system}.

But how do we actually describe these transformations? Before giving a more complete mathematical description, let's look at some examples, as shown in Tab.~\ref{tab:unitary-table}. The simplest transformation that we can think of is actually to do nothing. We call this the \textbf{\emph{identity operation}}\index{identity operation}, and it is usually represented by a capital $I$. When it's acting on a ket, it takes $\ket{0}$ to $\ket{0}$  (written $\ket{0}\rightarrow\ket{0}$) and $\ket{1}$ to $\ket{1}$ ($\ket{1}\rightarrow\ket{1}$). Classically, we can also do something similar by just not touching our classical bit. 0 remains 0 and 1 remains 1.

\begin{table}[h]
    \setcellgapes{5pt}
    \renewcommand\theadfont{}
    \makegapedcells
    \centering
    \begin{tabular}{cccc}
        \hline
        & \textbf{Notation} & \textbf{Quantum} & \textbf{Classical} \\
        \hline
        \thead{Identity \\ (do nothing)} & $I$ & {$\begin{aligned} |0\rangle & \rightarrow|0\rangle \\ |1\rangle & \rightarrow|1\rangle \end{aligned}$} & {$\begin{aligned} 0&\rightarrow0 \\ 1&\rightarrow1 \end{aligned}$} \\
        \thead{Pauli $X$ \\ (flip)} & $X$ & {$\begin{aligned} |0\rangle & \rightarrow|1\rangle \\ |1\rangle & \rightarrow|0\rangle \end{aligned}$} & {$\begin{aligned} 0&\rightarrow1 \\ 1&\rightarrow0 \end{aligned}$} \\
        \thead{Hadamard \\ (create superposition)} & $H$ & {$\begin{aligned} |0\rangle & \rightarrow \frac{|0\rangle + |1\rangle}{\sqrt{2}} \\ |1\rangle & \rightarrow \frac{|0\rangle - |1\rangle}{\sqrt{2}} \end{aligned}$} & \textcolor{darkred}{\xmark} \\
        \hline
    \end{tabular}
    \caption{Simple unitary transforms. The bit flip has both a classical and a quantum form, but the Hadamard gate that creates a quantum superposition has no classical equivalent.}
    \label{tab:unitary-table}
\end{table}

    
    

Another basic operation is the \textbf{\emph{Pauli X operation}}\index{Pauli X operation}, often called the ``flip''. We represent it by a capital $X$, and it does exactly what you would expect. It takes the input $\ket{0}$ into an output $\ket{1}$, and vice versa, $\ket{1}$ into a $\ket{0}$. Again, you have a corresponding classical operation as well, the NOT gate, which takes input 0 into classical bit 1 and input 1 into classical bit 0.

The third operation in Tab.~\ref{tab:unitary-table} can be used to create superpositions. It is known as the \textbf{\emph{Hadamard operation}}\index{Hadamard}, denoted by $H$. If the input is $\ket{0}$, it outputs an equal superposition of \ket{0} and \ket{1}, written $(\ket{0}+\ket{1})/\sqrt{2}$. If the input is \ket{1}, the output will be $(\ket{0}-\ket{1})/\sqrt{2}$. This is the first example of a quantum operation that doesn't really have a classical analog, because we cannot have superpositions of classical bits.

So what's the definition of a unitary operation? All of these examples that we just discussed are examples of unitary operations. Any unitary operation has the property of being \textbf{\emph{reversible}}\index{reversible operation}, meaning we can undo its effect on our data and return to the original input state. This is done by what's known as an \textbf{\emph{adjoint operator}}\index{adjoint operator}, denoted as $U^\dagger$ (read ``U dagger'') where $U$ is the unitary.
Let's see how that works. We start with a ket $\ket{\psi}$ and we apply a unitary that transforms it into a completely new ket $\ket{\psi'}$. Then, if we apply the adjoint (the operation which undoes the effect of the original unitary), we end up back again at the state $\ket{\psi}$. We can write $\ket{\psi'} = U\ket{\psi}$, or we can also write $\ket{\psi} = U^\dagger\ket{\psi'}$. 

Let's put these two equalities together. Use the above equality for \ket{\psi}, then replace \ket{\psi'} with $U\ket{\psi}$ and get
\begin{equation}
|\psi\rangle=U^{\dagger}\left|\psi^{\prime}\right\rangle=U^{\dagger} U|\psi\rangle.
\end{equation} 


We can also do similar operations starting from the other expression and get
\begin{equation}
\left|\psi^{\prime}\right\rangle=U|\psi\rangle=U U^{\dagger}\left|\psi^{\prime}\right\rangle.
\end{equation}
From these equations,  we can see that $UU^\dagger$ must be equal to the identity operator, and also $U^\dagger U$ must be equal to the identity.
In fact, this becomes precisely our definition of a unitary operation.
If we have that 
\begin{equation}
U U^{\dagger}=U^{\dagger} U=I,
\end{equation}
then $U$ is a unitary operator.

How can we represent this in matrix notation? So far we have been talking about states as kets, but in fact a ket is shorthand for a vector, and we know that in order to transform vectors we multiply them by matrices. Therefore, it should be the case that unitary operations can be represented by matrices. Let's look at some examples. 

First let's begin by seeing how kets for states become vectors. Usually we denote \ket{0} in vector notation as a column vector with 1 and 0,
\begin{equation}
\ket{0}\equiv\left(\begin{array}{l}
1 \\
0
\end{array}\right).
\end{equation}
The \ket{1}, on the other hand, is a column vector of 0 and 1,
\begin{equation}
\ket{1}\equiv\left(\begin{array}{l}
0 \\
1
\end{array}\right).
\end{equation}
You can see that any general state $\ket{\psi}$ can be represented as $\alpha$ times the vector $\left(\begin{array}{l}
1 \\
0
\end{array}\right)$ plus $\beta$ times $\left(\begin{array}{l}
0 \\
1
\end{array}\right)$, giving us the complex column vector $\left(\begin{array}{l}
\alpha \\
\beta
\end{array}\right)$. More formally, we have
\begin{equation}
\ket{\psi}=\alpha\left(\begin{array}{l}
1 \\
0
\end{array}\right)+\beta\left(\begin{array}{l}
0 \\
1
\end{array}\right)=\left(\begin{array}{l}
\alpha \\
\beta
\end{array}\right).
\end{equation}

Now let's look at examples of matrices representing unitary operations. The identity operator $I$ is represented by the matrix
\begin{equation}
I=\left(\begin{array}{ll}
1 & 0 \\
0 & 1
\end{array}\right).
\end{equation}
It's just a diagonal matrix with ones on the main diagonal and zeros everywhere else.

An important set of operations we will use many times is known as the set of \textbf{\emph{Pauli operators}}\index{Pauli operator}. We have encountered one Pauli operator already, the $X$ operator, which flips our ket from zero to one and from one to zero, but there are two other very important Pauli operators: the $Y$ and the $Z$. These three have the matrix representations 
\begin{equation}
    X=\left(\begin{array}{ll}
    0 & 1 \\
    1 & 0
    \end{array}\right), \quad
    Y=\left(\begin{array}{cc}
    0 & -i \\
    i & 0
    \end{array}\right), \quad
    Z=\left(\begin{array}{cc}
    1 & 0 \\
    0 & -1
    \end{array}\right).
\end{equation}

We also have the Hadamard operator~\footnote{In many physics books and papers, the symbol $H$ can also represent an operator known as the \emph{Hamiltonian} of a system.  In this book, we will not need the concept of the Hamiltonian. In general, it will be clear from context which one is meant.} that creates a superposition, written $H$.  The matrices corresponding to this operator is
\begin{equation}
H=\frac{1}{\sqrt{2}}\left(\begin{array}{cc}
1 & 1 \\
1 & -1
\end{array}\right).
\end{equation}

Let's see some examples just to give you a little bit of feeling for how this can actually work in practice. For the flip operation, you take the Pauli $X$, you apply it, or multiply it, by $\ket{0}$,
\begin{equation}
\begin{aligned}
X|0\rangle &=\left(\begin{array}{ll}
0 & 1 \\
1 & 0
\end{array}\right)\left(\begin{array}{l}
1 \\
0
\end{array}\right) \\
&=\left(\begin{array}{l}
0 \\
1
\end{array}\right)=|1\rangle.
\end{aligned}
\end{equation}


We can do the same thing for $\ket{1}$. Again, multiply the matrix representation of the Pauli $X$ operator with the column vector representation of state $\ket{1}$,
\begin{equation}
\begin{aligned}
X|1\rangle &=\left(\begin{array}{ll}
0 & 1 \\
1 & 0
\end{array}\right)\left(\begin{array}{l}
0 \\
1
\end{array}\right) \\
&=\left(\begin{array}{l}
1 \\
0
\end{array}\right)=|0\rangle
\end{aligned}
\end{equation}
and as expected we get $\ket{0}$. 

To create a superposition, take the Hadamard operator, apply it to the state $\ket{0}$. After going through the algebra, in the end we have an equal superposition of 0 and 1,
\begin{equation}
\begin{aligned}
H|0\rangle &=\frac{1}{\sqrt{2}}\left(\begin{array}{cc}
1 & 1 \\
1 & -1
\end{array}\right)\left(\begin{array}{l}
1 \\
0
\end{array}\right) \\
&=\frac{1}{\sqrt{2}}\left(\begin{array}{l}
1 \\
1
\end{array}\right)=\frac{1}{\sqrt{2}}(|0\rangle+|1\rangle)
\end{aligned}
\end{equation}

Please notice that this factor in the Hadamard operator, $1/\sqrt{2}$, ensures that the superposition vector at the end is properly normalized, as we saw in Eq.~\ref{eq:normalization-condition}. $|1/\sqrt{2}|^2 + |1/\sqrt{2}|^2 = 1$, therefore this vector is correctly normalized. You can follow the same process for the state 1, and again as we have seen in the previous step you get zero minus one,
\begin{equation}
\begin{aligned}
H|1\rangle &=\frac{1}{\sqrt{2}}\left(\begin{array}{cc}
1 & 1 \\
1 & -1
\end{array}\right)\left(\begin{array}{l}
0 \\
1
\end{array}\right) \\
&=\frac{1}{\sqrt{2}}\left(\begin{array}{c}
1 \\
-1
\end{array}\right)=\frac{1}{\sqrt{2}}(|0\rangle-|1\rangle).
\end{aligned}
\end{equation}
The modulus in Eq.~\ref{eq:normalization-condition} ensures that the negative coefficient still results in a non-negative probability and non-negative contribution to the normalization condition.

So far, the adjoint operator has been a rather abstract notion that can undo the effect of a unitary operation. How can we actually systematically find the adjoint 
of an operator, given the operator's unitary matrix? It's actually two very simple steps.
 
Step one: take the complex conjugate of the matrix. (Just to remind you, the \emph{complex conjugate}\index{complex conjugate} of a complex number $(x+iy)^*$ (read ``x plus i y star''), which is the notation for complete conjugation, is
\begin{equation}
(x+i y)^{*}=x-i y
\end{equation}
Wherever you see $i$, flip the sign of that term to obtain the complex conjugate.)

Step two: take the transpose. The transpose of the matrix is given as follows: for any matrix $U$ which has elements $U_{00}$, $U_{01}$, $U_{10}$ and $U_{11}$, transposing them will exchange the off-diagonal elements. Let's see how this unitary $U$ actually turns into an adjoint. First we apply the complex conjugation, so we take each element and we write a little star, denoting that this element is complex conjugated, and then we apply the transpose,
\begin{equation}
U=\left(\begin{array}{ll}
U_{00} & U_{01} \\
U_{10} & U_{11}
\end{array}\right) \rightarrow\left(\begin{array}{cc}
U_{00}^{*} & U_{01}^{*} \\
U_{10}^{*} & U_{11}^{*}
\end{array}\right) \longrightarrow\left(\begin{array}{ll}
U_{00}^{*} & U_{10}^{*} \\
U_{01}^{*} & U_{11}^{*}
\end{array}\right)=U^{\dagger}
\end{equation}
We have flipped the position of the off-diagonal elements, but have not moved the diagonal elements. If you look at an example given by the unitary Pauli $Y$ matrix, 

\begin{equation}
Y=\left(\begin{array}{cc}
0 & -i \\
i & 0
\end{array}\right) \longrightarrow\left(\begin{array}{cc}
0 & i \\
-i & 0
\end{array}\right) \longrightarrow\left(\begin{array}{cc}
0 & -i \\
i & 0
\end{array}\right)=Y.
\end{equation}
First we take the complex conjugate; the zero elements on the diagonal of the matrix remain zero but the signs of the off-diagonal elements change because they're pure imaginary. Second, we apply the transpose by flipping the off-diagonal elements, and we get back the same matrix $Y$ that we started with. When this happens, when the adjoint of a unitary matrix is equal to the unitary matrix, we say that that matrix is \textbf{\emph{self-adjoint}}\index{self-adjoint}.

Another very important class of unitary operations are \textbf{\emph{rotations}}\index{rotation}, and this is where the Bloch sphere representation will be extremely handy. The rotation can be written in this strange looking form (strange for now),
\begin{equation}
R_{\hat{n}}(\theta)=e^{-i \theta \hat{n} \cdot \hat{\sigma} / 2}.
\end{equation}
This notation means that we are rotating around some arbitrary axis by some arbitrary angle $\theta$. We write both the axis we will rotate around and the angle in the exponent. Here, $\hat{n}$ (read ``n hat'') is just a unit length vector given by coordinates $n_x$, $n_y$ and $n_z$, and the vector given by $\hat{\sigma}$ (``sigma hat'') is just a vector of our Pauli matrices $X$, $Y$ and $Z$.  The dot product expands out to be
\begin{equation}
      \hat{n} \cdot \hat{\sigma} = n_{x} X+n_{y} Y+n_{z} Z
\end{equation}
and when put it in the equation we get
\begin{equation}
e^{-i \theta \hat{n} \cdot \hat{\sigma} / 2 }=\cos \frac{\theta}{2} I-i \sin \frac{\theta}{2}\left(n_{x} X+n_{y} Y+n_{z} Z\right).
\end{equation}
It does look a little bit complicated, but in the Bloch sphere representation it becomes very clear what's going on: take the vector $\hat{n}$ and rotate the state vector around that by the angle $\theta$. When we need this rotation, we can simply write it as
\begin{equation}
    R_{\hat{n}}(\theta)|\psi\rangle=|\psi^{\prime}\rangle.
\end{equation}


\begin{figure}[H]
    \centering

    \def\r{4}
    \tdplotsetmaincoords{70}{115}
    
    \begin{tikzpicture}[tdplot_main_coords,scale=0.7,every node/.style={scale=0.7}]

    \tdplotsetcoord{O}{0}{0}{0}

    \shade[tdplot_screen_coords,ball color=gray!40,opacity=0.5] (0,0,0) circle (\r);    
    \tdplotCsDrawLonCircle[tdplotCsFill/.style={blue!50,opacity=0.1},tdplotCsBack/.style={black!50,dashed},thin]{\r}{90}
    \tdplotCsDrawLatCircle[tdplotCsBack/.style={black!50,dashed}]{\r}{0}

    \draw[blue!50] (O) -- (\r,0,0);
    \draw[thick] (O) -- (0,\r,0);
    \draw[blue!50] (O) -- (0,0,\r);
    \draw[dashed,black!50] (O) -- (-\r,0,0);
    \draw[dashed,black!50] (O) -- (0,-\r,0);
    \draw[dashed,black!50] (O) -- (0,0,-\r);

    \tdplotsetthetaplanecoords{0}
    \draw[tdplot_rotated_coords,-,blue!50,very thick,decoration={markings, mark=at position 0.5 with {\arrow{latex}}},postaction={decorate}] (0:\r) arc (0:90:\r);
    \draw[tdplot_rotated_coords,-,blue!50] (0:0.3*\r) arc (0:90:0.3*\r);
    
    \tdplotCsDrawCircle[tdplotCsFill/.style=blue!50,thick]{\r}{0}{0}{88.5}
    \tdplotCsDrawCircle[tdplotCsFill/.style=blue!50,thick]{\r}{0}{90}{88.5}

    \draw[-latex,thick] (\r,0,0) -- (1.3*\r,0,0) node[pos=1.5] {\Large $x$};
    \draw[-latex,thick] (0,\r,0) -- (0,1.3*\r,0) node[pos=1.4] {\Large $y$};
    \draw[-latex,thick] (0,0,\r) -- (0,0,1.3*\r) node[pos=1.38] {\Large $z$};
    
    \node[] at (\r,0.5,-0.4) {\Large $|+\rangle$};
    \node[] at (0,0.35,\r+0.6) {\Large $|0\rangle$};

    \node[] at (\r,-0.5,3.5) {\Large $R_y(\pi/2)$};
    \end{tikzpicture}
    
    \caption[Rotation about $y$ axis.]{Rotation about the $Y$ axis by an angle $\pi/2$.}
    
    \label{fig:y-rot}
\end{figure}

Let's consider the example shown in Fig.~\ref{fig:y-rot}: we want to rotate around the y-axis through an angle of $\pi/2$, and our initial state is given by $\ket{0}$.  Let's start at the point $\ket{0}$, at the north pole of the sphere, and rotate around the y-axis. Going down along the surface by angle $\pi/2$, you see that we reach the state \ket{+}, which is an equal superposition of \ket{0} and \ket{1}.  For this $Y$ rotation, $\hat{n}$ simply becomes the unit vector along the positive y-axis, which we will just write $y$. Written out, we have
\begin{equation}
R_y(\pi / 2)\ket{0}=\ket{+}.
\end{equation}
(Be careful here; although this rotation took us from \ket{0} to \ket{+}, just like the Hadamard gate we have already seen, $H$ is not a $\pi/2$ $Y$ axis rotation.  Instead, $H$ is a $\pi$ rotation about the axis $(X+Z)/2$. See the exercises for more.)

\begin{figure}[H]
    \centering

    \def\r{4}
    \def\alp{110}
    \def\bet{25}
    \def\eps{40}
    \tdplotsetmaincoords{70}{115}

    \begin{tikzpicture}[tdplot_main_coords,scale=0.7,every node/.style={scale=0.7}]

    \tdplotsetcoord{O}{0}{0}{0}
    \tdplotsetcoord{pointAxis}{\r}{\alp}{\bet}

    \shade[tdplot_screen_coords,ball color=gray!40,opacity=0.5] (0,0,0) circle (\r);    
    \tdplotCsDrawLatCircle[tdplotCsBack/.style={black!50,dashed}]{\r}{0}

    \pgfmathsetmacro\re {\r*cos(\eps)}
    \pgfmathsetmacro\ze {\r*sin(\eps)}
    \pgfmathsetmacro\coX{\ze*cos(\alp)*sin(\bet)}
    \pgfmathsetmacro\coY{\ze*sin(\alp)*sin(\bet)}
    \pgfmathsetmacro\coZ{\ze*cos(\bet)}
    \pgfmathsetmacro\length{sqrt(\coX*\coX+\coY*\coY+\coZ*\coZ)}
    \coordinate (pointAxis) at (\r*\coX/\length,\r*\coY/\length,\r*\coZ/\length);
    
    \draw[thick,-latex] (O) -- (pointAxis) node[anchor=east,yshift=4pt]{\Large $\hat{n}$};

    \draw[thick] (O) -- (\r,0,0);
    \draw[thick] (O) -- (0,\r,0);
    \draw[thick] (O) -- (0,0,\r);
    \draw[dashed,black!50] (O) -- (-\r,0,0);
    \draw[dashed,black!50] (O) -- (0,-\r,0);
    \draw[dashed,black!50] (O) -- (0,0,-\r);

    \draw[-latex,thick] (\r,0,0) -- (1.3*\r,0,0) node[pos=1.5] {\Large $x$};
    \draw[-latex,thick] (0,\r,0) -- (0,1.3*\r,0) node[pos=1.4] {\Large $y$};
    \draw[-latex,thick] (0,0,\r) -- (0,0,1.3*\r) node[pos=1.38] {\Large $z$};

    \tdplotCsDrawCircle[tdplotCsFill/.style={blue!50,opacity=0.1},tdplotCsBack/.style={black!50,dashed},thin]{\r}{\alp}{\bet}{\eps}

    \tdplotsetrotatedcoords{\alp}{\bet}{0}
    \coordinate (coffs) at (\coX,\coY,\coZ);
    \tdplotsetrotatedcoordsorigin{(coffs)}
    
    \draw[tdplot_rotated_coords,blue!50,very thick,decoration={markings, mark=at position 0.6 with {\arrow{latex}}},postaction={decorate}] (-160:\re) arc (-160:-30:\re) node[pos=0.6,below,black,yshift=-4pt] {\Large $R_{\hat{n}}(\theta)$};
    
    \draw[tdplot_rotated_coords,blue!50] (-160:0.3*\re) arc (-160:-30:0.3*\re);
    \draw[tdplot_rotated_coords,blue!50] (0,0,0) -- (-160:\re) node[black,anchor=east,xshift=-1pt,yshift=-4pt]{\Large $|\psi\rangle$};
    \draw[tdplot_rotated_coords,blue!50] (0,0,0) -- (-30:\re) node[anchor=north,black,yshift=-3pt]{\Large $|\psi'\rangle$};

    \tdplotCsDrawCircle[tdplotCsFill/.style=blue!50,thick]{\r}{-34}{27.9}{88.5}
    \tdplotCsDrawCircle[tdplotCsFill/.style=blue!50,thick]{\r}{85.5}{72.3}{88.5}
    \end{tikzpicture}
    
    \caption{Rotation about an arbitrary axis $\hat{n}$.}
    
    \label{fig:arb-rot}
\end{figure}

More generally, it is not necessary to rotate about one of these orthogonal axes; you can rotate about any axis that you want, as in Fig.~\ref{fig:arb-rot}. You are rotating in the plane defined by your chosen axis and parameterized by the angle $\theta$. Take some initial state $\ket{\psi}$ that's on the surface of the Bloch sphere, follow the surface and end at some new state $\ket{\psi'}$.  (Note that the plane a particular state is rotating in does not have to pass through the center of the Bloch sphere, but the axis of rotation is normal to the plane.) 

So you can see that you can work in the matrix representation or you can work also in the Bloch sphere representation for single-qubit operations.


\section{Measurement}
\label{sec:measurement}

\begin{figure}[H]
    \centering
    \includegraphics[width=0.8\textwidth]{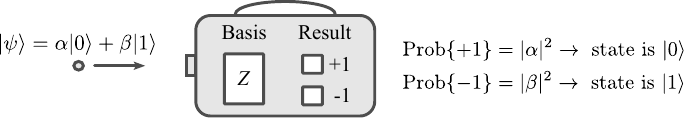}
    \caption{Measurement in the $Z$ basis.}
    \label{fig:z-measure}
\end{figure}

Now, let's consider measurements and how they can extract information from qubits~\footnote{The type of measurements we consider in this book are \emph{projective measurements}, as discussed more in the next chapter. There are other types of measurement that you will learn about in advanced quantum classes.}. Basically a measurement asks the question: is my qubit in the state $\ket{0}$ or is it in the state $\ket{1}$? You can think of a measurement as a big box and you feed it your prepared state, as shown in Fig.~\ref{fig:z-measure}. Here, we consider a general state given by $\ket{\psi} = \alpha\ket{0}+\beta\ket{1}$, and the measurement operation tells you which state it is in. Usually there are only two possible values that can come out of a measurement device, which we will assign the values $+1$ and $-1$.

The probabilities of these measurement outcomes, corresponding to state \ket{0} and state \ket{1}, are given by their probability amplitudes. The probability of getting a $+1$ outcome is given by $|\alpha|^2$ and the probability of a $-1$ outcome is given by $|\beta|^2$. Immediately after the measurement, the state \textbf{\emph{collapses}}\index{state collapse} either onto \ket{0} or onto \ket{1}, so if you get a $+1$ outcome you can be sure that immediately after the measurement the state is in the state \ket{0}, or if you get a $-1$ then the state of the system has changed from the initial $\ket{\psi}$ and is \ket{1},
\begin{align}
\operatorname{Prob}\{+1\}=|\alpha|^2 \rightarrow \textrm{state is } |0\rangle \\
\operatorname{Prob}\{-1\}=|\beta|^2 \rightarrow \textrm{state is } |1\rangle.
\end{align}

(Mathematically speaking, the $\pm 1$ outcomes of the measurement are the \emph{eigenvalues} corresponding to the \textbf{\emph{eigenvectors}}\index{eigenvector} of the Pauli $Z$ operator.  One way to remember which value corresponds to which vector is that $+1$ corresponds to \ket{0}, and $(-1)^0 = +1$, while $-1$ corresponds to \ket{1}, and $(-1)^1 = -1$.)

\begin{figure}[H]
    \centering
    \includegraphics[width=0.85\textwidth]{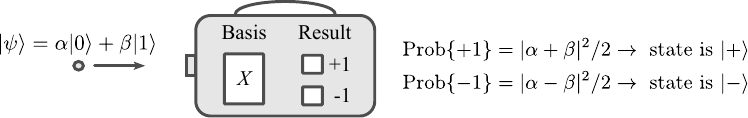}
    \caption{Measurement in the Pauli $X$ basis.}
    \label{fig:x-meas}
\end{figure}

We refer to this particular measurement as a measurement in the \textbf{\emph{computational basis}\index{computational basis}} or in the Pauli $Z$ basis. This already hints at the fact that this is not the only measurement we can do. We can also ask the following question: is the state in the $\ket{+}$ state or in the $\ket{-}$ state? We can answer this question by rewriting our original state
\begin{equation}
    \psi\rangle=\alpha|0\rangle+\beta|1\rangle
    \label{eq:superposition_basisZ}
\end{equation}
using the algebraic identities
\begin{equation}
|0\rangle=\frac{1}{\sqrt{2}}(|+\rangle+|-\rangle) \quad|1\rangle=\frac{1}{\sqrt{2}}(|+\rangle-|-\rangle).
\end{equation}
Notice that $\ket{0}$ is given by an equal superposition of the $\ket{+}$ state with the $\ket{-}$ state and the state $\ket{1}$ is also an equal superposition of $\ket{+}$ and $\ket{-}$, but this time with a minus phase in front of the $\ket{-}$ state. We can rewrite our original state $\ket{\psi}$ in terms of these \ket{+} and \ket{-} states, but this time the probability amplitudes are
\begin{equation}
    \ket{\psi}=\frac{\alpha+\beta}{\sqrt{2}}\ket{+}+\frac{\alpha-\beta}{\sqrt{2}}\ket{-}.
    \label{eq:superposition_basisX}
\end{equation}
Again, we can feed our qubit into the measurement device and it will give us an answer $\ket{+}$ or $\ket{-}$ with the following probabilities,
\begin{align}
\label{eq:plus-measurement}
\operatorname{Prob}\{+1\}=\frac{|\alpha+\beta|^2}{2} \rightarrow \textrm{state is } |+\rangle \\
\label{eq:minus-measurement}
\operatorname{Prob}\{-1\}=\frac{|\alpha-\beta|^2}{2} \rightarrow \textrm{state is } |-\rangle.
\end{align}
You can see that the probabilities still depend on $\alpha$ and $\beta$, but they're not just $|\alpha|^2$ and $|\beta|^2$.
Because we were asking a different question (whether the state state is in $\ket{+}$ or $\ket{-}$), the state changes from $\ket{\psi}$ and goes to $\ket{+}$ or $\ket{-}$ after the measurement, depending on the measurement outcome. This measurement is known as measurement in the Pauli $X$ basis, and is shown in Fig.~\ref{fig:x-meas}.

In fact, you can measure in any basis that you can think of.  This is explored more in the exercises.

We have been using this word \textbf{\emph{basis}}\index{basis}, but what is a ``basis''? Maybe you remember from your linear algebra class that a basis is a set of vectors that allows you to write down any vector that you can think of. If we have a general vector $\ket{\psi}$, we can express it in terms of the basis states $\ket{0}$ and \ket{1}, as shown in Eq.~(\ref{eq:superposition_basisZ}).
But the same state $|\psi\rangle$ can be expressed in terms of the states $|\pm\rangle$, as we have seen in Eq.~(\ref{eq:superposition_basisX}).
We do not have to use only the Pauli $Z$ or the Pauli $X$ bases, however.
Let's say you want to write the state $|\psi\rangle$ in terms of the states $\ket{i}$ and $\ket{-i}$.
This is known as the Pauli $Y$ basis, and their vector representations are given as
\begin{align}
    |i\rangle=\frac{1}{\sqrt{2}}\left(\begin{array}{l}1 \\ i\end{array}\right) \quad|-i\rangle=\frac{1}{\sqrt{2}}\left(\begin{array}{c}1 \\ -i\end{array}\right).
    \label{eq:Pauli_Y_basis}
\end{align}
Then the state $|\psi\rangle$ can be rewritten as
\begin{equation}
    |\psi\rangle=\frac{\alpha-i \beta}{\sqrt{2}}|i\rangle+\frac{\alpha+i \beta}{\sqrt{2}}|-i\rangle.
    \label{eq:superposition_basisY}
\end{equation}
The expressions in Eqs.~(\ref{eq:superposition_basisZ}), (\ref{eq:superposition_basisX}), and (\ref{eq:superposition_basisY}) look very different but they represent the same state: the state has not changed.
It's just that our description of the state is a different because we have chosen a different basis to express it.

Another important concept is the \textbf{\emph{inner product}}\index{inner product}. The inner product is basically a dot product between two vectors.
It tells us the ``overlap'' between two quantum states, which is a measure of how similar the states are. To write down the inner product in the notation we use in this book, first we have to define a \textbf{\emph{bra}}\index{bra}. Given a state $\ket{\psi}$, the corresponding bra is its adjoint. The bra of state $\ket{\psi}$ is given as the dagger of the original column vector,
\begin{equation}
\langle\psi|=(|\psi\rangle)^{\dagger}=\left(\begin{array}{l}
\alpha \\
\beta
\end{array}\right)^{\dagger}=\left(\begin{array}{ll}
\alpha^{*} & \beta^{*}
\end{array}\right).
\end{equation}
We applied the same trick we used for defining the adjoint of unitary operations: take the complex conjugate of $\alpha$ and $\beta$ and then transpose the column vector, which turns it into a row vector. Now we can find the inner product between two arbitrary states, $\ket{\psi} = \alpha\ket{0} + \beta\ket{1}$ and $\ket{\phi} = \gamma\ket{0} + \delta\ket{1}$. We write it as 
\begin{equation}
\langle\phi \mid \psi\rangle=\left(\begin{array}{ll}
\gamma^{*} & \delta^{*}
\end{array}\right)\left(\begin{array}{c}
\alpha \\
\beta
\end{array}\right)=\alpha \gamma^{*}+\beta \delta^{*}.
\end{equation}
This should be no surprise if you remember your dot products from your linear linear algebra class.
Unlike the dot product from your linear algebra class, the inner product can be a complex number.
This means that in general $\langle\psi|\phi\rangle \neq \langle\phi|\psi\rangle$.

Also notice that if we take the inner product of the state with itself, we get this expression:
\begin{equation}
\langle\psi \mid \psi\rangle=|\alpha|^{2}+|\beta|^{2}=1.
\end{equation}
Mod alpha squared plus mod beta squared will be equal to one, if the state is properly normalized.

For two different states, we can also come across the scenario where the inner product between two states is zero,
\begin{equation}
\langle\phi \mid \psi\rangle=0.
\end{equation}
In that case, we say that the two states are \textbf{\emph{orthogonal}}.

Being equipped with the knowledge of the inner product, we can address the following question: how can we compute the probabilities of different outcomes when measuring the state in a particular basis?
Given our general state $\ket{\psi}$, let's say that we want to measure in some arbitrary basis given by the states $\ket{b_{+1}}$ and $\ket{b_{-1}}$.
State $\ket{b_{+1}}$ represents the +1 outcome of the measurement, while the state $\ket{b_{-1}}$ represents the -1 outcome of the measurement.
We are asking the question: is my state $|\psi\rangle$ in the state $|b_{+1}\rangle$ or is my state in the state $|b_{-1}\rangle$?
To answer this, we take the inner product between $\ket{b_{+1}}$ and $\ket{\psi}$, take the modulus and we square it, and that gives us the probability of a $+1$ outcome. Correspondingly, the probability of the $-1$ outcome is given by the inner product between $\ket{b_{-1}}$ and $\ket{\psi}$ modulus squared,
\begin{equation}
\begin{aligned}
    \operatorname{Prob}\{+1\}&=\left|\left\langle b_{0}\mid \psi\right\rangle\right|^{2}, \\
    \operatorname{Prob}\{-1\}&=\left|\left\langle b_{1}\mid \psi\right\rangle\right|^{2}.
\end{aligned}
\end{equation}

Let's illustrate this with a familiar example. We start by measuring in the Pauli Z basis. This means that our basis states are $|b_{+1}\rangle=|0\rangle$, and $|b_{-1}\rangle=|1\rangle$. The probabilities of the two outcomes are then given by
\begin{equation}
\begin{aligned}
    \operatorname{Prob}\{+1\} &=|\langle 0 \mid \psi\rangle|^2=|\alpha|^2, \\
    \operatorname{Prob}\{-1\} &=|\langle 1 \mid \psi\rangle|^2=|\beta|^2.
\end{aligned}
\end{equation}

We can now apply this procedure to a new scenario. Let's measure our state $|\psi\rangle$ in the Pauli $Y$ basis.
Our basis states are given by $|b_{+1}\rangle=|i\rangle$, and $|b_{-1}\rangle=|-i\rangle$.
Using the definition of the Pauli $Y$ basis states in Eq.~(\ref{eq:Pauli_Y_basis}), we can compute the probabilities as follows,
\begin{equation}
\begin{aligned}
    \operatorname{Prob}\{+1\} &=|\langle i \mid \psi\rangle|^2=\frac{|\alpha-i \beta|^2}{2}, \\
    \operatorname{Prob}\{-1\} &=|\langle-i \mid \psi\rangle|^2=\frac{|\alpha+i \beta|^2}{2}.
\end{aligned}
\end{equation}
Notice that the for the case of measuring in the Pauli $Y$ basis, the inner product is a complex number, but the probability obtained from the inner product is a real number, as one would expect.


There is something very important to keep in mind: the measurement outcome is only $+1$ or $-1$. A single measurement does not reveal any information about the probability amplitudes $\alpha$ and $\beta$.
In the next section, we will consider multiple measurements, which can be used to obtain information about the probabilities of measurement outcomes.

\section{Probabilities, expectation, variance}

\begin{figure}[H]
    \centering
    \includegraphics[width=0.75\textwidth]{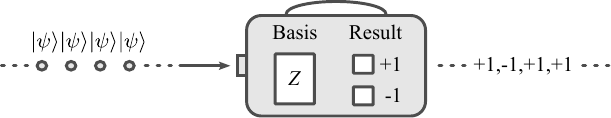}
    \caption[Repeated measurements.]{Repeated measurements of many copies of the state $|\psi\rangle$.}
    \label{fig:many-copies}
\end{figure}

We have seen how the measurement process can be described mathematically, and how the inner product is used to compute the probabilities of various measurement outcomes.
In this section, we will see how we can also compute the expectation value of a measurement outcomes as well its variance.
Up to this point, we have only measured the qubit once, as in Figs.~\ref{fig:z-measure} and \ref{fig:x-meas}.
This is known as \textbf{\emph{single-shot measurement}}\index{single-shot measurement}.
Now let's consider the scenario where we have many copies of the same state \ket{\psi} and we keep measuring them again and again by feeding them to our measurement device, as shown in Fig.~\ref{fig:many-copies}. What comes out of the measurement device is a string of $+1$ and $-1$ results.

We can count the number of $+1$ outcomes, which we denote by $N(+1)$, and we can count the number of $-1$ outcomes, which we denote by $N(-1)$. Intuitively, the ratio between the number of $+1$ outcomes to the total number of measurements will be approximately equal to the probability of obtaining the plus one outcome, which is $|\alpha|^2$, and similarly for the fraction of the number of $-1$ outcomes, which will be very close to $|\beta|^2$,
\begin{equation}
    \frac{N(+1)}{N(+1)+N(-1)} \approx|\alpha|^{2}, \qquad \frac{N(-1)}{N(+1)+N(-1)} \approx|\beta|^{2}.
\end{equation}
The approximation becomes more accurate as we increase the number of times the measurement is repeated.

In order to extract information about the probabilities of measurement outcomes, we have to repeat the same measurement on a fresh copy $|\psi\rangle$. Because these measurements are random, we're not going to get the same answer every time. However, we can ask the question: what is the expected result when we measure in the Pauli $Z$ basis? Maybe you remember from your probability course that the \textbf{\emph{expectation value}}\index{expectation value} of some random variable is the sum of the probability of each possible outcome times its value. Let's call our variable $Z$ because we are measuring in the Pauli $Z$ basis.  Its expectation value is given by the probability of the $+1$ outcome times $+1$ (the value of the outcome), plus the probability of $-1$ times the value of the outcome, which is in this case $-1$.  The probability of $+1$ remains $|\alpha|^2$, and the probability of the $-1$ outcome is given by $|\beta|^2$, but this time we are multiplying it by $-1$. So the expectation value when we measure in the Pauli Z basis is given by the expression
\begin{equation}
\begin{aligned}
    \mathbb{E}[Z] &=\operatorname{Prob}{+1} \cdot(+1)+\operatorname{Prob}{-1} \cdot(-1) \\
    &=|\alpha|^{2}-|\beta|^{2}.
    \label{eq:expectation_Z}
\end{aligned}
\end{equation}

In Dirac notation, we write it as follows. We write the operator in which basis we are measuring, for example the Pauli $Z$, and we enclose it in the angle brackets. This is a shorthand for saying that we are multiplying the bra of state $\bra{\psi}$ times the $Z$ operator times the ket of state $\ket{\psi}$,

\begin{equation}
    \langle Z\rangle=\langle\psi|Z| \psi\rangle.
\end{equation}

Let's check that this produces the expected result. Take the Pauli $Z$ operator and sandwich it between the bra $\langle\psi|$ and the ket $|\psi\rangle$. When we multiply this out, we get the following expression,
\begin{equation}
    \begin{aligned}
    \langle Z\rangle &=\left(\begin{array}{ll}
    \alpha^{*} & \beta^{*}
    \end{array}\right)\left(\begin{array}{cc}
    1 & 0 \\
    0 & -1
    \end{array}\right)\left(\begin{array}{l}
    \alpha \\
    \beta
    \end{array}\right) \\
    &=\left(\begin{array}{ll}
    \alpha^{*} & \beta^{*}
    \end{array}\right)\left(\begin{array}{c}
    \alpha \\
    -\beta
    \end{array}\right) \\
    &=|\alpha|^{2}-|\beta|^{2}.
    \end{aligned}
    \label{eq:z-expectation}
\end{equation}
This agrees with our expression from Eq.~(\ref{eq:expectation_Z}).

Because the outputs of the measurements are random we expect some degree of variance. The variance of a classical random variable $A$ is given by the expectation value of the square of that variable minus the square of the expectation value of that variable,
\begin{equation}
    \operatorname{Var}[A] =\mathbb{E}\left[A^{2}\right]-\mathbb{E}[A]^{2}.
\end{equation}
We can compute the variance of the Pauli $Z$ operator.
We have already computed the expectation value of $Z$ in Eq.~\ref{eq:z-expectation}. We therefore have to compute the expectation value of $Z^2$. That's the probability of getting a $+1$ outcome times $(+1)^2$ which is just $+1$.  The probability of the $-1$ outcome times $(-1)^2$, which is $1$, so actually what we get is
\begin{equation}
\begin{aligned}
    \mathbb{E}\left[Z^{2}\right] &=\operatorname{Prob}\{+1\} \cdot(+1)^{2}+\operatorname{Prob}\{-1\} \cdot(-1)^{2} \\
    &=|\alpha|^{2}+|\beta|^{2}.
\end{aligned}
\end{equation}
If the state is properly normalized, and it should be, then $\mathbb{E}[Z^{2}]=1$.
In fact, we could have guessed this without the explicit calculation by realizing that $Z^2=I$.
So that variance of the Pauli $Z$ operator can be written as
\begin{equation}
    \operatorname{Var}[Z] = 1 - \langle Z \rangle^2.
\end{equation}

Often in physics you will see that the variance is referred to as \textbf{\emph{fluctuations}}\index{fluctuations}, denoted by $(\Delta Z)^2$. In the Dirac notation, we again have this expression,
\begin{equation}
    (\Delta Z)^{2} \equiv \operatorname{Var}[Z] = \left\langle Z^{2}\right\rangle-\langle Z\rangle^{2}.
\end{equation}

\section{Multiple Qubits}
\label{sec:multi-qubit}

Now, let's consider how to write down the quantum state of multiple qubits. Let's start with the very simple task of writing down the state of two qubits. If we have two classical bits, they can be in four different states: $00$, $01$, $10$, or $11$. So what's the case for quantum bits? We need to switch to the Dirac notation, and we still have four possible states: $\ket{00}$, $\ket{01}$, $\ket{10}$, and $\ket{11}$. Any general state of two qubits can be written as a superposition of these four states weighted by their respective probability amplitudes $\alpha$, $\beta$, $\gamma$ and $\delta$, 
\begin{equation}
    |\psi\rangle=\alpha|00\rangle+\beta|01\rangle+\gamma|10\rangle+\delta|11\rangle.
    \label{eq:superposition_2qubits}
\end{equation}
The state $|\psi\rangle$ must be normalized, just like in the case of single-qubit states. The normalization condition is as follows,
\begin{equation}
    |\alpha|^2+|\beta|^2+|\gamma|^2+|\delta|^2 = 1,
\end{equation}
the moduli of all the probability amplitudes squared must sum it up to one.
In vector notation, we can write these kets as follows,
\begin{equation}
|00\rangle=\left(\begin{array}{l}
1 \\
0 \\
0 \\
0
\end{array}\right), \quad |01\rangle=\left(\begin{array}{l}
0 \\
1 \\
0 \\
0
\end{array}\right), \quad |10\rangle=\left(\begin{array}{l}
0 \\
0 \\
1 \\
0
\end{array}\right), \quad |11\rangle=\left(\begin{array}{l}
0 \\
0 \\
0 \\
1
\end{array}\right).
\end{equation}
We can easily check that these states are orthogonal to each other, and that they form a basis for the space of two-qubit states.
We can rewrite our general two-qubit superposition in Eq.~(\ref{eq:superposition_2qubits}) in the vector form as follows,
\begin{equation}
\ket{\psi}=\alpha\left(\begin{array}{l}
1 \\
0 \\
0 \\
0
\end{array}\right) + \beta\left(\begin{array}{l}
0 \\
1 \\
0 \\
0
\end{array}\right)+\gamma\left(\begin{array}{l}
0 \\
0 \\
1 \\
0
\end{array}\right)+\delta\left(\begin{array}{l}
0 \\
0 \\
0 \\
1
\end{array}\right)=\left(\begin{array}{l}
\alpha \\
\beta \\
\gamma \\
\delta
\end{array}\right).
\end{equation}

The mathematical concept behind being able to write the basis states in this fashion is the \textbf{\emph{tensor product}}\index{tensor product}~\footnote{There are actually multiple types of tensor products; by one definition, a tensor is like an $n$-dimensional matrix instead of the customary two-dimensional one.  In this book, we will use tensors in a fashion that expands the vector or matrix size, instead of making it into a multidimensional object.}.
The tensor product allows us to construct the vector representation of two qubits, given the states of the individual qubits.
Let's say that the first qubit is in the state $|a\rangle=(a_1,a_2)^T$, while the second qubit is in a different state $|b\rangle=(b_1,b_2)^T$.
The two-qubit state can then be expressed as
\begin{equation}
|a\rangle \otimes|b\rangle=\left(\begin{array}{l}
a_{1} \\
a_{2}
\end{array}\right) \otimes\left(\begin{array}{l}
b_{1} \\
b_{2}
\end{array}\right) \equiv\left(\begin{array}{l}
a_{1}\left(\begin{array}{l}
b_{1} \\
b_{2}
\end{array}\right) \\
a_{2}\left(\begin{array}{l}
b_{1} \\
b_{2}
\end{array}\right)
\end{array}\right)=\left(\begin{array}{l}
a_{1} b_{1} \\
a_{1} b_{2} \\
a_{2} b_{1} \\
a_{2} b_{2}
\end{array}\right),
\label{eq:tensor_product_state}
\end{equation}
where the symbol $\otimes$ denotes the tensor product.

To calculate the product, take the first probability amplitude $a_1$ times the whole column vector for $\ket{b}$. That will give us the first two elements in our new four dimensional vector.  $a_1$ times $b_1$ goes into the first entry in the resulting vector, $a_1$ times $b_2$ goes in the second entry. The bottom two elements are given by $a_2$ multiplying the whole column vector of $\ket{b}$, giving us $a_2$ times $b_1$ and $a_2$ times $b_2$.
Note that the tensor product symbol is often omitted to save space, and two-qubit state is written in a number of equivalent ways,
\begin{equation}
    |a\rangle\otimes|b\rangle = |a\rangle|b\rangle = |a,b\rangle = |ab\rangle.
\end{equation}

For example, if we have two qubits and the first qubit is in the state $\ket{0}$, and the second qubit is in the state $|1\rangle$, we write the two-qubit state as
\begin{equation}
|0\rangle \otimes|1\rangle=\left(\begin{array}{l}
1 \\
0
\end{array}\right) \otimes\left(\begin{array}{l}
0 \\
1
\end{array}\right)=\left(\begin{array}{l}
0 \\
1 \\
0 \\
0
\end{array}\right).
\end{equation}
We can the same logic to more complicated states. Let's say we have the tensor product of $\ket{1}$ with $\ket{-}$,
\begin{equation}
|1\rangle \otimes|-\rangle=\left(\begin{array}{l}
0 \\
1
\end{array}\right) \otimes \frac{1}{\sqrt{2}}\left(\begin{array}{c}
1 \\
-1
\end{array}\right)=\frac{1}{\sqrt{2}}\left(\begin{array}{c}
0 \\
0 \\
1 \\
-1
\end{array}\right).
\end{equation}

The tensor product is very general and goes beyond the case of two qubits. Given two vectors with $m$ and $n$ entries, their tensor product will be a vector with $mn$ entries in it. When dealing with qubits, our vectors will generally have a length that is a power of two since each qubit added to the system doubles the number of entries in the vector when we take the tensor product.

Notice that if the individual quantum states are normalized, their tensor product is automatically normalized as well.
And finally, be careful: just as with ordinary matrix products, the order of vectors (kets) written in a tensor product matters. This will be explored in detail in the exercises.

Having seen how we can describe multi-qubit states, it is time to learn how we can write down the operations that act on such states.
The logic is identical to our previous discussion and relies on the use of tensor products.
Consider operator $A$ acting on the first qubit, and operator $B$ acting on the second qubit.
The two-qubit operator is then described by the tensor product of the two operators,
\begin{equation}
\begin{aligned}
A \otimes B &=\left(\begin{array}{ll}
A_{11} & A_{12} \\
A_{21} & A_{22}
\end{array}\right) \otimes\left(\begin{array}{ll}
B_{11} & B_{12} \\
B_{21} & B_{22}
\end{array}\right) \\
&=\left(\begin{array}{lll}
A_{11}\left(\begin{array}{ll}
B_{11} & B_{12} \\
B_{21} & B_{22}
\end{array}\right) & A_{12}\left(\begin{array}{ll}
B_{11} & B_{12} \\
B_{21} & B_{22}
\end{array}\right) \\
A_{21}\left(\begin{array}{ll}
B_{11} & B_{12} \\
B_{21} & B_{22}
\end{array}\right) & A_{22}\left(\begin{array}{ll}
B_{11} & B_{12} \\
B_{21} & B_{22}
\end{array}\right)
\end{array}\right) \\
&=\left(\begin{array}{llll}
A_{11} B_{11} & A_{11} B_{12} & A_{12} B_{11} & A_{12} B_{12} \\
A_{11} B_{21} & A_{11} B_{22} & A_{12} B_{21} & A_{12} B_{22} \\
A_{21} B_{11} & A_{21} B_{12} & A_{22} B_{11} & A_{22} B_{12} \\
A_{21} B_{21} & A_{21} B_{22} & A_{22} B_{21} & A_{22} B_{22}
\end{array}\right)
\end{aligned}
\end{equation}
In the upper left, we have $A_{11}$ times the matrix representation of $B$. In the upper right square, we have the element $A_{12}$ times the whole matrix $B$. In the lower left, we have  $A_{21}$ times matrix $B$, and finally in the lower right $A_{22}$ times matrix $B$.

Having to multiply the individual matrix representations of the operators is quite cumbersome.
It is often more convenient to work in the Dirac notation to find out the action of the multi-qubit operator on the state.
Let's apply the two-qubit operator $A\otimes B$ to the two-qubit state $|a\rangle\otimes|b\rangle$ in Eq.~(\ref{eq:tensor_product_state}),
\begin{equation}
    (A\otimes B) |a\rangle\otimes|b\rangle = (A|a\rangle) \otimes (B|b\rangle).
\end{equation}
We can first compute the effect $A$ on $|a\rangle$, and of $B$ on $|b\rangle$ individually, and then take the tensor product.

As an example, consider applying a Pauli $X$ to the first qubit of the state $|\psi\rangle$ in Eq.~(\ref{eq:superposition_2qubits}), while leaving the second qubit alone. Leaving a qubit alone just means that we apply the identity operator to it. This is written as
\begin{equation}
\begin{aligned}
    (X \otimes I)|\psi\rangle &=X \otimes I(\alpha|00\rangle+\beta|01\rangle+\gamma|10\rangle+\delta|11\rangle) \\
    &=\alpha|10\rangle+\beta|11\rangle+\gamma|00\rangle+\delta|01\rangle
\end{aligned}
\end{equation}
If we write out the tensor of the operators, we have
\begin{equation}
X \otimes I=\left(\begin{array}{ll}
0 & 1 \\
1 & 0
\end{array}\right) \otimes\left(\begin{array}{ll}
1 & 0 \\
0 & 1
\end{array}\right)=\left(\begin{array}{llll}
0 & 0 & 1 & 0 \\
0 & 0 & 0 & 1 \\
1 & 0 & 0 & 0 \\
0 & 1 & 0 & 0
\end{array}\right).
\end{equation}

Note that, unlike ordinary matrix multiplication, two matrices being tensored do not have to have a dimension in common. They can be any size. If $A$ is an $l\times m$ matrix and $B$ is an $n \times p$ matrix, the result will be an $ln \times mp$ matrix.

Similarly to tensor product of two kets, where we often drop the twensor product symbol, we can do something the same with tensoring operators.
Though we have to be careful to keep in mind that doing ordinary matrix multiplication.
A good strategy is to add indices to the operators as a reminder that they act on different qubits,
\begin{equation}
    A \otimes B \equiv A_1B_2.
\end{equation}

Finally, by now you have probably recognized that one of the great advantages of the Dirac notation is its compactness. Two qubits can have four possible states, three qubits can have eight possible states, and so on; for $n$ qubits, the state vector is $2^n$ elements long when written out. But most of the time, we don't need to write out the entire state.  You have seen that the individual basis states are more compactly written, such as writing $\ket{00}$ instead of the whole four-element vector, let alone writing out the whole sixteen elements of the four-qubit state $\ket{0000}$. But more generally, even when dealing with an arbitrary state $\ket{\psi}$, we only need to write out the \textbf{\emph{nonzero}} elements, allowing us to ignore the zero elements.  Thus, a four-qubit superposition of the all-zeroes and all-ones states only requires us to write $\ket{\psi} = (\ket{0000} + \ket{1111})/\sqrt{2}$, saving our hands, our pens, and ultimately being easier cognitively as well~\footnote{If you are familiar with the execution of matrix math in high-performance computing, the Dirac and tensor notations are equivalent to a \emph{sparse representation}, keeping a list of nonzero elements instead of all of them.}.

\newpage
\begin{exercises}
\exer{Consider the following quantum state:}
\begin{equation*}
\ket{\psi} = \frac{\sqrt{3}}{2}\ket{0} + \frac{1}{2}\ket{1}
\end{equation*}
\subexer{Find the probability of measuring a zero.}
\subexer{Find the probability of measuring a one.}
\exer{Write out the vectors corresponding to the following tensor products. Confirm that the vectors remain normalized. Note that the order of writing the qubits matters, but the order of taking the products does not.} \subexer{$\ket{+}\otimes \ket{1}$.}
\subexer{$\ket{1}\otimes \ket{+}$.}
\subexer{$\ket{+}\otimes \ket{+}$.}
\subexer{$\ket{-}\otimes \ket{+}$.}
\subexer{$\ket{+}\otimes \ket{-}$.}
\subexer{$\ket{-}\otimes \ket{-}$.}
\subexer{$\ket{0}\otimes \ket{0} \otimes\ket{0}$.}
\subexer{$\ket{0}\otimes \ket{0} \otimes\ket{1}$.}
\subexer{$\ket{0}\otimes \ket{1} \otimes\ket{0}$.}
\subexer{$\ket{1}\otimes \ket{0} \otimes\ket{0}$.}
\subexer{$\ket{1}\otimes \ket{1} \otimes\ket{1}$.}
\subexer{$\ket{1}\otimes \ket{+} \otimes\ket{0}$.}

\exer{Write out the matrices corresponding to the following tensor products.} \subexer{$X\otimes X$.}
\subexer{$Z\otimes Z$.}
\subexer{$I\otimes X$.}
\subexer{$I\otimes Z$.}
\subexer{$X\otimes Z$.}
\subexer{$Z\otimes X$.}

\exer{Find the expectation value $\langle Z\rangle$ for the following single-qubit states.}
\subexer{$\ket{0}$}
\subexer{$\ket{1}$}
\subexer{$\ket{+}$}
\subexer{$\ket{-}$}
\subexer{$\frac{\sqrt{3}}{2}\ket{0} + \frac{1}{2}\ket{1}$}

\exer{Since we can find the expectation value of $Z$, you might suspect that we can do the same for $X$. Find the expectation value $\langle X\rangle$ for the following single-qubit states.}
\subexer{$\ket{0}$}
\subexer{$\ket{1}$}
\subexer{$\ket{+}$}
\subexer{$\ket{-}$}
\subexer{$\frac{\sqrt{3}}{2}\ket{0} + \frac{1}{2}\ket{1}$}

\exer{We said above that the Hadamard gate is not just a y-axis rotation.  To explore this further, calculate the following. \\
Use $\ket{\lambda_{+}} = \frac{\sqrt{2 + \sqrt{2}}}{2}\ket{0}+\frac{1}{\sqrt{2(2 + \sqrt{2})}}\ket{1}$ and $\ket{\lambda_{-}}=\frac{\sqrt{2 - \sqrt{2}}}{2}\ket{0} - \frac{1}{\sqrt{2(2 - \sqrt{2})}}\ket{1}$.
\subexer{$H\ket{i}$}
\subexer{$R_Y(\pi/2)\ket{i}$}
\subexer{$R_Y(\pi)\ket{i}$}
\subexer{$H(\sqrt{3}\ket{0}+\ket{1})/\sqrt{2}$}
\subexer{$R_Y(\pi/2)(\sqrt{3}\ket{0}+\ket{1})/\sqrt{2}$}
\subexer{$R_Y(\pi)(\sqrt{3}\ket{0}+\ket{1})/\sqrt{2}$}
\subexer{$H\ket{\lambda_+}$}
\subexer{$H\ket{\lambda_-}$}
\subexer{$R_Y(\pi/2)\ket{\lambda_+}$}
\subexer{$R_Y(\pi/2)\ket{\lambda_-}$}
\subexer{$R_Y(\pi)\ket{\lambda_+}$}
\subexer{$R_Y(\pi)\ket{\lambda_-}$}
}
\end{exercises}

\chapter[Pure and Mixed States]{Pure and Mixed States}
\label{sec:3_pure_mixed}

This chapter will expand on what we learned in the previous chapter.
It will deal with pure and mixed states, and finally we will get to the description of real noisy quantum states.
The notation and mathematical tools that we develop in this chapter will be extremely useful in the rest of these lecture notes. If you have taken an introductory course in quantum computation, your course may or may not have covered these topics already, depending on whether hardware or algorithmic concepts were emphasized.

\section{Noisy world}
\label{sec:3-1_noisy_world}

Up until now, we have only discussed quantum states which were noiseless, meaning they
were always in the exact state that we wanted them to be.
There was no uncertainty in our knowledge of the description of the state.
The operations that we used to transform them were perfect too, resulting in perfect outputs.
The world, however, is a noisy place and this description is not quite sufficient for real-world applications.
Let's consider three examples demonstrating why this is the case.

\begin{figure}[t]
    \centering
    \includegraphics[width=\textwidth]{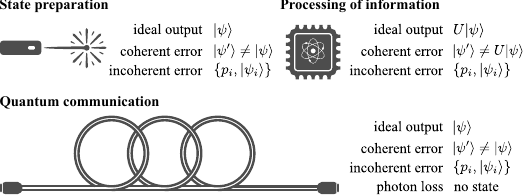}
    \caption[Noisy world.]{Real quantum states are noisy.}
    \label{fig:3-1_noise}
\end{figure}

The first example is that of \textbf{\emph{state preparation}}\index{state preparation} as shown in Fig.~\ref{fig:3-1_noise}.
You may go to your friend who works in a quantum laboratory and ask him or her to prepare
a state $\ket{\psi}$.
In a real lab, the prepared state is not the desired state $\ket{\psi}$.
If you are lucky it might be some other pure state $|\psi'\rangle$ which is close to the desired state $|\psi\rangle$.
You would still obtain a pure state, albeit not exactly the one you asked for, but at least you would have full knowledge of this state $|\psi'\rangle$.
A much more likely scenario is that your friend in the lab can only produce a distribution of pure states.
This means that we obtain some pure state $|\psi_1\rangle$ with probability $p_1$, or a different pure state $|\psi_2\rangle$ with probability $p_2$, and so on and so forth.
This distribution is often written as $\{p_i,|\psi_i\rangle\}$.

The second example is that of \textbf{\emph{processing of information}}\index{information processing}.
In this case, we wish to perform a quantum computation represented by some unitary operator $U$.
Even if the input to the computation is an ideal pure state $|\psi\rangle$, the output will not be the desired $U|\psi\rangle$.
The output might contain coherent errors, meaning that the applied operation was some other unitary $U'$, giving us the output $U'|\psi\rangle$.
The output might be affected by incoherent errors, such as probabilistic Pauli errors, or even relaxation errors.

The last example deals with the central scenario of these lecture notes, namely \textbf{\emph{quantum communication}}\index{quantum communication}.
Consider the case where you prepare a state $|\psi\rangle$, which you would like to send to a friend through a long optical fiber.
Even though we do not process the prepared state in any way, optical fibers themselves are sources of noise.
The state received by your friend will be affected by coherent and incoherent errors.
On top of these, the state may even get lost and never arrive due to various processes such as absorption or scattering in the fiber.
Photon loss is a huge headache in quantum communication and we will discuss it in great detail in later chapters.

\section{Outer product}
\label{sec:3-2_outer_product}

We have seen in the previous chapter that the inner product of two vector states $|a\rangle$ and $|b\rangle$ is formed by transforming the first ket into a bra and then multiplying them together,
\begin{equation}
    \langle a | b\rangle = \begin{pmatrix} a_0^* & a_1^* \end{pmatrix} \begin{pmatrix} b_0 \\ b_1 \end{pmatrix} = a_0^* b_0 + a_1^* b_1.
\end{equation}
What happens when we change the order of multiplication?
Let's multiply the ket $|b\rangle$ from the right by the bra $\langle a|$,
\begin{equation}
    | b\rangle \langle a | = \begin{pmatrix} b_0 \\ b_1 \end{pmatrix} \begin{pmatrix} a_0^* & a_1^* \end{pmatrix} = \begin{pmatrix} a_0^*b_0 & a_1^*b_0 \\ a_0^*b_1 & a_1^*b_1 \end{pmatrix}.
\end{equation}
By changing the order of multiplication, we obtained a complex matrix.
The above product of a ket with a bra, $|b\rangle\langle a|$, is called an \textbf{\emph{outer product}}\index{outer product}. It is necessary in order to describe measurements as well as noisy quantum states.

In order to see what the action of an outer product on a state vector is, we will begin with a few simple examples.
Let's consider the following outer product, $|0\rangle\langle0|$.
The matrix representation of $|0\rangle\langle0|$ is
\begin{equation}
    |0\rangle\langle0| = \begin{pmatrix} 1 \\ 0 \end{pmatrix} \begin{pmatrix} 1 & 0 \end{pmatrix} = \begin{pmatrix} 1 & 0 \\ 0 & 0 \end{pmatrix}.
\end{equation}
We also know from basic vector algebra that matrices transform vectors into other vectors.
Let's apply our example outer product to a general qubit state vector $|\psi\rangle = \alpha|0\rangle + \beta|1\rangle$, where $|\alpha|^2+|\beta|^2=1$,
\begin{align}
    |0\rangle\langle0|\psi\rangle & = |0\rangle\langle0| \left( \alpha|0\rangle + \beta|1\rangle \right) \label{eq:projector_00}\\
    & = \alpha |0\rangle\underbrace{\langle0|0\rangle}_{=1} + \beta |0\rangle\underbrace{\langle0|1\rangle}_{=0} \nonumber\\
    & = \alpha|0\rangle. \nonumber
\end{align}
In the second line of Eq.~(\ref{eq:projector_00}), we used the fact that the inner product of a normalized vector with itself is unity, and the orthogonality property of kets $|0\rangle$ and $|1\rangle$.
We see that our example outer product induces an interesting transformation of the initial general single-qubit state vector $|\psi\rangle$.
It changes the initial state vector into $|0\rangle$, up to a complex constant $\alpha$.
We did not assume any particular values for the initial probability amplitudes $\alpha$ and $\beta$, meaning that any initial state vector, provided $\alpha\neq0$, is transformed into a vector proportional to $|0\rangle$.
We say that $|0\rangle\langle0|$ \textit{\textbf{projects}} onto the state $|0\rangle$, and that the outer product is an example of a \textbf{\emph{projector}}\index{projector}.
The only time $|0\rangle\langle0|$ does not project onto $|0\rangle$ is when the initial state is orthogonal to $|0\rangle$, that is, when the initial state is $|1\rangle$.

Let's have a quick look at the action of a different outer product, namely $|1\rangle\langle1|$.
In matrix form, we have
\begin{equation}
    |1\rangle\langle1| = \begin{pmatrix} 0 \\ 1 \end{pmatrix} \begin{pmatrix} 0 & 1 \end{pmatrix} = \begin{pmatrix} 0 & 0  \\ 0 & 1 \end{pmatrix}.
\end{equation}
Since $|0\rangle\langle0|$ projects onto the state $|0\rangle$, we expect that $|1\rangle\langle1|$ projects onto $|1\rangle$.
Let's confirm this by acting with $|1\rangle\langle1|$ on a general state vector $|\psi\rangle$,
\begin{align}
    |1\rangle\langle1|\psi\rangle & = |1\rangle\langle1| \left( \alpha|0\rangle + \beta|1\rangle \right) \label{eq:projector_11}\\
    & = \alpha |1\rangle\underbrace{\langle1|0\rangle}_{=0} + \beta |1\rangle\underbrace{\langle1|1\rangle}_{=1} \nonumber\\
    & = \beta|1\rangle. \nonumber
\end{align}
Indeed, $|1\rangle\langle1|$ does project onto the $|1\rangle$ state.

Recall from Sec.~\ref{sec:measurement} that measurement in the $Z$ basis of a general qubit $|\psi\rangle = \alpha |0\rangle + \beta |1\rangle$ has two possible outcomes.
Outcome +1 is obtained with probability $|\alpha|^2$ and the post-measurement state is $|0\rangle$.
We just saw that this post-measurement state is obtained after application of the projector $|0\rangle\langle0|$, up to a factor of $\alpha$.
Similarly, he -1 outcome of the measurement is obtained with probability $|\beta|^2$ by application of the projector $|1\rangle\langle1|$, up to a factor of $\beta$.
The post-measurement state is $|1\rangle$ resulting from the application of the projector $|1\rangle\langle1|$.
The outer products $|0\rangle\langle0|$ and $|1\rangle\langle1|$ capture the effect of \textbf{\emph{projective measurements}}\index{projective measurements}.
The different probabilities of the measurement outcomes can also be readily calculated.
The probabilities are simply the expectation values of the projectors,
\begin{align}
    \langle \psi | 0 \rangle \langle 0 | \psi \rangle & = \left( \alpha^*\langle0| + \beta^*\langle1| \right) |0\rangle\langle0| \left( \alpha |0\rangle + \beta |1\rangle \right) = |\alpha|^2, \\
    \langle \psi | 1 \rangle \langle 1 | \psi \rangle & = \left( \alpha^*\langle0| + \beta^*\langle1| \right) |1\rangle\langle1| \left( \alpha |0\rangle + \beta |1\rangle \right) = |\beta|^2.
\end{align}

Can you guess the form of the projectors corresponding to measurement in the Pauli $X$ basis?
We can follow the same logic that we have used for the case of $Z$-basis measurement.
Post-measurement state after a $\pm1$ outcomes is $|\pm\rangle$.
The corresponding projectors are
\begin{align}
    |+\rangle\langle+| & = \frac{1}{\sqrt{2}} \begin{pmatrix} 1 \\ 1 \end{pmatrix} \frac{1}{\sqrt{2}} \begin{pmatrix} 1 & 1 \end{pmatrix} = \frac{1}{2} \begin{pmatrix} 1 & 1 \\ 1 & 1 \end{pmatrix}, \\
    |-\rangle\langle-| & = \frac{1}{\sqrt{2}} \begin{pmatrix} 1 \\ -1 \end{pmatrix} \frac{1}{\sqrt{2}} \begin{pmatrix} 1 & -1 \end{pmatrix} = \frac{1}{2} \begin{pmatrix} 1 & -1 \\ -1 & 1 \end{pmatrix}.
\end{align}
We can also convince ourselves that these operators really do project onto the correct states by applying them to a general qubit state,
\begin{align}
    |+\rangle\langle+|\psi\rangle & = |+\rangle \frac{1}{\sqrt{2}} \left( \langle0| + \langle1| \right) \left( \alpha |0\rangle + \beta|1\rangle \right) = \frac{\alpha+\beta}{\sqrt{2}} |+\rangle, \\
    |-\rangle\langle-|\psi\rangle & = |-\rangle \frac{1}{\sqrt{2}} \left( \langle0| - \langle1| \right) \left( \alpha |0\rangle + \beta|1\rangle \right) = \frac{\alpha-\beta}{\sqrt{2}} |-\rangle.
\end{align}
Similarly, projectors corresponding to the $\pm1$ outcomes of a measurement in the Pauli Y basis are
\begin{align}
    |i\rangle\langle i| & = \frac{1}{\sqrt{2}} \begin{pmatrix} 1 \\ i \end{pmatrix} \frac{1}{\sqrt{2}} \begin{pmatrix} 1 & -i \end{pmatrix} = \frac{1}{2} \begin{pmatrix} 1 & -i \\ i & 1 \end{pmatrix}, \\
    |-i\rangle\langle-i| & = \frac{1}{\sqrt{2}} \begin{pmatrix} 1 \\ -i \end{pmatrix} \frac{1}{\sqrt{2}} \begin{pmatrix} 1 & i \end{pmatrix} = \frac{1}{2} \begin{pmatrix} 1 & i \\ -i & 1 \end{pmatrix}.
\end{align}
The action of these projectors can be verified by applying them to an arbitrary qubit state,
\begin{equation}
    |i\rangle\langle i|\psi\rangle = \frac{\alpha-i\beta}{\sqrt{2}} |i\rangle, \quad
    |-i\rangle\langle-i|\psi\rangle = \frac{\alpha+i\beta}{\sqrt{2}} |-i\rangle.
\end{equation}

Before moving on, it is worth settling on the notation regarding projectors.
Sometimes, rather than writing projectors as outer products, we denote them by a single letter.
Usual choice is $\Pi$, the capital Greek letter ``pi'',
\begin{equation}
    \Pi_{\pm}^B \equiv |b_{\pm}\rangle\langle b_{\pm}|,
\end{equation}
where $B$ denotes the basis in which we are measuring and $|b_{\pm}\rangle$ denote the two orthogonal state vectors corresponding to the two possible measurement outcomes $\pm1$.
Table~\ref{tab:3-2_projectors} contains all six projectors that we have so far encountered written in both the short notation as well as outer products.
\begin{table}[h]
    \setcellgapes{3pt}
    \setlength{\tabcolsep}{15pt}
    \renewcommand\theadfont{}
    \makegapedcells
    \centering
    \begin{tabular}{ccc}
        \hline
        & \textbf{+1 Outcome} & \textbf{-1 Outcome} \\
        \hline
        \thead{Pauli $X$ basis} & $\Pi^X_+=|+\rangle\langle+|$ & $\Pi^X_-=|-\rangle\langle-|$ \\
        \thead{Pauli $Y$ basis} & $\Pi^Y_+=|i\rangle\langle i|$ & $\Pi^Y_-=|-i\rangle\langle-i|$ \\
        \thead{Pauli $Z$ basis} & $\Pi^Z_+=|0\rangle\langle0|$ & $\Pi^Z_-=|1\rangle\langle1|$ \\
        \hline
    \end{tabular}
    \caption{Projectors onto the eigenstates of the Pauli matrices.}
    \label{tab:3-2_projectors}
\end{table}

We have represented quantum states in the form of vectors.
For example,
\begin{equation}
    |0\rangle = \begin{pmatrix} 1 \\ 0 \end{pmatrix}, \quad |+\rangle = \frac{1}{\sqrt{2}} \begin{pmatrix} 1 \\ 1 \end{pmatrix}, \quad |\psi\rangle = \begin{pmatrix} \alpha \\ \beta \end{pmatrix}.
    \label{eq:3-2_vectors}
\end{equation}
We have also seen that outer products can project onto these states.
This suggests that we can equally \textbf{\emph{represent quantum states as matrices}} as well written in the form of outer products.
For example, the states in Eq.~(\ref{eq:3-2_vectors}) can be written as projectors in the following way,
\begin{equation}
    |0\rangle\langle0| = \begin{pmatrix} 1 & 0 \\ 0 & 0 \end{pmatrix}, \quad |+\rangle\langle+| = \frac{1}{2} \begin{pmatrix} 1 & 1 \\ 1 & 1 \end{pmatrix}, \quad |\psi\rangle\langle\psi| = \begin{pmatrix} |\alpha^*| & \alpha\beta^* \\ \alpha^*\beta & |\beta|^2 \end{pmatrix}.
\end{equation}
The state vector $|\psi\rangle$ and the projector $|\psi\rangle\langle\psi|$ represent the same state of a physical system.
You might ask yourselves, why complicate things with outer product description of quantum states if vectors seem to work just fine?
The vector description is sufficient only for a limited number of cases when the quantum systems are not affected by any imperfections.
Such states are called \textbf{\emph{pure states}}\index{pure states}.
So far we have been working exclusively with pure states.
In order to describe more realistic quantum states, we have to resort to description that relies on the outer product, as we will see in the next Section.

\section{Density matrices}
\label{sec:3-3_density_matrices}

\begin{figure}[t]
    \centering
    \includegraphics[width=\textwidth]{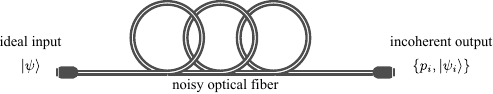}
    \caption[Noisy quantum communication.]{Ideal quantum input state becomes a probabilistic mixture at the output as a result of the noise introduced by the optical fiber.}
    \label{fig:3-3_noisy_communication}
\end{figure}

Let's consider the scenario of sending an ideal quantum state $|\psi\rangle$ through an optical fiber as shown in Fig.~\ref{fig:3-3_noisy_communication}.
The state of the quantum system inevitably changes as it travels through the optical fiber.
The fiber affects the state of the quantum system probabilistically.
The system remains unaffected with some probability, but also with some probability it suffers from unwanted transformations.
The incoherent output becomes a distribution of states $\{p_i, |\psi_i\rangle\}$.
With probability $p_i$ the output is described by the state vector $|\psi_i\rangle$.

How can we describe such an incoherent output mathematically?
The output is not a single state that we know can be represented by a vector, rather it is a probabilistic distribution of such states.
Therefore we cannot represent it as a superposition.
Superposition of various state vectors is still a state that represents perfect knowledge of the quantum system, which we do not have.
The correct way of describing the incoherent output in Fig.~\ref{fig:3-3_noisy_communication} is by using the outer product that we learned about in the previous section.
The output state is a projector $|\psi_i\rangle\langle\psi_i|$ with probability $p_i$, and the whole output state can be described as a sum of all such projectors,
\begin{equation}
    p_1 |\psi_1\rangle\langle\psi_1| + p_2 |\psi_2\rangle\langle\psi_2| + \ldots + p_n |\psi_n\rangle\langle\psi_n|.
    \label{eq:3-3_mixture}
\end{equation}
The state represented by the sum in Eq.~(\ref{eq:3-3_mixture}) is called a \textbf{\emph{mixed state}}\index{mixed state}.
It is important to remember is that the $p_i$'s are probabilities and therefore they must sum to unity, that is $\sum_{i=1}^n p_i=1$.

\begin{figure}[t]
    \centering
    \includegraphics[width=0.9\textwidth]{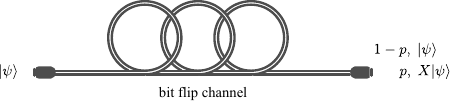}
    \caption[Bit-flip channel.]{Bit-flip channel. Pauli $X$ is applied to the input with probability $p$. The input is unaffected with probability $1-p$.}
    \label{fig:3-3_bit_flip_channel}
\end{figure}
In order to illustrate this on a simple example, we consider the  \textbf{\emph{bit-flip channel}}\index{bit-flip channel}, depicted in Fig.~\ref{fig:3-3_bit_flip_channel}.
The input is some general state $|\psi\rangle$.
The input is either affected by a Pauli $X$ transformation with probability $p$, or it is transmitted without errors with probability $1-p$.
We will denote the output by the Greek letter $\rho$ (rho),
\begin{equation}
    \rho = (1-p) |\psi\rangle\langle\psi| + p X|\psi\rangle\langle\psi|X.
    \label{eq:3-3_bit_flip}
\end{equation}
Let's explore how this channel works by considering a few simple inputs.
When the input is $|\psi\rangle = |0\rangle$, the output is
\begin{equation}
    \rho = (1-p)|0\rangle\langle0| + p|1\rangle\langle1| = \begin{pmatrix} 1-p & 0 \\ 0 & p \end{pmatrix}.
\end{equation}
On the other hand, if the input is $|\psi\rangle=|1\rangle$, we get
\begin{equation}
    \rho = (1-p)|1\rangle\langle1| + p|0\rangle\langle0| = \begin{pmatrix} p & 0 \\ 0 & 1-p \end{pmatrix}
\end{equation}
as the output.

We know that the general state of a pure state of a qubit is $|\psi\rangle = \alpha|0\rangle + \beta|1\rangle$.
What is the general form of a mixed state?
We have seen it already in Eq.~(\ref{eq:3-3_mixture}),
\begin{equation}
    \rho = \sum_i p_i |\psi_i\rangle\langle\psi_i|.
    \label{eq:general_mixed_state}
\end{equation}
The representation for a state vector we have used so far is known as a ket, while this more general representation using a sum of outer products is called a \textbf{\emph{density matrix}}\index{density matrix}.
Note that the density matrix can describe both pure and mixed states, whereas the state vector can only describe pure states.
When all $p_i$ values vanish except for one, that is $\rho=\ketbra{\psi}{\psi}$, the state $\rho$ is pure.
On the other hand, if more than one $p_i$ is non-zero, the state $\rho$ is mixed.  (The \ket{\psi} is not restricted to a single computational basis state; it can be any superposition.  That is, \ket{\psi} is not restricted to a state like \ket{0} or \ket{1}; $\ket{\psi} = \ket{+}$ is valid too.)

We saw in the previous chapter that all vectors representing quantum states must be normalized.
What does this normalization condition look like when we write the state as a density matrix?
To answer this question, let's first define the \textit{\textbf{trace}} of a matrix $A$,
\begin{equation}
    \text{Tr} \{ A \} = \sum_i A_{ii},
\end{equation}
as the sum of all the diagonal elements.
For example,
\begin{equation}
    A = \begin{pmatrix} A_{11} & A_{12} & A_{13} \\ A_{21} & A_{22} & A_{23} \\ A_{31} & A_{32} & A_{33} \end{pmatrix}, \qquad \text{Tr} \{ A \} = A_{11} + A_{22} + A_{33}.
\end{equation}
Density matrix $\rho$ is normalized when $\text{Tr}\{\rho\}=1$.
Let's test this on a general pure state $|\psi\rangle = \alpha |0\rangle + \beta |1\rangle$.
The density matrix  corresponding to $|\psi\rangle$ is
\begin{equation}
    \rho = |\psi\rangle\langle\psi| = \begin{pmatrix} |\alpha|^2 & \alpha\beta^* \\ \alpha^*\beta & |\beta|^2 \end{pmatrix}.
\end{equation}
We see that $\text{Tr}\{\rho\} = |\alpha|^2 + |\beta|^2 = 1$ if the state is normalized.

Computing the trace of a general density matrix can be made straightforward by the following realization.
The action of a trace on an outer product is to swap the places of the bra and the ket, changing the outer product into an inner product,
\begin{equation}
    \text{Tr} \{|b\rangle\langle a|\} = \langle a|b\rangle.
\end{equation}
Therefore for any pure state $|\psi\rangle$, we have
\begin{equation}
    \text{Tr}\{|\psi\rangle\langle\psi|\} = \langle\psi|\psi\rangle=1.
\end{equation}
How about a general mixed state $\rho$ given by the Eq.~(\ref{eq:general_mixed_state})?
The calculation is very similar,
\begin{align}
    \text{Tr}\{\rho\} & = \text{Tr}\left\{\sum_i p_i |\psi_i\rangle\langle\psi_i|\right\} \\
    & = \sum_i p_i \text{Tr}\left\{ |\psi_i\rangle\langle\psi_i| \right\} \nonumber\\
    & =\sum_i p_i \underbrace{\langle\psi_i|\psi_i\rangle}_{=1} \nonumber\\
    & = \sum_i p_i \nonumber\\
    & = 1.
\end{align}

\begin{figure}[t]
    \centering
    \def\r{4}
    \tdplotsetmaincoords{70}{115}
    \def\azimuthP{40}
    \def\polarP{60}
    \def\circleSize{88.5}
    
    \begin{tikzpicture}[tdplot_main_coords,scale=0.35,every node/.style={scale=0.35}]

    \tdplotsetcoord{O}{0}{0}{0}
    \tdplotsetcoord{P}{\r}{\azimuthP}{\polarP}
    \tdplotsetcoord{Q}{\r*sin(\azimuthP)}{90}{\polarP}

    \shade[tdplot_screen_coords,ball color=gray!40,opacity=0.5] (0,0,0) circle (\r); 
    \tdplotCsDrawLonCircle[tdplotCsBack/.style={black!50,dashed},thin]{\r}{\polarP-90}
    \tdplotCsDrawLatCircle[tdplotCsBack/.style={black!50,dashed},very thin]{\r}{90-\azimuthP}
    \tdplotCsDrawLatCircle[tdplotCsBack/.style={black!50,dashed}]{\r}{0}

    \draw[-latex,semithick] (O) -- (1.3*\r,0,0) node[pos=1.15] {\Huge $x$};
    \draw[-latex,semithick] (O) -- (0,1.3*\r,0) node[pos=1.07] {\Huge $y$};
    \draw[-latex,semithick] (O) -- (0,0,1.3*\r) node[pos=1.06] {\Huge $z$};
    
    \draw[-latex,thick,blue!50] (O) -- (P);
    
    \tdplotCsDrawCircle[tdplotCsFill/.style=blue!50,thick]{\r}{\polarP}{\azimuthP}{1.05*\circleSize}
    \node[] at (4, 3.8, 5.1) {\Huge $|\psi\rangle$};
    \node[] at (0, 0, 6.5) {\Huge Pure state};

    \begin{scope}[xshift=10cm]
    
    \tdplotsetcoord{O}{0}{0}{0}
    \tdplotsetcoord{P}{0.4*\r}{\azimuthP}{\polarP}
    \tdplotsetcoord{Q}{0.4*\r*sin(\azimuthP)}{90}{\polarP}

    \shade[tdplot_screen_coords,ball color=gray!40,opacity=0.5] (0,0,0) circle (\r);
    \tdplotCsDrawLatCircle[tdplotCsBack/.style={black!50,dashed}]{\r}{0}

    \shade[tdplot_screen_coords,ball color=blue!30,opacity=0.5] (0,0,0) circle (0.4*\r);
    \tdplotCsDrawLonCircle[tdplotCsBack/.style={black!50,dashed},thin]{0.4*\r}{\polarP-90}
    \tdplotCsDrawLatCircle[tdplotCsBack/.style={black!50,dashed},thin]{0.4*\r}{90-\azimuthP}
    \tdplotCsDrawLatCircle[tdplotCsBack/.style={black!50,dashed}]{0.4*\r}{0}

    \draw[-latex,semithick] (O) -- (1.3*\r,0,0) node[pos=1.15] {\Huge $x$};
    \draw[-latex,semithick] (O) -- (0,1.3*\r,0) node[pos=1.07] {\Huge $y$};
    \draw[-latex,semithick] (O) -- (0,0,1.3*\r) node[pos=1.06] {\Huge $z$};
    
    \draw[-latex,thick,blue!50] (O) -- (P);
    
    \tdplotCsDrawCircle[tdplotCsFill/.style=blue!50,thick]{0.4*\r}{\polarP}{\azimuthP}{1.09*\circleSize}
    \node[] at (1, 1.8, 1.3) {\Huge $\rho$};
    \node[] at (0, 0, 6.5) {\Huge Mixed state};
    
    \end{scope}

    \begin{scope}[xshift=20cm]
    
    \tdplotsetcoord{O}{0}{0}{0}
    \tdplotsetcoord{P}{0.4*\r}{\azimuthP}{\polarP}
    \tdplotsetcoord{Q}{0.4*\r*sin(\azimuthP)}{90}{\polarP}

    \shade[tdplot_screen_coords,ball color=gray!40,opacity=0.5] (0,0,0) circle (\r);
    \tdplotCsDrawLatCircle[tdplotCsBack/.style={black!50,dashed}]{\r}{0}

    \draw[-latex,semithick] (O) -- (1.3*\r,0,0) node[pos=1.15] {\Huge $x$};
    \draw[-latex,semithick] (O) -- (0,1.3*\r,0) node[pos=1.07] {\Huge $y$};
    \draw[-latex,semithick] (O) -- (0,0,1.3*\r) node[pos=1.06] {\Huge $z$};

    \filldraw[blue!50,draw=black,thick] (0,0) circle (5pt);
    \node[] at (1, 1.5, 1.2) {\Huge $I/2$};
    \node[] at (0, 0, 6.5) {\Huge Maximally mixed state};
    
    \end{scope}  
    \end{tikzpicture}

    \caption[Bloch sphere and mixed states.]{Bloch sphere representation of a pure state $|\psi\rangle$, a mixed state $\rho$ and the maximally mixed state $I/2$.}
    \label{fig:3-3_bloch}
\end{figure}

Mixed states can be visualized using the Bloch sphere as well.
Pure states are represented by the points on the surface of the sphere.
Mixed states, on the other hand, are represented by all the points \textbf{\emph{inside the Bloch sphere}}, as shown in Fig.~\ref{fig:3-3_bloch}.
The closer we get to the centre of the sphere, the more mixed the quantum state becomes.
The center of the sphere is where the \textbf{\emph{maximally mixed state}}\index{maximally mixed state},
\begin{equation}
    \frac{1}{2} |0\rangle\langle0| + \frac{1}{2} |1\rangle\langle1| = \frac{1}{2} \begin{pmatrix} 1 & 0 \\ 0 & 1 \end{pmatrix} = \frac{I}{2},
\end{equation}
is located.
This point represents zero knowledge about the quantum state.
If we measure this state in the Pauli $Z$ basis, both outcomes are equally likely.
We can think about this state as being ``half-way between'' the pure states $|0\rangle$ and $|1\rangle$.
However, there is nothing sacred about the Pauli $Z$ basis.
We can also think about the maximally mixed state being ``half-way between'' the pure states $|+\rangle$ and $|-\rangle$, and write it as
\begin{equation}
    \frac{I}{2} = \frac{1}{2} |+\rangle\langle+| + \frac{1}{2} |-\rangle\langle-|.
\end{equation}
This immediately suggests that if we measure the maximally mixed state in the Pauli $X$ basis then both outcomes are equally likely again.
You may have guessed that we can also use the Pauli $Y$ basis and write the maximally mixed state as $I/2 = ( |i\rangle\langle i| + |-i\rangle\langle-i| ) / 2$.

\section{Pure vs mixed states}
\label{sec:3-4_pure_vs_mixed}

Now that we have seen how pure and mixed states differ in terms of their representations, it is a good time to pause and consider the difference between them.
We mentioned that pure states represent states of full knowledge whereas mixed states are necessary if there is some uncertainty about the quantum state.
How can we see this difference in real life when we do measurements?
Let's consider two quantum states, one is an equal superposition of $|0\rangle$ and $|1\rangle$, and the other is an equal mixture of $|0\rangle$ and $|1\rangle$,
\begin{align}
    |\psi\rangle & = \frac{1}{\sqrt{2}} (|0\rangle + |1\rangle, \label{eq:3-4_superposition}\\
    \rho & = \frac{1}{2} (|0\rangle\langle0| + |1\rangle\langle1|). \label{eq:3-4_mixture}
\end{align}
We can immediately see that when we measure both states in the Pauli $Z$ basis, we obtain the same statistics for the measurement outcomes,
\begin{align}
    \text{superposition} \; |\psi\rangle & : \text{Prob}\{+1\} = \text{Prob}\{-1\} = \frac{1}{2},  \\
    \text{mixture} \; \rho & : \text{Prob}\{+1\} = \text{Prob}\{-1\} = \frac{1}{2}.
\end{align}
This means that we cannot tell these two states from each other by performing just a Pauli $Z$ measurement.
It also seems to go counter to our intuition that pure states represent full knowledge of the state. After all, we cannot predict the outcome of a measurement with certainty.
We will see that this is not quite true.

Let's measure the two states again but this time in the Pauli $X$ basis.
We can rewrite the two states in Eq.~(\ref{eq:3-4_superposition})-(\ref{eq:3-4_mixture}) in the Pauli $X$ basis,
\begin{align}
    |\psi\rangle & = |+\rangle, \\
    \rho & = \frac{1}{2} (|+\rangle\langle+| + |-\rangle\langle-|).
\end{align}
Probabilities of the measurement outcomes are now very different,
\begin{align}
    \text{superposition} \; |\psi\rangle & : \text{Prob}\{+1\} = 1, \; \text{Prob}\{-1\} = 0,  \\
    \text{mixture} \; \rho & : \text{Prob}\{+1\} = \text{Prob}\{-1\} = \frac{1}{2}.
\end{align}
We see that measuring the superposition $|\psi\rangle$ in the Pauli $X$ basis always results in a positive measurement outcome.
On the other hand, when measuring the mixture, both outcomes are still equally likely.
We can see that the equal superposition and equal mixture can be distinguished from each other when measured in the Pauli $X$ basis.
Now the meaning of having perfect knowledge of the states becomes more clear.
The outcome of measuring this state in the $X$ basis is deterministic: it always gives the same outcome. More generally, for any pure state, there is always some choice of measurement we could make that will give a deterministic outcome. For mixed states, on the other hand, the measurement outcomes in any basis will always be probabilistic.

The above procedure allows us to tell pure states from mixed states, but it is rather complicated.
We have to check the measurement statistics for various bases, which can be quite laborious.
Fortunately, there is a very simple way of checking if a state $\rho$ is pure or mixed using the \textbf{\emph{purity}}\index{purity} $\gamma$, pronounced ``gamma''.
Purity is defined as the trace of the square of the density matrix,
\begin{equation}
    \gamma = \text{Tr} \{ \rho^2 \} \longrightarrow \begin{cases}
        = 1, \quad \text{when } \rho \text{ is pure}, \\
        < 1, \quad \text{when } \rho \text{ is mixed}.
    \end{cases}
\end{equation}
When the quantum state is pure, the purity is $\gamma=1$.
On the other hand, when the state is mixed, purity is strictly less than 1.

\begin{table}[t]
    \setcellgapes{5pt}
    \renewcommand\theadfont{}
    \makegapedcells
    \centering
    \begin{tabular}{ccc}
        \hline
        & \textbf{Pure states} & \textbf{Mixed states} \\
        \hline
        \textbf{Notation} & $|\psi\rangle$ & $\rho$ \\
        \textbf{Represented by} & vectors or matrices & matrices only \\
        \textbf{Normalization} & $\langle\psi|\psi\rangle=1$ & $\text{Tr}\{\rho\}=1$ \\
        \textbf{Knowledge of the state} & perfect & imperfect \\
        \textbf{Purity} & 1 & $<1$ \\
        \hline
    \end{tabular}
    \caption[Pure versus mixed states.]{Summary of differences between pure and mixed states.}
    \label{tab:3-4_pure_vs_mixed}
\end{table}

This brings our discussion of the differences between pure and mixed states to an end.
Table~\ref{tab:3-4_pure_vs_mixed} provides a summary of the important points.

\section{Fidelity}
\label{sec:3-5_fidelity}

One thing you should always keep in mind is that it is impossible to exactly prepare the desires pure state.
Sometimes the prepared state is pure but not quite the target state we were aiming for.
Other times it can be affected by incoherent noise resulting in a mixed state.
It is crucial that we have a way of quantifying the difference between the real and the desired, or target, state.

A convenient tool that tells us how close the real state is to the target state is the \textbf{\emph{fidelity}}\index{fidelity}.
Consider the case where the target state is a pure state $|\psi\rangle$.
This state could be the desired output of a quantum computer, or a state distributed by a quantum network.
However, due to noise and imperfections in the hardware and our control of it, the actual state is inevitably some mixture $\rho$.
The deviation of this state from the target state is given by the fidelity $F$, defined as
\begin{equation}
    F(\rho,|\psi\rangle) = \langle\psi|\rho|\psi\rangle.
    \label{eq:3-5_fidelity}
\end{equation}
Note that other more general definitions for fidelity are sometimes used as well, particularly when the target state is not pure but mixed.
However, we will only consider scenarios where the target state is pure and therefore Eq.~(\ref{eq:3-5_fidelity}) will be sufficient.

Let's consider the case where the real state is equal to the target state, $\rho=|\psi\rangle\langle\psi|$.
The fidelity is then
\begin{equation}
    F(\rho,|\psi\rangle) = \bra{\psi}\rho\ket{\psi} = \underbrace{\langle\psi|\psi\rangle}_{=1} \underbrace{\langle\psi|\psi\rangle}_{=1} = 1.
\end{equation}
On the other hand, when the real state $|\phi\rangle$ is pure but orthogonal to the target state,
\begin{equation}
    F(\rho,|\psi\rangle) = \langle\psi|\rho|\psi\rangle = \underbrace{\langle\psi|\phi\rangle}_{=0} \underbrace{\langle\phi|\psi\rangle}_{=0} = 0.
\end{equation}
Orthogonal states are as far from each other as can be and this is signified by a vanishing fidelity.
In fact, the two cases considered above are the two extreme ones, and for general mixed state,
\begin{equation}
    0 \leq F(\rho,|\psi\rangle) \leq 1.
\end{equation}

Let's continue with another example.
Consider the case where we would like to prepare the state $|0\rangle$, but something goes very wrong with our hardware and the real state that we obtain after the preparation procedure is the maximally mixed state $\rho = (|0\rangle\langle0| + |1\rangle\langle1|)/2$.
The fidelity can be computed as follows,
\begin{align}
    F(\rho,|0\rangle) & = \langle 0 | \left[\frac{1}{2}(|0\rangle\langle0| + |1\rangle\langle1|)\right] | 0 \rangle \nonumber\\
    & = \frac{1}{2} \underbrace{\langle0|0\rangle}_{=1} \underbrace{\langle0|0\rangle}_{=1} + \frac{1}{2} \underbrace{\langle0|1\rangle}_{=0} \underbrace{\langle1|0\rangle}_{=0} \nonumber\\
    & = \frac{1}{2}.
\end{align}
It is interesting to note that the fidelity does not vanish.
We said that the maximally mixed state is where we have maximum uncertainty whether it is \ket{0} or \ket{1}, yet the fidelity with a pure state only drops to a half.

We can extend this example to two qubits and consider the fidelity between the pure target state $|00\rangle$ and the maximally mixed state $\rho = (|00\rangle\langle00|+|01\rangle\langle01|+|10\rangle\langle10|+|11\rangle\langle11|)/4=I/4$,
\begin{equation}
    F(\rho, |00\rangle) = \langle00| \rho | 00\rangle = \frac{1}{4} \langle 00| I |00\rangle = \frac{1}{4}.
\end{equation}
This calculation can be quickly extended to $N$ qubits.
The fidelity between the state $|0\rangle^{\otimes N}$ and a maximally mixed state of $N$ qubits, given by $\rho=I/2^N$.
The fidelity is
\begin{equation}
    F(\rho,|0\rangle^{\otimes N}) = \frac{1}{2^N}.
\end{equation}
We see that the fidelity scales with the number of qubits but never quite reaches zero.

As our last example, we will consider the bit-flip channel of Eq.~(\ref{eq:3-3_bit_flip}).
Let's again consider the initial state to be $|0\rangle$ and compute how the fidelity changes depending on the probability of error $p$.
We have
\begin{align}
    F(\rho,|\psi\rangle) & = \langle0| \rho | 0\rangle \\
    & = \langle0| \left[ (1-p)|0\rangle\langle0| + p|1\rangle\langle1| \right] |0\rangle \nonumber\\
    & = (1-p) \langle0|0\rangle \langle0|0\rangle + p \langle0|1\rangle \langle1|0\rangle \nonumber\\
    & = 1 - p.
\end{align}
The fidelity does depend on the probability of error, as we anticipated.
When $p=0$, meaning there is no bit-flip noise affecting the input state, the fidelity of the output state is 1.
On the other hand, if the bit-flip channel is maximally noisy, $p=1/2$, the output state is the maximally mixed state.
This is also evidenced by the fidelity dropping to $F(\rho,|0\rangle)=1/2$.

We mentioned that the fidelity is an important tool in quantifying the quality of a quantum state with respect to the ideal noiseless case.
It can be interpreted as the probability that real state is the target state.
Many protocols in quantum communication as well as quantum computation do not work when the real state is too noisy.
Often this condition is expressed in the form of a \textbf{\emph{threshold fidelity}}\index{threshold fidelity} $F_{crit}$.
If the fidelity of the real state is above this critical value, $F > F_{crit}$, then the protocol can be successfully executed.
On the other hand, if $F < F_{crit}$, the protocol cannot be used.
An example of such a protocol is entanglement purification, where the fidelity of noisy entangled pairs can be boosted provided the initial pairs are not too noisy.
Another example is fault-tolerant quantum computation, where scalable universal quantum computation can be achieved even with noisy components, provided the noise is below a certain threshold.

\newpage
\begin{exercises}

\exer{
\label{ex:dirac-notation-measurement}
\emph{Single-qubit measurements}.
Consider a general pure state $|\psi\rangle=\alpha|+\rangle+\beta|-\rangle$. Calculate the probabilities $\text{Pr}\{+1\}$ and $\text{Pr}\{-1\}$ in the following cases:
\subexer{
When the state is measured in the $X$ basis.
}
\subexer{
When the state is measured in the $Y$ basis.
}
\subexer{
When the state is measured in the $Z$ basis.
}
}

\exer{
\emph{Post-measurement state.} We have seen in Section~\ref{sec:3-2_outer_product} that a projector transforms the initial state into a state that resembles the expected state after the measurement. In order to obtain the correct post-measurement state $|\psi'\rangle$, we have to renormalize by the square root of the probability of the measurement outcome,
\begin{equation*}
    |\psi'\rangle = \frac{\Pi^B_i|\psi\rangle}{\sqrt{\langle\psi|\Pi^B_i|\psi\rangle}},
\end{equation*}
where index $i\in\{+1,-1\}$ denotes the measurement outcome. Obtain the post-measurement states in the following scenarios:
\subexer{
Initial state is $|\psi\rangle=\alpha|0\rangle + \beta|1\rangle$, it is measured in the $Z$ basis, the outcome is +1.
}
\subexer{
Initial state is $|\psi\rangle=|0\rangle$, it is measured in the $Y$ basis, the outcome is -1.
}
\subexer{
Initial state is $|\psi\rangle=|+\rangle$, the state is measured in the $X$ basis, can we obtain the -1 outcome?
}
}

\exer{
\emph{Measurement in rotated basis.} We have described projector operators in terms of outer products. Projectors can also be represented in the following way,
\begin{equation}
    \Pi^B_{\pm} = \frac{1}{2} (I \pm B),
    \label{eq:projector_new_form}
\end{equation}
where $B$ is the measurement basis.
Eq.~(\ref{eq:projector_new_form}) allows us to easily write down projectors onto rotated basis.
\subexer{
Verify Eq.~(\ref{eq:projector_new_form}) is true for the three Pauli operators.
}
\subexer{
Write down the projectors corresponding to $\pm 1$ outcomes of a measurement in $B = (Z+X)/\sqrt{2}$ basis.
}
\subexer{
Write down the projectors corresponding to $\pm 1$ outcomes of a measurement in $B = r_x X + r_y Y + r_z Z$ basis, where $(r_x,r_y,r_z)$ is a unit vector.
}
}

\exer{
\emph{Mixed states}. We have a quantum machine that prepares a particular quantum state. By using our machine many times, we have determined that 60\% of the time, the machine prepares the state $|0\rangle$, 20\% of the time, it prepares the state $|1\rangle$, and 20\% of the time, it prepares the state $|+\rangle$.
\subexer{
Write down the state the machine prepares using Dirac notation.
}
\subexer{
What is the matrix representation of this state?
}
}

\exer{
\emph{Maximally mixed state}. We have seen that, considered in the Z basis, the density matrix for a completely mixed state of a single qubit has $1/2$ in each of the two entries on the diagonal,
\begin{equation*}
    \rho=\left(\begin{array}{cc}
    1/2 & 0 \\
    0 & 1/2
    \end{array}\right).
\end{equation*}
\subexer{
Prove that when measuring $\rho$ in the $X$ basis, the probability of finding each outcome is also $1/2$.
}
\subexer{
Prove that a completely mixed state of $\ket{+}$ and $\ket{-}$ is equal to the fully mixed state of $\ket{0}$ and $\ket{1}$.
}
}

\exer{
\emph{Bit-flip channel}. We have introduced the bit-flip channel,
\begin{equation*}
    \rho = (1-p)|\psi\rangle\langle\psi| + p X|\psi\rangle\langle\psi|X,
\end{equation*}
where an initial pure state $|\psi\rangle$ is transformed into a final mixed state by a probabilistic application of the Pauli $X$ operator.
\subexer{
Check that the final state $\rho$ is properly normalized.
}
\subexer{
Show that the purity of the output state $\rho$ is given by
\begin{equation}
    \gamma = (1-p)^2 + p^2 + 2p(1-p) \langle X\rangle^2.
\end{equation}
(\emph{Hint:} you might find the following property of the trace useful,
\begin{equation*}
    \text{Tr}\{|\psi\rangle\langle\psi|U\} = \text{Tr}\{U|\psi\rangle\langle\psi|\} = \langle\psi|U|\psi\rangle.)
\end{equation*}
}
\subexer{
If the input state is $|\psi\rangle=|0\rangle$, what is the purity of the output state $\rho$?
}\subexer{
If the input state is $|\psi\rangle=|+\rangle$, what is the purity of the output state $\rho$? Does this answer make sense?
}
}

\exer{
\emph{Phase-flip channel}. Another important noise model is one where the Pauli $Z$ operator is applied to the input state $|\psi\rangle$ with probability $p$.
\subexer{
Write down the output state in terms of the input state.
}
\subexer{
What input state remains unaffected by the phase-flip channel?
}
\subexer{
Compute the purity of the output state $\rho$ for a general input state $|\psi\rangle$.
}
}

\end{exercises}

\chapter{Entanglement}
\label{sec:4_entanglement}

In this chapter, we will finally learn about \textbf{\emph{entanglement}}: a special and unique feature of quantum mechanics.
We will begin by demonstrating how strange and counter-intuitive entanglement can be.
Then, we will move on to the definition of entangled states.
Following that, we will discuss a particular class of entangled states called ``Bell states'', which play an important role in quantum communication.
Finally, we will finish by framing entanglement as a resource for computational and communication tasks.

\section{CHSH Game}
\label{sec:chsh-game}
\index{CHSH game}

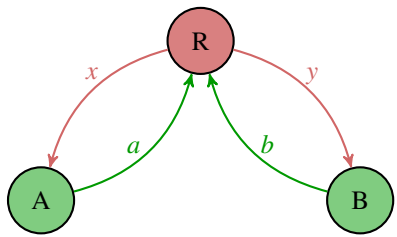
\begin{figure}[t]
    \centering
    \begin{tikzpicture}[->, >=stealth', auto, semithick, node distance=3cm]
    \tikzstyle{every state}=[draw=black,thick,text=black,scale=1]

    \node[state,fill=myred!50]    (R)                     {R};
    \node[state,fill=mygreen!50]   (A)[below left of=R]    {A};
    \node[state,fill=mygreen!50]   (B)[below right of=R]   {B};

    \path
    (R) edge[bend right,above,myred!60,thick]	node[yshift=1pt]{$x$}	(A)
    (R) edge[bend left,above,myred!60,thick]	node[yshift=-1pt,xshift=1pt]{$y$}	(B)
    (A) edge[bend right,left,mygreen,thick]	node[yshift=1pt]{$a$}	(R)
    (B) edge[bend left,right,mygreen,thick]	node[yshift=2pt]{$b$}	(R);

    \end{tikzpicture}
    
    \caption[CHSH game.]{CHSH game view. The players A and B attempt to collaborate to beat the referee R.}
    \label{fig:chsh-game}
\end{figure}

We begin with a game that will demonstrate just how wonderful and strange entangled states are.
The game is related to a proposal of four scientists named Clauser, Horn, Shimony, and Holt, though it is normally referred to as the \textbf{\emph{CHSH game}}\index{CHSH game}.
It consists of two players, labelled $A$ and $B$, and a referee denoted by $R$, as shown in Fig.~\ref{fig:chsh-game}.
The rules of the game are the following:
\begin{itemize}
    \item Referee $R$ generates two random bits, $x,y\in\{0,1\}$. Bit $x$ is sent to player $A$, and bit $y$ is sent to player $B$.
    \item Upon receiving the referee's bits, players $A$ and $B$ reply with their own bits. Player $A$ sends back a bit $a$, and player $B$ sends back bit $b$.
    \item Referee $R$ checks whether the players' bits satisfy the winning condition,
    \begin{equation}
        x\cdot y = a \oplus b,
    \end{equation}
    where $\oplus$ represents binary addition. This concludes a single round of the game.
    \item The game continues to see what percentage of the rounds the players can win.
    \item IMPORTANT: The players are not allowed to communicate once the game has started.
\end{itemize}

The players cannot communicate in order to make the game interesting, otherwise they would win every single round.
This does not mean that the players are not allowed to communicate at all.
Before the game starts, they can agree on a common strategy in order to maximize their winning chances.

Before discussing what an optimal strategy would look like, let's have a look at a few rounds of the CHSH game to become more comfortable with the rules.
\begin{table}[h]
    \setcellgapes{5pt}
    \renewcommand\theadfont{}
    \makegapedcells
    \centering
    \begin{tabular}{cccccccc}
        \hline
        & $x$ & $y$ & $a$ & $b$ & $x\cdot y$ & $a\oplus b$ \\
        \hline
        \textbf{Round 1} & 0 & 1 & 0 & 1 & 0 & 1 & \textcolor{myred}{Loss} \\
        \textbf{Round 2} & 0 & 1 & 1 & 1 & 0 & 0 & \textcolor{mygreen}{Win} \\
        \textbf{Round 3} & 1 & 1 & 1 & 0 & 1 & 1 & \textcolor{mygreen}{Win} \\
        \hline
    \end{tabular}
    \caption[CHSH game example.]{Three example rounds of the CHSH game.}
    \label{tab:4-1_chsh_rounds}
\end{table}
Table~\ref{tab:4-1_chsh_rounds} shows three example rounds of the game.
In Round 1, the referee $R$ generates the input bits with values $x=0$ and $y=1$, which are sent to the players.
The players generate their answers according to a pre-agreed strategy, and reply with $a=0$ and $b=1$.
The product of the input bits is $x\cdot y=0$, while the binary sum of the output bits is $a\oplus b=1$.
The referee determines that winning condition is not satisfied, and the players lose this round.
In Round 2, the product of input qubits is $x\cdot y=0$, and so is the binary sum of the outputs, $a\oplus b=0$. The players win this round.
Similarly for Round 3 where $x\cdot y=a\oplus b=1$.

So what is the optimal strategy that players $A$ and $B$ should follow in order to maximize their chances of winning?
We will now address this important question.
Let's begin by looking at the possible input bit pairs that the referee $R$ can generate.
There are four possible pairs, as shown in Tab.~\ref{tab:4-2_chsh_strategy}, and they are $(0,0)$, $(0,1)$, $(1,0)$, and $(1,1)$.
\begin{table}[h]
    \setcellgapes{5pt}
    \renewcommand\theadfont{}
    \makegapedcells
    \centering
    \begin{tabular}{cccccc}
        \hline
        $x$ & $y$ & $x\cdot y$ & $a=b$ & $a\oplus b$ & \\
        \hline
        0 & 0 & 0 & 0 & 0 & \textcolor{newgreen}{Win} \\
        0 & 1 & 0 & 0 & 0 & \textcolor{newgreen}{Win} \\
        1 & 0 & 0 & 0 & 0 & \textcolor{newgreen}{Win} \\
        1 & 1 & 1 & 0 & 0 & \textcolor{darkred}{Loss} \\
        \hline
    \end{tabular}
    \caption[CHSH game classical strategy.]{An example strategy for the CHSH game. Players $A$ and $B$ always reply with 0, regardless of the values of the input bits sent by the referee $R$.}
    \label{tab:4-2_chsh_strategy}
\end{table}
In the first three cases, the product of the input bits is $x\cdot y=0$. Only in the last case, when $x=y=1$, is the product equal to 1.
Realizing this, the players can agree on a strategy where they always output $a=b=0$, regardless of the value of the input bits sent by the referee.
This strategy results in the output bit binary sum $a\oplus b=0$, meaning the players win the CHSH game in three out of the four possible cases.
Since the input bits are generated uniformly at random, the probability of of a particular input pair $(x,y)$ is 1/4.
Therefore, the players have a 75\% chance of winning every round.
Not bad for such a simple strategy.

\begin{figure}[t]
    \centering
    \begin{tikzpicture}[auto, semithick, node distance=3cm]
    \tikzstyle{every state}=[draw=black,thick,text=black,scale=1]

    \node[state,fill=red!50]    (R)                     {R};
    \node[state,fill=newgreen!50]   (A)[below left of=R]    {A};
    \node[state,fill=newgreen!50]   (B)[below right of=R]   {B};

    \path
    (R) edge[bend right,above,red!60,thick,->,>=stealth']	node[yshift=1pt]{$x$}	(A)
    (R) edge[bend left,above,red!60,thick,->,>=stealth']	node[yshift=-1pt,xshift=1pt]{$y$}	(B)
    (A) edge[bend right,left,newgreen,thick,->,>=stealth']	node[yshift=1pt]{$a$}	(R)
    (B) edge[bend left,right,newgreen,thick,->,>=stealth']	node[yshift=2pt]{$b$}	(R);

    \coordinate (qubitA) at ($(A)+(0,-1)$);
    \coordinate (qubitB) at ($(B)+(0,-1)$);

    \draw[decorate,decoration={coil,aspect=0},thick] (qubitA)  -- (qubitB);
    \filldraw[black!30,draw=black,thick] (qubitA) circle (5pt);
    \filldraw[black!30,draw=black,thick] (qubitB) circle (5pt);

    \node[] at (-3, -3.15) {qubit A};
    \node[] at (3, -3.15) {qubit B};
    \node[] at (0, -3.5) {$|\Phi^+\rangle$};

    \end{tikzpicture}
    
    \caption[Optimal quantum CHSH game strategy.]{Optimal quantum strategy for the CHSH game requires the players to pre-share an entangled state of two qubits.}
    \label{fig:chsh-game_entanglement}
\end{figure}
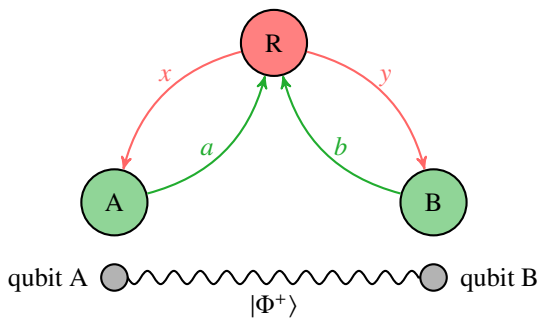

But can the players do better than that?
The answer is no, if the players are restricted to using only classical strategies.
However they can do better if they use pre-shared \textbf{\emph{entangled states}}\index{entangled state}.
A particular quantum strategy is depicted in Fig.~(\ref{fig:chsh-game_entanglement}).
Players $A$ and $B$ now share an entangled state,
\begin{equation}
    |\Phi^+\rangle = \frac{1}{\sqrt{2}} (|00\rangle + |11\rangle),
    \label{eq:4_1-phiPlus}
\end{equation}
where the first qubit corresponds to qubit $A$, while the second qubit to corresponds to qubit $B$.
\footnote{We will discuss what it means for a quantum state to be entangled in the following section.
For now, we will just accept that Eq.~(\ref{eq:4_1-phiPlus}) is entangled.
In fact, it is good to get used to this state because we will keep encountering it again and again.}
The capital Greek letter $\Phi$ is pronounced ``Phi''.
If the referee's input bit is $x=0$, player $A$ measures qubit $A$ in the $Z$ basis.
If $x=1$, player $A$ measures in the $X$ basis.
For +1 measurement outcome, player $A$ replies with $a=0$.
If the measurement outcome is -1, player $A$ replies with $a=1$.
Similar procedure applies to player $B$.
Only difference is that player $B$ measures in a \textbf{\emph{rotated basis}}\index{rotated basis}.
If the input bit is $y=0$, player $B$ measures in the $(Z+X)/\sqrt{2}$ basis.
For $y=1$, player $B$ measures in the $(Z-X)/\sqrt{2}$ basis.

The above quantum strategy is a lot more complicated that the optimal classical one.
Actions of the players now depend on the inputs received from the referee.
They also need to pre-share one entangled for every round they play.
All of this extra work is well worth it however.
By following this quantum strategy, the players have probability of winning a round of 85\%.
This is an increase of 10\% compared to the optimal classical strategy.
We do not yet have the necessary tools to analyze the quantum strategy quantitatively in order to derive the winning probability.
We are about to remedy that in the remainder of this chapter.

\section{Entangled states}
\label{sec:4_2-entagnled_states}

Having seen how entangled states can be useful in the CHSH game, it is time to dicuss them in more detail.
To keep things simple, we will consider a system of two qubits.
We will use the term \textbf{\emph{local state}}\index{local state} to refer to the state of either of two qubits alone, and the term \textbf{\emph{global state}}\index{global state} to refer to the two qubits together.~\footnote{These terms can easily be extended to discuss more than one qubit in each location.}

We saw in Sec.~\ref{sec:multi-qubit} that in order to describe states of many qubits, we have to use the tensor product. For the case of the two qubits $A$ and $B$, we have the local state $\ket{\psi}_A$ of the qubit $A$, and a local state $\ket{\psi}_B$ of qubit $B$.
For concreteness, let's say that they are $\ket{0}$ and $\ket{0}$.
In order to write the global state of both qubits, we form the tensor product and we call this state $\ket{\psi}_{AB}$, which in this case is simply $\ket{00}$.
We can also consider a different state where $A$ is still in $\ket{0}$ but $B$ is now in the $\ket{+}$, a superposition of zero and one.
Again, the global state is straightforward: form the tensor product of $\ket{0}$ and $\ket{+}$,
\begin{equation}
\ket{\psi}_{AB} = \ket{0}\ket{+} = \ket{0}\left(\frac{\ket{0} + \ket{1}}{\sqrt{2}}\right) = \frac{\ket{00} + \ket{01}}{\sqrt2}.
\end{equation}

The above two examples both started with local states, which were then used to formulate the global state.
We can also ask the reverse question: given the global state, how do we write the local states of the qubits?
We can consider the global state of two qubits that we have encountered in the CHSH game in the previous section,
\begin{equation}
    \ket{\Phi^+}_{AB} = \frac{1}{\sqrt2}(\ket{00} + \ket{11})
    \label{eq:entangled_state}
\end{equation}
Identifying the local states is not that straightforward in this case.
By looking at Eq.~(\ref{eq:entangled_state}), we can see that the local state of either qubit is not quite $|0\rangle$, and it is also not quite $|1\rangle$.

At this point, it is better to get a bit more rigorous.
Let's write the local states as general pure states,
\begin{equation}
    |\psi\rangle_A = a_0|0\rangle + a_1|1\rangle, \quad |\psi\rangle_B = b_0|0\rangle + b_1|1\rangle.
\end{equation}
Taking the tensor product of $\ket{\psi}_A$ with $\ket{\psi}_B$, we arrive at the following two-qubit state, 
\begin{align}
    |\psi\rangle_{A} \otimes|\psi\rangle_{B} & = \left(a_{0}|0\rangle+a_{1}|1\rangle\right) \otimes\left(b_{0}|0\rangle+b_{1}|1\rangle\right) \nonumber\\
    & = a_0 b_0\ket{00} + a_0 b_1\ket{01} + a_1 b_0\ket{10} + a_1 b_1\ket{11}.
    \label{eq:two_qubit_separable}
\end{align}
Our goal is to find suitable probability amplitudes $a_i$ and $b_j$ such that $|\psi\rangle_{A} \otimes|\psi\rangle_{B} = |\Phi^+\rangle_{AB}$.
By comparing Eq.~(\ref{eq:entangled_state}) with Eq.~(\ref{eq:two_qubit_separable}), we see that the probablity amplitudes must satisfy two conditions,
\begin{equation}
    a_0 b_0 = a_1 b_1 = \frac{1}{\sqrt{2}}, \quad \text{ and }\quad a_0 b_1 = a_1 b_0 = 0.
\end{equation}
Let's look at the second condition first.
Either $a_0$ or $b_1$ is zero.
But setting either of them to 0, then we see that $a_0 b_0$ or $a_1 b_1$ will also be 0, meaning the first condition cannot be satisfied.
This is a remarkable discovery.
It says that \emph{not all global states can be written as a tensor product of local states}.

This leads us to the realization that there are two major classes of states.
There are \textbf{\emph{product states}}\index{product state}.
A product state is one that \textbf{\emph{can}} be written as a tensor product of local states.
Given two local states $\ket{\psi}_A$ and $\ket{\psi}_B$, we can easily find the global state by forming the tensor product.
Since we can write down the state vectors for both the global state and the local states, we say that we perfect knowledge of the local states.

The other class of quantum states are \textbf{\emph{entangled states}}\index{entangled state}.
An entangled state is a state whose global state \textbf{\emph{cannot}} be written as the tensor product of local states.
In the previous example, we have demonstrated this for an equal superposition of $\ket{00}$ and $\ket{11}$, this is in fact the case.
This is very interesting because it implies that we have perfect knowledge of the global state, but we have imperfect knowledge of local states.

\section{Bell states}
\label{sec:4_3-bell_states}

After defining what entangled states are, it is time to take a look at a very useful class of entangled states, called the \textbf{\emph{Bell states}}\index{Bell states}.
There are a total of four Bell states, 
\begin{equation}
    \begin{aligned}
        |\Phi^+\rangle & = \frac{1}{\sqrt{2}} \left( |00\rangle + |11\rangle \right), & 
        |\Phi^-\rangle = \frac{1}{\sqrt2} \left( |00\rangle - |11\rangle \right), \\
        |\Psi^+\rangle & = \frac{1}{\sqrt{2}} \left( |01\rangle + |10\rangle \right), & 
        |\Psi^-\rangle = \frac{1}{\sqrt2} \left(|01\rangle - |10\rangle \right).
    \end{aligned}
\end{equation}
We have already encountered the equal superposition of $|\Phi^+\rangle$.
Its counterpart is the $\ket{\Phi^-}$ (``phi minus'') state, where the $\ket{00}$ and $\ket{11}$ have opposite phase, as shown represented by the minus sign.
There are two more states, $\ket{\Psi^+}$ (``psi plus'') and $\ket{\Psi^-}$ (``psi minus''), which are superpositions of \ket{01} and \ket{10}, with \ket{\Psi^-} having opposite phase between the two terms. 
We will encounter these states countless number of times in the context of quantum communication.~\footnote{In some texts and papers, more commonly but not exclusively older ones, you will see these states written using the lower-case Greek letters $\phi$ (phi) and $\psi$ (psi).  In this book, we will use the capital Greek letters when referring to the Bell pairs and the lower-case letters when referring to specific variables.} 

The Bell states form an orthogonal basis for the space of two qubits.
This means that any state vector of two qubits can be written in terms of the four Bell states.
We begin with rewriting the computational basis states of two qubits $\{\ket{00},\ket{01},\ket{10},\ket{11}\}$ in terms of the Bell states.
It is not too difficult to verify that the following equalities are indeed true,
\begin{equation}
    \begin{aligned}
    |00\rangle &=\frac{1}{\sqrt{2}}\left(\left|\Phi^{+}\right\rangle+\left|\Phi^{-}\right\rangle\right), & 
    \ket{01} = \frac{1}{\sqrt2}(\ket{\Psi^{+}} + \ket{\Psi^{-}}), \\
    |10\rangle &=\frac{1}{\sqrt{2}}\left(\left|\Psi^{+}\right\rangle-\left|\Psi^{-}\right\rangle\right), & 
    \ket{11} = \frac{1}{\sqrt2}(\ket{\Phi^{+}} - \ket{\Phi^{-}}).
    \label{eq:comp_to_bell_basis}
    \end{aligned}
\end{equation}
An arbitrary two-qubit pure state can be written in the computational basis as follows,
\begin{equation}
    |\psi\rangle=\alpha|00\rangle+\beta|01\rangle+\gamma|10\rangle+\delta|11\rangle,
\end{equation}
where the probability amplitude satisfy the usual normalization condition, $|\alpha|^2+|\beta|^2+|\gamma|^2+|\delta|^2=1$.
Using Eq.~(\ref{eq:comp_to_bell_basis}), this pure state can be readily written in the Bell basis,
\begin{equation}
    |\psi\rangle=\frac{\alpha+\delta}{\sqrt{2}}\left|\Phi^{+}\right\rangle+\frac{\alpha-\delta}{\sqrt{2}}\left|\Phi^{-}\right\rangle+\frac{\beta+\gamma}{\sqrt{2}}\left|\Psi^{+}\right\rangle+\frac{\beta-\gamma}{\sqrt{2}}\left|\Psi^{-}\right\rangle.
    \label{eq:psi_bell_basis}
\end{equation}
Note that since we have changed the basis the probability amplitudes have transformed accordingly.

We have so far been dealing with single-qubit measurements.
Such measurements only had two possible outcomes.
Measuring in the Pauli $Z$ basis would project the initial state onto either $\ket{0}$ or $\ket{1}$.
If the basis of measurement is the Pauli $X$, then the post-measurement state will be either $\ket{+}$ or $\ket{-}$.
However, there is no reason to stop at single-qubit measurements.
We can extend our previous discussion to two-qubit measurements.
Not only that, we can consider two-qubit measurement in an entangled basis.
An example of this is \textbf{\emph{measurement in the Bell basis}}\index{Bell-basis measurement}.
Measuring a general two-qubit state $\ket{\psi}$ in the Bell basis has now four possible outcomes.
The post-measurement state can be any of the four Bell states $\{\ket{\Phi^+}, \ket{\Phi^-}, \ket{\Psi^+}, \ket{\Psi^-}\}$.
Using Eq.~(\ref{eq:psi_bell_basis}), we can easily determine the probabilities of these outcomes,
\begin{equation}
    \begin{aligned}
    &\operatorname{Prob}\left\{|\Phi^{+}\rangle\right\}=\frac{|\alpha+\delta|^{2}}{2} \quad \operatorname{Prob}\left\{|\Phi^{-}\rangle\right\}=\frac{|\alpha-\delta|^{2}}{2} \\
    &\operatorname{Prob}\left\{|\Psi^{+}\rangle\right\}=\frac{|\beta+\gamma|^{2}}{2} \quad \operatorname{Prob}\left\{|\Psi^{-}\rangle\right\}=\frac{|\beta-\gamma|^{2}}{2}
    \end{aligned}
\end{equation}
This will be again very important because measurements in the Bell basis are crucial in many protocols in quantum communication, especially teleportation and entanglement swapping, which we will look at quite closely later in this book.

We mentioned in Sec.~\ref{sec:4_2-entagnled_states} that entangled states have the curious property that the global state is known perfectly, while it is not possible to express the state of the local qubits using pure states.
This suggests that we only have imperfect knowledge of the local states.
We will make this notion more concrete.

Consider the two-qubit state to be the $|\Phi^+\rangle$ Bell state.
Measuring this state in the Bell basis will yield the $|\Phi^+\rangle$ outcome with unit probability.
This is consistent with our interpretation of having full knowledge of the global state.
Let's see what happens if we perform a single-qubit measurement on the state $|\Phi^+\rangle$.
For example, we measure qubit $A$ in the Pauli $Z$ basis.
Since this is a single-qubit measurement, we can only two possible outcomes.
The projectors associated with these two outcomes are
\begin{equation}
    \Pi^A_{Z,+} = |0\rangle_A\langle0| \otimes I_B, \quad \text{ and } \Pi^A_{Z,-} = |1\rangle_A\langle1| \otimes I_B.
    \label{eq:local_projectors}
\end{equation}
Note the slight but necessary change of notation in Eq.~(\ref{eq:local_projectors}).
The superscript $A$ labels which local qubit is being measured, while the subscript now labels both the basis and the measurement outcome.
The probability of obtaining the +1 outcome is
\begin{align}
    \operatorname{Pr}\{+1\} & = \operatorname{Tr}\left\{\Pi^A_{Z,+} |\Phi^+\rangle_{AB}\langle\Phi^+|\right\} \\
    & = \langle\Phi^+|_{AB} \left( |0\rangle_A\langle0| \otimes I_B \right) |\Phi^+\rangle_{AB} \nonumber\\
    & = \frac{1}{2} \left( \langle00|_{AB} + \langle11|_{AB} \right) \left( |0\rangle_A\langle0| \otimes I_B \right) \left( |00\rangle_{AB} + |11\rangle_{AB} \right) \nonumber\\
    & = \frac{1}{2} \left( \langle0|0\rangle_A^2 \langle0|0\rangle_B + \langle0|0\rangle_A \langle0|1\rangle_A \langle0|1\rangle_B \right. \nonumber\\
    & + \left. \langle1|0\rangle_A \langle0|0\rangle_A \langle1|0\rangle_B + \langle1|0\rangle_A \langle0|1\rangle_A \langle1|1\rangle_B \right) \nonumber\\
    & = \frac{1}{2}. \nonumber
\end{align}
Similar calculation can be repeated for the case of the -1 measurement outcome,
\begin{equation}
    \operatorname{Pr}\{-1\} = \operatorname{Tr}\left\{\Pi^A_{Z,-} |\Phi^+\rangle_{AB}\langle\Phi^+|\right\} = \frac{1}{2}.
\end{equation}

\label{page:plus-is-pure}
At this point, you may be a little puzzled; didn't we see this same fifty-fifty behavior with a single qubit in the $\ket{+}$ state?  What is different with entanglement?  When measuring in the Z basis, it is true that we always see the fifty-fifty statistics.  However, if we measure a $\ket{+}$ in the X basis, we will always find that the state is in the $+1$ state.  We can also choose to apply a Hadamard gate to the $\ket{+}$ state and return it to a known $\ket{0}$ state.  Thus, although the qubit is in superposition, we have complete information about that superposition.  In contrast, when we hold only one qubit of an entangled pair and attempt to measure it, our results will differ.


If we change the measurement basis to a Pauli $Y$ basis, again we find that the probability of the $+1$ outcome is the same as the probability of the $-1$ outcome. It is the same for the $X$ basis, and in fact in any basis that you measure qubit $A$, you will get the same result: both outcomes are fifty-fifty. In fact, the local measurement results are uniformly random in any basis.
Remember that the global state of the system is pure.
We know exactly what it is, yet no matter in what basis we measure the states locally we are getting 50 percent one outcome and 50 percent the other outcome.
We can say that we have no knowledge of the local states of qubit $A$ and qubit $B$.
This goes back to our previous discussion about the difference between entangled states and product states. 
Since this state is entangled, the correct description of the local qubits must be given in terms of density matrices, and we can write that the state of qubit $A$ is a maximally mixed state.
The same holds for qubit $B$.
The local states of the two qubits are therefore
\begin{equation}
    \rho_A=\frac{1}{2}(|0\rangle\langle 0|+| 1\rangle\langle 1|), \quad
    \rho_B=\frac{1}{2}(|0\rangle\langle 0|+| 1\rangle\langle 1|).
\end{equation}

Just to stress how strange this is that we have a full knowledge of the global state yet we have zero knowledge of the local states, let's look at this again: qubit $A$ is a maximally mixed state. We have no knowledge about its state. The same is true for qubit $B$ considered in isolation. Yet, somehow, when we look at the state globally, we have perfect knowledge of the entire state. This distinction is addressed more in the chapter exercises.

\section{Spontaneous parametric down-conversion}
\label{sec:4-4_spdc}

In this section, we will learn about one possible way of producing the entangled states that we have been discussing so far.
The physical systems that will encode the qubits are photons, and the process generates entangled photon pairs is called \textbf{\emph{spontaneous parametric down-conversion}}\index{spontaneous parametric down-conversion}, or SPDC for short. There are many physical processes that can generate entangled pairs of photons, but this one is common in laboratories and will serve as our example.

The basic idea behind SPDC is illustrated in Fig.~\ref{fig:spdc}.
A high-energy \textbf{\emph{pump photon}}\index{pump photon}, pictured in green, is incident on a nonlinear crystal.
A popular choice of material for the crystal is \textbf{\emph{beta barium borate}}\index{BBO}, or BBO for short.
The pump photon interacts with atoms inside the nonlinear crystal and gets transformed into two lower-energy photons.
Often these photons are referred to as the \textbf{\emph{signal}}\index{signal photon} and the \textbf{\emph{idler}}\index{idler photon}.

\begin{figure}[t]
    \centering
    \begin{tikzpicture}[scale=2, every node/.style={scale=1.3},wave/.style={semithick,color=#1,smooth}]
        
        \begin{scope}
            \draw[wave=black, fill=black!10, variable=\x, samples at={-0.3,-0.29,...,0.3}] plot (\x-0.6, {0.2*e^(-\x*\x/0.005)+0.5});
            \draw[thick, -latex, line cap=round] (-1.1,0.5) -- (-0.1, 0.5);
            \node[black] at (-0.6,0.8) {\scriptsize pump};
        \end{scope}

        \begin{scope}[rotate around={30:(0.0,0.5)}]
            \draw[wave=black, fill=black!10, variable=\x, samples at={-0.3,-0.29,...,0.3}] plot (\x+0.6, {0.2*e^(-\x*\x/0.005)+0.5});
            \draw[thick, -latex] (0.0,0.5) -- (1.1,0.5);
        \end{scope}

        \begin{scope}[rotate around={-30:(0.0,0.5)}]
            \draw[wave=black, fill=black!10, variable=\x, samples at={-0.3,-0.29,...,0.3}] plot (\x+0.6, {0.2*e^(-\x*\x/0.005)+0.5});
            \draw[thick, -latex] (0.0,0.5) -- (1.1,0.5);
        \end{scope}
        
        \draw[thick, fill=gray!20] (-0.1,0) rectangle (0.1,1);

        \node[] at (0,-0.2) {\scriptsize BBO};
        \node[] at (1.2,1.05) {\scriptsize signal};
        \node[] at (1.2,-0.05) {\scriptsize idler};
    \end{tikzpicture}
    \caption[Spontaneous parametric down conversion (SPDC).]{Symmetric parametric down conversion (SPDC) turns one high-energy pump photon into two lower-energy photons, with some very low probability.}
    \label{fig:spdc}
\end{figure}

``Spontaneous'' means that the process is not stimulated.
In a spontaneous process, no other input is needed, only the incoming photon of high energy that gets converted into two photons of lower energies.
(In contrast, a \emph{stimulated} process requires both the input photon and some other photon as well; a good example would be lasing.) “Parametric” means the whole process does not depend only on the intensity of the electric field but also on the phase of the field. In the context of the SPDC, it means there is a relationship between the input and output electric fields.

\begin{figure}[t]
    \centering
    \includegraphics[width=0.8\textwidth]{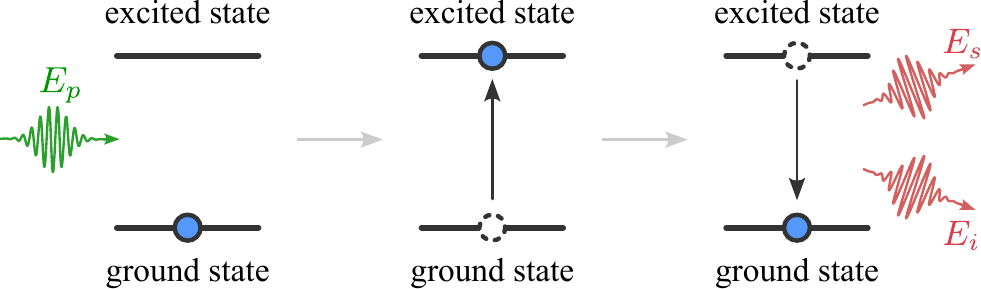}
    \caption[Pump photon exciting an atom.]{A representation of the pump photon of energy $E_p$ and the energy levels of the atom. Energy level diagrams such as these are an abstraction that allows us to describe absorption and emission of photons by an atom. The circle indicates that there is a single electron in the ground state.
    The atom absorbs the energy provided by the pump photon and transitions to the excited state. The atom decays spontaneously after some time, releasing energy in the form of two photons with energies $E_s$ and $E_i$. Due to conservation of energy, we have $E_p = E_s + E_i$.}
    \label{fig:spdc-energy-levels}
\end{figure}

We can zoom into the nonlinear crystal to see what happens at the level of individual atoms interacting with single photons of the pump, as shown in Fig.~\ref{fig:spdc-energy-levels}.
We can tune the energy of the pump photons to be resonant with a particular transition frequency of the atoms inside the BBO crystal.
Therefore, to a good approximation the pump laser will only affects those two energy levels.
The lower level is called the \textbf{\emph{ground state}}\index{ground state} and the upper level is call the \textbf{\emph{excited state}}\index{excited state}.
The atom in the BBO crystal initially starts in the ground state.
The pump photon carries the right amount of energy to excite the atom.
After a short time the atom de-excites and transitions back into the ground state.

The vast majority of the time, this de-excitation is coupled with the emission of a single, high-energy photon.
With some very low probability, however, the atom emits two photons with energy $E_s$ for the signal photon and energy $E_i$ for the idler photon.
Because energy is conserved in this transition, we must have that the energy of the pump photon be equal to the sum of the energies of the outgoing photons, $E_p = E_s + E_i$.
This is the basic process of spontaneous parametric down-conversion.

Now that we have covered the basic idea of the SPDC process it is time to discuss how we can use photons encode qubits.
Qubits have two distinguishable states.
They are represented by the abstract orthogonal kets, for example $\{|0\rangle,|1\rangle\}$ or $\{|+\rangle,|-\rangle\}$.
A general state of the qubit is then written as an arbitrary superposition of these two orthogonal states.
Therefore, we require the photons to be able to possess two distinguishable \textbf{\emph{degrees of freedom}}\index{degree of freedom} which can also form superposition.
A natural choice for the photons is their \textbf{\emph{polarization}}\index{polarization}.

\begin{figure}[t]
    \centering
    \begin{tikzpicture}
        \draw[semithick,dashed,black!50] (0,0) circle (1.5cm);
        \draw[thick,latex-latex,newgreen] (-1.5,0) -- (1.5,0);
        \draw[thick,latex-latex,darkred] (0,-1.5) -- (0,1.5);
        \draw[thick,latex-latex,blue!50] (-1.06,-1.06) -- (1.06,1.06);

        \draw[-latex] (0:0.75) arc (0:45:0.75);
        \node[] at (1.0, 0.35) {\small $45\degree$};

        \node[newgreen] at (2.5,0.2) {horizontal};
        \node[newgreen] at (2.5,-0.2) {polarization};

        \node[darkred] at (0, 2.2) {vertical};
        \node[darkred] at (0, 1.8) {polarization};

        \node[blue!50] at (1.9, 1.4) {diagonal};
        \node[blue!50] at (2.15, 1.05) {polarization};
    \end{tikzpicture}
    \caption[Linear polarization of light.]{Linear polarization of light. Horizontally polarized light can be superposed with vertically polarized one to form diagonal polarization.}
    \label{fig:polarization}
\end{figure}

Natural light is composed of many electromagnetic waves, each oscillating in a different direction.
It is this direction of oscillations that we call polarization\footnote{We will not cover polarization in depth as this would require delving into electromagnetism. We will address electromagnetic theory and Maxwell's equations in great detail in our next book}.
If the electromagnetic waves are oscillating in the horizontal plane we say the light is \textbf{\emph{horizontally polarized}}\index{horizontally polarized}, as shown in Fig.~\ref{fig:polarization}.
And if the oscillations are in the vertical direction, you guessed it, the light is \textbf{\emph{vertically polarized}}\index{vertically polarized}.
These two polarizations are just two possible ones out of infinitely many.
Another example that we will prove very useful in this Section is \textbf{\emph{diagonal polarization}}\index{diagonal polarization}.
It can be viewed as a superposition of horizontally and vertically polarized light.

\begin{figure}[t]
    \centering
    \begin{tikzpicture}[x={(0.866cm,-0.5cm)}, y={(0.866cm,0.5cm)}, z={(0cm,1cm)}, scale=0.8, >=stealth, axis/.style={thick,->}, wave/.style={thick,color=#1,smooth}, polaroid/.style={fill=black!30!white, opacity=0.7}]

    \begin{scope}
    \coordinate (O) at (0,0,0);

    \draw[thick] (-3.25,0,0) -- (O);

    \draw[wave=newgreen, fill=newgreen!60, fill opacity=0.2, variable=\x,samples at={-3.14,-3.13,...,0}] plot (\x,{sin(2*\x r)},0);

    \draw[wave=lightred, fill=lightred!60, fill opacity=0.2, variable=\x,samples at={-3.14,-3.13,...,0}]
        plot (\x,0,{sin(2*\x r)});

    \filldraw[polaroid] (0,-2,-1.5) -- (0,-2,1.5) -- (0,2,1.5) -- (0,2,-1.5) -- (0,-2,-1.5) node[below, sloped, near end]{polarizer};
    \draw[thick,latex-latex] (0,-1.75,-1) -- (0,-0.75,-1);

    \draw[wave=newgreen, fill=newgreen!60, fill opacity=0.2, variable=\x,samples at={0,0.01,...,3.14}] plot (\x,{sin(2*\x r)},0);

    \draw[axis] (O) -- (4,0,0) node [right] {};
    \draw[axis] (O) -- (0,2.5,0) node [right,newgreen] {$H$};
    \draw[axis] (O) -- (0,0,2) node [above,lightred] {$V$};
    \end{scope}

    \begin{scope}[xshift=7cm]
    \coordinate (O) at (0,0,0);
    \draw[axis] (O) -- (4,0,0) node [right] {};
    \draw[axis] (O) -- (0,2.5,0) node [right,newgreen] {$H$};
    \draw[axis] (O) -- (0,0,2) node [above,lightred] {$V$};

    \draw[thick] (-3.25,0,0) -- (O);

    \draw[wave=newgreen, fill=newgreen!60, fill opacity=0.2, variable=\x,samples at={-3.14,-3.13,...,0}] plot (\x,{sin(2*\x r)},0);

    \draw[wave=lightred, fill=lightred!60, fill opacity=0.2, variable=\x,samples at={-3.14,-3.13,...,0}]
        plot (\x,0,{sin(2*\x r)});

    \filldraw[polaroid] (0,-2,-1.5) -- (0,-2,1.5) -- (0,2,1.5) -- (0,2,-1.5) -- (0,-2,-1.5) node[below, sloped, near end]{polarizer};
    \draw[thick, <->] (0, -1.5,-1) -- (0, -1.5, 0);

    \draw[axis] (O) -- (0,2.5,0) node [right,newgreen] {$H$};
    \draw[axis] (O) -- (0,0,2) node [above,lightred] {$V$};

    \draw[wave=lightred, fill=lightred!60, fill opacity=0.2, variable=\x,samples at={0,0.01,...,3.14}] plot (\x,0,{sin(2*\x r)});

    \draw[axis] (O) -- (4,0,0) node [right] {};
    \end{scope}
    \end{tikzpicture}
    
    \caption[Polarizing filter.]{Horizontal polarizer only lets through light of horizontal polarization as shown in the left image. Similarly, only vertically polarized light can pass through a vertical polarizer.}
    \label{fig:polarizer}
\end{figure}
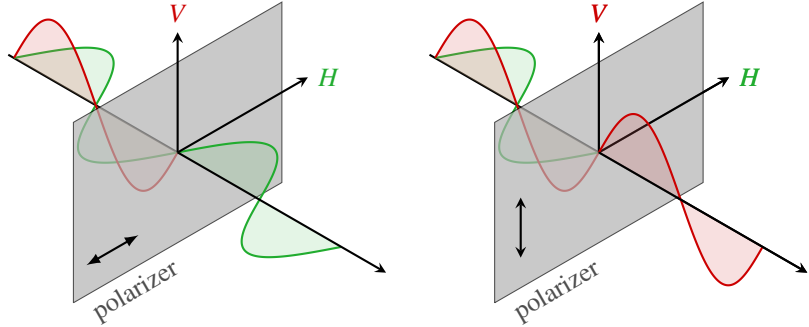

Distinguishing different orthogonal polarization states of light can be done easily with the help of \textbf{\emph{polarizers}}\index{polarizer}.
Polarizer is an optical element that lets through light of only certain polarization and blocks all other light.
Sunglasses are a good example of polarizing filters that are used to reduce the intensity of sunlight in order to protect our eyes.
Figure~\ref{fig:polarizer} shows what happens when light of different polarization is incident onto a polarizer.
Horizontal polarizer lets through only horizontally polarized light, while a vertical polarizer lets through only vertically polarized light.

All of this discussion carries through to the level of individual photons as well.
Photons can be polarized too.
However, individual photons are quantum states and therefore we have to describe then with kets.
We label a horizontally polarized photon by the ket $|H\rangle$, and a vertically polarized photon with the ket $|V\rangle$.
Now the idea of a photon encoding a qubit becomes more clear.
We can use the horizontal polarization state $|H\rangle$ to correspond to the $|0\rangle$ state of our qubit, and the vertical polarization state $|V\rangle$ to correspond to the $|1\rangle$ state.
General pure state of the qubit $\alpha|0\rangle+\beta|1\rangle$ can be then encoded into the photon by a general superposition of the photon's polarization,
\begin{equation}
    |\psi\rangle = \alpha|H\rangle + \beta|V\rangle.
\end{equation}
For example, the equal superposition $|+\rangle$ is encoded using the diagonal polarization, that is $|D\rangle=(|H\rangle+|V\rangle)/\sqrt{2}$.
Table~\ref{tab:4-4_polarization_encoding} summarizes how the polarization degree of freedom of individual photons is used to encode a qubit.
\begin{table}[h]
    \setcellgapes{5pt}
    \renewcommand\theadfont{}
    \makegapedcells
    \centering
    \begin{tabular}{cccc}
        \hline
         & \textbf{Basis states} & \textbf{Equal superposition} & \textbf{General state} \\
        \hline
        Qubits & $\{|0\rangle,|1\rangle\}$ & $|+\rangle$ & $\alpha|0\rangle+\beta|1\rangle$ \\
        Polarized photons & $\{|H\rangle,|V\rangle\}$ & $|D\rangle$ & $\alpha|H\rangle+\beta|V\rangle$ \\
        \hline
    \end{tabular}
    \caption[Polarization encoding.]{Mapping between qubit states and their corresponding polarization-encoded photonic states.}
    \label{tab:4-4_polarization_encoding}
\end{table}

Now we are in a position to discuss how the polarization of the pump photons affects the polarization of the signal and idler photons in the SPDC.
When a pump photon of a particular polarization is incident onto a BBO crystal, it has a chance to be down-converted to a pair of photons with equal polarizations which are orthogonal to that of the pump photon\footnote{This is known as Type-I down-conversion. It is possible for the signal and idler photons to have opposite polarization to each other. This is known as Type-II down-conversion.}.
Left side of Fig.~\ref{fig:poalrization_BBO} shows a horizontally polarized pump photon being down-converted into two vertically polarized photons,
\begin{equation}
    |H\rangle_p \rightarrow |V\rangle_s|V\rangle_i.
\end{equation}
The figure also displays the \textbf{\emph{optic axis}}\index{optic axis} indicating the orientation of the BBO crystal\footnote{We do not go into much detail about the significance of the optic axis. Briefly, if light propagates parallel to the optic axis it suffers no bending or refraction. We will cover this topic in great detail in our next book.}.
Rotating the whole setup by 90\degree, as shown on the right side of Fig.~\ref{fig:poalrization_BBO}, we can down-convert a vertically polarized pump photon into a pair of horizontally polarized photons,
\begin{equation}
    |V\rangle_p \rightarrow |H\rangle_s|H\rangle_i.
\end{equation}

\begin{figure}[t]
    \centering
    \begin{tikzpicture}[x={(0.866cm,-0.5cm)}, y={(0.866cm,0.5cm)}, z={(0cm,1cm)}, scale=0.9, >=stealth, wave/.style={thick,color=#1,smooth}]

    \def\h{1}
    \def\w{1.2}
    \def\d{0.1}

    \begin{scope}
    \coordinate (O) at (0,0,0);
    
    \draw[wave=newgreen, fill=newgreen!10, variable=\x, samples at={-0.35,-0.34,...,0.35}, thin] plot (\x-2,{0.5*e^(-\x*\x/0.02)},0);
    \node[newgreen] at (-2,0.89,0) {\footnotesize $|H\rangle_p$};
    \draw[newgreen,thick,line cap=round] (-3,0,0) -- (O);

    \draw[] (-\d,-\w,-\h) -- (-\d,-\w,\h) -- (-\d,\w,\h) -- (-\d,\w,-\h) -- cycle;
    \draw[] (-\d,\w,-\h) -- (\d,\w,-\h) -- (\d,-\w,-\h) -- (-\d,-\w,-\h) -- cycle;
    \draw[pattern=north east lines] (-\d,-\w,-\h) -- (-\d,-\w,\h) -- (\d,-\w,\h) -- (\d,-\w,-\h) -- cycle;
    \draw[pattern=north east lines] (-\d,\w,\h) -- (-\d,\w,-\h) -- (\d,\w,-\h) -- (\d,\w,\h) -- cycle;
    \draw[] (-\d,-\w,\h) -- (-\d,\w,\h) -- (\d,\w,\h) -- (\d,-\w,\h) -- cycle;
    \draw[] (\d,-\w,-\h) -- (\d,-\w,\h) -- (\d,\w,\h) -- (\d,\w,-\h) -- cycle;
    \draw[-latex,thick,line cap=round] (0,-\w,0) -- (0,{-1.5*\w},0);

    \begin{scope}[rotate around z=20]
        \draw[wave=lightred, fill=lightred!10, variable=\x, samples at={-0.35,-0.34,...,0.35}, thin] plot (\x+1.7,0,{0.5*e^(-\x*\x/0.02)});
    \draw[-latex,lightred,thick,line cap=round] (O) -- (2.3,0,0);
    \end{scope}
    \node[lightred] at (1.3,1.4,0) {\footnotesize $|V\rangle_s$};

    \begin{scope}[rotate around z=-20]
        \draw[wave=lightred, fill=lightred!10, variable=\x, samples at={-0.35,-0.34,...,0.35}, thin] plot (\x+1.7,0,{0.5*e^(-\x*\x/0.02)});
    \draw[-latex,lightred,thick,line cap=round] (O) -- (2.3,0,0);
    \end{scope}
    \node[lightred] at (1.6,-0.1,0) {\footnotesize $|V\rangle_i$};

    \node[] at (0.6,\w,\h/2) {\footnotesize BBO};
    \node[] at (-0.6,-\w-0.5,0) {\footnotesize optic};
    \node[] at (-0.6,-\w-0.5,-0.3) {\footnotesize axis};
    \end{scope}

    \begin{scope}[xshift=5.5cm]
    \coordinate (O) at (0,0,0);
    
    \draw[wave=lightred, fill=lightred!10, variable=\x, samples at={-0.35,-0.34,...,0.35}, thin] plot (\x-2,0,{0.5*e^(-\x*\x/0.02)});
    \node[lightred] at (-2.2,0.8,0) {\footnotesize $|V\rangle_p$};
    \draw[lightred,thick,line cap=round] (-3,0,0) -- (O);

    \draw[] (-\d,-\w,-\h) -- (-\d,-\w,\h) -- (-\d,\w,\h) -- (-\d,\w,-\h) -- cycle;
    \draw[pattern=north east lines] (-\d,\w,-\h) -- (\d,\w,-\h) -- (\d,-\w,-\h) -- (-\d,-\w,-\h) -- cycle;
    \draw[] (-\d,-\w,-\h) -- (-\d,-\w,\h) -- (\d,-\w,\h) -- (\d,-\w,-\h) -- cycle;
    \draw[] (-\d,\w,\h) -- (-\d,\w,-\h) -- (\d,\w,-\h) -- (\d,\w,\h) -- cycle;
    \draw[pattern=north east lines] (-\d,-\w,\h) -- (-\d,\w,\h) -- (\d,\w,\h) -- (\d,-\w,\h) -- cycle;
    \draw[] (\d,-\w,-\h) -- (\d,-\w,\h) -- (\d,\w,\h) -- (\d,\w,-\h) -- cycle;
    \draw[-latex,thick,line cap=round] (0,0,\h) -- (0,0,{1.5*\h});

    \begin{scope}[rotate around z=20]
        \draw[wave=newgreen, fill=newgreen!10, variable=\x, samples at={-0.35,-0.34,...,0.35}, thin] plot (\x+1.7,{0.5*e^(-\x*\x/0.02)},0);
    \draw[-latex,newgreen,thick,line cap=round] (O) -- (2.3,0,0);
    \end{scope}
    \node[newgreen] at (1.3,1.4,0) {\footnotesize $|H\rangle_s$};

    \begin{scope}[rotate around z=-20]
        \draw[wave=newgreen, fill=newgreen!10, variable=\x, samples at={-0.35,-0.34,...,0.35}, thin] plot (\x+1.7,{0.5*e^(-\x*\x/0.02)},0);
    \draw[-latex,newgreen,thick,line cap=round] (O) -- (2.3,0,0);
    \end{scope}
    \node[newgreen] at (2,0,0) {\footnotesize $|H\rangle_i$};

    \node[] at (0.6,\w,\h/2) {\footnotesize BBO};
    \node[] at (0.2,0.2,1.8*\h) {\footnotesize optic axis};
    \end{scope}
    \end{tikzpicture}
    \caption[Photon pair via SPDC.]{Horizontally polarized pump photons can be down-converted to a pair of vertically polarized photons by the BBO crystal as shown on the left side. Rotating the crystal by 90\degree makes it possible to convert a vertically polarized pump photon into a pair of horizontally polarized photons.}
    \label{fig:poalrization_BBO}
\end{figure}

\begin{figure}[t]
    \centering
    \begin{tikzpicture}[x={(0.866cm,-0.5cm)}, y={(0.866cm,0.5cm)}, z={(0cm,1cm)}, scale=0.9, wave/.style={thick,color=#1,smooth}]

    \begin{scope}
    \def\h{1}
    \def\w{1.2}
    \def\d{0.3}
    \coordinate (O) at (0,0,0);
    
    \draw[wave=blue!50, fill=blue!10, variable=\x, samples at={-0.35,-0.34,...,0.35}, thin, rotate around x=45] plot (\x-2,{0.5*e^(-\x*\x/0.02)},0);
    \node[blue!50] at (-2.1,0.89,0.2) {\footnotesize $|D\rangle_p$};
    \draw[blue!50,line cap=round,thick] (-3,0,0) -- (2*\d,0,0);

    \draw[] (-2*\d,-\w,-\h) -- (-2*\d,-\w,\h) -- (-2*\d,\w,\h) -- (-2*\d,\w,-\h) -- cycle;
    \draw[pattern=horizontal lines] (-2*\d,\w,-\h) -- (0,\w,-\h) -- (0,-\w,-\h) -- (-2*\d,-\w,-\h) -- cycle;
    \draw[] (-2*\d,-\w,-\h) -- (-2*\d,-\w,\h) -- (0,-\w,\h) -- (0,-\w,-\h) -- cycle;
    \draw[] (-2*\d,\w,\h) -- (-2*\d,\w,-\h) -- (0,\w,-\h) -- (0,\w,\h) -- cycle;
    \draw[pattern=horizontal lines] (-2*\d,-\w,\h) -- (-2*\d,\w,\h) -- (0,\w,\h) -- (0,-\w,\h) -- cycle;
    \draw[] (0,-\w,-\h) -- (0,-\w,\h) -- (0,\w,\h) -- (0,\w,-\h) -- cycle;
    \draw[-latex,thick,line cap=round] (-\d,0,\h) -- (-\d,0,{1.6*\h});

    \draw[] (0,\w,-\h) -- (2*\d,\w,-\h) -- (2*\d,-\w,-\h) -- (0,-\w,-\h) -- cycle;
    \draw[pattern=horizontal lines] (0,-\w,-\h) -- (0,-\w,\h) -- (2*\d,-\w,\h) -- (2*\d,-\w,-\h) -- cycle;
    \draw[pattern=horizontal lines] (0,\w,\h) -- (0,\w,-\h) -- (2*\d,\w,-\h) -- (2*\d,\w,\h) -- cycle;
    \draw[] (2*\d,\w,\h) -- (2*\d,-\w,\h);
    \draw[-latex,thick,line cap=round] (\d,-\w,0) -- (\d,{-1.6*\w},0);

    \begin{scope}[rotate around z=20]
        \draw[wave=newgreen, fill=newgreen!10, variable=\x, samples at={-0.35,-0.34,...,0.35}, thin] plot (\x+2.5,{0.5*e^(-\x*\x/0.02)},0);
    \draw[-latex,newgreen,thick] (O) -- (3.3,0,0);
    \end{scope}
    \node[newgreen] at (2.5,1.5,0.2) {\footnotesize $|H\rangle_s$};

    \begin{scope}[rotate around z=-20]
        \draw[wave=newgreen, fill=newgreen!10, variable=\x, samples at={-0.35,-0.34,...,0.35}, thin] plot (\x+2.5,-{0.5*e^(-\x*\x/0.02)},0);
    \draw[-latex,newgreen,thick] (O) -- (3.3,0,0);
    \end{scope}
    \node[newgreen] at (2.5,-1.5,0) {\footnotesize $|H\rangle_i$};

    \begin{scope}[rotate around z=-20,transform canvas={xshift=12,yshift=-7}]
    \draw[wave=lightred, fill=lightred!10, variable=\x, samples at={-0.35,-0.34,...,0.35}, thin] plot (\x+1.8,0,{0.5*e^(-\x*\x/0.02)});
    \draw[-latex,lightred,thick,line cap=round] (O) -- (3,0,0);
    \end{scope}
    \node[lightred] at (1.6,1.38,0.05) {\footnotesize $|V\rangle_s$};

    \begin{scope}[rotate around z=20,transform canvas={xshift=12,yshift=-7}]
    \draw[wave=lightred, fill=lightred!10, variable=\x, samples at={-0.35,-0.34,...,0.35}, thin] plot (\x+1.8,0,{0.5*e^(-\x*\x/0.02)});
    \draw[-latex,lightred,thick,line cap=round] (O) -- (3,0,0);
    \end{scope}
    \node[lightred] at (2.2,-0.1,0) {\footnotesize $|V\rangle_i$};

    \node[] at (0,0,2.3) {\footnotesize Product photon pair};

    \end{scope}

    \begin{scope}[xshift=6cm]
    \def\h{1}
    \def\w{1.2}
    \def\d{0.1}
    \coordinate (O) at (0,0,0);
    
    \draw[wave=blue!50, fill=blue!10, variable=\x, samples at={-0.35,-0.34,...,0.35}, thin, rotate around x=45] plot (\x-2,{0.5*e^(-\x*\x/0.02)},0);
    \node[blue!50] at (-2.1,0.89,0.2) {\footnotesize $|D\rangle_p$};
    \draw[blue!50,line cap=round,thick] (-3,0,0) -- (0,0,0);

    \draw[] (-2*\d,-\w,-\h) -- (-2*\d,-\w,\h) -- (-2*\d,\w,\h) -- (-2*\d,\w,-\h) -- cycle;
    \draw[pattern=horizontal lines] (-2*\d,\w,-\h) -- (0,\w,-\h) -- (0,-\w,-\h) -- (-2*\d,-\w,-\h) -- cycle;
    \draw[] (-2*\d,-\w,-\h) -- (-2*\d,-\w,\h) -- (0,-\w,\h) -- (0,-\w,-\h) -- cycle;
    \draw[] (-2*\d,\w,\h) -- (-2*\d,\w,-\h) -- (0,\w,-\h) -- (0,\w,\h) -- cycle;
    \draw[pattern=horizontal lines] (-2*\d,-\w,\h) -- (-2*\d,\w,\h) -- (0,\w,\h) -- (0,-\w,\h) -- cycle;
    \draw[] (0,-\w,-\h) -- (0,-\w,\h) -- (0,\w,\h) -- (0,\w,-\h) -- cycle;
    \draw[-latex,thick,line cap=round] (-\d,0,\h) -- (-\d,0,{1.6*\h});

    \draw[] (0,\w,-\h) -- (2*\d,\w,-\h) -- (2*\d,-\w,-\h) -- (0,-\w,-\h) -- cycle;
    \draw[pattern=horizontal lines] (0,-\w,-\h) -- (0,-\w,\h) -- (2*\d,-\w,\h) -- (2*\d,-\w,-\h) -- cycle;
    \draw[pattern=horizontal lines] (0,\w,\h) -- (0,\w,-\h) -- (2*\d,\w,-\h) -- (2*\d,\w,\h) -- cycle;
    \draw[] (2*\d,\w,\h) -- (2*\d,-\w,\h);
    \draw[-latex,thick,line cap=round] (\d,-\w,0) -- (\d,{-1.6*\w},0);

    \begin{scope}[rotate around z=20]
    \draw[wave=black, fill=black!10, densely dotted, variable=\x, samples at={-0.35,-0.34,...,0.35}, thin] plot (\x+2,{0.5*e^(-\x*\x/0.02)},0);
    \draw[wave=black, fill=black!10, densely dotted, variable=\x, samples at={-0.35,-0.34,...,0.35}, thin] plot (\x+2,0,{0.5*e^(-\x*\x/0.02)});
    \draw[-latex,thick] (O) -- (2.7,0,0);
    \end{scope}

    \begin{scope}[rotate around z=-20]
    \draw[wave=black, fill=black!10, densely dotted, variable=\x, samples at={-0.35,-0.34,...,0.35}, thin] plot (\x+2,-{0.5*e^(-\x*\x/0.02)},0);
    \draw[wave=black, fill=black!10, densely dotted, variable=\x, samples at={-0.35,-0.34,...,0.35}, thin] plot (\x+2,0,{0.5*e^(-\x*\x/0.02)});
    \draw[-latex,thick] (O) -- (2.7,0,0);
    \end{scope}
    \node[] at (0,0,2.3) {\footnotesize Entangled photon pair};
    \end{scope}
\end{tikzpicture}
    \caption[Entangled photon pairs via SPDC.]{If the crystals are too thick (left image) then the propagation paths for horizontal and vertical photons are distinguishable, resulting in a product state. The two BBO crystals must be sufficiently thin in order to erase the information about where the signal-idler pair were produced (right iamge). The gray color of the photon pair represents the fact that we cannot assign definite ket states to the individual photons since they are entangled.}
    \label{fig:entanglement_BBO}
\end{figure}

So far, we have discussed production of pairs of photons which are not entangled.
With a simple change of this setup, we can use SPDC to produce entangled pairs of photons as well.
The first step is to use two identical BBO crystals back-to-back, where the second crystal is rotated by 90\degree.
The second step is to choose an appropriate polarization for the pump photon.
If we use $|H\rangle_p$, then down-conversion always occurs in only one of the BBO crystals.
If we use $|V\rangle_p$, then down-conversion always occurs in the other crystal.
The trick is to use a diagonally polarized pump photon, $|D\rangle_p$.
With probability of $1/2$, such a pump photon will be down-converted in the first crystal, and with equal probability it will be down-converted in the second crystal.
The final step is to ensure that it is not possible to determine whether the down-conversion occurred in the first or in the second BBO crystal.
This can be achieved by using thin enough BBO crystals, resulting in very nearly overlapping propagation paths for the down-converted photons coming from either crystal, as shown in Fig.~\ref{fig:entanglement_BBO}.
Only when this \textbf{\emph{indistinguishability condition}}\index{indistinguishability condition} is satisfied, do we obtain an entangled pair of photons,
\begin{equation}
    |D\rangle_p \rightarrow \frac{|H\rangle_s|H\rangle_i + |V\rangle_s|V\rangle_i}{\sqrt{2}}.
\end{equation}

SPDC is an extremely rare process.
On average, we can expect to obtain one photon pair per $10^6$ pump photons.
Even this low rate can be achieved only in state-of-the-art experiments.
With a more common laboratory setup, it takes several orders of magnitude more pump photons to successfully create an entangled pair.
The photons that are not down-converted simply pass through the crystal unaffected and are discarded.

\section{Entanglement as a resource}

\begin{figure}[t]
   \centering
    \begin{tikzpicture}[auto, semithick, node distance=3cm]
    \tikzstyle{every state}=[draw=black,thick,text=black,scale=1]

    \node[state,fill=red!50]    (R)                     {R};
    \node[state,fill=newgreen!50]   (A)[below left of=R]    {A};
    \node[state,fill=newgreen!50]   (B)[below right of=R]   {B};

    \path
    (R) edge[bend right,above,red!60,thick,->,>=stealth']	node[yshift=1pt]{$x$}	(A)
    (R) edge[bend left,above,red!60,thick,->,>=stealth']	node[yshift=-1pt,xshift=1pt]{$y$}	(B)
    (A) edge[bend right,left,newgreen,thick,->,>=stealth']	node[yshift=1pt]{$a$}	(R)
    (B) edge[bend left,right,newgreen,thick,->,>=stealth']	node[yshift=2pt]{$b$}	(R);

    \coordinate (qubitA) at ($(A)+(0,-1)$);
    \coordinate (qubitB) at ($(B)+(0,-1)$);

    \draw[decorate,decoration={coil,aspect=0},thick] (qubitA)  -- (qubitB);
    \filldraw[black!30,draw=black,thick] (qubitA) circle (5pt);
    \filldraw[black!30,draw=black,thick] (qubitB) circle (5pt);

    \node[] at (-3.6, -3.15) {measure qubit A};
    \node[] at (3.6, -3.15) {measure qubit B};
    \node[darkred] at (0, -3.15) {\xmark};

    \end{tikzpicture}
    
    \caption[CHSH game consumes entanglement.]{In the CHSH game, players consume one Bell pair every time they play with the optimal strategy of winning.}
    \label{fig:chsh-broken}
\end{figure}


Let's go back and revisit the CHSH game from the first section of this chapter.
We have seen that entanglement can help players $A$ and $B$ win the game more often.
The optimal strategy is to share a Bell pair, which is then locally measured by the players.
As shown in Fig.~\ref{fig:chsh-broken}, players $A$ and $B$ destroy the entanglement in the process.
If they want to play another round of the CHSH game, and still have the optimal 85\% probability of winning, they have to find a way of sharing a Bell pair again.
We can see that usingthis strategy, one Bell pair is consumed in every round of the game.
Therefore, if the players want to play $n$ rounds with optimal strategy of winning, they must have access to at least $n$ shared Bell pairs.

\begin{figure}[t]
    \centering
    \includegraphics[width=\textwidth]{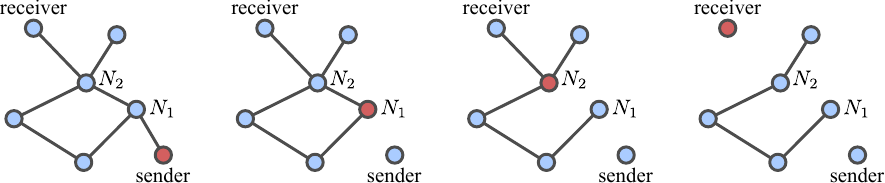}
    \caption[Consumption of entanglement in a quantum network.]{Entanglement is consumed as quantum information is propagated from the sender to the receiver.}
    \label{fig:4-5_resource_network}
\end{figure}

Abstract scenarios, such as the CHSH game, are not the only place where entanglement is useful.
Quantum networks is one such important example where entanglement plays a vital role.
Figure~\ref{fig:4-5_resource_network} shows a simple quantum network, where the blue circles represent quantum nodes of the network.
They could be individual qubits or even small quantum computers; for now, the distinction is not important.
The lines joining the quantum nodes represent shared entangled pairs.
There is a sender node in possession of a qubit, represented by the red circle, that it is trying to communicate to the receiver node.
The usual way to send quantum information is to use the teleportation protocol, which we will discuss in detail in Chapter~\ref{sec:8_teleportation}.
For now, all we need to know is that the protocol consumes a single Bell pair to transfer the quantum information from one node to its neighbor.
By executing the teleportation protocol first at the sender's node, and then at nodes $N_1$ and $N_2$, the quantum information will be delivered to the receiver node.
We can see that delivering the quantum information has consumed all of the entangled pairs along the path connecting the sender with the receiver.
In order for the quantum network to be fully functional again, we must re-establish the destroyed entangled links.

We can think of entanglement as the fuel that drives many quantum technologies. Entanglement offers improved, and sometimes completely new, functionality that is not seen in either classical networks or classical computation.
One of the key takeaways is that entanglement is a \textbf{\emph{resource}}, and is consumed just like fuel in your car or battery charge in your phone.
It is the job of quantum networks to distribute entanglement in order to satisfy the demand for this precious resource.

\newpage
\begin{exercises}
\exer{\emph{Classical strategies for the CHSH game.} We have seen the optimal classical strategy. For each strategy described below, determine the probability of winning a round of the CHSH game.
\subexer{Player $A$ and player $B$ reply with bits that are completely random, regardless of the bits received from the referee $R$.}
\subexer{Reply bits from player $A$ and player $B$ are the same as the bits received from referee $R$. This means $a=x$ for player $A$, and $b=y$ for player $B$.
}
}

\exer{\emph{Local measurements on two qubits.}
Eq.~(\ref{eq:local_projectors}) describes local measurement in Pauli $Z$ basis performed by player $A$.
If player $B$ also measures in the Pauli $Z$ basis, the full projector is written as the tensor product of the two local projectors,
\begin{equation}
    \Pi^A_{Z,\pm} \otimes \Pi^B_{Z,\pm}.
\end{equation}
\subexer{
Consider a shared entangled state between the players $A$ and $B$ to be $|\Phi^+\rangle$.
If players $A$ and $B$ both measure in the Pauli $Z$ basis, what are the possible outcomes of their measurements, and their corresponding probabilities?
}
\subexer{
Measurements in other bases can be described in a straightforward fashion. If player $A$ measures in the $O_A$ basis and player $B$ measures in the $O_B$ basis, the two-qubit projector has the following form,
\begin{equation}
    \Pi^A_{O_A,\pm} \otimes \Pi^B_{O_B,\pm}.
\end{equation}
Consider the case where the players still share $|\Phi^+\rangle$. $A$ measures in the Pauli $Z$ basis, but this time player $B$ measures in the Pauli $X$ basis.
What are the possible measurements outcomes and their corresponding probabilities?
}
}

\exer{\emph{Quantum strategy for the CHSH game.} Consider a particular instance of the CHSH game, where the referee generates the bits $x=0$ and $y=0$.
\subexer{
What values of $a$ and $b$ will result in a win for the players?
}
\subexer{
Assuming the players use the optimal quantum strategy, show that the probability of winning this round is $\frac{1+\sqrt{2}}{2\sqrt{2}}\approx0.85$.
}
}

\exer{
\emph{Bell basis.} We have seen that the four Bell pairs form a convenient basis when we consider two qubits.
\subexer{
Prove that the four Bell states are orthogonal to each other.
}
\subexer{
Express an arbitrary pure two-qubit state,
\begin{equation}
    |\psi\rangle = \alpha |00\rangle + \beta|01\rangle + \gamma|10\rangle + \delta|11\rangle,
\end{equation}
in the Bell basis.
}
\subexer{
The two-qubit maximally mixed state is the tensor product of two local maximally mixed states, $I/2 \otimes I/2$.
Write down this state in the Bell basis.
}
}

\exer{\emph{Singe-qubit vs two-qubit measurement statistics.}
On p.~\pageref{page:plus-is-pure}, we stated without proof that measuring the state of a single qubit provides different outcome statistics than measuring one qubit that is part of a Bell pair.
Let's prove this now.
\subexer{
Consider single qubit in $|+\rangle$. Compute outcome probabilities when this state is measured in the Pauli $X$, $Y$, and $Z$ bases.
}
\subexer{
Now consider a shared two-qubit Bell pair, say $|\Psi^+\rangle$.
Again, compute the outcome probabilities when the first qubit is measured in the Pauli $X$, $Y$, and $Z$ bases.
}
}

\exer{\emph{Rate of production of down-converted photons.}
Consider a high power UV diode laser at wavelength 405 nm. The pump laser power is 15 mW.
Consider that only 1 photon pair is produced on average per $10^{8}$ pump photons.
\subexer{
How many pump photons are incident on the BBO crystal per second? \\
(\emph{Hint:} Energy of a photon is given by $E_{\text{photon}}=hc/\lambda$, where $h$ is Planck's constant, $c$ is speed of light in free space, $\lambda$ is the wavelength of light.)
}
\subexer{
What is the rate at which down-converted photon pairs are produced?
}
\subexer{
What is the wavelength of the down-converted photons if both have the same energy?
}
}

\end{exercises}

\newpage
\section*{Quiz}
  \addcontentsline{toc}{section}{Quiz}

The online version of this course includes a quiz for this block of chapters. Discussion of the quiz questions will be provided there.

\section*{Further reading for chapters 1-4}
\addcontentsline{toc}{section}{Further reading for chapters 1-4}

\textbf{Chapter 1}\\

The first chapter serves as a gentle introduction to the area of quantum communication. Its goal is to provide a qualitative overview and discuss ideas behind communication and its evolution rather than focusing on any quantitative discussion.

An excellent popular science book on this topic is James Gleick, \emph{The Information: A History, The Theory, A Flood}~\cite{gleick2012information}. It covers the evolution of communication and spends a sizeable portion on discussing the switch from analog to digital. It does not cover quantum communication.

There are also a number of great science YouTube channels with excellent, concise introductions on the topic of quantum communication and computation:
\begin{enumerate}
    \item Veritasium
    \item Science Girl
    \item PBS Space Time
\end{enumerate}

Understanding the role that encryption plays in achieving the overall goals of computer security is valuable. We recommend Matt Bishop's authoritative textbook, \emph{Computer Security: Art and Science}~\cite{bishop2002art}. \\

\textbf{Chapter 2}\\

The second chapter is more technical and introduces fundamental concepts such as qubits and measurements, which we will encounter throughout the entire curriculum. Being comfortable with these concepts and knowing how to describe them mathematically is crucial.

An article called ``Quantum Computing as a High School Module''~\cite{perry2019quantum} is a great extension to the content of this and following chapters. It introduces basic mathematical descriptions and contains many short exercises designed to check your understanding.

A more advanced introduction to the mathematical concepts in this chapter can be found in the classic textbook by Michael A. Nielsen and Isaac L. Chuang, \emph{Quantum Computation and Quantum Information}~\cite{nielsen-chuang:qci}. Almost all serious quantum researchers have a copy of this book, but its learning curve is steep and we recommend attacking it after a more introductory book or course (which you presumably are gaining through this learning module and other courses). This book, known as "Mike and Ike", covers key ideas in computer science theory, the basic quantum information ideas and algorithms, and the principles of quantum hardware. Its descriptions of hardware and error correction are now rather out of date, and the description of algorithms is limited to a few important cases, but the principles are foundational and the explanation clear.

An alternative to Mike and Ike is John Preskill's lecture notes, which are available chapter by chapter as open access PDF files.~\cite{preskill1998lecture}.

A new textbook that covers the basics of quantum information and their role in quantum computation and communication is Robert Sutor's \emph{Dancing with Qubits}~\cite{sutor19:dancing}. This textbook spends great effort in explaining the basics and goes over the fundamental calculations in great detail. The first half of the book covers the basic mathematics you will need, including complex numbers, probability and linear algebra (vectors and matrices). We strongly recommend this book for all beginning students, especially those who are worried about the math required. We recommend it to our own beginning undergraduates.

An alternative to Sutor that focuses primarily on basic algorithms is Eleanor Rieffel and Wolfgang Polak's \emph{Quantum Computing: a Gentle Introduction}~\cite{rieffel2011quantum}.\\

\textbf{Chapter 3}\\

The third chapter continues with our exposition of the basic mathematical formalism underpinning quantum communication as well as quantum computation. The primary focus is description of noisy quantum states using density matrices.

Chapter 2 of “Mike \& Ike” and Perry’s article mentioned above are great for this chapter as well. Another fantastic book on quantum information (though primarily targeted at graduate students so parts of it might be too technical to follow) is Mark Wilde's \emph{Quantum Information Theory}~\cite{wilde2013quantum}.

Chapter 3 of Wilde’s textbook deals with noisy states. Discussion on fidelity of quantum states can be found in Chapter 9 of “Mike \& Ike” and in Chapter 9 of Wilde’s textbook.\\

\newpage

\textbf{Chapter 4}\\

This chapter is the conclusion of the introductory chapters and deals with the scenario when we have multiple qubits which naturally leads to entanglement. Perry’s article, “Mike \& Ike” as well as Wilde’s textbook remain great introductions and ordered in this way represent the ramping up difficulty of mathematical rigour. 

The CHSH game is discussed in Chapter 3 of Wilde’s textbook~\cite{wilde2013quantum}. Another excellent exposition of this game is by Umesh Vazirani as part of his lecture series on YouTube.

A great introduction to SPDC can be found in Betchart’s bachelor's thesis~\footnote{\url{https://etd.ohiolink.edu/apexprod/rws_olink/r/1501/10?clear=10&p10_accession_num=oberlin1206296667}}.

\part{Fundamentals of Optics}

\begin{partintro}
\partintrotitle{Introduction to the second chapter block}

In this second block of chapters, we cover the basic elements of optics that are used in both quantum and classical communication systems. Coherent light and lasers that can generate it are addressed in Chapter 5. Chapter 6 presents interference and the concepts of phase and group velocities; here, interference is a specific concept in wave mechanics, rather than the use of the term to refer to e.g. noise that disrupts radio communications. Constructive and destructive interference patterns appear in light, and are also the heart of quantum computing algorithms, so this chapter may deepen your understanding of quantum algorithms as well.  Chapter 7 presents the geometric optics analysis of waveguides.  The most common form of waveguides, of course, is optical fiber, which forms the backbone of today's classical Internet.
\end{partintro}

\chapter{Coherent Light and Single Photons}
\label{sec:5_coherent_light_single_photons}

In this chapter, we will look at why light is such an integral part of modern day communication.
We will discuss the difference between coherent and incoherent light, and outline the basic principle behind producing coherent light using lasers.
Finally, we will transition to talking about light in the context of quantum communication that relies on single photons as information carriers.

\section{Introduction}
\label{sec:5-1_intoduction}

Why do we want to encode information as optical signals?
The first reason behind light being a good carrier of information is its incredible \textit{\textbf{speed}}.
Table~\ref{tab:5-1_speed_light} summarizes the speed of light in various media.
\begin{table}[b!]
    \setcellgapes{5pt}
    \renewcommand\theadfont{}
    \makegapedcells
    \centering
    \begin{tabular}{ccc}
        \hline
        \textbf{vacuum} & \textbf{air} & \textbf{silica fiber} \\
        \hline
        $c$ & $c/1.0003$ & $c/1.47$ \\
        $2.998\times10^{8} \; \text{ms}^{-1}$ & $2.998\times10^{8} \; \text{ms}^{-1}$ & $2.039\times10^{8} \; \text{ms}^{-1}$ \\
        \hline
    \end{tabular}
    \caption[Speed of light.]{Speed of light in various media.}
    \label{tab:5-1_speed_light}
\end{table}
In vacuum, the speed of light is $c=2.998\times 10^8$ $\text{ms}^{-1}$.
Here on Earth, most of the time we do not send light through vacuum.
Even in air the light slows by only a small factor of 1.0003.
Most often, we use fiber optic cables made from pure silica glass with refractive index of 1.47.
This decreases the speed of light somewhat but it still remains very fast.

Apart from being fast, light is also relatively \textit{\textbf{easy to produce}}.
Even in the early days of civilization, a reliable source of light was fire.
We saw an example of this when we learned about optical telegraphy used for rapid communication on the Great Wall of China in Chapter~\ref{sec:1_Introduction}.
Today, for long-distance communication, we mainly use lasers and send the optical signals through fibers.
Due to their ability to produce highly coherent light, lasers have had a transformative impact, not only on the way we communicate, but on many other aspects of our lives. 

The third reason why light is so useful in communication is that photons do not interact easily with each other.
Once in flight, photons will continue speeding to their destination nearly unaffected.
This makes optical signals \textit{\textbf{robust to noise}}.
Compare this with copper wires carrying electric signals.
The moving electrons are susceptible to external electric and magnetic noise and require thorough shielding to protect the integrity of the signal.
Furthermore, the moving electrons themselves produce electromagnetic fields which may affect other nearby carriers of electric signals.
This means that copper wires cannot be packed too closely to each other.
In contrast, this is not the case for optical fibers.

\begin{figure}[t]
    \centering
    \includegraphics[width=0.8\textwidth]{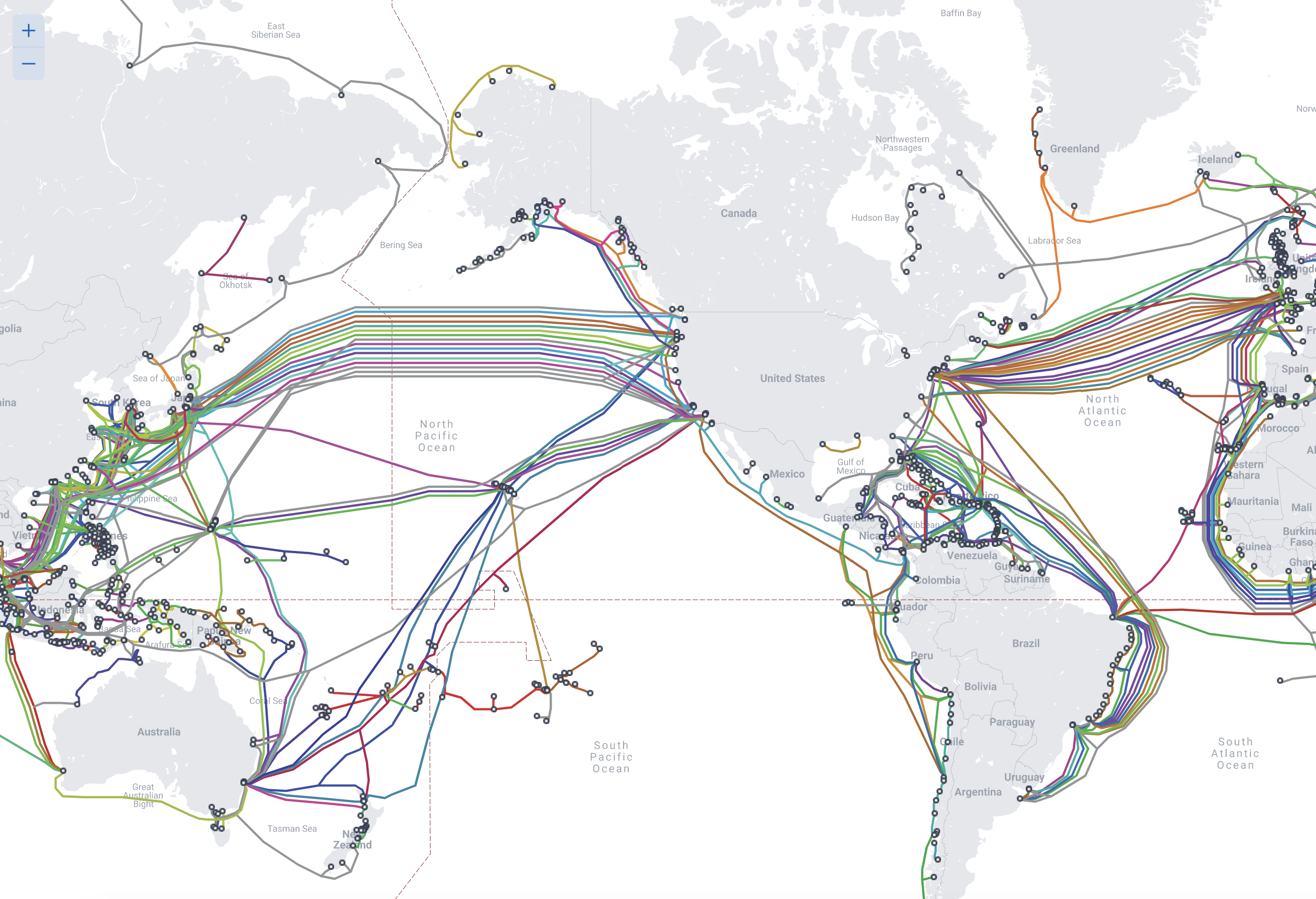}
    \caption[Underwater cable map.]{Map of submarine cables.}
    \label{fig:5-1_underwater_cable_map}
\end{figure}

Optics has always played an important role in communication.
We saw a couple of examples of that already in the Great Wall of China and Napoleon's semaphore in chapter~\ref{sec:1_Introduction}.
These methods were limited in the sense that you had to have a direct visual path between the sender and the receiver, you needed good weather conditions, and in the case of Napoleon's semaphore, it only worked during the day.
\textit{\textbf{Waveguides}}\index{waveguides} circumvent all these problems.
Use of electric wires and optical fibers sparked a rapid expansion in our ability to communicate fast and far.
Figure~\ref{fig:5-1_underwater_cable_map} shows a map of submarine cables \footnote{This map was obtained from TeleGeography at \href{https://www.submarinecablemap.com/}{https://www.submarinecablemap.com/} under the \href{https://creativecommons.org/licenses/by-sa/4.0/}{CC BY-SA 4.0} license.}, connecting the continents.
It is these cables that allow seamless global communication at incredible speeds.

In this chapter, we are going to be concerned with how to produce three types of light.
We will begin with \textit{\textbf{incoherent light}}\index{incoherent light}, which can be produced by burning fuel or heating a gas.
We will explain in what sense this light is incoherent in Section~\ref{sec:5-2_coherent_vs_incoherent}.
This type of light is known as a classical state of light as it does not manifest any quantum behavior.

We will compare incoherent light with \textit{\textbf{coherent light}}\index{coherent light} produced by lasers. The main mechanism behind producing this light is known as stimulated emission, which we will discuss in Sec.~\ref{sec:5-3_lasers1} and Sec.~\ref{sec:5-4_lasers2}.
Lasers sparked the first information revolution and therefore played an important historical role.
Despite its coherent nature, light produced by lasers is still not fully quantum. 
Developments in laser technology over the last decades have led to great proliferation of available sources of coherent light.
Lasers range from large and powerful ones found in high-tech laboratories to small pointer devices hanging from our keychains.

We will conclude this chapter by looking at \textit{\textbf{single photon sources}} in Sec~\ref{sec:5-5_single_photons}.
We will discuss three ways to produce single photons.
The first one is by attenuation of laser light.
The second method is by using heralded photons.
The third way is to use genuine emitters of single photons, such as nitrogen vacancy centres in diamond.
Compared to the previous two types of light, single photons are very difficult to make.
They can only be produced in laboratories or devices that meet very stringent requirements, but they can display quantum behavior, which is why they are crucial in quantum communication.

\section{Coherent vs incoherent light}
\label{sec:5-2_coherent_vs_incoherent}

\begin{figure}
    \centering
    \includegraphics[width=\textwidth]{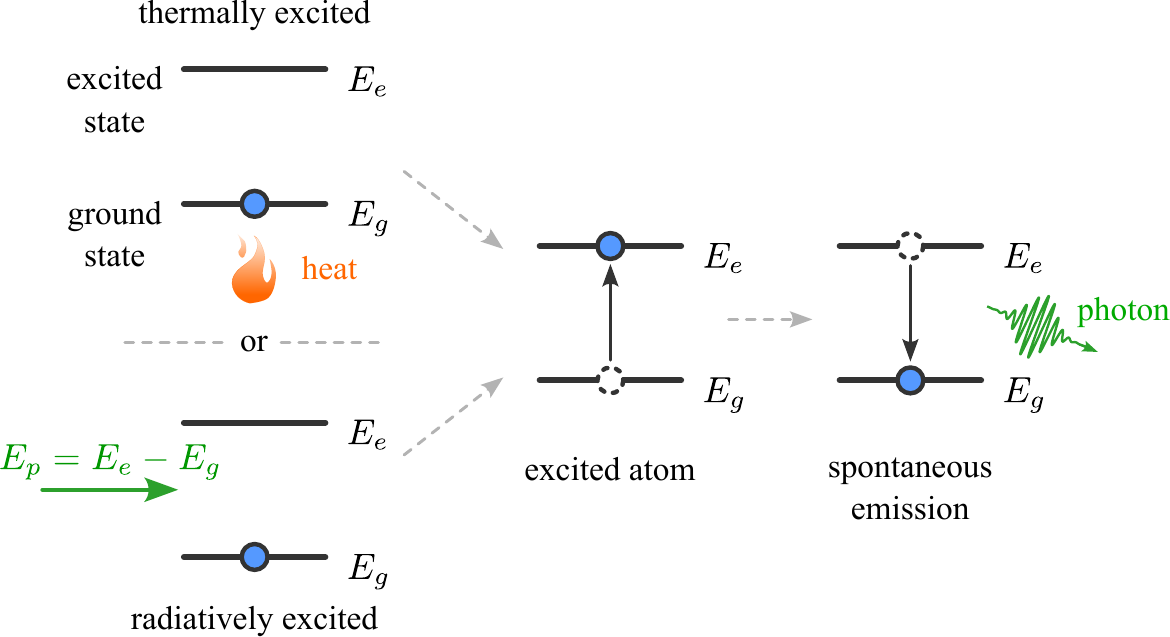}
    \caption[Spontaneous emission.]{A two-level atom is excited either thermally or radiatively. After some time it deexcites via spontaneous emission producing a photon of light.}
    \label{fig:5-2_spontaneous_emission}
\end{figure}

Be begin by understanding how matter can radiate light.
Let's consider a model of a simple two-level atom as shown in~Fig.~\ref{fig:5-2_spontaneous_emission}.
The atom has two distinguishable states given by the ground state with energy $E_g$, and the excited state with energy $E_e$.
The state of the atom is pictured by the blue circle.
Atoms are most stable when they are in the lowest energy state, in this case the ground state.
Atoms in the ground state do not radiate any light.
In order to produce light, the atom needs to first receive enough energy.
One way to do that is through heat.
If this is the case, we say the atom becomes \textbf{\emph{thermally excited}}\index{thermal excitation}.
This means that the thermal energy received by atom forced it to transition to the excited state, as shown in the middle panel of~Fig.~\ref{fig:5-2_spontaneous_emission}.
Different way of exciting the atom is to irradiate it with light of the right energy.
The atom can absorb radiation of energy equal to the difference of energies between the excited and ground states, $E_p = E_e - E_g$.
When this happens, we say the atom has been \textbf{\emph{radiatively excited}}\index{radiative excitation}.
After some time, the atom in the excited staet releases the stored energy in the form of a photon of light.
This process happens without an external stimulus at a random time and is called \textit{\textbf{spontaneous emission}}\index{spontaneous emission}.

\begin{figure}
    \centering
    \includegraphics[width=0.8\textwidth]{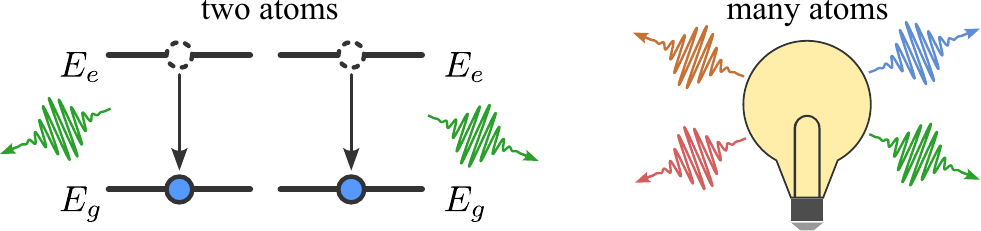}
    \caption[Incoherent emission.]{Two or more atoms emitting incoherently.}
    \label{fig:5-2_incoherent_emission}
\end{figure}

Let's consider two such excited atoms as shown in Fig.~\ref{fig:5-2_incoherent_emission}.
The atoms eventually undergo the process of spontaneous emission producing one photon each.
The direction of emission is random and generally different for both atoms.
There is a possibility that they will emit in the same direction but the chance of this happening is very small.
Furthermore, both photons have random and different phases.
We say that the emitted photons are incoherent.

We can go a step further and consider a large number of atoms emitting photons.
To have a more specific image in mind, we can consider an incandescent light bulb, as shown in the right panel of~Fig.~\ref{fig:5-2_incoherent_emission}.
The filament is heated up by an electric current running through it.
This in turn excites the atoms in the filament, which eventually undergo spontaneous emission producing a large amount of photons.
Not only are these photons travelling in all possible directions and are out of phase, their energies are different as well.
This is because the atoms in the filament have a much more complicated energy level structure than our simple two-level model.
In conclusion, incoherent light is composed of components with different energies travelling in different directions, each having a different phase.

Producing incoherent light is very easy.
The question that you might be asking yourself is, what does it take to produce coherent light?
How can we make light that has a single component of the same energy, travelling in the same direction and with the same phase?
We will answer this question in the next two sections.

\section{Lasers I: Stimulated emission}
\label{sec:5-3_lasers1}

In this section, we begin to answer the question that was raised at the end of the previous section.
What are the basic ingredients to make coherent light?
Such light is in-phase, monochromatic (meaning each photon has the same frequency), and travels in the same direction.
Typical example of a source that produces light with these properties is the \textbf{\emph{laser}}\index{laser}.
Laser stands for ``light amplification via stimulated emission of radiation''.
Let's have a look at what these individual terms mean.

\begin{figure}[t]
    \centering
    \includegraphics[width=0.55\textwidth]{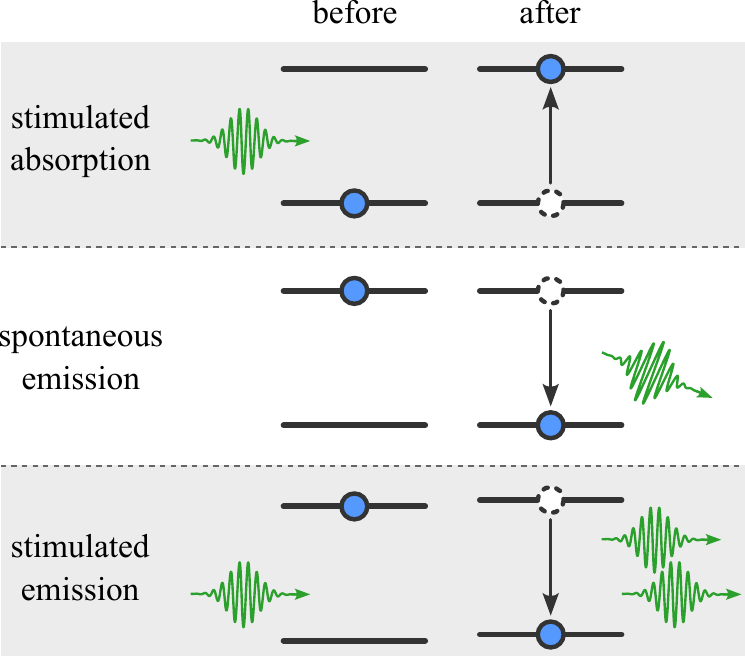}
    \caption[Light-matter interactions.]{Three fundamental types of light-matter interactions.}
    \label{fig:5-3_light_matter_interaction}
\end{figure}

Let's begin with \textbf{\emph{stimulated emission}}\index{stimulated emission}, which is the physical process behind lasing.
We have encountered two of the three fundamental ways in which light interacts with matter, namely stimulated absorption and spontaneous emission, also shown in Fig.~\ref{fig:5-3_light_matter_interaction}.
Stimulated absorption is when an atom, initially in the ground state, interacts with an incoming photon. If the frequency of the photon is just right, the atom may absorb this photon, and the atom becomes excited.
Spontaneous emission is when an initially excited atom emits a photon of light without any external stimulus.
Important property of spontaneous emission, which we mentioned in the previous section, is that this photon is emitted in a random direction.
Stimulated emission on the other hand is when an initially excited atom interacts with an incoming photon.
This causes the atom to emit a photon of light.
But this time, the emitted photon of light has the same energy as the external photon, same phase and, it is emitted in the same direction.
In other words, the two photons are coherent.

We can see from Fig.~\ref{fig:5-3_light_matter_interaction} that stimulated emission starts with a single photon and finishes with two coherent photons.
This opens up the possibility of amplifying light, which brings us to the \textbf{\emph{light amplification}}\index{light amplification} part of the laser.
Imagine having a large number of atoms in the excited state.
A single photon of light can stimulate the first atom to emit a photon.
Both photons (the initial one and the newly emitted one) can now stimulate further atoms to emit, triggering a cascade of emissions and producing a coherent beam of highly intense light. 
There is however one catch to the above scheme, not all of the atoms are usually found in the excited state.
When left alone, the atoms are much more likely to be in the ground state.
Getting all of them into an excited state is no easy task.

Let's first consider a single atom interacting with a single photon and do some simple accounting to better understand light amplification.
This time, we do not assume that the atom is in the excited state.
There are three possibilities that are summarized in Table~\ref{tab:5-3_three_possibilities}.
\begin{table}[h]
    \setcellgapes{5pt}
    \renewcommand\theadfont{}
    \makegapedcells
    \centering
    \begin{tabular}{ccccc}
        \hline
        & \textbf{Photons IN} & \textbf{Interaction} & \textbf{State of the atom} & \textbf{Photons OUT} \\
        \hline
        \textbf{Case 1} & 1 & No & ground/excited & 1 \\
        \textbf{Case 2} & 1 & Yes & ground & 0 \\
        \textbf{Case 3} & 1 & Yes & excited & 2 \\
        \hline
    \end{tabular}
    \caption[Stimulated emission accounting.]{Number of photons initially and after the process of stimulated emission.}
    \label{tab:5-3_three_possibilities}
\end{table}
The first case is the trivial one, the photon does not interact with the atom.
The atom's state has no effect on the final photon count.
This is represented by the first row of Table~\ref{tab:5-3_three_possibilities}.
The second possibility is that the photon interacts with the atom while it is in the ground state.
The atom absorbs the energy of the photon meaning the number of photons after the process drops to zero, as seen in the second row of Table~\ref{tab:5-3_three_possibilities}.
The last possibility is that the photon interacts with an atom in the excited state, the atom is stimulated to emit a photon, resulting in two coherent photons after the interaction as seen in the last row of Table~\ref{tab:5-3_three_possibilities}.
Assuming that each of the three cases occurs with the same probability, the average number of photons before and after the interaction remains unchanged.
This means a single atom is not able to amplify light purely via stimulated emission.

\begin{figure}[t]
    \centering
    \includegraphics[width=\textwidth]{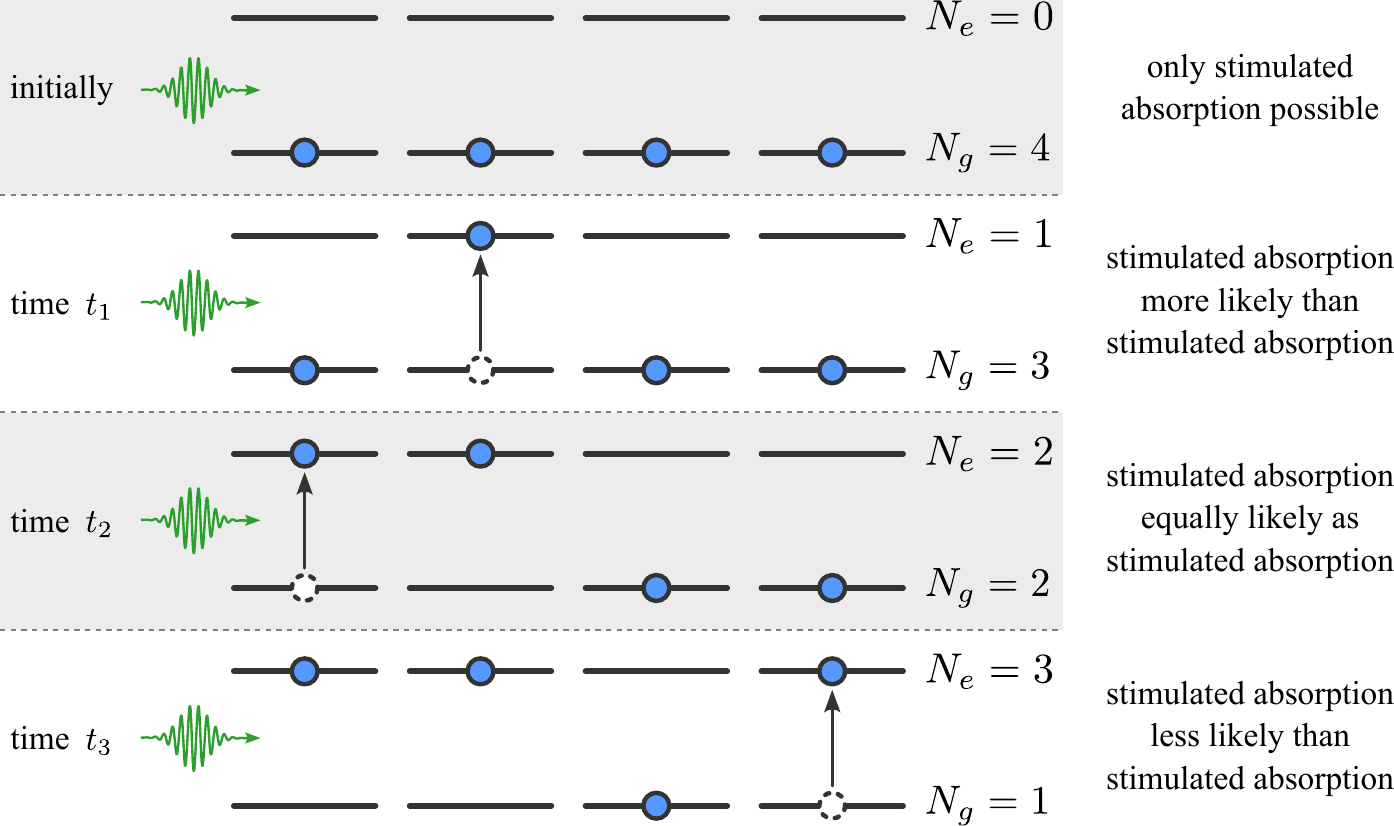}
    \caption[Population inversion.]{Light amplification can occur only if stimulated emission is more likely than stimulated absorption. This can occur only when more atoms are in the excited state than in the ground state.}
    \label{fig:5-3_population_inversion}
\end{figure}

Let's now consider multiple atoms to see how the situation changes.
Figure~\ref{fig:5-3_population_inversion} depicts four two-level atoms.
For simplicity, we assume that the incoming photon has equal probability to interact with any of the four atoms.
The probability of spontaneous absorption taking place is proportional to the number of atoms in the ground state, $N_g$.
Not surprisingly, the probability of stimulated emission is proportional to the number of atoms in the excited state, $N_e$.
All atoms are initially in the ground state, $N_g=4$.
The only interaction that is possible for an incoming photon is to get absorbed by one of the atoms.
Let's assume that atom 2 is the one that absorbs the photon and transitions to an excited state.
At time $t_1$, there is a new incoming photon.
This time $N_e=1$ and $N_g=3$, meaning both stimulated absorption and emission are possible.
However, stimulated absorption is more likely because more atoms are in the ground state.
Let's say that the photon is absorbed by atom 1.
At time $t_2$, there is a new incoming photon, and both stimulated absorption and emission are equally likely since $N_e=N_g=2$.
For the sake of this example, let's say that this photon is absorbed by atom 4, bring the totally tally to $N_e=3$ and $N_g=1$.
Finally, when another photon at time $t_3$ comes along, it has a higher chance of stimulating an emission from one of the excited atoms.
This example demonstrates that if we want to achieve light amplification, we require
\begin{equation}
    N_e > N_g.
\end{equation}
This condition is known as \textit{\textbf{population inversion}}.

It seems that we now have a way of producing an intense and highly coherent light.
There is however one final obstacle that needs to be overcome.
We have seen in Fig.~\ref{fig:5-3_population_inversion} that if $N_g > N_e$, the incoming photon is more likely to be absorbed and contribute to the population of the excited state.
On the other hand, when $N_g < N_e$, then the incoming photon is more likely to stimulate an emission from one of the excited atoms, contributing to the population of the atoms in the ground state.
This means that in the long-time limit, the population of atoms approaches an equal distribution where $N_g = N_e$, where the population is not inverted.
This demonstrates that population inversion is not possible for an ensemble of two-level atoms.

\section{Lasers II: Population inversion}
\label{sec:5-4_lasers2}

\begin{figure}[t]
    \centering
    \includegraphics[width=\textwidth]{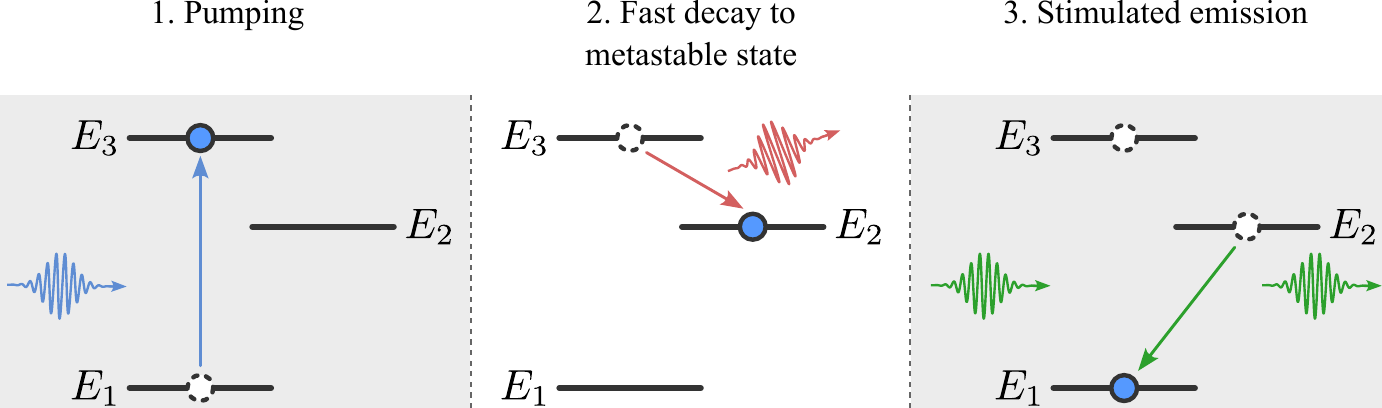}
    \caption[Laser cycle.]{Three-stage cycle consisting of pumping via stimulated absorption followed by a quick decay to the middle level via spontaneous emission. The cycle is completed by stimulated emission $E_2 \rightarrow E_1$ contributing to lasing.}
    \label{fig:5-4_three_level_atom}
\end{figure}

We finished the previous section by concluding that we cannot achieve population inversion with a ensemble of two-level atoms.
Three-level atoms on the other hand are suitable for this task as we will learn in this section.
Figure~\ref{fig:5-4_three_level_atom} shows an example of such a three-level atom and demonstrates the basic working principle of a laser.
The ground state is labelled with $E_1$, the middle excited state with $E_2$, and the top excited level with $E_3$.
We assume that the new level $E_3$ is \textbf{\emph{unstable}}\index{unstable state}, meaning that whenever the atom is excited to level $E_3$, it very quickly decays via spontaneous emission to the middle level $E_2$.
We also assume that level $E_2$ is \textbf{\emph{metastable}}\index{metastable state}, meaning the atom does not quickly decay to the ground state.

Before addressing the issue of population inversion with an ensemble of three-level atom, let's have a look at the lasing cycle.
It consists of the following three stages:
\newline
\textit{\textbf{1. Pumping:}}
We assume that the atom is initially in the ground state $E_1$.
The atom is then pumped to the excited level $E_3$ by a strong pump laser represented by the blue arrow in Fig.~\ref{fig:5-4_three_level_atom}.
Provided that the energy levels $E_2$ and $E_3$ are well separated, the pump laser has negligible probability of exciting the atom to level $E_2$.
\newline
\textit{\textbf{2. Fast decay:}}
Due to the instability of level $E_3$, the atom quickly decays to the metastable level $E_2$ represented by the red arrow in Fig.~\ref{fig:5-4_three_level_atom}.
We have learned that photons produced via spontaneous emission travel in a random direction.
These photons do not contribute to the amplification of light.
\newline
\textit{\textbf{3. Stimulated emission:}}
Finally, a photon of energy $E_2-E_1$, interacts with the atom causing it to decay to the ground state via stimulated emission.
The transition $E_2\rightarrow E_1$ is known as the \textbf{\emph{lasing transition}}\index{lasing transition}.
Photons produced via this transition do contribute to light amplification.

The lasing cycle demonstrates why we need the pump.
Its role is to make sure there are enough atoms populating the metastable level $E_2$.
Now we see how population inversion can be achieved using three-level atoms.
By using an intense pump, we decrease the population of the ground state $N_1$.
The instability of level $E_3$ and metastability of $E_2$ ensure that
\begin{equation}
    N_2 > N_1,
\end{equation}
leading to population inversion required for lasing.

\begin{figure}[t]
    \centering
    \includegraphics[width=\textwidth]{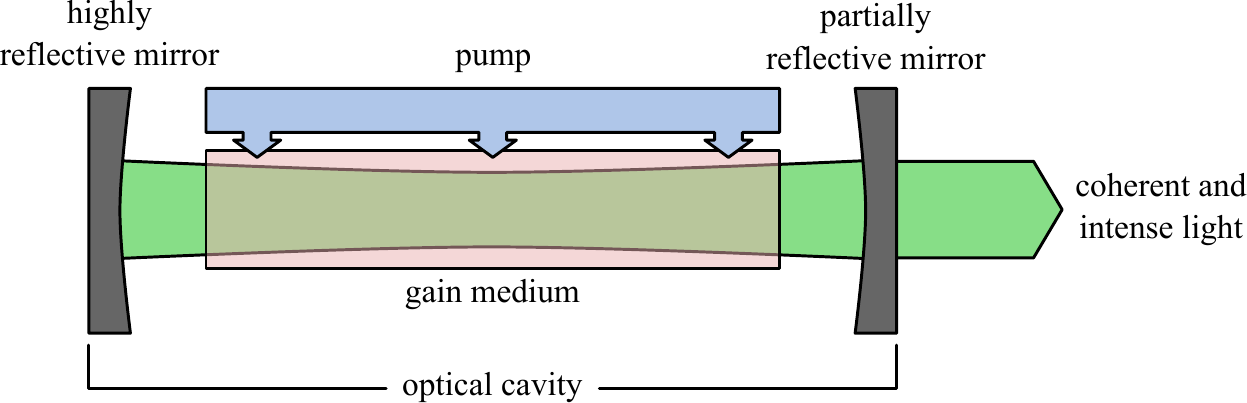}
    \caption[Laser construction.]{Basic construction of a laser.}
    \label{fig:5-4_laser_construction}
\end{figure}

Having described the basic working principle of a laser, let's see what are the main its components.
Figure~\ref{fig:5-4_laser_construction} depicts the basic construction of a laser.
The \textbf{\emph{gain medium}}\index{gain medium} is an ensemble of our three-level atoms.
Left to their devices, the atoms will mostly be found in the ground state.
Turning on the pump will begin the lasing cycle.
Initially, the excited atoms will decay via transitions $E_3 \rightarrow E_2 \rightarrow E_1$ through spontaneous emission and the produced photons will mostly be incoherent.
Some of the photons emitted from the transition $E_2 \rightarrow E_1$ will be reflected by the mirrors and remain in the \textbf{\emph{optical cavity}}\index{optical cavity}.
These photons will move through the gain medium again where they will encounter a significant population in the middle level $E_2$.
Now the atoms will decay through stimulated emission, producing coherent photons with the ones that were reflected by the mirrors.
Each time the photons get reflected from the mirrors, they stimulate further emission from the lasing transition leading to very intense coherent light inside the optical cavity.
In order to extract this light, one of the mirrors is only partially reflective and a portion of the coherent light is transmitted resulting in an intense and coherent beam of light outside the optical cavity.
Eventually, the number of photons inside the optical cavity reaches a steady state where the rate of production of new photons via stimulated emission balances the rate of loss of photons due to the partially reflective mirror.

Having learned about the fundamental working principles and basic construction of the laser, it is time to make our discussion a bit more quantitative.
In particular, we would like to have a better understanding of the following behavior.
If the pump is weak, no lasing is observed.
Once the rate at which the gain medium is pumped reaches a certain \textbf{\emph{threshold}}, lasing takes place.
In the remainder of this section, we discuss a simplified nonlinear dynamical model that captures this behavior.

The two variables of interest are the number of photons inside the optical cavity, denoted by $n$, and the number of atoms in the state $E_2$, denoted by $N_2$.
During the lasing process, the number of photons rapidly increases resulting in a large positive rate of change of the number of photons.
This rate of change is given by the time derivative of the number of photons, written $dn/dt$.
Time derivatives are often written with a dot above the changing variable, $\dot{n} \equiv dn/dt$, notation which we will use in the rest of this section as well.
This rate of change is given by the following difference,
\begin{equation}
    \dot{n} = \text{gain} - \text{loss}.
    \label{eq:5-4_n_dot_basic}
\end{equation}
The gain represents processes that contribute to the number of photons $n$.
In our case, this captures the effect of stimulated emission.
The amount of gain depends on both the number of photons $n$ as well as the population of excited atoms $N_2$.
In the absence of either, the gain vanishes, therefore we can write
\begin{equation}
    \text{gain} = G n  N_2.
    \label{eq:5-4_gain}
\end{equation}
We have introduced the gain coefficient $G>0$.
The loss represents photons escaping the optical cavity.
More photons inside the optical cavity lead to larger loss term,
\begin{equation}
    \text{loss} = k n,
    \label{eq:5-4_loss}
\end{equation}
where $k>0$ is the loss coefficient.
Substituting Eqs.~(\ref{eq:5-4_gain}) and (\ref{eq:5-4_loss}) into Eq.~(\ref{eq:5-4_n_dot_basic}), we obtain
\begin{equation}
    \dot{n} = G n N_2 - k n.
    \label{eq:5-4_n_dot_detailed}
\end{equation}

Next, we make the crucial observation that stimulated emission decreases the population $N_2$.
The more photons present in the optical cavity the more likely that stimulated emission takes place.
Equally important, we must realize that in the absence of lasing, the pump maintains a constant population $N_2 = N_0$.
This allows us to write the population of excited atoms as
\begin{equation}
    N_2 = N_0 - \alpha n,
    \label{eq:5-4_population}
\end{equation}
where $\alpha$ is the rate of stimulated emission.
Substituting Eq.~(\ref{eq:5-4_population}) into Eq.~(\ref{eq:5-4_n_dot_detailed}) leads to our final expression for the rate of change of photons inside the optical cavity,
\begin{equation}
    \dot{n} = (G N_0 - k) n - \alpha G n^2.
    \label{eq:5-4_n_dot_final}
\end{equation}

\begin{figure}
    \centering
    \begin{tikzpicture}[scale=0.8,every node/.style={scale=0.8}]
    \begin{scope}
        \draw[] (-0.2,0) -- (3,0) node[below] {$n$};
        \draw[] (0,-2) -- (0,2) node[left] {$\dot{n}$};
        \draw[blue!50, variable=\x, samples at={0,0.01,...,2.5}, very thick] plot (\x,{-0.3*\x*\x});
        \filldraw[black, draw=black] (0,0) circle (1.4pt) node[above right] {stable};
        \draw[-stealth'] (2,0) -- (1.5,0);
        \node[] at (1.5,2.5) {$N_0<k/G$};
        \node[] at (1.5,2) {no lasing};
    \end{scope}

    \begin{scope}[xshift=4.5cm]
        \draw[] (-0.2,0) -- (3,0) node[below] {$n$};
        \draw[] (0,-2) -- (0,2) node[left] {$\dot{n}$};
        \draw[blue!50, variable=\x, samples at={0,0.01,...,2.8}, very thick] plot (\x,{2.2*\x-\x*\x});
        \filldraw[white, draw=black, thick] (0,0) circle (1.4pt) node[below right, black] {unstable};
        \filldraw[black, draw=black] (2.2,0) circle (1.4pt) node[above right] {stable};
        \draw[-stealth'] (1,0) -- (1.2,0);
        \draw[-stealth'] (3,0) -- (2.6,0);
        \node[] at (1.5,2.5) {$N_0>k/G$};
        \node[] at (1.5,2) {lasing};
    \end{scope}

    \begin{scope}[xshift=9cm]
        \draw[] (-0.2,0) -- (3,0) node[below] {$N_0$};
        \draw[] (0,-2) -- (0,2) node[left] {$n^*$};
        \draw[blue!50, variable=\x, samples at={0,0.01,...,1.5}, very thick] plot (\x,0);
        \draw[blue!50, variable=\x, samples at={1.5,1.51,...,3}, very thick, dashed] plot (\x,0);
        \draw[blue!50, variable=\x, samples at={1.5,1.51,...,3}, very thick] plot (\x,\x-1.5);
        \draw[black!50, dashed] (1.5,0) -- (1.5,2);
        \node[] at (1.55,-0.4) {$k/G$};
        \node[] at (0.75,2.4) {light};
        \node[] at (0.75,2) {bulb};
        \node[] at (2.5,2) {lasing};
    \end{scope}
  
    \end{tikzpicture}
    \caption[Dynamical model of a laser.]{Solutions to the dynamical model of a laser in Eq.~(\ref{eq:5-4_n_dot_final}). In the weak-gain regime, there is only a single fixed point. }
    \label{fig:5-4_lasing}
\end{figure}

Eq.~(\ref{eq:5-4_n_dot_final}) is not easily solved analytically.
Frankly, such a solution would not be very enlightening anyway.
To gain better understanding of the dynamics between the number of photons in the optical cavity $n$ and the number of excited atoms $N_2$, we proceed by using graphical methods of analysis.
We would like to determine how the rate of change of the photon number $\dot{n}$ changes as a function of $n$ in different parameter regimes.
The main object of interest are the \textbf{\emph{fixed points}}\index{fixed points} of the dynamical system in Eq.~(\ref{eq:5-4_n_dot_final}).
Fixed points, denoted by $n^*$, are states of the dynamical system that do not change in time,
\begin{equation}
    \left.\dot{n}\right|_{n^*} = 0.
\end{equation}

In some cases, fixed points can be good approximations to the long-term solution of the dynamical system, that is $n^* = \lim_{t\rightarrow\infty} n$.
Such fixed points are known as \textbf{\emph{stable}}\index{stable fixed point} fixed points.
In other cases, the fixed point may satisfy the property of vanishing rate of change, but any small deviation from the fixed point pushes the state of the system further away from it.
Such fixed points are known as \textbf{\emph{unstable}}\index{unstable fixed points}\footnote{There are fixed points that are both stable and unstable. However, we will not encounter such exotic cases in this book.}.
We can get good intuition for these two types of fixed points by considering a ball rolling down a hill.
If the ball starts exactly at the top of the hill it will not move.
However, any small deviation from the hill's top will result in the ball starting to roll down a hill.
The top of the hill represents an unstable fixed point.
The ball will continue rolling down until it finds a valley, where it will stop and settle down.
Even if we give the ball a little push, the ball will simply roll back into the valley.
The valley represents a stable fixed point.
We will see that depending on the parameters of our simplified lasing model, Eq.~(\ref{eq:5-4_n_dot_final}) can have one or two fixed points.

\emph{Case 1: Weak pumping.}
We observe from Eq.~(\ref{eq:5-4_n_dot_final}) that when the pump is such that $N_0 < k / G$, the right-hand-side of the equation is negative, $\dot{n} < 0$, for any $n$ because the number of photons can only be non-negative.
This means that regardless of how many photons we start with, eventually they will all leak from the cavity and $n$ will always tend to zero.
The left panel of Fig.~\ref{fig:5-4_lasing} shows a plot of $\dot{n}$ versus $n$ in this weak-gain regime.
We see that the rate of change of the photon number is indeed negative as predicted.
The arrow on the horizontal axis represents the flow of $n$ as time progresses.
It always decreases and asymptotically approaches the stable fixed point $n^* = 0$, represented by the solid circle.
There is no lasing in this weak-gain regime.

\emph{Case 2: Strong pumping.}
Now let's look at the strong-gain regime when $N_0 > k / G$.
In this case, the right-hand-side of Eq.~(\ref{eq:5-4_n_dot_final}) may be positive as well as negative.
This means that for some starting values of $n(t)$ the photon number will decrease, and for others it will increase.
The middle panel of Fig.~\ref{fig:5-4_lasing} depicts this regime.
This time, $\dot{n} = 0$ has two solutions, suggesting there are two fixed points.
One of them is our old fixed point $n^*=0$, represented by the white circle in Fig.~\ref{fig:5-4_lasing}.
Unlike before, this fixed point is now unstable, meaning that any small deviation from it will result in an increase of $n$, which will flow towards a new stable fixed point at a finite value.
The unstable fixed point is represented by an empty circle in Fig.~\ref{fig:5-4_lasing} while the stable fixed point is solid.
We can observe amplification of the photon number which is a clear sign of lasing.

The last panel on the right of Fig.~\ref{fig:5-4_lasing} summarizes our analysis of Eq.~(\ref{eq:5-4_n_dot_final}) by plotting the fixed points $n^*$ and their stability.
When $N_0 < k / G$, there is only a single fixed point at $n^* = 0$.
In this regime the atoms are weakly pumped and decay via spontaneous emission producing incoherent light, just like a light bulb.
When $N_0 > k / G$, two fixed points exist.
The stable fixed point is represented by the solid line, while the unstable one is represented by the dashed line.
It is in this regime where lasing takes place.

\section{Single photons}
\label{sec:5-5_single_photons}

Lasers are excellent sources of intense coherent light and are indispensable in modern fiber-optic communication.
Pulses of laser light can be used to encode classical bits.
Presence of a pulse can encode a bit value of 1 and absence of a pulse can encode a bit value of 0.
Such encoding has a number of desired properties making it suitable for classical communication.
It is robust against \textbf{\emph{attenuation}}\index{attenuation}.
Due to the large number of photons constituting a single pulse, loss of a few of them along the fiber presents no issue, and the message remains legible to a decoder at the destination.
Even if the pulse travels a large distance along the fiber and attenuation becomes a problem, the message can be read and \textbf{\emph{amplified}}\index{amplification} along the way.
Finally, producing these pulses of light is relatively easy and therefore \textbf{\emph{reliable}}.

On the other hand, intense pulses of laser light cannot be easily put in a quantum superposition and they cannot be entangled with other systems.
This makes the above encoding scheme entirely unsuitable for quantum communication.
In order to exploit the full toolbox of quantum mechanics, we have to use single photons.
In this section, we outline the three basic methods of producing single photons.

\begin{figure}
    \centering
    \includegraphics[width=0.9\textwidth]{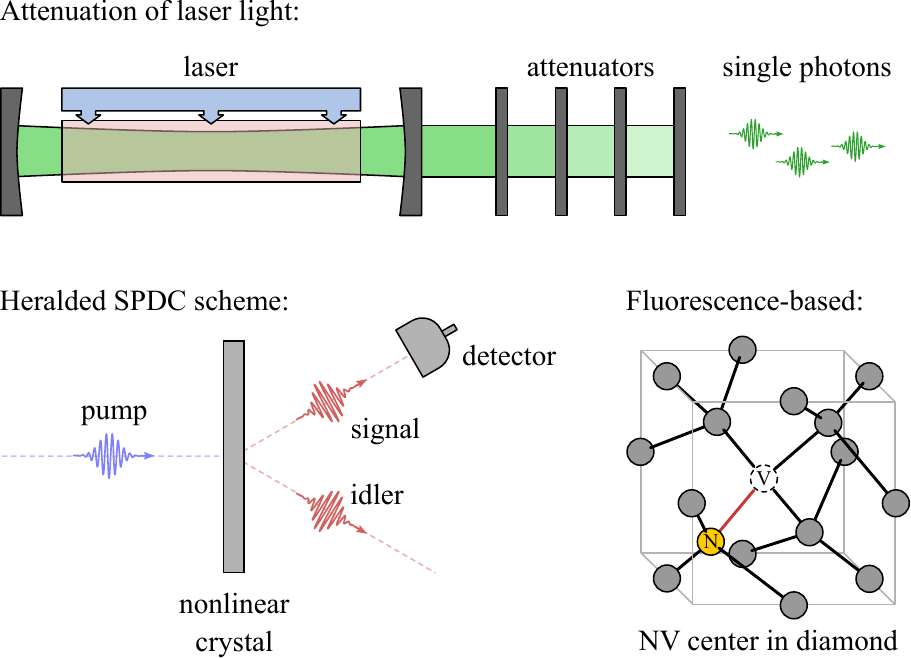}
    \caption[Single photon sources.]{Various approaches to producing single photons.}
    \label{fig:5-5_single_photons}
\end{figure}

The first method relies on gradual \textbf{\emph{attenuation of laser light}} as shown in the top panel of Fig.~\ref{fig:5-5_single_photons}.
The intense laser pulse exiting the optical cavity is passed through a series of attenuator plates.
Each plate transmits only a portion of the laser light and therefore decreases the pulses intensity.
This process is repeated until the average energy of the pulse, $\lambda$, is less than that of a single photon.

This approach to producing single photons is conceptually simple but suffers from a number of serious issues.
Firstly, this source of single photons is probabilistic.
The number of photons contained in the pulse after the attenuation process follows a Poissonian distribution.
If the average number of photons per pulse is $\lambda = 0.1$, then the probability of there being zero photons is 90.5\%, probability of a single photon is 9.1\%, and probability of two photons is 0.4\%.
Most of the time, the pulse becomes completely attenuated and the process of generation of single photons fails.
There is also a finite probability that this method generates more than one photon, which is highly undesirable in quantum communication.

The second issue is quite technical but very important, and has to do with something called the \textbf{\emph{second-order correlation function}}\index{second-order correlation function} $g^{(2)}(\tau)$.
The full derivation of this function is well beyond the scope of this book but its meaning is not so complicated.
The second-order correlation function $g^{(2)}(\tau)$ tells us how likely it is to detect two photons, one at time $t$, and the other one at time $t+\tau$.
Particular case of interest in our discussion is when the time interval between the detection events vanishes, $\tau = 0$.
For ideal single photons, the second-order correlation function $g^{(2)}(0) = 0$.
This is quite intuitive.
If there is only a single photon, then it should be impossible to detect a second photon when we detect the first one.
We say that a single-photon source produces light which is \textbf{\emph{anti-bunched}}\index{anti-bunched light}.
For realistic sources, as long as $g^{(2)}(0) < 1$, we say the light is anti-bunched and possesses quantum properties.
On the other hand, for classical light $g^{(2)}(0) \geq 1$, and when it is strictly larger than unity, we say the light is \textbf{\emph{bunched}}\index{bunched light}.
Laser light is neither bunched nor anti-bunched as $g^{(2)}(0)=1$.
Since the single photons produced by attenuation started as a laser they are not anti-bunched.
This is a problem because numerous protocols in quantum computation require anti-bunched light.

The second method of producing single photons is via \textbf{\emph{heralded spontaneous parametric down-conversion}}\index{heralded single-photon source}, shown in the lower left panel of Fig.~\ref{fig:5-5_single_photons}.
We have discussed the basics of SPDC in Section~\ref{sec:4-4_spdc}, where we used it to produce entangled pairs of photons.
Conversion of a single high energy photon into two photons of lower energy can be also used as a source of single photons by using the fact that the two produced photons have a well-defined direction of travel.
We can detect one of these photons in order to herald presence of a single photon in the other mode.
This scheme is still probabilistic as the SPDC is a very rare process, meaning we have to try many times before we produce a heralded single photon.
In some cases this rarity can be an advantage because once we detect the signal photon we have a very high probability that the idler mode contains only a single photon.
The properties of the detector also affect this scheme.
Real detectors are not 100\% efficient, meaning sometimes a signal photon goes undetected.
This might be annoying but really it just means that we have to try the whole process again.
More importantly, real detectors have a finite \textbf{\emph{dark count rate}}\index{dark counts}, meaning occasionally they register a detection event even in the absence of a signal photon.
This ``heralds'' a non-existent idler photon and has a deleterious effect on any quantum communication protocol relying on single photons produced by this scheme.

The last approach to producing single photons is through \textit{\textbf{fluorescence}} of atoms and molecules.
The idea is basically the same as the one we have been discussing throughout this Chapter.
A physical system with discrete levels is first excited to a higher energy level and later transitions back to a lower energy level by emitting a single photon.
One promising source of single photons that is currently under intense research focus is \textbf{\emph{nitrogen-vacancy centers in diamond}}\index{nitrogen-vacancy}, shown in the lower right panel of Fig.~\ref{fig:5-5_single_photons}.
The NV center consists of a nitrogen atom $N$ located next to a vacant site $V$ of a diamond lattice.
This vacancy is used to trap an electron whose \textbf{\emph{spin}}\index{spin} then acts as a qubit~\footnote{You should have already studied the basics of atomic structure, but for the record, electrons generally have two states that can be described as ``spin up'' and ``spin down'', which we write as \ket{\uparrow} and \ket{\downarrow}.}.
The electron qubit can be manipulated optically, can retain its quantum properties even at room temperatures, and acts as an excellent source of single photons with nearly vanishing second-order correlation function $g^{(2)}(0)$.
All these properties make NV centers in diamond very promising physical systems for quantum communication.

\newpage
\begin{exercises}

\exer{
\emph{Laser model.}
We introduced our mathematical toy model of a laser, where the rate of change of the number of photons inside the optical cavity, $n$, is given by
\begin{equation}
    \dot{n} = (G N_0 - k) n - \alpha G n^2,
\end{equation}
where $G$ is the gain coefficient, $N_0$ is the number of excited atoms maintain by the pump in the absence of lasing, $k$ is photon loss coefficient, and $\alpha$ is the rate of stimulated emission.
We gained intuition into the dynamics of this model via graphic tools.
Here, let's solve it mathematically.
\subexer{
By solving $\dot{n} = 0$, show that the there are two fixed points,
\begin{equation}
    n_1^* = 0, \qquad \text{and }\qquad n_2^*=\frac{GN_0 - k}{\alpha}.
\end{equation}
Determine under what conditions are both of these fixed points physical.
}
\subexer{
Stability of the fixed points can be determined by \emph{linearizing} around the fixed points\footnote{This is the same idea used in calculus to determine whether a stationary point of a function is a maximum or a minimum.}.
The trick is to look at the value of the derivative of the rate of change with respect to the photon number at the fixed point.
If it is positive, the fixed point unstable.
If it is negative, the fixed point is stable,
\begin{equation}
    \left.\frac{d\dot{n}}{dn}\right|_{n=n^*}\longrightarrow\begin{cases}
        <0, \quad\text{when } n^*\text{ is stable}, \\
        >0, \quad\text{when } n^*\text{ is unstable}.
    \end{cases}
\end{equation}
Determine the stability of all fixed points.\\
(\emph{Hint}: For one of the fixed points you will find that $d\dot{n}/dn|_{n=n^*}=0$. This means that linearization around the fixed point is not enough to determine the stability of the fixed point. Use graphical methods instead.)
}
}

\exer{
\emph{Single photons via attenuation.}
How many photons per second are in one milliwatt of laser power at a wavelength of 1550nm?  In order to solve this problem, you will need to know that the energy in a single photon is
\begin{equation}
    E = \frac{hc}{\lambda}
\end{equation}
where $\lambda$ is the wavelength of the light and $h$ is Planck's constant,
\begin{equation}
    h = 6.62607015\times 10^{-34} \frac{\text{J}}{\text{Hz}}.
\end{equation}
(You may need to look up the relationship between energy in joules (J) and power in watts.)
\subexer{
Assume an attenuator plate blocks 99\% of the incoming photons.
Find the average number of photons per second after the light passes through (i) 1 attenuator plate, (ii) 2 attuantor plates, (iii) $m$ attenuator plates.
}
\subexer{
How many attenuator plates are needed in order for the average number of photons per second to drop below 1?
}
}

\end{exercises}

\chapter{Interference}
\label{sec:6_interference}

In this chapter, we will discuss the fundamental phenomenon of interference.
Most of you have very likely encountered interference of waves in either water, light or even sound.
We begin this chapter with a mathematical description of what happens when two waves of different frequencies interfere together.
After that, we will move to interference at the quantum level, using single photons.
Finally, we will conclude this chapter by learning that interference can occur at the level of individual qubits as well.

\section{Constructive and destructive interference}
\label{sec:6-1_constructive_and_destructive_interference}

Good starting point is to settle on the notation that will be used in this chapter.
Let's consider a wave oscillating at a single frequency, and which is propagating in time along a single dimension denoted by coordinate $x$,

\begin{equation}
    \Psi(x,t) = A \sin [\omega t-(k x+\phi_0)].
    \label{eq:simple_wave}
\end{equation}

We denote the wave with $\Psi(x,t)$.
The \textbf{\emph{amplitude}}\index{amplitude}, denoted by $A$, is the maximum value that $\Psi(x,t)$ takes.
It signifies how how much the wave is displaced from its rest state.
The symbol $\omega$ (small Greek letter ``omega'') is the \textbf{\emph{angular frequency}}\index{angular frequency}, and it determines how quickly the wave propagates in time.
Time is denoted by $t$.
The symbol $k$ is known as the \textbf{\emph{wave number}}\index{wave number}.
It tells us how quickly the wave oscillates in space, and therefore is related to the wavelength of the oscillations.
The wave number also tells us about the direction in which the wave propagates.
Finally, $\phi_0$ (small Greek letter ``phi'') is called the initial phase of oscillations.
The angular frequency $\omega$, and the wave number $k$, are related to the period of oscillations $T$ and the wavelength $\lambda$, respectively,
\begin{equation}
    \omega=\frac{2 \pi}{T}, \qquad k=\frac{2 \pi}{\lambda},
    \label{eq:freq-period_k-wavelength}
\end{equation}

Now is a good time to have a look at some simple examples of the simple wave described by Eq.~(\ref{eq:simple_wave}).
Let's freeze the wave at time $t=0$, and for convenience we also set $\phi_0 = 0$, and only vary the wave number $k$.
The blue wave in Fig.~\ref{fig:two-waves} represents the case when $k=1$, whereas the orange one represents the case when $k=0.5$.
The wavelength is the shortest distance between two points of the where it begins to repeat itself.
We can see that halving the wave number $k$ doubles the wavelength.
In other words, increasing the wave number ``compresses the oscillations'', while decreasing the wave number ``stretches'' them.

\begin{figure}[t]
    \centering
    \resizebox {0.6\textwidth} {!} {
    \begin{tikzpicture}
        \begin{axis}[xtick=\empty,
                    ytick={-1,1},
                    yticklabels={$-A$,$A$},
                    axis lines=middle,
                    enlarge x limits={abs=0.2},
                    enlarge y limits={abs=0.3},
                    x label style={anchor=north},
                    y label style={anchor=east},
                    xlabel={$x$},
                    ylabel={$\Psi(x,0)$},
                    width=9cm,
                    height=6cm]
        \addplot[domain=-1:18,samples=100,smooth,myblue,thick] {sin(deg(-x))};
        \addlegendentry{$k=1$}
        \addplot[domain=-1:18,samples=100,smooth,myorange,thick] {sin(deg(-0.5*x))};
        \addlegendentry{$k=0.5$}
        \node[circle,fill,inner sep=1.5pt,myblue] at (axis cs:3*pi/2,1) {};
        \node[circle,fill,inner sep=1.5pt,myblue] at (axis cs:7*pi/2,1) {};
        \draw[dashed, blue!80, thick] (axis cs:3*pi/2,1) -- node[above] {$\lambda=2\pi/1.0$} (axis cs:7*pi/2,1);
        \node[circle,fill,inner sep=1.5pt,myorange] at (axis cs:pi,-1) {};
        \node[circle,fill,inner sep=1.5pt,myorange] at (axis cs:5*pi,-1) {};
        \draw[dashed, orange!80, thick] (axis cs:pi,-1) -- node[below] {$\lambda=2\pi/0.5$} (axis cs:5*pi,-1);
        \end{axis}
    \end{tikzpicture}
    }
    \caption[Same frequency, different wave numbers.]{Two independent waves with $k=1$ (blue), and $k=0.5$ (orange). The time is frozen at $t=0$, and we set the initial phase to be $\phi_0=0$ for convenience.}    
    \label{fig:two-waves}
\end{figure}

Let's consider what happens when we fix the wave number but we vary the initial phase, as shown in Fig.~\ref{fig:phase-diff-waves}.
In this case, we have shifted the second wave by an initial phase of $\phi_0=\pi/2$.
Varying the initial phase has the effect of shifting the wave along the $x$ coordinate.
The wavelength of the wave remains unaffected.

\begin{figure}[t]
    \centering
    \resizebox {0.6\textwidth} {!} {
    \begin{tikzpicture}
        \begin{axis}[xtick=\empty,
                    ytick={-1,1},
                    yticklabels={$-A$,$A$},
                    axis lines=middle,
                    enlarge x limits={abs=0.2},
                    enlarge y limits={abs=0.3},
                    x label style={anchor=north},
                    y label style={anchor=east},
                    xlabel={$x$},
                    ylabel={$\Psi(x,0)$},
                    width=9cm,
                    height=6cm]
        \addplot[domain=-1:18,samples=100,smooth,myblue,thick] {sin(deg(-x))};
        \addlegendentry{$\phi_0=0$}
        \addplot[domain=-1:18,samples=100,smooth,myorange,thick] {sin(deg(-x+pi/2))};
        \addlegendentry{$\phi_0=\pi/2$}
        \node[circle,fill,inner sep=1.5pt,black!80] at (axis cs:3*pi/2,1) {};
        \node[circle,fill,inner sep=1.5pt,black!80] at (axis cs:2*pi,1) {};
        \draw[dashed, black!80, thick] (axis cs:3*pi/2,1) -- node[above] {$\pi/2$} (axis cs:2*pi,1);
        \end{axis}
    \end{tikzpicture}
    }
    \caption[Same wavelength, different initial phase.]{Two waves with the same wavelength and frequency but different initial phases, $\phi_0 = 0$ (blue), $\phi_0=\pi/2$ (orange).}
    \label{fig:phase-diff-waves}
\end{figure}

Finally, let's add time dependence into the picture, as shown in Fig.~\ref{fig:propagating-waves}, where we have set the wave number $k=1$, and the initial phase $\phi=0$.
The angular frequency of the blue wave is set to $\omega=0.1$.
Figure~\ref{fig:propagating-waves} plots the wave at three different times $t_1$, $t_2$, $t_3$, where $t_1<t_2<t_3$.
The angular frequency of the orange wave is set to $\omega=0.2$.
We observe that the wave is propagating faster as it covers longer distance in the same amount of time.
We can compare Fig.~\ref{fig:propagating-waves} with Fig.~\ref{fig:two-waves} to better appreciate the difference between the angular frequency $\omega$ and the wave number $k$.
The wave number $k$ controls the ``shape'' of the wave.
More specifically, it stretches or contracts the oscillations by determining their wavelength.
On the other hand, the angular frequency $\omega$ controls how much the whole wave shifts in time.
We will see in Sec~\ref{sec:6-2_phase_group_velocity} that both $\omega$ and $k$ are related to the velocity at which the wave propagates.

\begin{figure}[t]
   \centering
    \resizebox {0.6\textwidth} {!} {
    \begin{tikzpicture}
        \begin{axis}[name=plotA,
                    xtick=\empty,
                    ytick={-1,1},
                    yticklabels={$-A$,$A$},
                    axis lines=middle,
                    enlarge x limits={abs=0.2},
                    enlarge y limits={abs=0.3},
                    x label style={anchor=north},
                    y label style={anchor=east},
                    xlabel={$x$},
                    ylabel={$\Psi(x,t)$},
                    width=9cm,
                    height=6cm]
        \addplot[domain=-1:18,samples=100,smooth,myblue,thick] {sin(deg(0.1*15-x))};
        \addplot[domain=-1:18,samples=100,smooth,myblue,thick,opacity=0.2] {sin(deg(0.1*0-x))};
        \addplot[domain=-1:18,samples=100,smooth,myblue,thick,opacity=0.6] {sin(deg(0.1*7.5-x))};
        \addlegendentry{$\omega=0.1$}
        \node[myblue,opacity=0.2] at (axis cs:3*pi/2,1.15) {$t_1$};
        \node[myblue,opacity=0.6] at (axis cs:3*pi/2+0.1*7.5,1.15) {$t_2$};
        \node[myblue] at (axis cs:3*pi/2+0.1*15,1.15) {$t_3$};
        \end{axis}
    
        \begin{axis}[name=plotB,
                    at=(plotA.outer south west),
                    anchor=outer north west,
                    xtick=\empty,
                    ytick={-1,1},
                    yticklabels={$-A$,$A$},
                    axis lines=middle,
                    enlarge x limits={abs=0.2},
                    enlarge y limits={abs=0.3},
                    x label style={anchor=north},
                    y label style={anchor=east},
                    xlabel={$x$},
                    ylabel={$\Psi(x,t)$},
                    width=9cm,
                    height=6cm]
        \addplot[domain=-1:18,samples=100,smooth,myorange,thick] {sin(deg(0.2*15-x))};
        \addplot[domain=-1:18,samples=100,smooth,myorange,thick,opacity=0.2] {sin(deg(0.2*0-x))};
        \addplot[domain=-1:18,samples=100,smooth,myorange,thick,opacity=0.6] {sin(deg(0.2*7.5-x))};
        \addlegendentry{$\omega=0.2$}
        \node[myorange,opacity=0.2] at (axis cs:3*pi/2,1.15) {$t_1$};
        \node[myorange,opacity=0.6] at (axis cs:3*pi/2+0.2*7.5,1.15) {$t_2$};
        \node[myorange] at (axis cs:3*pi/2+0.2*15,1.15) {$t_3$};
        \end{axis}
    \end{tikzpicture}
    }
    \caption[Propagation of waves in time.]{Propagation of waves in time. We observe that the wave with angular frequency $\omega=0.1$ (blue) propagates more slowly compared to the wave with angular frequency $\omega=0.2$ (orange).}
    \label{fig:propagating-waves}    
\end{figure}

Having gained some intuition how the various parameters in Eq.~(\ref{eq:simple_wave}) affect the behavior of the wave, it is time to discuss what happens when two waves meet.
Let's consider a special case of two waves, $\Psi_1(x,t)$ and $\Psi_2(x,t)$, each with a different amplitude and initial phase, but both having the same frequency,
\begin{align}
    \Psi_1(x,t) & = A_1 \sin \left(\omega t+\alpha_1\right), \quad\text{where } \alpha_1=-\left(k x+\phi_1\right), \\
    \Psi_2(x,t) & = A_2 \sin \left(\omega t+\alpha_2\right), \quad\text{where } \alpha_2=-\left(k x+\phi_2\right).
    \label{eq:superposition}
\end{align}
These two waves produce a new wave given by their sum,
\begin{equation}
    \Psi(x,t) = \Psi_1(x,t) + \Psi_2(x,t).
    \label{eq:wave_superposition}
\end{equation}
This is known as the \textbf{\emph{principle of superposition}}\index{principle of superposition},
and it should look familiar.
We have in fact encountered this principle in Chapter~\ref{sec:2_quantum_states}, where we talked about superposition of two quantum state vectors.

The superposition in Eq.~(\ref{eq:wave_superposition}) will have the same form as its constituent waves,
\begin{equation}
    \Psi(x,t) = A \sin (\omega t-\alpha).
    \label{eq:resulting_wave}
\end{equation}
The question that we would like to answer is to determine how the amplitudes and initial phases of $\Psi_1(x,t)$ and $\Psi_2(x,t)$ affect the amplitude and the phase of the superpostion.
Using the trigonometric identity $\sin (a + b)=\sin a \cos b +\cos a \sin b$, we can rewrite Eq.~\ref{eq:wave_superposition} as
\begin{align}
    \Psi(x,t) & = A_1 \left( \sin\omega t \cos\alpha_1 + \cos\omega t \sin \alpha_1 \right) \nonumber\\
    & + A_2 \left( \sin\omega t \cos\alpha_2 + \cos\omega t \sin\alpha_2 \right).
\end{align}
We can group the time dependent terms together to obtain the following, 
\begin{equation}
    \Psi(x,t) = ( \underbrace{A_1 \cos\alpha_1 + A_2 \cos\alpha_2}_{\equiv A\cos\alpha} ) \sin \omega t + ( \underbrace{A_1 \sin\alpha_1 + A_2 \sin\alpha_2}_{\equiv A\sin\alpha} ) \cos \omega t.
    \label{eq:resulting_wave_2}
\end{equation}
We observe that the resulting wave $\Psi(x,t)$ does indeed oscillate with angular frequency $\omega$ as we anticipated in Eq.~(\ref{eq:resulting_wave}).
The coefficients in front of the time-dependent terms are functions of the original two waves as they depend on $A_1$, $A_2$, and $\alpha_1$, $\alpha_2$.

It would be nice to obtain a cleaner expression for the amplitude of the resulting wave $A$, and its phase $\alpha$.
We can do that by using the trigonometric identity $\cos^2\theta + \sin^2\theta=1$,
\begin{equation}
    A^2 = A^2 (\cos^2\alpha + \sin^2\alpha) = A^2\cos^2\alpha + A^2 \sin^2\alpha.
\end{equation}
We can now substitute the definition for $A\cos\alpha$ and $A\sin\alpha$ from Eq.~(\ref{eq:resulting_wave_2}), which after some rearranging yields the following expression for the square of the amplitude $A$,
\begin{equation}
    A^2 = A_1^2 + A_2^2 + 2 A_1 A_2 \cos(\alpha_2-\alpha_1).
    \label{eq:superposition-amplitude}
\end{equation}
The phase $\alpha$ can be obtained in the following way,
\begin{equation}
    \tan \alpha = \frac{A\sin\alpha}{A\cos\alpha} = \frac{A_1\sin\alpha_1 + A_2\sin\alpha_2}{A_1\cos\alpha_1 + A_2\cos\alpha_2}.
    \label{eq:superposition-alpha}
\end{equation}

We started with two initial waves that only differed in their amplitudes and initial phases.
Using the principle of superposition, we determined the mathematical description of the resulting wave $\Psi(x,t)$.
From Eq.~(\ref{eq:superposition-amplitude}), we can see that the amplitude of $\Psi(x,t)$ is determined by the amplitudes and phases of $\Psi_1(x,t)$ and $\Psi_2(x,t)$.
The last term in Eq.~(\ref{eq:superposition-amplitude}) is known as the \textbf{\emph{interference term}}\index{interference term}.
It oscillates as a function of the phase difference $\delta\equiv\alpha_2 - \alpha_1$, resulting in either positive or negative contribution to the amplitude of $\Psi(x,t)$.
When the phase difference is such that $\cos\delta>1$, the interference term contributes by increasing the amplitude $A$, which is known as \textbf{\emph{constructive interference}}\index{constructive interference}.
On the other hand, when $\cos\delta<1$, the interference term contributes negatively, situation known as \textbf{\emph{destructive interference}}\index{destructive interference}.

Let's look at the example, where the amplitudes of the initial two waves are equal, $A_1=A_2$, but their wave numbers are different.
This difference results in a phase difference, allowing us to observe both constructive and destructive interference.
The two waves have the following form,
\begin{equation}
    \Psi_1(x,t) = \sin (\omega t - k_1x), \qquad \Psi_2(x,t) = \sin (\omega t - k_2 x).
\end{equation}
The interference of these two waves is plotted in Fig.~\ref{fig:interference_example}, where we set $k_1=1.0$ (blue) and $k_2=0.8$ (orange).
The resulting wave $\Psi(x,t)$ is shown in green.
We observe that when the phase difference $\delta$ is small, which occurs in the region where $x$ is small, the two waves interfere constructively.
The amplitude of the resulting wave almost reaches all the way to $2A$.
In the region where $x$ is large however, the phase difference also increases.
We also say that the waves are ``out of phase''.
In this region, the interference is destructive.
\begin{figure}[t]
    \centering
    \resizebox {0.6\textwidth} {!} {
    \begin{tikzpicture}
        \begin{axis}[xtick=\empty,
                    ytick={-2,-1,1,2},
                    yticklabels={$-2A$,$-A$,$A$,$2A$},
                    axis lines=middle,
                    enlarge x limits={abs=0.2},
                    enlarge y limits={abs=0.8},
                    x label style={anchor=north},
                    y label style={anchor=east},
                    xlabel={$x$},
                    ylabel={$\Psi(x,t)$},
                    width=9cm,
                    height=6cm]
        \addplot[domain=-1:18,samples=100,smooth,myblue,thick,opacity=0.5] {sin(deg(-x))};
        \addlegendentry{$k_1=1.0$}
        \addplot[domain=-1:18,samples=100,smooth,myorange,thick,opacity=0.5] {sin(deg(-0.8*x))};
        \addlegendentry{$k_2=0.8$}
        \addplot[domain=-1:18,samples=100,smooth,mygreen,very thick] {sin(deg(-x))+sin(deg(-0.8*x))};
        \end{axis}
    \end{tikzpicture}
    }
    \caption[Superposition of two waves.]{Interference of two waves with different wave numbers.}
    \label{fig:interference_example}    
\end{figure}

\section{Phase and group velocities}
\label{sec:6-2_phase_group_velocity}

In this section, we will investigate the speed at which a wave propagates.
We will begin with case of waves with single frequency before moving onto more complicated scenario resulting from interference of multiple waves.

\emph{Phase velocity.}
First, let's consider a single wave of a single frequency $\omega$ propagating through space.
We saw in the previous section that we can write down such a wave as
\begin{equation}
    \Psi(x,t) = A \sin(\omega t - kx - \phi_0),
\end{equation}
where the phase at a particular point in time $t$ and space $x$ is $\theta(x,t)=\omega t-k x-\phi_0$.
The speed of propagation can be determined by inspecting any point on the wave.
This point is characterized by constant phase $\theta(x,t)$.
How fast does this point of constant phase propagate?
We need to differentiate the expression for phase with respect to time and set it equal to zero (since the phase does not change in time),
\begin{equation}
    \frac{d\theta(x,t)}{d t} = \omega - k\frac{dx}{dt} = 0.
\end{equation}
The rate of change of the position, $dx/dt$, is exactly the speed of propagation of the point of constant phase.
This quantity is called the \textbf{\emph{phase velocity}}\index{phase velocity},
\begin{equation}
    v_p \equiv \frac{dx}{dt}, \longrightarrow v_p = \frac{\omega}{k}.
    \label{eq:6-2_phase_velocity}
\end{equation}
We see that in the simple case of a single-frequency wave, the phase velocity is given as the ratio of the angular velocity $\omega$ and the wave number $k$.

\begin{figure}[t]
    \centering
    \resizebox {0.6\textwidth} {!} {
    \begin{tikzpicture}
        \begin{axis}[xtick=\empty,
                    ytick={-2,-1,1,2},
                    yticklabels={$-2A$,$-A$,$A$,$2A$},
                    axis lines=middle,
                    enlarge x limits={abs=0.2},
                    enlarge y limits={abs=0.3},
                    x label style={anchor=north},
                    y label style={anchor=east},
                    xlabel={$x$},
                    ylabel={$\Psi(x,t)$},
                    width=9cm,
                    height=6cm]
        \addplot[domain=-1:18,samples=100,smooth,myblue,thick] {sin(deg(x))};
        \node[circle,fill,inner sep=1.5pt,black!80] at (axis cs:5*pi/2,1) {};
        \node[black!80] at (axis cs:5*pi/2,1.15) {$P_1$};
        \node[circle,fill,inner sep=1.5pt,black!80] at (axis cs:9*pi/2,1) {};
        \node[black!80] at (axis cs:9*pi/2,1.15) {$P_2$};
        \draw[dashed, black!80, thick] (axis cs:5*pi/2,1) -- node[above] {$\lambda$} (axis cs:9*pi/2,1);
        \end{axis}
    \end{tikzpicture}
    }   
    \caption[Phase velocity.]{The black points represent points of same phase and are separated by the wavelength $\lambda$.}
    \label{fig:6-2_phase_velocity}
\end{figure}

We can check that the expression for phase velocity in Eq.~(\ref{eq:6-2_phase_velocity}) makes sense.
Consider one of the peaks on the wave in Fig.~\ref{fig:6-2_phase_velocity}, say point $P_1$.
This point will move to point $P_2$, covering the distance of one wavelength $\lambda$.
This represents one full oscillation of the wave, therefore the time taken to cover this distance is one period $T$.
The wave propagates at constant speed $v$, given by the ratio $\lambda/T$.
Using Eq.~(\ref{eq:freq-period_k-wavelength}), we can express this ratio in terms of the angular frequency and the wave number,
\begin{equation}
    v = \frac{\frac{2\pi}{k}}{\frac{2\pi}{\omega}} = \frac{\omega}{k} = v_p,
\end{equation}
which agrees with the phase velocity of Eq.~(\ref{eq:6-2_phase_velocity}).

\emph{Group velocity.}
How is the speed of the wave affected when the wave itself is a superposition of multiple single-frequency waves?
Let's take things slow at first by considering only two interfering waves.,
\begin{equation}
    \Psi_1(x,t) = A \sin (\omega_1 t - k_1 x), \qquad \Psi_2(x,t) = A \sin (\omega_2 t - k_2 x).
\end{equation}
We have assumed that the two waves have same amplitude, and we set the initial phase to $\phi_0=0$ for convenience.
\begin{figure}[t]
    \centering
    \resizebox {0.9\textwidth} {!} {
    \begin{tikzpicture}
        \begin{axis}[xtick=\empty,
                    ytick={-2,2},
                    yticklabels={$-2A$,$2A$},
                    axis lines=middle,
                    enlarge x limits={abs=2},
                    enlarge y limits={abs=0.6},
                    x label style={anchor=north},
                    y label style={anchor=east},
                    xlabel={$x$},
                    ylabel={$\Psi(x,t)$},
                    width=12cm,
                    height=6cm]
        \addplot[domain=-1:100,samples=500,smooth,myblue,thick] {sin(deg(-2*x))+sin(deg(-2.1*x))};
        \addplot[domain=-1:100,samples=500,smooth,myorange,thick,dashed] {2*cos(deg(-0.5*(2.0-2.1)*x))};
        \addplot[domain=-1:100,samples=500,smooth,myorange,thick,dashed] {-2*cos(deg(-0.5*(2.0-2.1)*x))};
        \end{axis}
    \end{tikzpicture}
    }   
    \caption[Group velocity.]{Superposition of two sinusoidal waves. The fast oscillations (blue solid line) are modulated by a slowly-varying envelope (orange dashed line).}
    \label{fig:6-2_superposition}
\end{figure}
Interference of these two waves results in the following superposition, 
\begin{align} 
    \Psi(x,t) & = \Psi_1(x,t) + \Psi_2(x,t) \nonumber\\
     & = A\left[\sin \left(\omega_{1} t-k_{1} x\right)+\sin \left(\omega_{2} t-k_{2} x\right)\right] \nonumber\\
    & = 2 A \underbrace{\sin \left( \frac{\omega_{1}+\omega_{2}}{2} t-\frac{k_{1}+k_{2}}{2} x \right)}_{\text{fast oscillations}} \underbrace{ \cos \left( \frac{\omega_{1}-\omega_{2}}{2} t-\frac{k_{1}-k_{2}}{2} x \right)}_{\text{slow oscillations}}.
    \label{eq:fast_slow_oscillations}
\end{align}
We see that the superposition can be split into two terms.
The first term captures the fast oscillations of the superposition at frequency $(\omega_1+\omega_2)/2$.
These fast oscillations are pictured in blue in Fig.~\ref{fig:6-2_superposition}.
Similar to the case of a single-frequency wave, we can ask the question at what speed a point of constant phase travels to obtain the \textit{\textbf{phase velocity}} for the superposition.
This information is contained in the fast-oscillation term of Eq.~(\ref{eq:fast_slow_oscillations}),
\begin{equation}
    v_p = \frac{\omega_1 + \omega_2}{k_1 + k_2}.
\end{equation}
These fast oscillations are modulated by a slowly-varying envelope as seen in by the oranged dashed line in Fig.~\ref{fig:6-2_superposition}.
This is captured by the second term in Eq.~(\ref{eq:fast_slow_oscillations}).
The frequency of these slow oscillations is $(\omega_1-\omega_2)/2$.
This term also gives rise to \textit{\textbf{group velocity}},
\begin{equation}
    v_g = \frac{\omega_1 - \omega_2}{k_1 - k_2}.
\end{equation}
The group veocity tells us how fast the entire envelope is propagating, rather than just a single point of constant phase.
The phase and group velocities of a superposition are, in general, different.
There are scenarios where one is larger than the other but also where they are equal,
\begin{equation}
    v_p > v_g, \quad v_p = v_g, \quad v_p < v_g.
\end{equation}
Real wave packets and signals are never single-frequency waves but rather results of a superposition of a number of single-frequency components.
When we talk about the speed of such a wave, we usually refer to the group velocity.
It is also the group velocity that tells us how quickly a signal carries information.

\begin{figure}[t]
    \centering
    \resizebox {0.9\textwidth} {!} {
    \begin{tikzpicture}
        \begin{axis}[xtick=\empty,
                    ytick={-11,11},
                    yticklabels={$-11A$,$11A$},
                    axis lines=middle,
                    enlarge x limits={abs=2},
                    enlarge y limits={abs=4},
                    x label style={anchor=north},
                    y label style={anchor=east},
                    xlabel={$x$},
                    ylabel={$\Psi(x,0)$},
                    width=12cm,
                    height=6cm]
        \addplot[domain=-1:100,samples=500,smooth,myblue,thick] {sin(deg(-2*x))+sin(deg(-2.1*x))+sin(deg(-2.2*x))+sin(deg(-2.3*x))+sin(deg(-2.4*x))+sin(deg(-2.5*x))+sin(deg(-2.6*x))+sin(deg(-2.7*x))+sin(deg(-2.8*x))+sin(deg(-2.9*x))+sin(deg(-3.0*x))};
        \end{axis}
    \end{tikzpicture}
    }   
    \caption[Superposition of more than two waves.]{Superposition of more than two sinusoidal waves resulting in a pulse.}
    \label{fig:6-2_pulse}
\end{figure}
To complete our discussion, let's look at the case when we have more than two single-frequency components interfering together.
Let's say we have a superposition of the following form,
\begin{equation}
    \Psi(x,t) = A [\sin(-2.0x) + \sin(-2.1x) + \ldots + \sin(-3.0x)].
    \label{eq:6-2_pulse}
\end{equation}
We have assumed that we are looking at the superposition frozen in time at $t=0$, hence the simple form.
The resultant superposition is pictured in Fig.~\ref{fig:6-2_pulse}.
Unlike before, we can observe long regions where destructive interference results in almost no disturbance.
Then the constructive interference kicks in and we observe a short \textbf{\emph{pulse}}\index{pulse}.
It is these pulses that can be used to carry information at the group velocity of the superposition.

\section{Interference with single photons}
\label{sec:6-3_interference_single_photons}

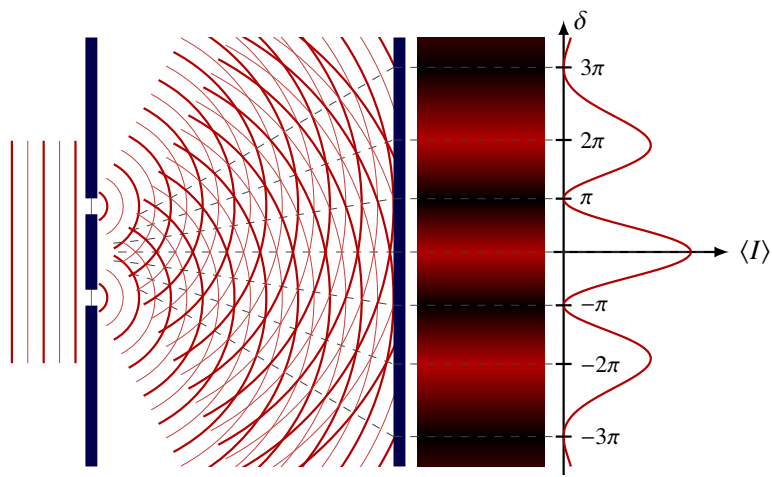
\begin{figure}[H]
   \centering

    \resizebox {0.95\textwidth} {!} {
    \begin{tikzpicture}[nodal/.style={black!75,dashed,very thin},
            declare function={
            xnode(\n,\dn,\lam,\f) = \lam/\f*sqrt( \n^2*(\f^2-\dn^2)+\n*\dn*(\f^2-\dn^2)+\dn^2*\f^2/2-(\f^4+\dn^4)/4 );
            ynode(\n,\dn,\lam,\a) = (2*\n*\dn+\dn^2)*\lam/(2*\f);
            intensity(\y,\lam,\a,\L) = exp(-\y*\y/5)*cos(180*\a*\y/(2*\lam*sqrt(\L*\L+\y*\y)))^2;}]
  
    \def\L{3.8}       
    \def\H{5.4}       
    \def\h{2.8}       
    \def\t{0.15}      
    \def\a{1.15}      
    \def\d{0.20}      
    \def\N{21}        
    \def\lambd{0.20}  
    \def\R{\N*\lambd} 
    \def\Nlines{3}    
    \def\A{1.6}       
    \def\nsamples{100}
    \def\ang{62}
  
    \begin{scope}
    \clip (-\t/2,-\H/2) rectangle (\L,\H/2);
    
    \draw[nodal] (0.08*\N*\lambd,0) -- (1.06*\R,0);
    \coordinate (NP0) at (\L,0);  
    \foreach \dn [evaluate={
                   \f=\a/\lambd;
                   \nmin=2.5+0.2*\dn; 
                   \nmax=10; 
                   \c=int(\dn<\f);
                   \y=\L/sqrt((\a/(\lambd*\dn))^2-1);
                 }] in {1,...,\Nlines}{
      \coordinate (NP+\dn) at (\L,\y);  
      \coordinate (NP-\dn) at (\L,-\y); 
      \ifnum\c=1
        \draw[nodal,variable=\n,samples=\nsamples,smooth]
          plot[domain=\nmin:\nmax] ({xnode(\n,\dn,\lambd,\f)},{ynode(\n,\dn,\lambd,\f)})
          -- (NP+\dn);
        \draw[nodal,variable=\n,samples=\nsamples,smooth]
          plot[domain=\nmin:\nmax] ({xnode(\n,\dn,\lambd,\f)},{-ynode(\n,\dn,\lambd,\f)})
          -- (NP-\dn);
      \fi
    }
    
    \foreach \i [evaluate={\R=\i*\lambd;}] in {1,...,\N}{
      \ifodd\i
        \draw[myred,line width=0.8] (0,\a/2)++(\ang:\R) arc (\ang:-\ang:\R);
        \draw[myred,line width=0.8] (0,-\a/2)++(\ang:\R) arc (\ang:-\ang:\R);
      \else
        \draw[myred!75,line width=0.1] (0,\a/2)++(\ang:\R) arc (\ang:-\ang:\R);
        \draw[myred!75,line width=0.1] (0,-\a/2)++(\ang:\R) arc (\ang:-\ang:\R);
      \fi
    }
  \end{scope}
  
  \foreach \i [evaluate={\x=-\i*\lambd;}] in {0,...,5}{
    \ifodd\i
      \draw[myred,line width=0.8] (\x,-\h/2) -- (\x,\h/2);
    \else
      \draw[myred!75,line width=0.1] (\x,-\h/2) -- (\x,\h/2);
    \fi
  }
  
  \fill[wall]
    (\t/2,\a/2-\d/2) rectangle (-\t/2,-\a/2+\d/2)
    (\t/2,\a/2+\d/2) rectangle (-\t/2,\H/2)
    (\t/2,-\a/2-\d/2) rectangle (-\t/2,-\H/2)
    (\L,-\H/2) rectangle (\L+\t,\H/2);
  
  \begin{scope}[shift={(1.08*\L,0)}]
    \def\yz{\L/sqrt((\a/\lambd)^2-1)} 
    \def\yZ{\L/sqrt((\a/\lambd/2)^2-1)} 
    \clip (0,-\H/2) rectangle (1.1*\A,\H/2);
    \fill[white] (0,-\H/2) rectangle++ (\A,\H); 
    \foreach \i [evaluate={\n=0.5*\i;\yn=\L/sqrt((\a/(2*\lambd*\n))^2-1);
                 }] in {1,...,\Nlines}{
      \ifodd\i 
        \fill[myred] (0,{-\yn-0.1}) rectangle++ (\A,0.2); 
        \fill[myred] (0,{ \yn-0.1}) rectangle++ (\A,0.2); 
      \fi
    }
    \path[left color=black,right color=black,middle color=myred,shading angle={180}]
      (0,{-\yz}) rectangle (\A,{\yz});
    \foreach \i [evaluate={
                  \n=0.5*\i;
                  \m=0.5*(\i+1);
                  \yn=\L/sqrt((\a/(2*\lambd*\n))^2-1);
                  \ym=\L/sqrt((\a/(2*\lambd*\m))^2-1);
                  \dang=mod(\i,2)*180;
                 }] in {1,...,\Nlines}{
      \path[left color=black,right color=myred,shading angle={\dang}]
        (0,\yn) rectangle (\A,\ym);
      \path[left color=black,right color=myred,shading angle={180+\dang}]
        (0,-\yn) rectangle (\A,-\ym);
    }
  \end{scope}
  
  \begin{scope}[shift={(1.1*\L+1.1*\A,0)}]
    \draw[->,thick] (-0.08*\A,0) -- (1.3*\A,0) node[right=0] {$\langle I\rangle$}; 
    \draw[->,thick] (0,-0.52*\H) -- (0,0.54*\H) node[right] {$\delta$}; 
    \draw[nodal] (NP0) --++ (0.15*\L+2.1*\A,0); 
    \foreach \i in {1,...,\Nlines}{ 
      \draw[nodal] (NP+\i) --++ (2,0);
      \draw[nodal] (NP-\i) --++ (2,0);
    }
    \draw[myred,thick,variable=\y,samples=\nsamples,smooth,domain=-\H/2:\H/2]
      plot({\A*intensity(\y,\lambd,\a,\L)},\y);
    \tick{0,{\L/sqrt((\a/(\lambd*1))^2-1)}}{180} node[right=0,scale=0.85] {$\pi$};
    \tick{0,{\L/sqrt((\a/(\lambd*2))^2-1)}}{180} node[right=0,scale=0.85] {$2\pi$};
    \tick{0,{\L/sqrt((\a/(\lambd*3))^2-1)}}{180} node[right=0,scale=0.85] {$3\pi$};
    \tick{0,{-\L/sqrt((\a/(\lambd*1))^2-1)}}{180} node[right=0,scale=0.85] {$-\pi$};
    \tick{0,{-\L/sqrt((\a/(\lambd*2))^2-1)}}{180} node[right=0,scale=0.85] {$-2\pi$};
    \tick{0,{-\L/sqrt((\a/(\lambd*3))^2-1)}}{180} node[right=0,scale=0.85] {$-3\pi$};
    \end{scope}
    \end{tikzpicture}
    }
    \caption[Double-slit experiment.]{Double-slit experiment with coherent light.}
    \label{fig:two-slit}
\end{figure}

Perhaps you have encountered the scenario of a double-slit experiment from one of your physics classes.
In Fig.~\ref{fig:two-slit}, coherent light produced by a laser is incident on a screen with two narrow slits.
The peaks of the waves are represented by thick lines while the troughs are represented by the thin lines.
The two slits diffract the incident light and act as two sources of spherical waves.
The waves propagate towards a second screen, where we observe an interference pattern of alternating bright and dark fringes.
Bright fringes are a result of constructive interference, where two peaks or two troughs meet.
Dark fringes on the other hand are a result of destructive interference where a peak is cancelled by a trough.
The dashed lines mark the points of constructive and destructive interference.

For example, consider the point right in the middle of the screen in Fig.~\ref{fig:two-slit}.
The distances travelled by light passing through the upper and lower slits are equal, therefore there is no phase difference between the two waves, $\delta=0$.
This leads to constructive interference and the average intensity $\langle I\rangle$ reaches its maximum.
This is also true for the cases when the phase difference is an even integer multiple of $\pi$.
We observe repeating bright fringes when the phase difference is $\delta=\pm 2\pi n$, for $n\in\{0,1,2,\ldots\}$.
On the other hand, when the phase difference is an odd integer multiple of $\pi$, that is when $\delta=\pm \pi n$, for $n\in\{0,1,2,\ldots\}$, the two waves cancel each other due to destructive interference.
This results in vanishing average intensity $\langle I \rangle$.
The bright fringes become fainter as we move away from the center of the screen due to diffraction effects modulating the average intensity.

Let's consider the scenario where we attenuate the laser light to such low levels that only single photons are incident at the first screen with the two slits.
In order to gain some intuition into the pattern that is observed on the second screen, let's analyze the case when one of the slits is covered, as shown in Fig.~\ref{fig:single-slit-attenuated}.
The incident photons can pass only through the open slit where they get diffracted.
Most photons pass through the slit without too much diffraction and hit the second screen at a point that is in line with the open slit's position.
Few photons do get diffracted significantly, resulting in an intensity distribution that shows a bring peak in the middle but decreases as we move away from this peak.
We can swap which slit is open and which is closed, and the observed pattern will be nearly the same.
The only difference will be that the peak of the average intensity would shift to be in line with the open slit.

\begin{figure}[t]
    \centering
    \resizebox {0.95\textwidth} {!} {
    \begin{tikzpicture}[
    nodal/.style={black!75,dashed,very thin},
    declare function={
      xnode(\n,\dn,\lam,\f) = \lam/\f*sqrt( \n^2*(\f^2-\dn^2)+\n*\dn*(\f^2-\dn^2)+\dn^2*\f^2/2-(\f^4+\dn^4)/2 );
      ynode(\n,\dn,\lam,\a) = (2*\n*\dn+\dn^2)*\lam/(2*\f);
      intensity(\y,\a) = exp(-(\y-\a/2)^2);
    }
  ]
  
  \def\L{3.8}       
  \def\H{5.4}       
  \def\h{2.8}       
  \def\t{0.15}      
  \def\a{1.15}      
  \def\d{0.20}      
  \def\N{21}        
  \def\lambd{0.20}  
  \def\R{\N*\lambd} 
  \def\Nlines{3}    
  \def\A{1.6}       
  \def\nsamples{100}
  \def\ang{62}
  
  \begin{scope}
    \clip (-\t/2,-\H/2) rectangle (\L,\H/2);
    \newcommand*\Angles{{5,2,-3,20,-1,-10,0,4,-7,6,1,2,-3,1,15,-8,-6,2,-4,5,2,3}}
    \foreach \i [evaluate={\R=\i*\lambd;}] in {1,...,\N}{
      \ifodd\i
        \filldraw[fill=myred,rotate around={\Angles[\i]:(0,\a/2)}] (\R,0\a/2) circle (0.08);
      \fi
    }
  \end{scope}
  
  \foreach \i [evaluate={\x=-\i*\lambd;}] in {0,...,5}{
    \ifodd\i
      \filldraw[fill=myred] (\x,0) circle (0.08);
    \fi
  }
  
  \fill[wall]
    (\t/2,\a/2-\d/2) rectangle (-\t/2,-\a/2+\d/2)
    (\t/2,\a/2+\d/2) rectangle (-\t/2,\H/2)
    (\t/2,-\a/2-\d/2) rectangle (-\t/2,-\H/2)
    (\L,-\H/2) rectangle (\L+\t,\H/2)
    (\t,-\a/2+\d/2) rectangle (\t/2,-\a/2-\d/2);
  
  \begin{scope}[shift={(1.08*\L,0)}]
    \clip (0,-\H/2) rectangle (1.1*\A,\H/2);
    \fill[black] (0,-\H/2) rectangle++ (\A,\H); 
    \path[left color=black,right color=black, middle color=myred, shading angle=0] (0,-\H/2+\a/2) rectangle (\A,\H/2+\a/2);
  \end{scope}
  
  \begin{scope}[shift={(1.1*\L+1.1*\A,0)}]
    \draw[->,thick] (-0.08*\A,0) -- (1.3*\A,0) node[right=0] {$\langle I\rangle$};
    \draw[->,thick] (0,-0.52*\H) -- (0,0.54*\H) node[right] {$\delta$}; 
    \draw[myred,thick,variable=\y,samples=\nsamples,smooth,domain=-\H/2:\H/2]
      plot({\A*intensity(\y,\a)},\y);
  \end{scope}

    \end{tikzpicture}
    }
    \caption[Single-slit experiment with single photons.]{Single-slit experiment with single photons.}
    \label{fig:single-slit-attenuated}
\end{figure}

What happens when both slits are open?
Naive expectation would suggest that since the photons can pass through either slit, we should add the two intensity distributions obtained from the photons passing through the top slit and the photons passing through the bottom slit.
The resulting pattern should look like the one displayed in Fig.~\ref{fig:two-slit-expectation}.
Individual intensity distributions are shown in thin pink lines, while the total expected distribution is their sum, shown in thick red.
Our naive expectation does not display any interference pattern, rather it is a simple consequence of diffraction of photons from the two slits.
However, this is \emph{\textbf{not}} what is observed when the experiment is performed in a laboratory.

\begin{figure}[H]
    \centering
    \resizebox {0.95\textwidth} {!} {
    \begin{tikzpicture}[
    nodal/.style={black!75,dashed,very thin},
    declare function={
      xnode(\n,\dn,\lam,\f) = \lam/\f*sqrt( \n^2*(\f^2-\dn^2)+\n*\dn*(\f^2-\dn^2)+\dn^2*\f^2/2-(\f^4+\dn^4)/4 );
      ynode(\n,\dn,\lam,\a) = (2*\n*\dn+\dn^2)*\lam/(2*\f);
      intensity_single(\y,\a) = exp(-(\y+\a/2)^2)/1.5;
      intensity_total(\y,\a) = (exp(-(\y-\a/2)^2)+exp(-(\y+\a/2)^2))/1.5;
    }
  ]
  
  \def\L{3.8}       
  \def\H{5.4}       
  \def\h{2.8}       
  \def\t{0.15}      
  \def\a{1.15}      
  \def\d{0.20}      
  \def\N{21}        
  \def\lambd{0.20}  
  \def\R{\N*\lambd} 
  \def\Nlines{3}    
  \def\A{1.6}       
  \def\nsamples{100}
  \def\ang{62}
  
  \begin{scope}
    \clip (-\t/2,-\H/2) rectangle (\L,\H/2);
    \newcommand*\Angles{{5,2,-3,20,-1,-10,0,4,-7,6,1,2,-3,1,15,-8,-6,2,-4,5,2,3}}
    \foreach \i [evaluate={\R=\i*\lambd;}] in {1,...,\N}{
        \ifodd\i
        \filldraw[fill=myred,rotate around={\Angles[\i]:(0,\a/2)}] (\R,\a/2) circle (0.08);
        \else
        \filldraw[fill=myred,rotate around={\Angles[\i]:(0,\a/2)}] (\R,-\a/2) circle (0.08);
      \fi
    }
  \end{scope}
  
  \foreach \i [evaluate={\x=-\i*\lambd;}] in {0,...,5}{
    \ifodd\i
      \filldraw[fill=myred] (\x,0) circle (0.08);
    \fi
  }
  
  \fill[wall]
    (\t/2,\a/2-\d/2) rectangle (-\t/2,-\a/2+\d/2)
    (\t/2,\a/2+\d/2) rectangle (-\t/2,\H/2)
    (\t/2,-\a/2-\d/2) rectangle (-\t/2,-\H/2)
    (\L,-\H/2) rectangle (\L+\t,\H/2);
  
  \begin{scope}[shift={(1.08*\L,0)}]
    \clip (0,-\H/2) rectangle (1.1*\A,\H/2);
    \fill[black] (0,-\H/2) rectangle++ (\A,\H); 
    \path[left color=black,right color=black, middle color=myred, shading angle=0] (0,-4) rectangle (\A,4);
  \end{scope}
  
  \begin{scope}[shift={(1.1*\L+1.1*\A,0)}]
    \draw[->,thick] (-0.08*\A,0) -- (1.3*\A,0) node[right=0] {$\langle I \rangle$}; 
    \draw[->,thick] (0,-0.52*\H) -- (0,0.54*\H) node[right] {$\delta$}; 
    \draw[myred!50,thin,variable=\y,samples=\nsamples,smooth,domain=-\H/2:\H/2]
      plot({\A*intensity_single(\y,\a)},\y);
    \draw[myred!50,thin,variable=\y,samples=\nsamples,smooth,domain=-\H/2:\H/2]
      plot({\A*intensity_single(\y,-\a)},\y);
    \draw[myred,thick,variable=\y,samples=\nsamples,smooth,domain=-\H/2:\H/2]
      plot({\A*intensity_total(\y,\a)},\y);
    \end{scope}

    \end{tikzpicture}
    }
    \caption[Double-slit experiment with single photons, naive expectation.]{Naive expectation of the observed pattern for a double-slit experiment with single photons. This pattern is not observed in a real experiment.}
    \label{fig:two-slit-expectation}
\end{figure}

The observed pattern is that of bright and dark interference fringes as shown in Fig.~\ref{fig:two-slit-single-photon-pattern}.
These are nearly the same as the ones we observed when coherent light was incident on the screen with the two slits.
This may not be that surprising upon a quick reflection.
We have started with strong coherent light in Fig.~\ref{fig:two-slit}, where we observed the interference pattern.
All we have done was we attenuated the light to the level of single photons, therefore there is no reason to expect the observed pattern to change.
It may take longer for the pattern to emerge clearly on the screen due to the attenuation of the light, but that is the only difference.

The situation gets a lot more interesting if we attenuate the incident light even further, such that with high probability there is only a single photon travelling between the two screens.
After enough time, the clear interference pattern of Fig.~\ref{fig:two-slit-single-photon-pattern} emerges yet again.
This is a lot spookier than the previous scenario since it seems a single photon can interfere with itself.
The incident photon has a two possibilities when it reaches the first screen in the form of the two open slits.
It is these possibilities that further interfere between the two screens, resulting in some places on the second screen having zero probability of detecting a photon.

\begin{figure}[t]
    \centering
    \resizebox {0.95\textwidth} {!} {
    \begin{tikzpicture}[
    nodal/.style={black!75,dashed,very thin},
    declare function={
      xnode(\n,\dn,\lam,\f) = \lam/\f*sqrt( \n^2*(\f^2-\dn^2)+\n*\dn*(\f^2-\dn^2)+\dn^2*\f^2/2-(\f^4+\dn^4)/4 );
      ynode(\n,\dn,\lam,\a) = (2*\n*\dn+\dn^2)*\lam/(2*\f);
      intensity(\y,\lam,\a,\L) = exp(-\y*\y/5)*cos(180*\a*\y/(2*\lam*sqrt(\L*\L+\y*\y)))^2;
    }
  ]
  
  \def\L{3.8}       
  \def\H{5.4}       
  \def\h{2.8}       
  \def\t{0.15}      
  \def\a{1.15}      
  \def\d{0.20}      
  \def\N{21}        
  \def\lambd{0.20}  
  \def\R{\N*\lambd} 
  \def\Nlines{3}    
  \def\A{1.6}       
  \def\nsamples{100}
  \def\ang{62}
  
    \filldraw[fill=myred] (-2*\lambd,0) circle (0.08);
    \filldraw[fill=myred] (17*\lambd,0) circle (0.08);
  
  \fill[wall]
    (\t/2,\a/2-\d/2) rectangle (-\t/2,-\a/2+\d/2)
    (\t/2,\a/2+\d/2) rectangle (-\t/2,\H/2)
    (\t/2,-\a/2-\d/2) rectangle (-\t/2,-\H/2)
    (\L,-\H/2) rectangle (\L+\t,\H/2);
  
  \begin{scope}[shift={(1.08*\L,0)}]
    \def\yz{\L/sqrt((\a/\lambd)^2-1)} 
    \def\yZ{\L/sqrt((\a/\lambd/2)^2-1)} 
    \clip (0,-\H/2) rectangle (1.1*\A,\H/2);
    \fill[myred] (0,-\H/2) rectangle++ (\A,\H); 
    \foreach \i [evaluate={\n=0.5*\i;\yn=\L/sqrt((\a/(2*\lambd*\n))^2-1);
                 }] in {1,...,\Nlines}{
      \ifodd\i 
        \fill[myred] (0,{-\yn-0.1}) rectangle++ (\A,0.2); 
        \fill[myred] (0,{ \yn-0.1}) rectangle++ (\A,0.2); 
      \fi
    }
    \path[left color=black,right color=black,middle color=myred,shading angle={180}]
      (0,{-\yz}) rectangle (\A,{\yz});
    \foreach \i [evaluate={
                  \n=0.5*\i;
                  \m=0.5*(\i+1);
                  \yn=\L/sqrt((\a/(2*\lambd*\n))^2-1);
                  \ym=\L/sqrt((\a/(2*\lambd*\m))^2-1);
                  \dang=mod(\i,2)*180;
                 }] in {1,...,\Nlines}{
      \path[left color=black,right color=myred,shading angle={\dang}]
        (0,\yn) rectangle (\A,\ym);
      \path[left color=black,right color=myred,shading angle={180+\dang}]
        (0,-\yn) rectangle (\A,-\ym);
    }
  \end{scope}
  
    \coordinate (NP0) at (\L,0);  
    \foreach \dn [evaluate={
                   \f=\a/\lambd;
                   \nmin=2.5+0.2*\dn; 
                   \nmax=10; 
                   \c=int(\dn<\f);
                   \y=\L/sqrt((\a/(\lambd*\dn))^2-1);
                 }] in {1,...,\Nlines}{
      \coordinate (NP+\dn) at (\L,\y);  
      \coordinate (NP-\dn) at (\L,-\y); 
    }
    
  \begin{scope}[shift={(1.1*\L+1.1*\A,0)}]
    \draw[->,thick] (-0.08*\A,0) -- (1.3*\A,0) node[right=0] {$\langle I\rangle$}; 
    \draw[->,thick] (0,-0.52*\H) -- (0,0.54*\H) node[right] {$\delta$}; 
    \draw[myred,thick,variable=\y,samples=\nsamples,smooth,domain=-\H/2:\H/2]
      plot({\A*intensity(\y,\lambd,\a,\L)},\y);
  \end{scope}

    \end{tikzpicture}
    }
    \caption[Double-slit experiment with single photons.]{Interference pattern remains even if the photons arrive slowly enough that a single photon is present between the screens.}
    \label{fig:two-slit-single-photon-pattern}
\end{figure}

\section{Interference with qubits}
\label{sec:6-4_interference_single_qubits}

We have finished Sec.~\ref{sec:6-3_interference_single_photons} by saying that a single photon's all possible routes to the second screen interfere together, resulting in a pattern displaying bright and dark fringes.
In this Section, we will learn more about this concept in its simplest form.
But before we do that, let's consider interference from a more abstract perspective of a single qubit.

Consider the action of a Hadamard operation on a qubit.
We will consider two initial states, state $\ket{0}$ and state $\ket{1}$.
The initial states and the Hadamard operation can be represented in matrix form as follows,
\begin{equation}
    |0\rangle = \begin{pmatrix} 1 \\ 0 \end{pmatrix}, \quad
    |1\rangle = \begin{pmatrix} 0 \\ 1 \end{pmatrix}, \quad
    H = \frac{1}{\sqrt{2}} \begin{pmatrix} 1 & 1 \\ 1 & -1 \end{pmatrix}.
\end{equation}
If the initial state is $\ket{0}$, applying the Hadamard operation creates an equal superposition of the states $\ket{0}$ and $\ket{1}$,
\begin{equation}
    |0\rangle \longrightarrow H|0\rangle=\frac{1}{\sqrt{2}}(|0\rangle+|1\rangle).
\end{equation}
We can apply the Hadamard again, this time to the superposition,
\begin{equation}
    H\left[\frac{1}{\sqrt{2}}(|0\rangle+|1\rangle)\right] = \frac{1}{2}(|0\rangle+|1\rangle+|0\rangle-|1\rangle) = |0\rangle.
    \label{eq:qubit_interference}
\end{equation}
This may seem like a trivial calculation.
The Hadamard is a unitary operation, therefore applying it twice should return the initial state.
But it is worth pausing briefly and thinking about the calculation in Eq.~(\ref{eq:qubit_interference}).
We can observe that both $|0\rangle$ terms have the same probability amplitude, $+1$.
This is another example of constructive interference.
On the other hand, the terms $|1\rangle$ have probability amplitudes $+1$ and $-1$, leading to destructive interference.

We can draw a parallel between the double-slit experiment with single photons and the interference of a single qubit in Eq.~(\ref{eq:qubit_interference}).
In the double-slit experiment, the photon had infinitely many possibilities where to hit the second screen.
Due to interference, some of these possibilities had finite probability of occurring, while the probability vanished for others.
In the case of a single-qubit interference, there are only two possibilities, represented by states $|0\rangle$ and $|1\rangle$.
It is the \textbf{\emph{interference of probability amplitudes}} that determines whether the state of the qubit is $|0\rangle$ or $|1\rangle$.

Let's consider two transformations that are not hermitian conjugates of each other, denoted by BS1 and BS2,
\begin{equation}
    \text{BS}1 = \frac{1}{\sqrt{2}} \begin{pmatrix} 1 & 1 \\ 1 & -1 \end{pmatrix}, \quad
    \text{BS}2 = \frac{1}{\sqrt{2}}\begin{pmatrix} -1 & 1 \\ 1 & 1 \end{pmatrix}, \quad
    \text{BS}2 \cdot \text{BS1} \neq I.
    \label{eq:BS_transformations}
\end{equation}
The operation BS1 is none other than the Hadamard operation.
The choice of notation will become clear in a short while. 
BS2 appears to be similar to BS1, the difference being that -1 is not located in the bottom right, but instead is in the top left.
You can check for yourself that applying these gates in sequence is not the same thing as doing nothing.
In particular, the product of BS2 and BS1 is not equal to the identity.
To see this, let's apply BS1, followed by BS2, to the initial state $|1\rangle$,
\begin{align}
    \text{BS}2 \cdot \text{BS}1 |1\rangle & = \text{BS}2 \cdot \frac{1}{\sqrt{2}} \begin{pmatrix} 1 & 1 \\ 1 & -1 \end{pmatrix} \cdot \begin{pmatrix} 0 \\ 1 \end{pmatrix} \nonumber\\
    & = \text{BS}2 \cdot \frac{1}{\sqrt{2}} \begin{pmatrix} 1 \\ -1 \end{pmatrix} \nonumber\\
    & = \frac{1}{2} \begin{pmatrix} -1 & 1 \\ 1 & 1 \end{pmatrix} \begin{pmatrix} 1 \\ -1 \end{pmatrix} \nonumber\\
    & = \frac{1}{2} \begin{pmatrix} -2 \\ 0 \end{pmatrix} \nonumber\\
    & = -|0\rangle = |0\rangle.
    \label{eq:BS2-BS1}
\end{align}
The global phase in the last line can be ignored as it has no physically observable consequence.
Application of BS1 creates a superposition of states $|0\rangle$ and $|1\rangle$.
Subsequent application of BS2 results in destructive interference of probability amplitudes for $|1\rangle$.

\begin{figure}[t]
   \centering
    \includegraphics[width=0.8\textwidth]{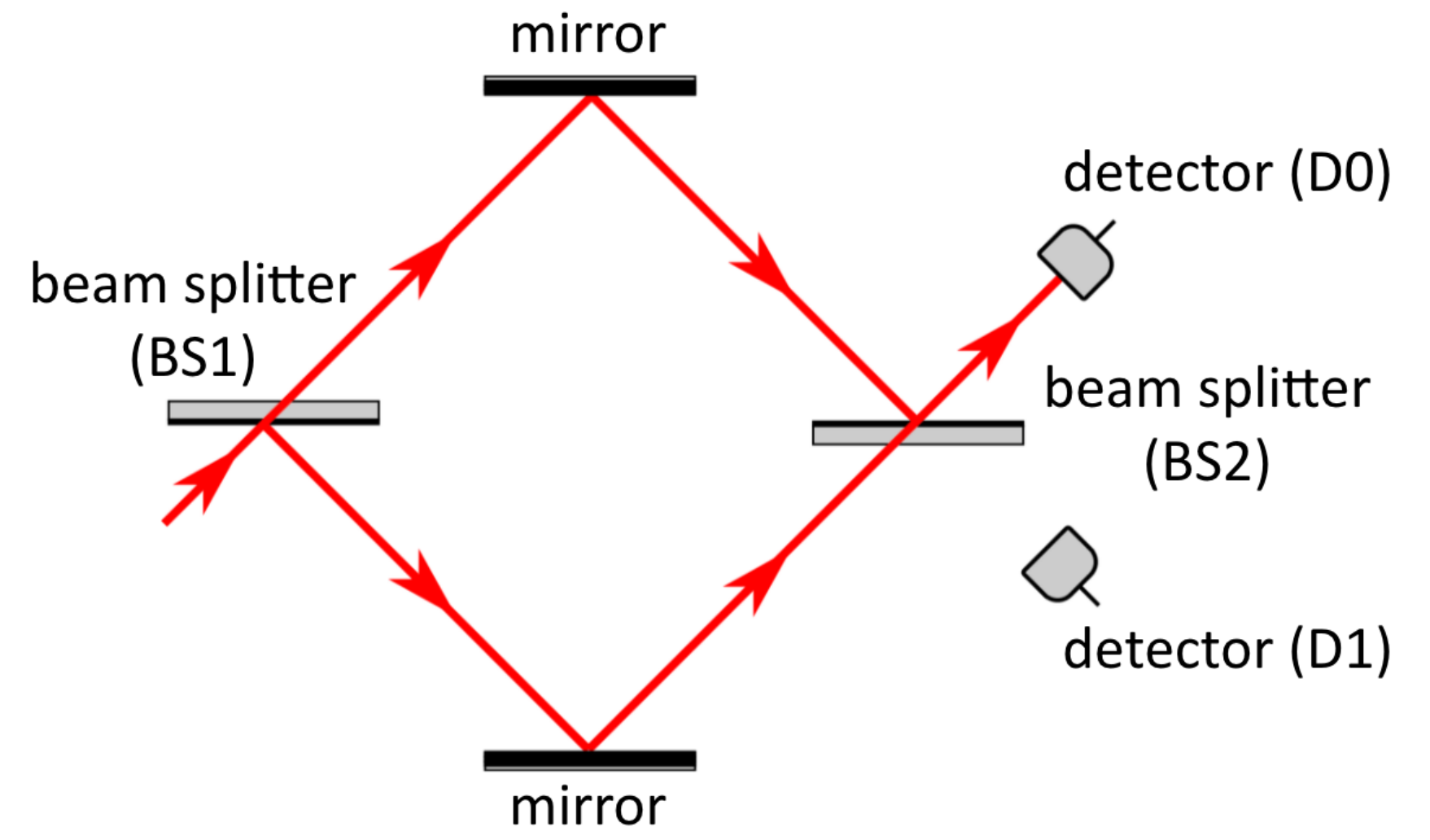}
    \caption{Mach-Zehnder interferometer.}
    \label{fig:mach-zehnder}  
\end{figure}

The previous discussion of qubit interference may seem a little abstract.
Now is the time to make more concrete by considering an optical instrument called the \textbf{\emph{Mach-Zehnder interferometer}}\index{Mach-Zehnder interferometer}\index{interferometer}, as shown in Fig.~\ref{fig:mach-zehnder}.
It consists of two beam splitters, BS1 and BS2, two mirrors, and two detectors.
The experiment that we would like to carry out is as follows.
We use a source of light as input for the beam splitter BS1.
The light can enter the beam splitter either from the top port or from the bottom one.
The detectors are placed at the two output ports of beam splitter BS2.
We are interested in the detection pattern of D0 and D1 based on the input pattern.
For the particular input in Fig.~\ref{fig:mach-zehnder}, only detector D0 registers a click, the light is never detected by D1.

\begin{figure}[t]
   \centering
    \includegraphics[width=0.8\textwidth]{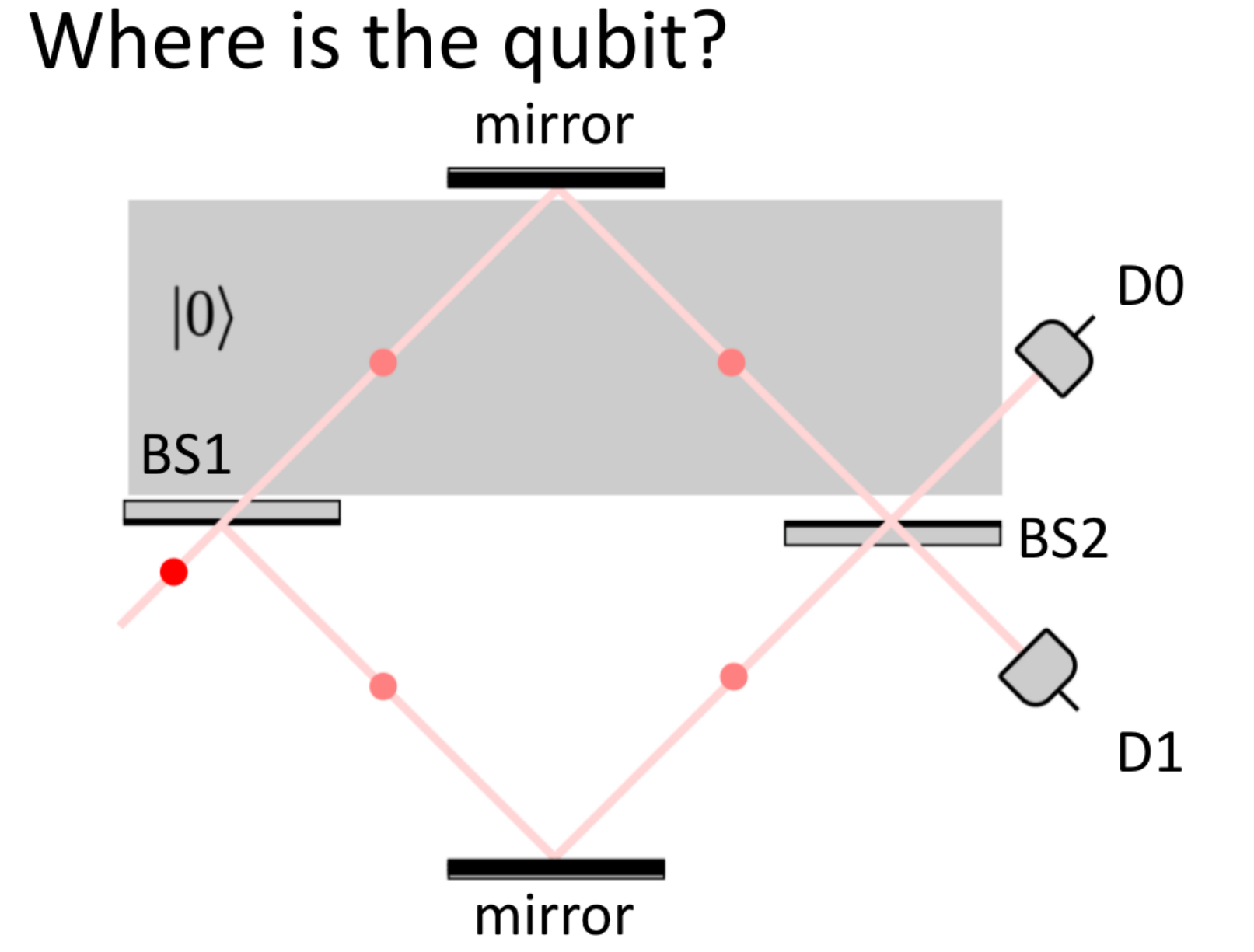}
    \caption[State $|0\rangle$ in the MI interferometer.]{State $\ket{0}$ is represented by a photon in the upper path.}
    \label{fig:m-z-upper}
\end{figure}

\begin{figure}[t]
   \centering
    \includegraphics[width=0.8\textwidth]{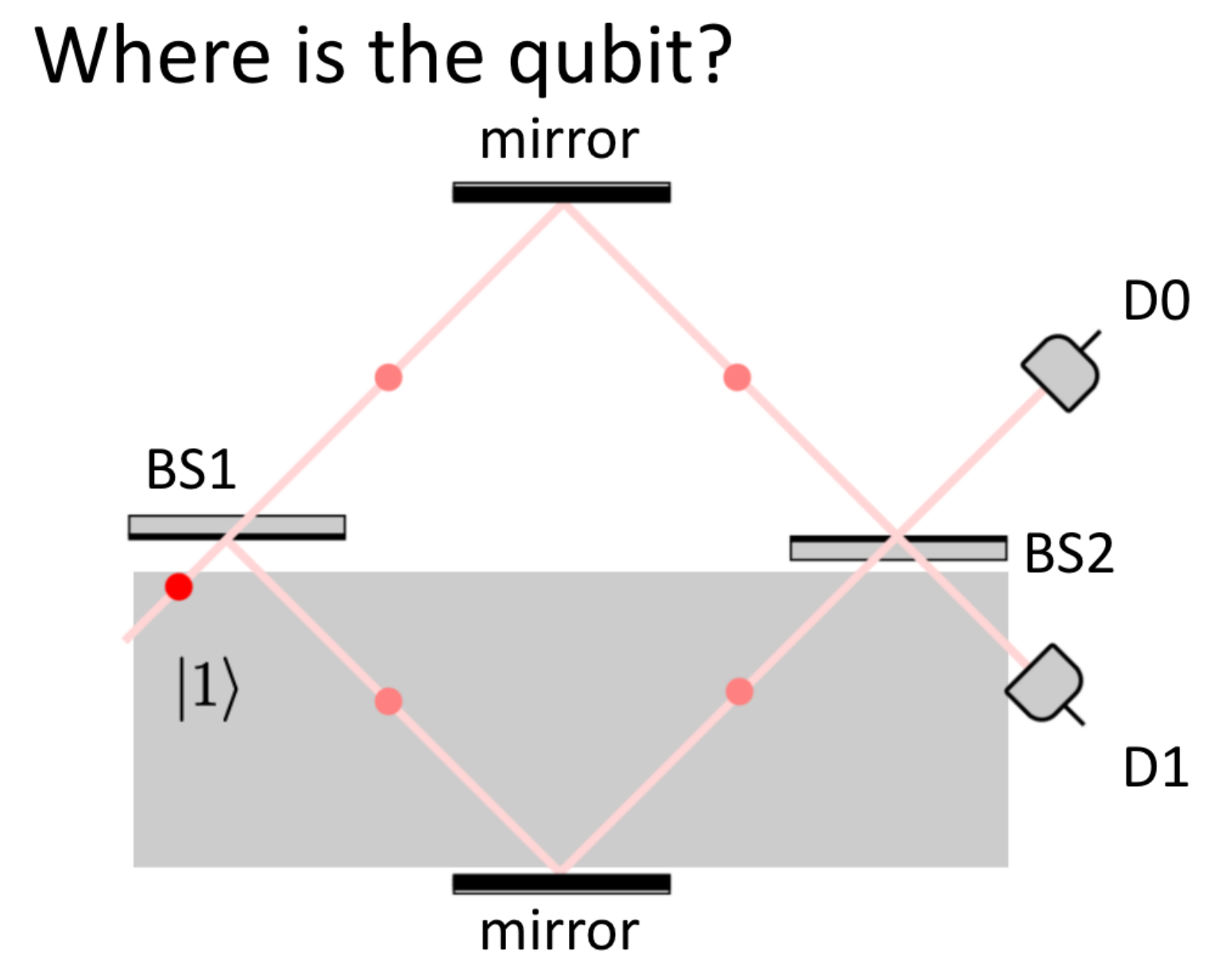}
    \caption[State $|1\rangle$ in the MI interferometer.]{State $\ket{1}$ is represented by a photon in the lower path.}
    \label{fig:m-z-lower}
\end{figure}

The Mach-Zehnder interferometer can be used to encode a qubit.
An incident photon, regardless of the input port, has two possible paths that it can take to BS2.
If it is in the upper half of the interferometer then we say the state of the qubit is $|0\rangle$, as shown in Fig.~\ref{fig:m-z-upper}.
If the photon is found in the bottom half of the interferometer, the state of the qubit is $|1\rangle$ as shown in Fig.~\ref{fig:m-z-lower}.
We say that the qubit is \textbf{\emph{spatially encoded}}\index{spatial encoding}.

Let's consider the initial state to be $\ket{1}$, meaning the photon enters the interferometer from the bottom port.
Now it is more clear why we have called those previous transformations in Eq.~(\ref{eq:BS_transformations}) BS1 and BS2.
They correspond to the mathematical description of how the beam splitters affect the probability amplitudes of our qubit~\footnote{In fact, the quantum mechanical description of a beam splitter always has to involve writing down the state of both of the input ports to the beam splitter; here, we are being a little loose with the description. BS2 in this figure shows the correct accounting for photons coming from above and below the beam splitter, but for BS1 we have ignored the fact that we need to count the number of photons coming in from both above and below it, as well.  In future courses, you will see a more rigorous treatment.}.
We proved in Eq.~(\ref{eq:BS2-BS1}) that if we first act on the input qubit $|1\rangle$ with BS1, and subsequently with BS2, then the final state of the qubit will be $|0\rangle$.
This means that a photon in the bottom input port of BS1 will be always detected by detector D0 after passing through the Mach-Zehnder interferometer.
This is one of the most bare-bones demonstrations of single-photon interference and also shows how the abstract calculation in Eq.~(\ref{eq:BS2-BS1}) connects to the real world.

\begin{figure}[t]
   \centering
    \includegraphics[width=0.8\textwidth]{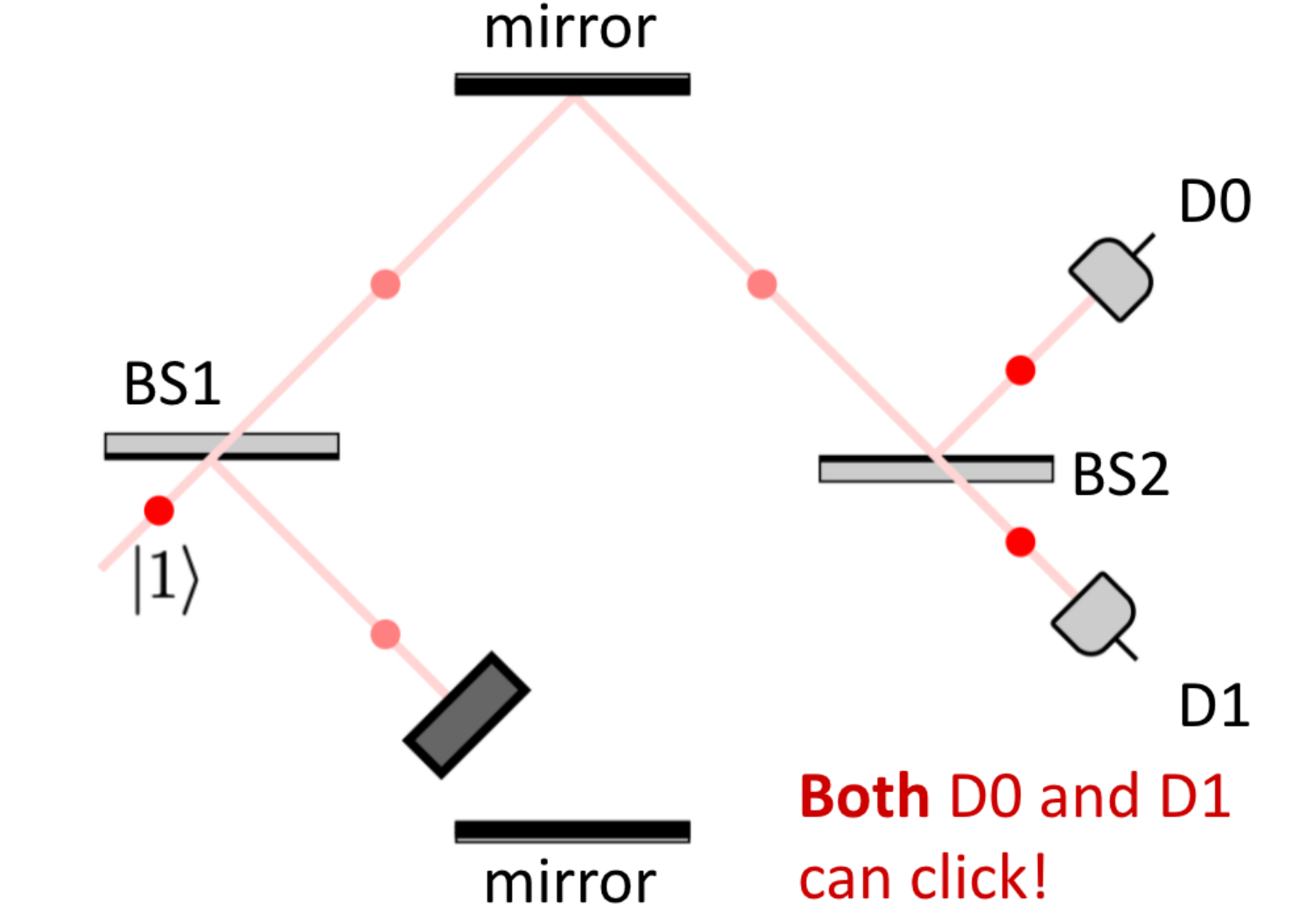}
    \caption[Lower path blocked.]{Lower path blocked. Because there is no photon coming from the lower path at BS2, no interference occurs, and any photon coming from above has a 50/50 chance of being directed to either detector.}
    \label{fig:m-z-blocked}
\end{figure}

Let's do a simple test to better understand the behavior of the Mach-Zehnder interferometer.
Let's put some absorbing material and block the possibility of the photon going through the lower half of the Mach-Zehnder Interferometer, as in Fig.~\ref{fig:m-z-blocked}.
An input photon at BS1 in the bottom port has a chance to get reflected at BS1.
If it does, it hits the block, gets absorbed and we don't get any clicks at either detector.
However, if it gets transmitted at BS1, it bounces off the upper mirror and it is incident onto the second beam splitter, where again, it has an equal probability of being reflected or passing through the beam splitter.
Therefore, it has a probability of being detected by both detector D0 and the detector D1.
Effectively, by blocking the bottom path of the Mach-Zehnder Interferometer, we have prevented interference from taking place at BS2.
That is why we see both possibilities D0 and D1.

\newpage
\begin{exercises}

\exer{
\emph{Visualization of wave propagation.}
Significant portions of this Chapter discuss propagation of waves in time. Naturally, this is can be only partially visualized in the form of a book. This exercise gives you the opportunity to write some basic code that will create an animation depicting interference of two waves in time. You can choose any package or library. If you are not familiar with creating animations, we suggest you look in to {\tt Python} library called {\tt matplotlib}.
\subexer{
Begin by replicating Fig.~\ref{fig:6-2_superposition}. Choose any two sinusoidal waves, but make sure you can clearly see the fast and slow oscillations.
}
\subexer{
Create an animation that clearly shows how the interference of the two waves propagates in time.
}
\subexer{
Using a small square, mark a point of constant phase on the interference pattern.
Using a small circle, mark a point on the envelope that modulates the oscillations.
For your chosen parameters, which point has higher velocity?
}
\subexer{
Change the parameters and make sure you can create animations that demonstrate the following three cases clearly: $v_p>v_g$, $v_p=v_g$, $v_p<v_g$.
}
\subexer{
Can you find parameters for the two waves, such that the phase velocity of the resulting wave is negative, $v_p<0$?
}
}

\exer{
\emph{Single-qubit interference.}
Consider the Mach-Zehnder interferometer of Fig.~\ref{fig:mach-zehnder}.
\subexer{
Show that BS1 and BS2 are not hermitian conjugates of each other, that $\text{BS}1\cdot\text{BS}2\neq I$.
}
\subexer{
What are the detection probabilities of the D0 and D1 if the input state is an equal superposition $(|0\rangle + |1\rangle)/\sqrt{2}$?
}
\subexer{
Can you think of a way to physically implement such an input state?
}
}

\exer{
\emph{Destroying the interference.}
We said that the interference at BS2 can be destroyed by placing an obstacle in one of the arms of the Mach-Zehnder interferometer, see Fig.~\ref{fig:m-z-blocked}. Let's better quantitative intuition into this scenario by replacing the obstacle with another detector, labelled by D2.
\subexer{
What is the state of the qubit immediately after BS1?
}
\subexer{
What is the probability of detector D2 detecting a photon?
}
\subexer{
What is the state of the qubit given that detector D2 did not click?
}
\subexer{
What are the individual click probabilities for the three detectors?
}
\subexer{
If we input a photon in the top port of BS1, where is it most likely going to be detected?
}
}

\end{exercises}

\chapter{Waveguides}
\label{sec:7_waveguides}

This chapter is dedicated to the question of how to guide light using optical fibers.
We begin by discussing why waveguides are necessary in long-distance communication before focusing on the main optical principles that allow us to confined light within optical fibers.
We conclude this chapter by looking at the basic construction of optical fibers and some of their basic types and properties.

\section{Brief history of guiding light}
\label{sec:7-1_history}

Light is an excellent information carrier.
In previous chapters, we have focused mainly on light's speed and robustness to noise as the characteristics that make it so suitable for communication.
We have not really discussed another important property of light, namely that is travels in \emph{\textbf{straight lines}}.
At first, this might seem like another advantage.
Coupled with light's speed and insensitivity to noise, it would seem that all we have to do is point at our intended target and light will carry our message to its destination unimpeded.
However, this is not quite true.
Due to the curvature of the Earth, the range of direct transmission is limited.
This range depends on the altitude at which the source of light as well as the receiver are found.
In order to increase the range, both need to be placed as high as possible.

\begin{figure}[H]
    \centering
    \includegraphics[width=0.75\textwidth]{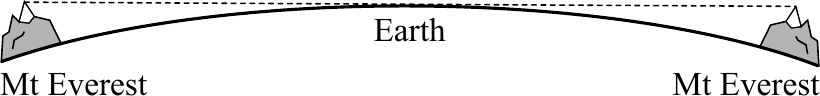}
    \caption[Two Everests.]{The disadvantage of direct transmission of light.}
    \label{fig:7-1_everest}
\end{figure}

Let's perform a little thought experiment in order to get an intuition for the maximum distance the transmitter and receiver can be separated by before they lose direct line of sight due to the curvature of the Earth.
Consider placing the light source at the highest point on Earth, the top of Mount Everest at 8849 meters above sea level, as shown in Fig.~\ref{fig:7-1_everest}.
Let's say that there is another Mount Everest with the receiver placed at its top.
How far apart can the two mountains be such that they are just able to maintain direct line of sight, assuming there are no obstacles between them?
With some basic trigonometry, the answer comes to a mere 672 kilometers.
And this is without worrying about absorption due to bad weather.

Being able to transmit light through a waveguide gets around some of these problems.
Weather conditions are no concern anymore.
Absorption is still an issue but not a fundamental obstacle, as we will learn in Chapter \ref{sec:11_long-distance}.
Before learning how optical fibers work, it is worth looking at some brief history of guiding light.

\begin{figure}[t]
    \centering
    \includegraphics[width=0.45\textwidth]{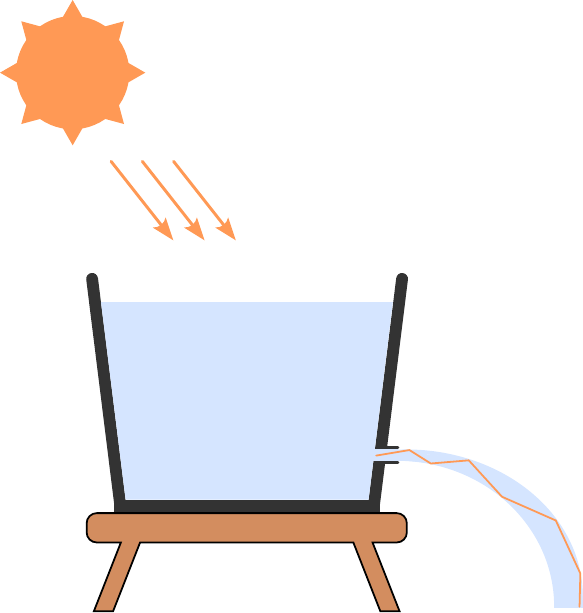}
    \caption[Tyndall's experiment.]{An early observation of light being guided by water pouring out of a bucket.}
    \label{fig:7-1_tyndall}
\end{figure}

One of the earliest demonstrations of guiding light in a waveguide was in an experiment by John Tyndall in 1870, depicted in Fig.~\ref{fig:7-1_tyndall}.
He filled a large vessel with water and let the water pour out from a hole in one of the sides.
He observed that light entering the water vessel did not exit the bucket through the hole in a straight line.
Remarkably, the light rays were guided by the water and seemingly followed the water's path.
Tyndall noticed that the rays were being reflected at the boundary between the water and air, confining them to stay within the streaming water.

This observation led many researchers to investigate fiber optics and how to guide light, particularly in glass.
But it was only in 1960 with the invention of the laser when people truly realized the potential of using laser light coupled to fiber optics.
They spent a lot of effort in researching how to do that, and in 1966 people managed to couple lasers with fiber optics.
This really sparked the first information revolution (and was rewarded with a Nobel Prize in Physics).

Just to give you some idea how far we've come, in 1970 it was possible to transmit about one percent of the original light over a distance of one kilometer.
If you put in light of some power at the beginning, after one kilometer, you only had one hundredth of the original signal remaining.
Twenty years later, in 1990, it was possible to transmit 96\% of the original power over the same distance of one kilometer.

Let's have a quick look at what we're going to discuss in this chapter.
First, we're going to begin with two basic phenomena of how light behaves when it hits the interface of two materials.
We're going to talk about \textbf{\emph{reflection}}\index{reflection} and \textbf{\emph{refraction}}\index{refraction}.
These two are crucial for understanding how fiber optics works.
Then we will talk about \textbf{\emph{total internal reflection}}\index{total internal reflection}, where we're going to combine reflection and refraction and derive the condition for total internal reflection.
Total internal reflection is what we have seen in Tyndall's experiment, where light was not escaping the water stream but was being reflected.
We will conclude with some basics of fiber optic cables, in particular how they are constructed and the differences between various types of optical fibers.

\section{Light at an interface}
\label{sec:7-2_light_at_interface}

In this section, we will look at how light behaves when it travels from one medium to another.
To make this quantitative, we need a way to describe light. 
There are three major descriptions we can use, all useful in different circumstances.
Light can be viewed as a particle traveling in straight lines.
This scenario is described by \textbf{\emph{geometric optics}}\index{geometric optics}.
All that is required to understand geometric description of light is some basic knowledge of trigonometry, making it fairly straightforward to understand and use.

However, light possesses wave properties as well.
We have seen this on the example of the double-slit experiment in Section~\ref{sec:6-3_interference_single_photons}. 
In order to correctly account for these phenomena, we have to abandon geometric optics and resort to the use of \textit{\textbf{wave optics}}.
This description is necessary in some cases when light is travelling down a very narrow fiber.
In such a case we need the full power of electromagnetism and Maxwell's equations.
This description is a lot harder and we're not going to use it here~\footnote{Maxwell's equations are covered in the second module in this series, ``From Classical to Quantum Light''.  They use partial differential equations (PDEs), which we briefly introduce in that module, but we recommend taking a math class that covers PDEs if you can. The more you have worked with PDEs the easier Maxwell's equations will be.}.

Finally, the third description is \textit{\textbf{quantum electrodynamics}}, because light is fundamentally a quantum field.
In this course, we will only use geometric optics.
There are a number of reasons for this.
Quantum electrodynamics describes interaction of light with matter at the atomic level.
We will not worry about such a detailed description in this Chapter.
Similarly, wave optics are really only necessary for single-mode fibers that have cross-section $d$ comparable to the wavelength of light $\lambda$, where wave effects become important.
We will mainly discuss multi-mode fibers with cross-sections much larger than the wavelength of light.
Such fibers have cross-sections at the order of tens of micrometers, and
they often carry light of wavelength $\lambda=1550$ nanometers.

\begin{figure}[t]
    \centering
    \includegraphics[width=0.65\textwidth]{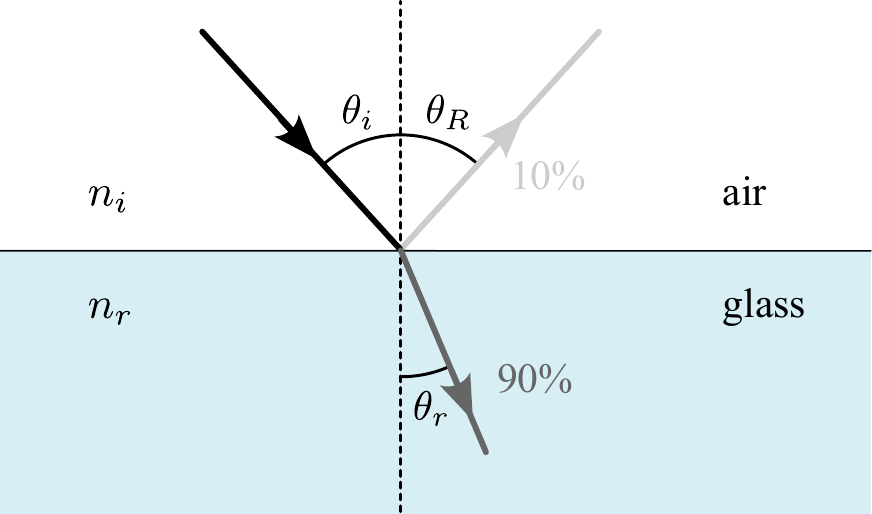}
    \caption[Angle of incidence.]{Light arriving at the interface between air and glass.}
    \label{fig:7-2_interface}
\end{figure}

Having settled on geometric optics as the means to describe how light behaves, let's see how to apply it to the case of light arriving at an interface of two different materials.
To be more concrete, let's consider the light ray travelling in air, and being incident onto a piece of glass.
The \textbf{\emph{angle of incidence}}\index{angle of incidence} $\theta_i$ is measured with respect to the normal to the surface as shown in Fig.~\ref{fig:7-2_interface}.
At the interface between air and glass, a portion of the incoming light is reflected back at the \textbf{\emph{angle of reflection}}\index{angle of reflection} $\theta_R$, again measured with respect to the normal.
The rest of the light enters the glass but the angle at which it travels changes to the \textbf{\emph{angle of refraction}} \index{angle of refraction}$\theta_r$.

The amount of light that is reflected and refracted is not important to our discussion right now, but we will return to it in our next module ``From Classical to Quantum Light''.
We are going to focus on the angles of reflection $\theta_R$ and refraction $\theta_r$.
The relationship between the angle of incidence and the angle of reflection is straightforward,
\begin{equation}
    \theta_R = \theta_i.
\end{equation}

The relationship between the angles of incidence and refraction is a little bit more involved.
We can observe from Fig.~\ref{fig:7-2_interface} that $\theta_r \neq \theta_i$.
An important quantity that helps with determining how much the angle of propagation changes as light travels from one medium into another is the \textbf{\emph{refractive index}}\index{refractive index} $n$, defined as the ratio of the speed of light in vacuum $c$ to the speed of light in the medium $v$,
\begin{equation}
    n = \frac{c}{v}.
    \label{eq:7-2_refractive_index}
\end{equation}
The refractive index tells us how much the speed of light changes when travelling through different media.
The speed of light is greatest in vacuum and therefore $n\geq 1$ for all materials.
The refractive index of air is slightly larger than one, meaning that light slows down when travelling in air only very slightly.
Glass, on the other hand, has a refractive index of $n_{\text{glass}} = 1.46$, meaning the light travels 1.46 times slower in the glass than in vacuum.
Light slows down even more in diamond, which has a refractive index of $n_{\text{diamond}} = 2.42$, giving diamond its unique and prized sparkling properties.

Let's return to the setting of Fig.~\ref{fig:7-2_interface}.
The light is initially travelling in some material, in our case air, at speed $v_i$, giving the refractive index $n_i = c / v_i$.
The refracted portion of light travels at a different speed $v_r$, resulting in refractive index $n_r = c / v_r$.
The angles of incident and refraction are related according to \textit{\textbf{Snell's law}},
\begin{equation}
    \frac{\sin\theta_i}{v_i} = \frac{\sin\theta_r}{v_r}.
\end{equation}
Using the relationship between the refractive index and the speed of light in a material in Eq.~(\ref{eq:7-2_refractive_index}), we can rewrite Snell's law in its more usual form,
\begin{equation}
    n_i \sin \theta_i = n_r \sin \theta_r.
    \label{eq:7-2_snell}
\end{equation}

\begin{figure}
    \centering
    \includegraphics[width=0.7\textwidth]{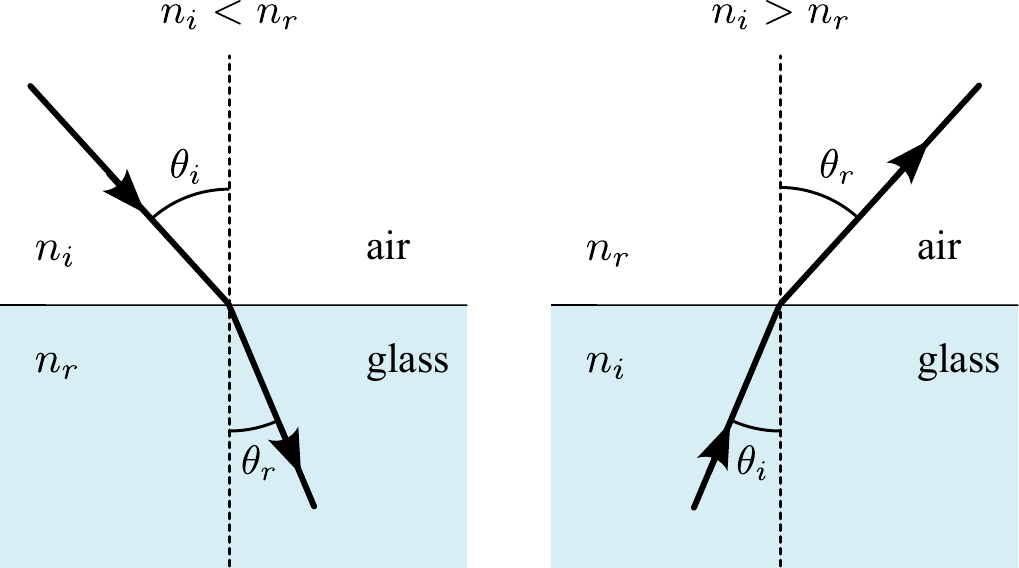}
    \caption[Snell's law.]{Angle of refraction can be larger or smaller than the angle of incidence.}
    \label{fig:7-2_snell_law}
\end{figure}

Using Snell's law (Eq.~\ref{eq:7-2_snell}), let's look at how light behaves when entering a new material.
We start with the case pictured in the left panel of Fig.~\ref{fig:7-2_snell_law}, where the light is travelling \textit{\textbf{from air to glass}}.
We can rearrange Eq.~(\ref{eq:7-2_snell}) to obtain
\begin{equation}
    \frac{n_i}{n_r} = \frac{\sin\theta_r}{\sin\theta_i}.
\end{equation}
Light travels faster in air than in glass, therefore we have $n_i / n_r < 1$.
This in turn means that $\sin\theta_r / \sin\theta_i < 1$ must be true.
This can be satisfied only when the angle of refraction is smaller than the angle of incidence,
\begin{equation}
    \theta_r < \theta_i, \qquad \text{when } \; n_i < n_r.
\end{equation}
We can also consider the opposite case when the light travels from \textit{\textbf{glass to air}}, as shown in the right panel of Fig.~\ref{fig:7-2_snell_law}.
This time, the refractive indices satisfy $n_i > n_r$.
Repeating the same argument as above, we discover that the light moves away from the normal,
\begin{equation}
    \theta_r > \theta_i, \qquad \text{when } \; n_i > n_r.
\end{equation}
This second case when light travels from a denser material to a less dense one will be studied in more detail in the next section.

\section{Total internal reflection}
\label{sec:7-3_total_internal_reflection}

\begin{figure}[t]
    \centering
    \includegraphics[width=0.4\textwidth]{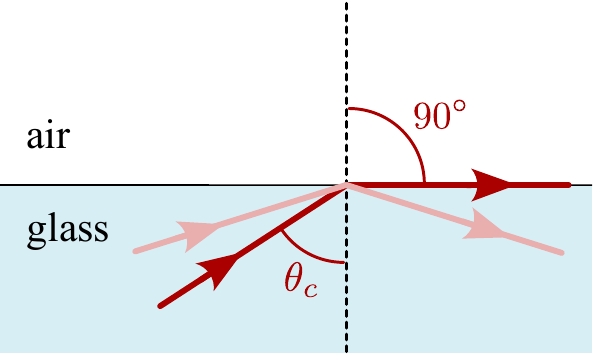}
    \caption[Total internal reflection.]{At the critical angle, reflected light travels just along the surface of the glass.}
    \label{fig:7-3_total_internal_reflection}
\end{figure}
We saw that if light is incident on an interface between two materials and traveling from a more dense material into a less dense one, then it gets refracted away from the normal.
The angle of refraction is larger than the angle of incidence, $\theta_r > \theta_i$.
Consider the scenario where we are steadily increasing the angle of incidence, as pictured in Fig.~\ref{fig:7-3_total_internal_reflection}.
The refracted beam is moving further away from the normal.
An interesting question to ask is at what angle of incidence will the refracted beam travel parallel to the surface between the two materials.
We can answer this question by setting $\theta_r=90^{\circ}$ in Eq.~(\ref{eq:7-2_snell}).
The \textbf{\emph{critical angle}}\index{critical angle} $\theta_c$ is the angle at which this is true, so we also replace the angle of incidence $\theta_i$ with $\theta_c$,
\begin{equation}
    n_i \sin \theta_c = n_r \sin 90^{\circ}.
    \label{eq:7-4_crit_angle}
\end{equation}
Using $\sin 90^{\circ} = 1$, we can obtain an expression for the critical angle,
\begin{equation}
    \theta_c=\sin ^{-1}\left(\frac{n_r}{n_i}\right).
    \label{eq:7-3_crit_angle}
\end{equation}

When light is at this angle $\theta_c$, then after reaching the interface, the light ray travels parallel to the surface.
We can keep increasing the angle of incidence so that $\theta_i > \theta_c$.
In this case, all of the light gets reflected back, undergoing \textbf{\emph{total internal reflection}}\index{total internal reflection}.
This is the basic working principle behind optical fibers and it is in fact responsible for the observations made by Tyndall in his experiment.

\begin{figure}
    \centering
    \includegraphics[width=0.9\textwidth]{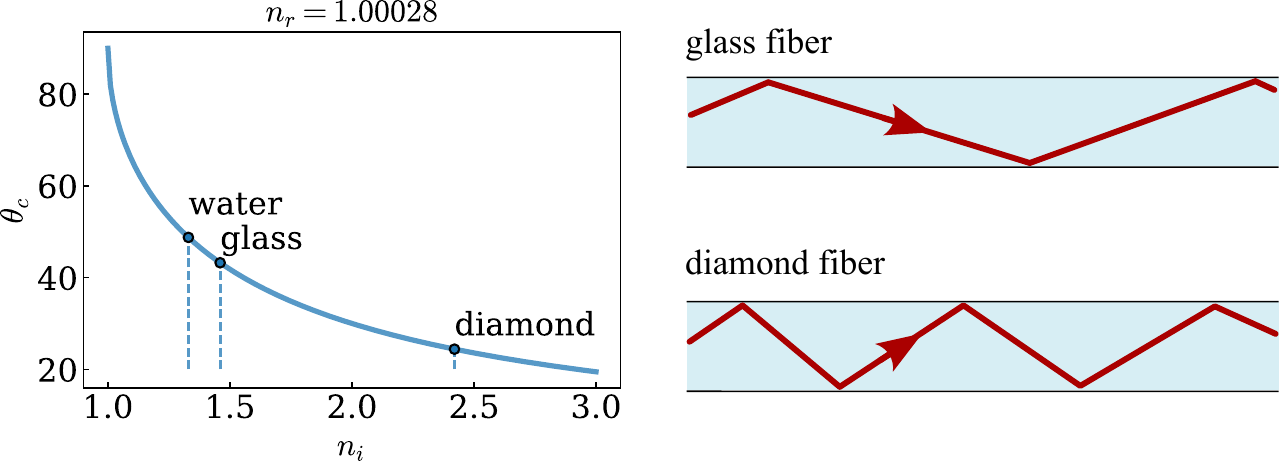}
    \caption[Critical angle.]{The critical angle gets smaller as the refractive index increases.}
    \label{fig:7-3_crit_angle}
\end{figure}

Left panel of Fig.~\ref{fig:7-3_crit_angle} shows a plot of how the critical angle $\theta_c$ changes with the refractive index $n_i$ of the material the waveguide is made from.
We assume that the material outside of the waveguide is still just air with a refractive index of $n_r = 1.00028$.
As the refractive index of the waveguide increases, the critical angle decreases.
This has important implications for the design of an optical fiber.
Fibers with smaller refractive indices must ensure that the light is incident on the surface at a larger angle, otherwise the light will escape.
On the other hand, fibers made from material with high refractive index make very good waveguides because they are capable of containing the light better, as shown in the right panel of Fig.~\ref{fig:7-3_crit_angle}.

\section{Optical fibers}
\label{sec:7-4_optical fibers}

\begin{figure}[t]
    \centering
    \includegraphics[width=0.7\textwidth]{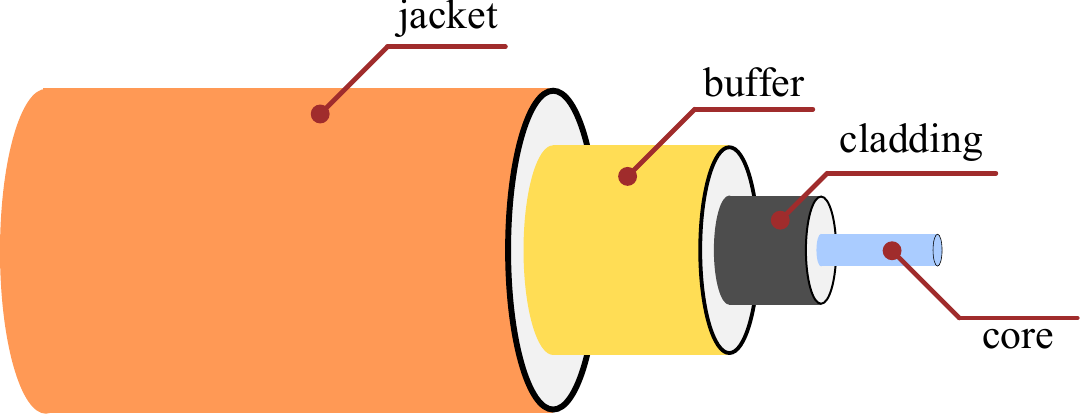}
    \caption[Anatomy of an optical fiber.]{Anatomy of an optical fiber.}
    \label{fig:7-4_fiber}
\end{figure}

Let's have a look at the composition of a typical optical fiber\index{optical fiber}.
Often, many fibers are bundled into a fiber-optic cable, but here we will look at an individual fiber as shown in Fig.~\ref{fig:7-4_fiber}. 
The outermost layer is the \textbf{\emph{jacket}}\index{jacket}.
Its function is to protect the inside components of the fiber from the environment.
Under the jacket is the \textbf{\emph{buffer}}\index{buffer}.
The buffer offers further protection, and sometimes bundles multiple optical fibers.
The next layer under the buffer is the \textbf{\emph{cladding}}\index{cladding}.
This is the material that is responsible for reflecting light back and keeping it contained within the innermost layer, which is the \textbf{\emph{core}}\index{core}.
Cladding serves as a form of protection and prevents cross-talk in the case of multiple fibers bundled within the buffer.
Having two cores next to each other would result in light leaking from one core to the other.
The refractive index of the cladding must be smaller than the refractive index of the core in order for the light signal to have a chance of undergoing total internal reflection.

\begin{figure}[t]
    \centering
    \includegraphics[width=0.7\textwidth]{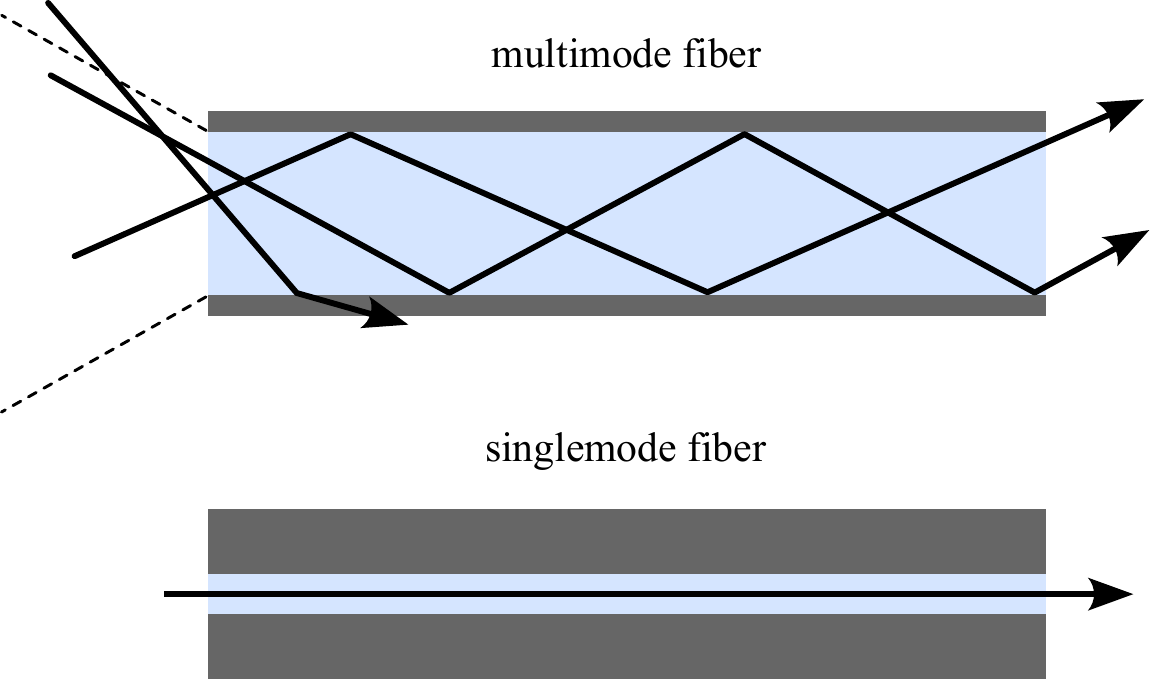}
    \caption[Multimode and singlemode fibers.]{Two main types of optical fibers, multimode and singlemode.}
    \label{fig:7-4_multimode_singlemode_fiber}
\end{figure}

Depending on the thickness of the core, optical fibers are classified into two types, as shown in Fig.~\ref{fig:7-4_multimode_singlemode_fiber}.
\textbf{\emph{Multimode fiber}}\index{multimode fiber} has cross-section diameter much larger than the wavelength of light.
The mode of light here refers to the spatial direction in which each beam of light travels.
The multimode fiber in Fig.~\ref{fig:7-4_multimode_singlemode_fiber} has three modes coupling to it but only two modes making the journey all the way to the end.
We will discuss why this is shortly.
\textbf{\emph{Singlemode fiber}} \index{single-mode fiber} has cross-section diameter comparable to the wavelength of light and is able to support only a single mode of light.
This mode is called the \textbf{\emph{transverse mode}}\index{transverse mode} of light.

Let's look at a multimode fiber first.
Figure~\ref{fig:7-4_multimode_singlemode_fiber} shows one of the modes of light being lost in the multimode fiber.
This mode leaks out of the fiber due to its angle of incidence onto the cladding being less than the critical angle.
Modes of light can couple to a multimode fiber only if they are within the \textbf{\emph{acceptance cone}}\index{acceptance angle}, represented by the dashed lines.
Any modes that arriving at the entrance of the fiber from outside of this cone will be refracted by the cladding and leak out of the fiber.
This suggests there is a relationship between the angle of acceptance and the critical angle.

\begin{figure}[t]
    \centering
    \includegraphics[width=0.6\textwidth]{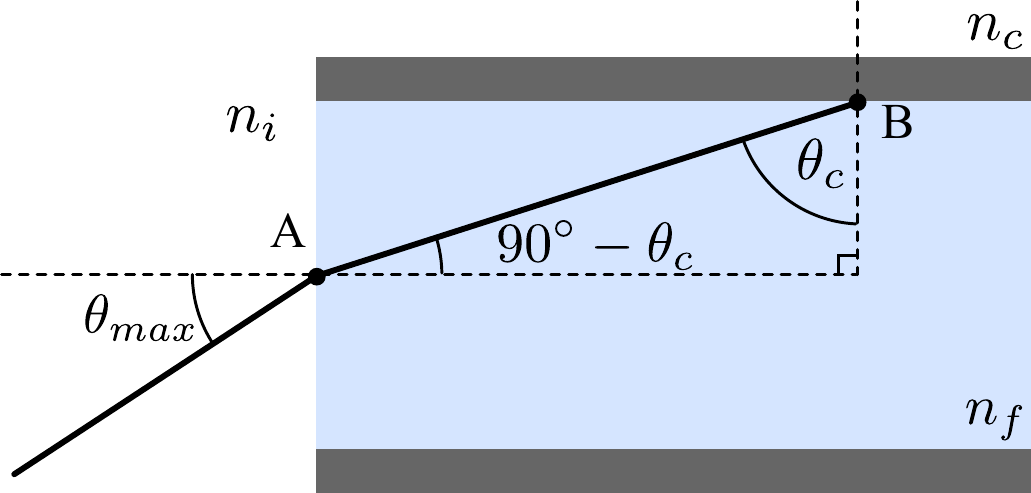}
    \caption[Acceptance cone.]{Maximum incidence angle of the light beam onto the fiber's core such that it hits the cladding at the critical angle.}
    \label{fig:7-4_acceptance_cone}
\end{figure}

Let's make this relationship more quantitative.
Consider the situation of Fig.~\ref{fig:7-4_acceptance_cone}, where a light beam is incident onto the core of a fiber at point $A$, gets refracted, and is later incident onto the cladding at point $B$.
Our aim is to find the maximum angle of incidence $\theta_{max}$ onto the core such that angle of incidence onto the cladding is the critical angle $\theta_c$.
We denote the refractive index of the material outside the fiber by $n_i$, the refractive index of the core by $n_f$, and the refractive index of the cladding by $n_c$.
Using Snell's law at point $A$, we have $n_i \sin \theta_i = n_f \sin \theta_r$.
We can use the right-hand triangle in Fig.~\ref{fig:7-4_acceptance_cone} to see that $\theta_r = 90^{\circ} - \theta_c$.
In that case we can set the incidence angle at point $A$ to $\theta_i = \theta_{max}$,
\begin{equation}
    n_i \sin \theta_{max} = n_f \sin (90^{\circ} - \theta_c).
\end{equation}
Using the trigonometric identity $\sin(\theta + \phi) = \sin\theta\cos\phi + \sin\phi\cos\theta$, we obtain
\begin{equation}
    n_i \sin \theta_{max} = n_f \cos\theta_c,
\end{equation}
where we used $\cos(-\theta_c) = \cos\theta_c$.
We express the cosine of the critical angle in terms of sine, $\cos\theta_c = \sqrt{1 - \sin^2\theta_c}$, and using Eq.~(\ref{eq:7-3_crit_angle}),
\begin{equation}
    n_i \sin \theta_{max} = n_f \sqrt{1 - \left( \frac{n_c}{n_f} \right)^2}.
\end{equation}
Bringing the refractive index of the fiber under square root, we obtain
\begin{equation}
    n_i \sin \theta_{max} = \sqrt{n_f^2 -  n_c^2}.
    \label{eq:7-4_theta_max}
\end{equation}
The expression on the left of Eq.~(\ref{eq:7-4_theta_max}), $n_i \sin \theta_{max}$, is known as the \textbf{\emph{numerical aperture}}\index{numerical aperture}.
A larger numerical aperture means that it is easier to couple to the fiber as the acceptance cone is wider.
Equation~(\ref{eq:7-4_theta_max}) allows us to calculate the maximum incidence angle given the three refractive indices $n_i$, $n_c$, and $n_f$.
The acceptance cone is given by twice this angle, $2 \theta_{max}$.

We said that the diameter of the core for a multimode fiber is much larger than the wavelength of light, leading to the core being able to support a number of modes.
The speed of light in the core is the same for all these modes, however the time it takes for individual modes to reach the end of the fiber is different.
Due to the different angles of incidence of different modes, they have to cover different total distances before reaching the end.
This is known as \textbf{\emph{mode dispersion}}\index{mode dispersion}.
We will cover mode dispersion in more detail in Section~\ref{sec:11-2_mode_dispersion}.
Because of this, multimode fibers are more suitable for short distances, such as in data centers, offices and campus networks.
On the other hand, because the core does not need to be extremely small, multimode fibers are easier to manufacture, leading to savings in production costs.

In contrast, singlemode fibers are much more difficult to produce, as their core diameter is typically less than $10$ micrometers.
For comparison, the average width of a human hair is larger than $20$ micrometers.
Singlemode fibers support only the transverse mode, therefore there is no mode dispersion.
Mode dispersion changes the shape of the signal received at the far end, so the rate at which a signal can be changed while maintaining a clean signal at the receiver is determined by the amount of this dispersion.
Singlemode fibers allow for high rate of transfer of data, making them suitable for long distance communications.
They are still susceptible to attenuation, and therefore the signal must be boosted approximately every 50 kilometers.
We will discuss how to combat attenuation in Section~\ref{sec:11-4_overcoming_losses}.

\begin{figure}[t]
    \centering
    \includegraphics[width=\textwidth]{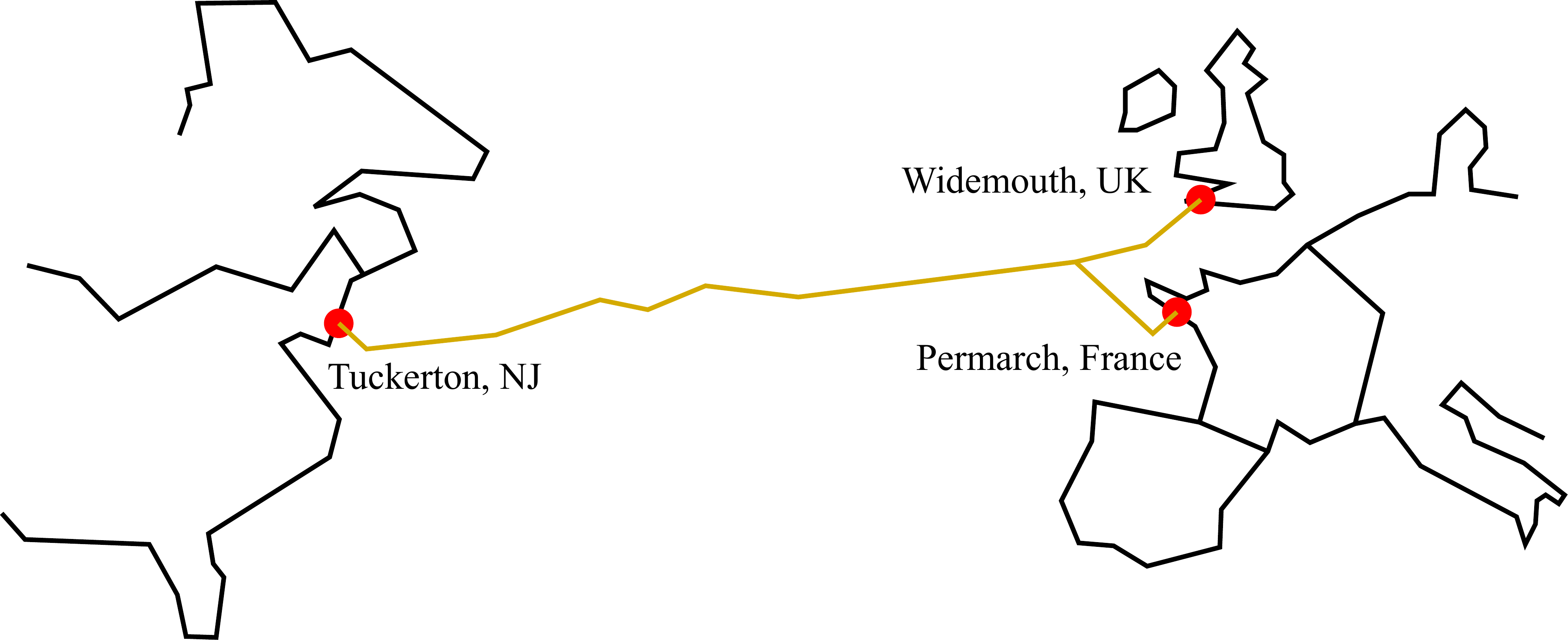}
    \caption[TAT-8.]{The first transatlantic communications cable, TAT-8.}
    \label{fig:7-4_TAT8}
\end{figure}

The first transatlantic fiber optic cable, called TAT-8, was laid in 1988 between Tuckerton, New Jersey, and the UK and France.
The submarine cable was constructed at a cost of approximately 335 million US dollars, and it was retired in 2002.
The bandwidth was 280 Mbits/s, which can carry around 40,000 phone circuits simultaneously. At the time of installation, it was predicted that the bandwidth of this fiber optic cable would be adequate to serve the demand for a long time and probably would never be reached.
Instead, the capacity was reached in eighteen months, necessitating the laying of more cables.

\newpage
\begin{exercises}
\exer{
\emph{How far is the horizon?}
We motivated the need for optical fibers by saying that direct line of sight transmission of information is limited in range.
How far is the horizon for an average human?
Assume spherical Earth, no dispersion due to atmosphere, and average human height of 1.7 meters.
}

\exer{
\emph{Light passing through a glass tube.}
Consider a ray of light being incident onto a long glass tube at angle $\theta_i$. The light gets refracted when entering the tube and when leaving the tube.
\subexer{
Draw this scenario. Using your drawing, what is the angle to the normal at which the light ray exits the glass tube.
}
\subexer{
Confirm your previous answer by calculating it from Snell's law. 
}
\subexer{
If the thickness of the glass tube is $h$, calculate the deviation $\Delta$ of the light ray with respect to its original path.
}
}

\exer{
\emph{Negative refractive index.}
We mentioned in this chapter that the refractive index is a positive number. This is true of conventional materials light air, water, or glass. However, metamaterials can be engineered to have a \textbf{\emph{negative refractive index}}\index{negative refractive index}.
\subexer{
Can you guess what happens when a light ray, travelling through material with refractive index $n$, is incident onto a material with refractive index $-n$?
}
\subexer{
Use Snell's law to check your answer.
}
}

\exer{
\emph{Acceptance angle for typical optical fibers.}
Consider an optical fiber with core refractive index of $1.522$, and cladding refractive index of $1.343$.
\subexer{
What is the critical angle $\theta_c$ for this optical fiber?
}
\subexer{
Compute the numerical aperture for this fiber?
}
\subexer{
What is the angle of acceptance $\theta_{max}$ for this fiber?
}
\subexer{
What does the angle of acceptance increase to if we use a diamond core?
}
}

\end{exercises}

\newpage
\section*{Quiz}
\addcontentsline{toc}{section}{Quiz}
The online version of this course includes a quiz for this block of chapters.
Discussion of the quiz questions will be provided there.

\section*{Further reading Chapters 5-7}
\addcontentsline{toc}{section}{Further reading Chapters 5-7}

{\bf Chapter 5}\\

This Chapter discusses the various types of light depending on their coherence properties. Coherent light produced by lasers is crucially important from the perspective of communication. A good qualitative as well as quantitative introduction to lasers can be found in Chapter 1 of Orazio Svelto's \emph{Principles of Lasers}~\cite{svelto2010principles}.

A great discussion of lasing behaviour from the perspective of a mathematician is in Chapter 3 of Steven Strogatz's \emph{Nonlinear Dynamics and Chaos: With Applications to Physics, Biology, Chemistry, and Engineering}~\cite{strogatz2018nonlinear}.\\

{\bf Chapter 6}\\

Following three books are great excellent introductions as well as references for the topics of geometric and wave optics:
\begin{enumerate}
    \item Eugene Hecht, \emph{Optics}~\cite{hecht2012optics}
    \item Grant R. Fowles, \emph{Introduction to Optics}~\cite{fowles1989introduction}
    \item Bahaa E. A. Saleh, Malven C. Teich, \emph{Fundamentals of Photonics}~\cite{saleh2019fundamentals}
\end{enumerate}

These books provide a good foundation for transitioning to quantum optics which we will do later.
The first is a common undergraduate textbook, while the last is more advanced and contains a lot of specialized information, perhaps best suited to those planning further study in optics.

Interference is an extremely important concept.
We encourage you to read and understand the discussion in Chapter 1 of
Richard P. Feynman, Robert P. Leighton, Matthew Sands, \emph{The Feynman Lectures on Physics}~\cite{feynman1971feynman}.\\

{\bf Chapter 7}\\

This Chapter relies mainly on the geometric description of light propagation. Section 4.3 and Section 5.6 of Hecht’s textbook contain great discussion of reflection and propagation in optical fibers, respectively.

\part{Quantum Communication Protocols}

\begin{partintro}
\partintrotitle{Introduction to the third chapter block}
With the basics in place, in this block of chapters we introduce quantum teleportation and two forms of QKD (single-photon BB84 and entanglement-based E91) as three of the intermediate-layer building blocks.  It is worth noting that teleportation and QKD are often described as "applications" of the Quantum Internet. In our terminology, however, teleportation is not an application protocol; it is instead a function that is used within systems to build more complete systems. In classical Internet terms, it is very roughly like the transport protocol TCP, whose job it is to see that data is correctly transmitted end-to-end but does not care about how that data is used. Even this analogy, however, is poor; teleportation does not assess or correct errors. For a more complete, robust system, then, it must be combined with an error correction or at least error detection method, as we will see in Sec.~\ref{sec:12-4_purification} in the following block of chapters, when we study purification.

QKD likewise provides a function that is of little use alone, but must be incorporated into larger systems to provide a service that is used by end users. QKD, whether single-photon or entanglement-based, requires the measurement of qubits and the collapse of superposition and/or entanglement. This bridging of the quantum and classical levels of a complete system is of particular interest, but the details of how the random numbers generated in QKD are incorporated into a complete system are deferred to Sec.~\ref{sec:classical-integration}.
\end{partintro}

\chapter{Teleportation}
\label{sec:8_teleportation}

Teleportation is one of the most wonderful and most fundamental protocols in quantum information processing.
It is used extensively in quantum computation and quantum communication.

\section{Introduction}
\label{sec:8-1_introduction}

This Chapter deals with the question of transmitting quantum information in a quantum network.
Before we deal with this question, let's see how we transmit classical information.
In Ch.~\ref{sec:1_Introduction}, we discussed the example of the Great Wall of China.
If a guard tower was under attack, the guards lit a fire to alert the other guard towers and ask for help.
The message, in the form or light, was generated at the place of attack and travelled to the neighboring guard towers.
Nowadays, if we want to download data from a server, we encode the request in the form of a light signal that is carried by optical fibers.
The physical systems that encode the information, in this case photons, are transmitted from from us to the server hop-by-hop using intermediate nodes of the classical network.
\begin{figure}[H]
    \centering
    \includegraphics[width=0.5\textwidth]{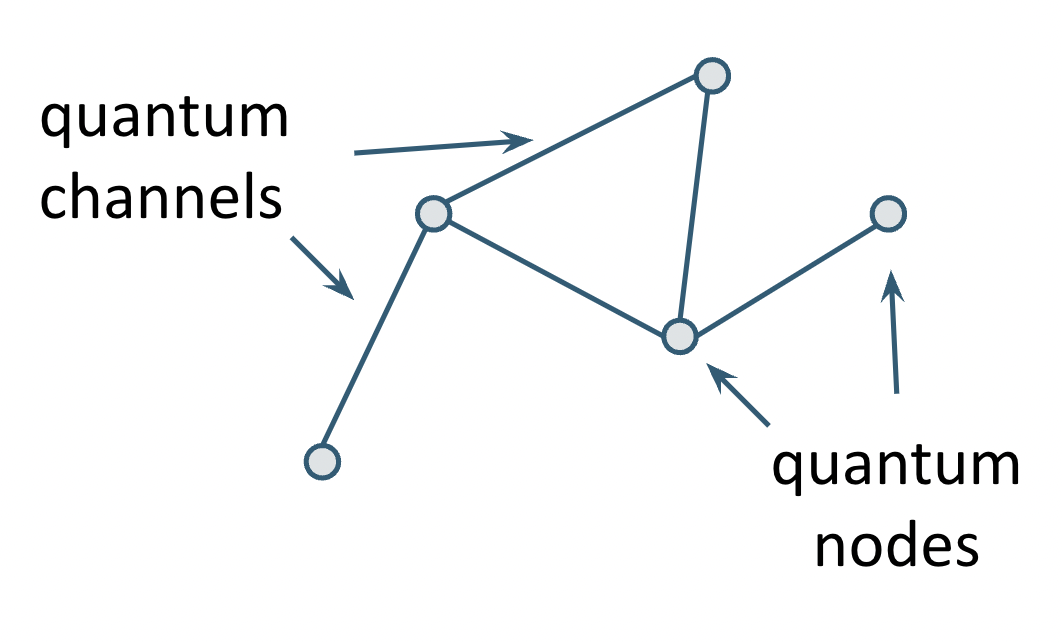}
    \caption[A quantum network of nodes and channels.]{A quantum network consists of quantum nodes, and quantum channels that we will call links.}
    \label{fig:quantum-network}
\end{figure}
\begin{figure}[H]
    \centering
    \includegraphics[width=0.5\textwidth]{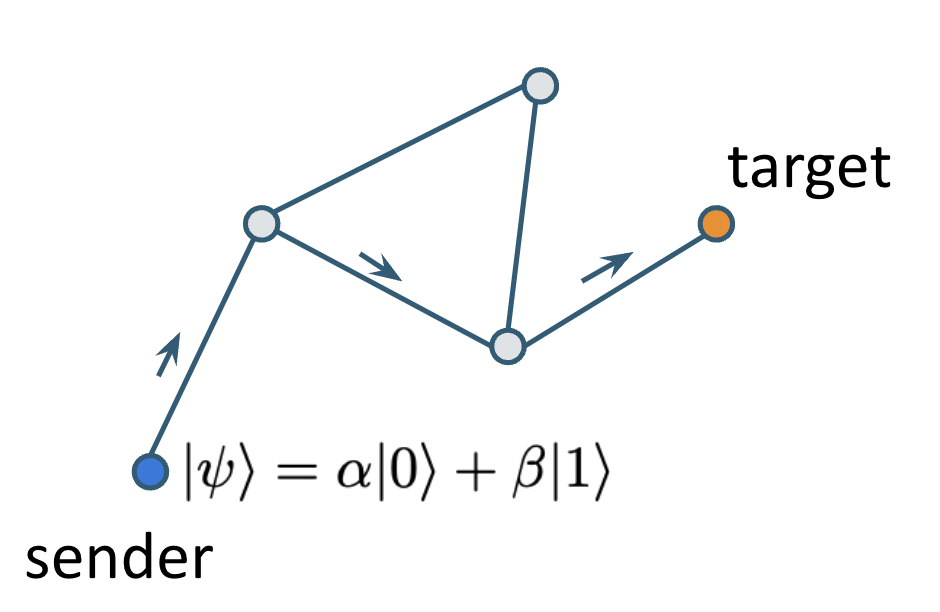}
        \caption[Hop-by-hop transmission.]{Physically moving a quantum message hop-by-hop through a network.}
    \label{fig:hop-by-hop}
\end{figure}
\begin{figure}[H]
    \centering
    \includegraphics[width=0.5\textwidth]{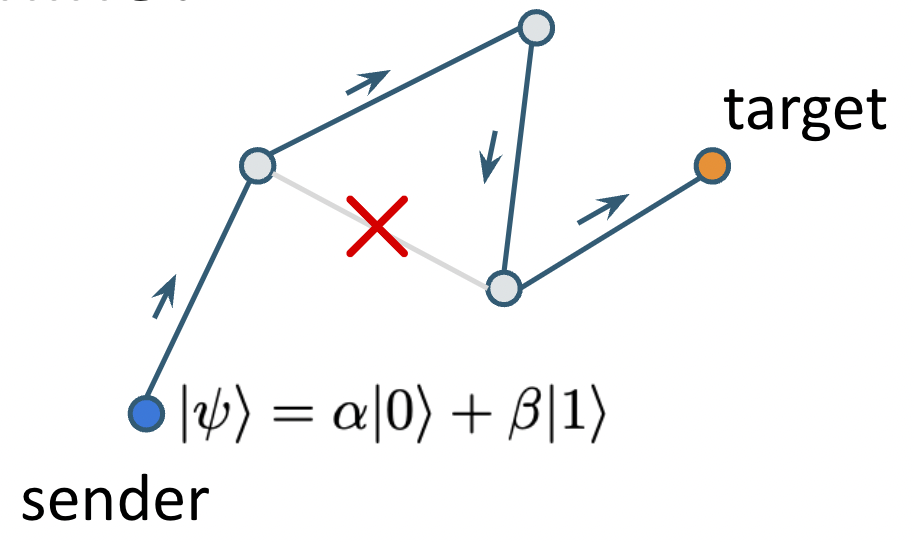}
        \caption{Rerouting when a link goes down.}
    \label{fig:link-down}
\end{figure}
\begin{figure}[H]
    \centering
    \includegraphics[width=0.5\textwidth]{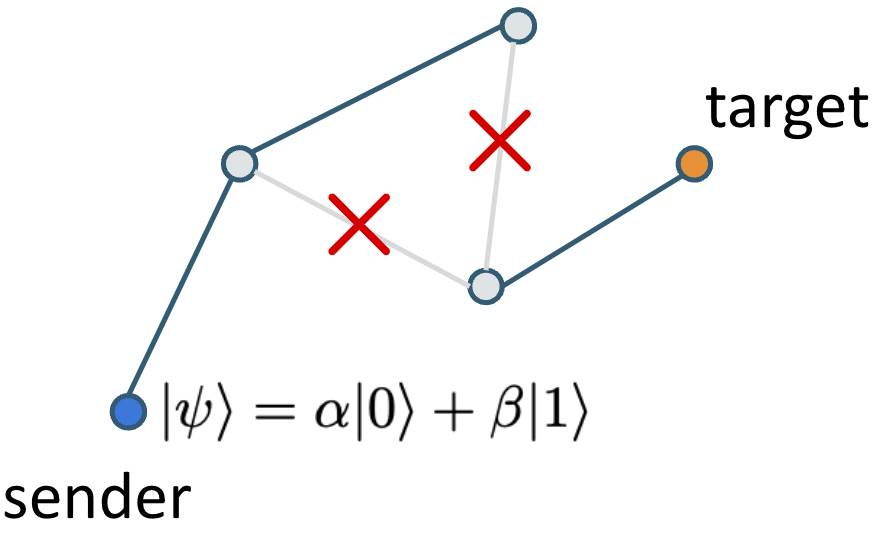}
        \caption{A partitioned network.}
    \label{fig:partition}
\end{figure}

However, quantum networks offer different ways of transmitting information.
Consider a small quantum network in Fig.~\ref{fig:quantum-network}.
The circles represent our nodes and the lines represent the links, or quantum channels, between the nodes.
The sender, represented by the blue node, is in possession of a pure state $\ket{\psi} = \alpha\ket{0}+\beta\ket{1}$, and  wants to send this state to the target given by the orange network node.
The transmission can be done hop-by-hop by encoding it in a physical system and sending that physical system along a path that connects to the target node.
The network may decide to send the encoded quantum message via the path represented by the arrows in Fig.~\ref{fig:hop-by-hop}.
What happens if one of the quantum links is down, as in Fig.~\ref{fig:link-down}?
In this particular case, it's not a big problem because the message can be simply rerouted to go around the damaged link and still reach the target node.
But what happens if yet another link is down, as in Fig.~\ref{fig:partition}?
In this case, the quantum portion of our network is \textbf{\emph{partitioned}}.
It may seem that we cannot transmit the message to the target...unless we use \textbf{\emph{teleportation}}\index{teleportation}\footnote{Here, we assume that the supporting classical network remains intact, only the quantum channels are going up or down.}
We can teleport this state from the sender to the target, provided that some initial conditions are met.
We can move the information itself \textbf{\emph{without moving the actual physical system}}.
That physical system remains stationary.

\begin{figure}[H]
    \centering
    \includegraphics[width=0.8\textwidth]{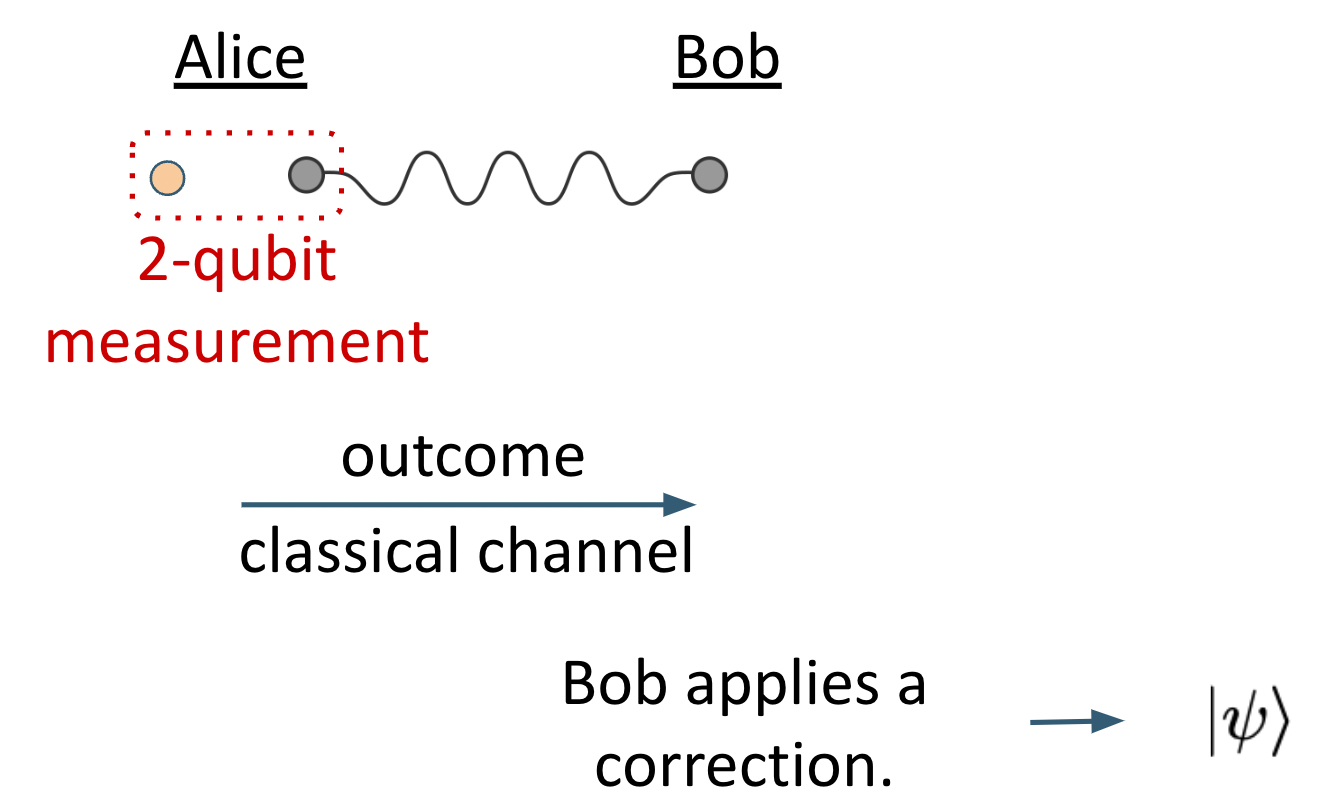}
        \caption{Outline of the teleportation protocol.}
    \label{fig:teleportation}
\end{figure}

This is the outline of the teloportation is the following, pictured in Fig.~\ref{fig:teleportation}.
\begin{enumerate}
    \item Let's call the sender ``Alice'', who wishes to communicate the quantum state $\ket{\psi}$ to her friend ``Bob''.
    \item Alice and Bob start by sharing an entangled pair of qubits.
    \item Alice performs a two-qubit measurement jointly on her qubits. That measurement provides her with two classical bits of information.
    Naturally, this means that there are four possible outcomes.
    \item Alice communicates the outcome of the measurement to Bob via a classical channel.
    \item Bob receives this classical message and applies appropriate local corrections.
    \item The state of Bob's qubit is the desired state $\ket{\psi}$.
\end{enumerate} 

Now you see the two conditions necessary to maintain our ability to teleport quantum data even when the quantum network is partitioned: we must have a supply of entanglement, and we must have the ability to exchange classical messages.

\section{Teleportation protocol}
\label{sec:8-2_teleportation_protocol}

In this Section, we will look at the teleportation protocol in more detail, and make the outline of the previous Section more mathematical.
First, consider the initial state in our protocol.
Alice has an arbitrary qubit given by
\begin{align}
    \ket{\psi} = \alpha\ket{0} + \beta\ket{1}.
\end{align}
Alice and Bob also share an entangled state.
It's one of the Bell states, $\ket{\Phi^+}$, which is an equal superposition of $\ket{00}$ and $\ket{11}$.
We label the qubits as follows: $A_1$ is Alice's first qubit, holding the state that she is trying to communicate to Bob. 
Her second qubit, $A_2$, is one half of the entangled Bell pair.\
We designate Bob's qubit, which is the other half of the entangled pair that he is sharing with Alice, as $B$.
If we write out the initial three-qubit state in its full form, we have
\begin{align}
    |\psi\rangle_{A_{1}}|\Phi^{+}\rangle_{A_{2} B} &=(\alpha|0\rangle+\beta|1\rangle)_{A_{1}} \frac{1}{\sqrt{2}}(|00\rangle+|11\rangle)_{A_{2} B} \nonumber\\
    &=\frac{1}{\sqrt{2}}(\alpha|000\rangle+\alpha|011\rangle+\beta|100\rangle+\beta|111\rangle)_{A_{1} A_{2} B}.
    \label{eq:teleportation_initial_state}
\end{align}

Alice then performs a two-qubit measurement in the Bell basis.
To refresh your memories, you may want to take a look at Sec.~\ref{sec:4_3-bell_states}, where we discussed Bell-state measurements.
Doing a measurement in the Bell basis is basically asking the question: which of the following states is Alice's state in? 
\begin{align}
\begin{array}{ll}
    \left|\Phi^{+}\right\rangle=\frac{1}{\sqrt{2}}(|00\rangle+|11\rangle), \quad & \left|\Phi^{-}\right\rangle=\frac{1}{\sqrt{2}}(|00\rangle-|11\rangle), \\
    \left|\Psi^{+}\right\rangle=\frac{1}{\sqrt{2}}(|01\rangle+|10\rangle), \quad & \left|\Psi^{-}\right\rangle=\frac{1}{\sqrt{2}}(|01\rangle-|10\rangle).
\end{array}
\end{align}
To answer that question, we begin by rewriting our initial state in the Bell basis.
We will use the following set of four identities relating computatinal basis to the Bell basis,
\begin{align}
\begin{array}{ll}
    |00\rangle=\frac{1}{\sqrt{2}}\left(\left|\Phi^{+}\right\rangle+\left|\Phi^{-}\right\rangle\right), & |01\rangle=\frac{1}{\sqrt{2}}\left(\left|\Psi^{+}\right\rangle+\left|\Psi^{-}\right\rangle\right), \\
    |10\rangle=\frac{1}{\sqrt{2}}\left(\left|\Psi^{+}\right\rangle-\left|\Psi^{-}\right\rangle\right), & |11\rangle=\frac{1}{\sqrt{2}}\left(\left|\Phi^{+}\right\rangle-\left|\Phi^{-}\right\rangle\right).
\end{array}
\end{align}

Going term by term of the initial state in Eq.~(\ref{eq:teleportation_initial_state}), we first look at the portion of Alice's state that is in $\ket{00}$.
We substitute for it our superposition of Bell states $\ket{\Phi^+}$ and $\ket{\Phi^-}$.
We do the same thing for the next term in the initial state $\ket{01}$, and we substitute for it the superposition of $\Psi^+$ and $\Psi^-$, and so on and so forth.
In full, we get the following superposition,
\begin{align}
    \begin{aligned}
        |\psi\rangle_{A_{1}}|\Phi^{+}\rangle_{A_{2} B} & = (\alpha|0\rangle+\beta|1\rangle)_{A_{1}} \frac{1}{\sqrt{2}}(|00\rangle+|11\rangle)_{A_{2} B} \\
        = & \frac{1}{\sqrt{2}}(\alpha|000\rangle+\alpha|011\rangle+\beta|100\rangle+\beta|111\rangle)_{A_{1} A_{2} B} \\
        = & \frac{1}{2}\left( \alpha\left(|\Phi^{+}\rangle+|\Phi^{-}\rangle\right)|0\rangle\right.\\
        & + \alpha\left(|\Psi^{+}\rangle+|\Psi^{-}\rangle\right)|1\rangle \\
        & + \beta\left(|\Psi^{+}\rangle-|\Psi^{-}\rangle\right)|0\rangle \\
        & \left. +\beta\left(|\Phi^{+}\rangle-|\Phi^{-}\rangle\right)|1\rangle\right). \\
\end{aligned}
\end{align}
Let's collect all the terms that have the same Bell state for Alice's qubits $A_1$ and $A_2$.
For example, the first term has a $\ket{\Phi^+}$, and also the last term has a $\ket{\Phi^+}$, but they differ in their amplitudes and also in the state of Bob's qubit.
For the first term, we have probability amplitude $\alpha$ and Bob's qubit in the state $\ket{0}$.
For the last term, we have probability amplitude $\beta$ and Bob's qubit is in the state $\ket{1}$.
We collect them together, and we get the expression
\begin{align}
\begin{aligned}
    |\psi\rangle_{A_{1}}|\Phi^{+}\rangle_{A_{2} B} & = \frac{1}{2}\left|\Phi^{+}\right\rangle_{A_{1} A_{2}}(\alpha|0\rangle+\beta|1\rangle)_{B} \\
    & + \frac{1}{2}|\Phi^{-}\rangle_{A_{1} A_{2}}(\alpha|0\rangle-\beta|1\rangle)_{B} \\
    & + \frac{1}{2}|\Psi^{+}\rangle_{A_{1} A_{2}}(\alpha|1\rangle+\beta|0\rangle)_{B} \\
    & + \frac{1}{2}|\Psi^{-}\rangle_{A_{1} A_{2}}(\alpha|1\rangle-\beta|0\rangle)_{B}.
\end{aligned}
\label{eq:teleport-alice-basis}
\end{align}

Remember, we are not really doing anything yet. 
We are still dealing with the initial state, just rewriting it in a more convenient form.
We are now in a position to answer the question that we asked earlier: in which of the Bell states are Alice's two qubits?
Alice performs the two-qubit measurement, and obtains two classical bits as the measurement outcome.
Probabilities of the four possible outcomes are all equal,
\begin{equation}
    \operatorname{Prob}\left(|\Phi^{\pm}\rangle\right)=\operatorname{Prob}\left(|\Psi^{\pm}\rangle\right)=\frac{1}{4}.
\end{equation}
Notice that the probabilities of the four possible outcomes are also independent of the probability amplitudes $\alpha$ and $\beta$, even though Alice's state was in an arbitrary superposition.

The outcome of Alice's measurement determines the state of Bob's qubit.
If Alice measures $\ket{\Phi^+}$, then we know that Bob has the desired output state $\alpha|0\rangle+\beta|1\rangle$, which was the initial state that Alice was trying to communicate to him.
So in that case, we can say, ``Great! The teleportation has succeeded.''

How about the other measurement outcomes?
Remember we said that all of these outcomes are equally likely, so Alice might get $\ket{\Phi^-}$, $\ket{\Psi^+}$ or $\ket{\Psi^-}$ as the outcome of her Bell state measurement.
For each of these measurement outcomes, Bob's qubit is different from the desired state $\alpha\ket{0}+\beta\ket{1}$.
Summarizing,

\hspace{2cm} If Alice measures $\left|\Phi^{+}\right\rangle \rightarrow$ Bob has $\alpha|0\rangle+\beta|1\rangle$.

\hspace{2cm} If Alice measures $\left|\Phi^{-}\right\rangle \rightarrow$ Bob has $\alpha|0\rangle-\beta|1\rangle$.

\hspace{2cm} If Alice measures $\left|\Psi^{+}\right\rangle \rightarrow$ Bob has $\alpha|1\rangle+\beta|0\rangle$.

\hspace{2cm} If Alice measures $\left|\Psi^{-}\right\rangle \rightarrow$ Bob has $\alpha|1\rangle-\beta|0\rangle$.

Does this mean that teleportation succeeds only with probability of 0.25?
No.
Luckily, all these other states can be transformed into $\ket{\psi}$ by a suitable unitary.
For example, notice that if we apply a Pauli $Z$ on the state $\ket{\psi}$, then we obtain $\alpha\ket{0}-\beta\ket{1}$, which happens to correspond to Bob's state when Alice's measurement outcome is $\ket{\Phi^-}$.
Applying a Pauli $X$ transforms \ket{\psi} to $\alpha\ket{1}+\beta\ket{0}$, while applying the Pauli product $XZ$ produces $\alpha|1\rangle-\beta|0\rangle$.
Summarizing,
\begin{align}
\begin{aligned}
    |\psi\rangle &=\alpha|0\rangle+\beta|1\rangle \\
    Z|\psi\rangle &=\alpha|0\rangle-\beta|1\rangle \\
    X|\psi\rangle &=\alpha|1\rangle+\beta|0\rangle \\
    X Z|\psi\rangle &=\alpha|1\rangle-\beta|0\rangle.
    \label{eq:teleportation_possible_states}
\end{aligned}
\end{align}
Note that all of those operations are unitary.  This means the states can be \textbf{\emph{returned}} to the original, desired state $\ket{\psi}$ using their adjoint operations.

Alice needs to let Bob know which measurement outcome she obtained, and she does so by sending two classical bits.
After receiving the classical bits, Bob knows exactly which correction he has to apply.
If Alice measures $\ket{\Phi^+}$, he does nothing because he already has the state $\ket{\psi}$.
If Alice measures $\ket{\Phi^-}$, then Bob knows he has to apply a Pauli $Z$ to his qubit.
If she tells Bob that she obtained the measurement outcome $\ket{\Psi^+}$, he simply applies Pauli $X$, and if she obtains a $\ket{\Psi^-}$ then he just needs to apply $ZX$ to his qubit.  Summarizing,

\hspace{2cm} If Alice measures $\left|\Phi^{+}\right\rangle \rightarrow$ Bob applies $I$.

\hspace{2cm} If Alice measures $\left|\Phi^{-}\right\rangle \rightarrow$ Bob applies $Z$.

\hspace{2cm} If Alice measures $\left|\Psi^{+}\right\rangle \rightarrow$ Bob applies $X$.

\hspace{2cm} If Alice measures $\left|\Psi^{-}\right\rangle \rightarrow$ Bob applies $Z X$.

Teleportation demonstrates the interchangeability of resources in quantum communication and quantum computation.
We have exchanged one qubit of communication for one entangled pair and two classical bits.
What does this mean?
It means that we can send the information about state $\ket{\psi}$ from Alice to Bob in two different ways.
We can send it directly, like in Fig.~\ref{fig:hop-by-hop}, which is counted as one qubit of communication.
Or, as we outlined with the teleportation protocol, if Alice and Bob share an entangled pair, \textbf{\emph{and}} are allowed to communicate two classical bits, then they can achieve the same task.

\section{No-cloning theorem and Faster-than-light communication}
\label{sec:8-3_no-cloning}

Let's go back to the teleportation protocol and ask a few important questions.
Alice communicated her state to Bob.
\textbf{\emph{Did she clone the qubit while teleporting it?}}
After all, she started with the state $\ket{\psi}$ and she did not send the physical qubit to Bob.
Yet at the end of the protocol, we saw that Bob does have the qubit $\ket{\psi}$.
What was going on there?
Initially Bob's qubit was part of a maximally entangled state.
He had one part of the entangled state and Alice had the other.
At the end of the protocol, Alice's qubit $A_1$ was part of a maximally entangled state.
She performed a measurement in the Bell basis that projected her two qubits onto one of the four possible Bell states, all of which are maximally entangled.~\footnote{Implementations commonly \emph{destructively} measure the two qubits at Alice as part of this process, in which case a more accurate description is, "The two qubits \emph{were} in Bell state..." rather than, "The two qubits \emph{are} in Bell state..." The mathematical implications for Bob's qubit and the no-cloning theorem remain the same.}
Initially, the state of the qubit $A_1$ was $\ket{\psi}$, and the state of the qubit that Bob had in his possession, considered alone, was a maximally mixed state.
After completing the protocol, the state of Alice's qubit $A_1$ became the maximally mixed state, whereas Bob's qubit became the state $\ket{\psi}$.
There was no cloning or copying of the state even though there was no direct physical transmission of the state $\ket{\psi}$ from Alice to Bob.
This leads to another interesting and important question.
\textbf{\emph{Is cloning possible?}}
Let's have a look at this question.

We will only consider cloning of pure states.
Fig.~\ref{fig:cloner} shows what a hypothetical cloning device.
It takes two qubits as inputs.
The first input is an arbitrary state $\ket{\psi}$ that we wish to clone, and the second one is initialized in \ket{0}.
The cloning operation is represented by a  unitary operation $U$.
The output is given by two copies of the arbitrary state \ket{\psi}.
The questions that we would like to naswer are teh following.
Is such a transformation possible? If so, what does the unitary look like that can achieve such a transformation?
\begin{figure}[t]
    \centering
    \begin{tikzpicture}
    \begin{yquant*}
    qubit {$|\psi\rangle$} in;
    qubit {$|0\rangle$} q;

    [shape=yquant-rectangle, rounded corners=.25em, fill=blue!10, y radius=10pt]
    box {cloning\\device\\$U$} (in, q);

    output {$|\psi\rangle$} in;
    output {$|\psi\rangle$} q;
    \end{yquant*}
    \end{tikzpicture}
    \caption[A hypothetical cloning device.]{A hypothetical cloning device would have to create an unentangled copy of the input state $\ket{\psi}$ on the second qubit initialized in $\ket{0}$.}
    \label{fig:cloner}
\end{figure}

For example if we start in the state $\ket{0}\ket{0}$, after applying the cloning unitary $U$, we should obtain state $\ket{0}\ket{0}$. If we start in the state $\ket{+}\ket{0}$ and apply our cloning unitary, we should obtain \ket{+} for the first qubit and \ket{+} for the second qubit.

Let's consider some candidate unitaries for the cloning unitary $U$.
As a first guess, let's consider a CNOT gate,
\begin{equation}
    \operatorname{CNOT} =|0\rangle\langle 0|\otimes I+| 1\rangle\langle 1| \otimes X = \begin{pmatrix}
        1 & 0 & 0 & 0 \\
        0 & 1 & 0 & 0 \\
        0 & 0 & 0 & 1 \\
        0 & 0 & 1 & 0 \\
    \end{pmatrix}
\end{equation}
If the state of the first qubit is zero, then apply the identity to the second qubit, meaning ``don't do anything''.
If the state of the first qubit is one, then apply Pauli $X$ to the second qubit.
Let's consider some inputs and their results,
\begin{equation}
\begin{aligned}
    \operatorname{CNOT} |0\rangle|0\rangle &=|0\rangle|0\rangle, \\
    \operatorname{CNOT} |1\rangle|0\rangle &=|1\rangle|1\rangle, \\
    \operatorname{CNOT} |+\rangle|0\rangle &=\frac{1}{\sqrt{2}}(|00\rangle+|11\rangle)\neq|+\rangle|+\rangle.
    \label{eq:cloning_examples_CNOT}
\end{aligned}
\end{equation}
First, let's try a very simple input, $\ket{0}\ket{0}$, as in the first equation above.
Applying the CNOT, you can check for yourself that the output is also $\ket{0}\ket{0}$.
In this case, the cloning of the input state is trivial since we do not have to actually do anything.
The second case is more interesting.
We wish to clone the state \ket{1}, and so the initial state is \ket{1}\ket{0}.
Since the first input qubit is \ket{1}, the CNOT gate is applied, flipping the state of the second qubit to \ket{1}.
We went from input $\ket{1}\ket{0}$ to output $\ket{1}\ket{1}$, cloning the qubit.

How about if we consider a superposition of zero and one as our input? The input is $\ket{+}\ket{0}$, and the desired cloned output is $\ket{+}\ket{+}$.
However, applying the CNOT, we obtain an entangled state $(\ket{0}\ket{0}+\ket{1}\ket{1})/\sqrt{2}$, which is not the product state that we were aiming for. Cloning has failed in this case.

The two cases in Eq.~(\ref{eq:cloning_examples_CNOT}) that were cloned successfully, states \ket{0} and \ket{1}, were orthogonal.
On the other hand, the one case that failed, state \ket{+}, was a superposition of these two.
Let's keep this observation in mind, and consider the case of a general cloning unitary applied to a general input state.

We start by assuming that the cloning unitary can be applied successfully to two different states \ket{\psi} and \ket{\phi}.
In other words applying the cloning unitary $U$ to the input states \ket{\psi}\ket{0} and \ket{\phi}\ket{0} produces the following outputs,
\begin{align}
    U(\ket{\psi}\ket{0}) & = \ket{\psi}\ket{\psi}, \label{eq:cloning_first_eq}\\
    U(\ket{\phi}\ket{0}) & = \ket{\phi}\ket{\phi} \label{eq:cloning_second_eq}
\end{align}
We next take the inner product between the two equations.
This may seem like a strange things to do, but it is a standard trick that you will see on numerous occasions in the future.
We begin by taking the adjoint of Eq.~(\ref{eq:cloning_second_eq}),
\begin{equation}
    (\bra{\phi}\bra{0}) U^{\dagger} = \bra{\phi}\bra{\phi}. 
\end{equation}
Next, we multiply the left-hand side Eq.~(\ref{eq:cloning_first_eq}) by the left-hand side of Eq.~(\ref{eq:cloning_second_eq}), and similarly for the right-hand sides of both equations,
\begin{equation}
    (\bra{\phi}\bra{0}) \underbrace{U^{\dagger} \cdot U}_{=I} \ket{\psi}\ket{0} = (\bra{\phi}\bra{\phi}) \cdot (\ket{\psi}\ket{\psi}).
\end{equation}
The cloning operation is assumed to be unitary, meaning $U^{\dagger}U=I$.
Therefore we can write
\begin{equation}
    \braket{\phi}{\psi} \cdot \underbrace{\braket{0}{0}}_{=1} = \braket{\phi}{\psi} \cdot \braket{\phi}{\psi}, \longrightarrow
    \braket{\phi}{\psi} = \braket{\phi}{\psi}^2.
    \label{eq:cloning_scalars}
\end{equation}
The inner product \braket{\phi}{\psi} is complex scalar.
Only scalars that satisfy Eq.~(\ref{eq:cloning_scalars}) are
\begin{equation}
    \braket{\phi}{\psi} = 0, \quad \text{and } \quad \braket{\phi}{\psi} = 1.
\end{equation}
This tells us something interesting about the initial states \ket{\psi} and \ket{\phi}.
Either the states are orthogonal, meaning $\braket{\phi}{\psi}=0$, or the states are the same, meaning $\braket{\phi}{\psi}=1$.
Any other state, for example a superposition of \ket{\psi} and \ket{\phi}, will not satisfy Eq.~(\ref{eq:cloning_scalars}), and therefore will not be cloned successfully by the unitary $U$.

This proves that our cloning really works \textbf{\emph{only}} for states that can be distinguished with certainty -- there is no ambiguity or overlap in their superpositions.
(Remember, orthogonal states are always distinguishable deterministically.)
So we can say that we cannot clone an arbitrary state in quantum mechanics, which is in stark contrast to classical physics.
This is the essence of \textbf{\emph{no-cloning theorem}}\index{no-cloning theorem}.

\begin{figure}[t]
    \centering
    \includegraphics[width=0.8\textwidth]{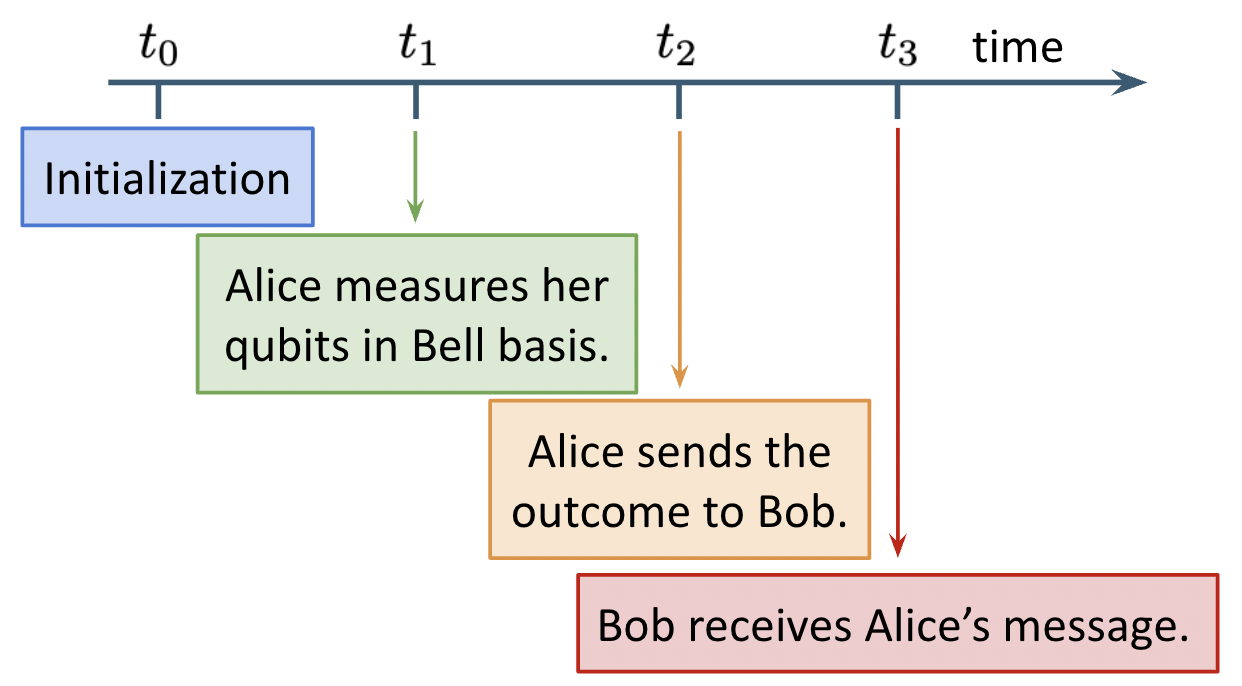}
        \caption[Event timing in teleportation.]{The timeline of necessary events and messages in teleportation makes it clear that the speed of light is not violated.}
    \label{fig:teleportation-timeline}
\end{figure}

The second interesting question that we would like to answer about teleportation is the following.
\textbf{\emph{Is teleportation instantaneous?}}
Does it allow us to communicate faster than the speed of light?
The short answer is no, otherwise teleportation would violate the special theory of relativity.
In order to understand why this is the case, we have to consider the timeline of events during the teleportation protocol.
This is pictured in Fig.~\ref{fig:teleportation-timeline}.
We start at time $t_0$, when we initialize our system.
At some later time $t_1$, Alice performs her measurement in the Bell basis on the two qubits that she holds.
At some later time $t_2$, Alice sends the outcome of the measurements to Bob.
He receives these outcomes at time $t_3$.

Let's go step by step and consider what the state of Bob's qubit at these different times.
At time $t_0$, we said that Bob's qubit is part of a maximally entangled pair, which means that the reduced density matrix of his qubit is a maximally mixed state,
\begin{equation}
    \left|\Phi^{+}\right\rangle_{A_2 B}=\frac{1}{\sqrt{2}}(|00\rangle+|11\rangle) \longrightarrow \rho_B=\frac{1}{2}\ketbra{0}{0} +\frac{1}{2}\ketbra{1}{1}.
\end{equation}
This means that if Bob measures his qubit in the Pauli $Z$ basis, he will get the outcome zero ($+1$ eigenvalue) with probability one half, or the outcome one ($-1$ eigenvalue) with probability one half. In fact, if he measures in any basis, he will get the outcome $+1$ or $-1$ with the same probability, so he's maximally unsure about his own state. That also means that he has no idea what state Alice is trying to send him.

At time $t_1$, Alice performs her measurement.
Just before the measurement, the total state of all three qubits is given by the expression in Eq.~(\ref{eq:teleport-alice-basis}).
Bob does not know anything about Alice's state $A_1$.
The initial state written in the form of Eq.~(\ref{eq:teleport-alice-basis}) might suggest that he knows something about the state of his own qubit.
This is however not true, Bob's qubit is still in a maximally mixed state.

Let's look at what is known immediately after Alice measures in the Bell basis at time $t_1$.
Just after the measurement, Bob knows that with probability one quarter, Alice measured the outcome $\ket{\Phi^+}$, and with the same probability she could have obtained each of the other three  Bell states.
So he knows that with probability $1/4$, he has the state $\ket{\psi}$, or with probability $1/4$ he has some equivalent state given by these following expressions,

\hspace{2cm} $\operatorname{Prob}\left(\left|\Phi^{+}\right\rangle\right)=1 / 4, \quad$ Bob has $\alpha|0\rangle+\beta|1\rangle$

\hspace{2cm} $\operatorname{Prob}\left(\left|\Phi^{-}\right\rangle\right)=1 / 4, \quad$ Bob has $\alpha|0\rangle-\beta|1\rangle$

\hspace{2cm} $\operatorname{Prob}\left(\left|\Psi^{+}\right\rangle\right)=1 / 4, \quad$ Bob has $\alpha|1\rangle+\beta|0\rangle$

\hspace{2cm} $\operatorname{Prob}\left(\left|\Psi^{-}\right\rangle\right)=1 / 4, \quad$ Bob has $\alpha|1\rangle-\beta|0\rangle$.
This is a distribution which we can write in the density matrix formalism.
The calculation is a bit lengthy but stick with it.
Bob's qubit can be written as follows:
\begin{equation}
\begin{aligned}
    \rho_B &=\frac{1}{4}(\alpha|0\rangle+\beta|1\rangle)\left(\alpha^*\langle 0|+\beta^*\langle 1|\right) \\
    &+\frac{1}{4}(\alpha|0\rangle-\beta|1\rangle)\left(\alpha^*\langle 0|-\beta^*\langle 1|\right) \\
    &+\frac{1}{4}(\alpha|1\rangle+\beta|0\rangle)\left(\alpha^*\langle 1|+\beta^*\langle 0|\right) \\
    &+\frac{1}{4}(\alpha|1\rangle-\beta|0\rangle)\left(\alpha^*\langle 1|-\beta^*\langle 0|\right) \\
    &=\frac{1}{2}\left(|\alpha|^2+|\beta|^2\right)\ketbra{0}{0} + \frac{1}{2}\left(|\alpha|^2+|\beta|^2\right)\ketbra{1}{1} \\
    &=\frac{1}{2}\ketbra{0}{0}+\frac{1}{2}\ketbra{1}{1}\\
    \label{eq:Bobs_qubit_t2}
\end{aligned}
\end{equation}
In the penultimate line, we used the fact that Alice's state to be teleported is normalized, meaning $|\alpha|^2+|\beta|^2=1$.
Alice's measurement breaks the entanglement between her and Bob's qubit.
Yet, Bob's qubit is still in a maximally mixed state, meaning he has no information about the state \ket{\psi}.

Let's move to the later time $t_2$ when Alice sends her measurement outcome to Bob.
Nothing really changes on Bob's side compared to the situation a time $t_1$.
His qubit is still in a maximally mixed state described by Eq.~(\ref{eq:Bobs_qubit_t2}).
Bob still has no information about the state \ket{\psi} because the information sent by Alice about her measurement outcome requires a finite amount of time to travel to Bob.

Finally, at time $t_3$, Bob receives Alice's message about the measurement outcome.
This changes the state of his qubit to one of the four possibilities in Eq.~(\ref{eq:teleportation_possible_states}).
He applies the appropriate correction unitary ensuring his qubit is in the desired state \ket{\psi}.

\newpage
\begin{exercises}

\exer{
\emph{Filling in the gaps.}
During our discussion of the teleportation we have skipped some details, which we will consider now.
\subexer{
Consider the input state of the three qubits $A_1,A_2$, and $B$ in Eq.~(\ref{eq:teleportation_initial_state}).
Write down, both in Dirac and matrix notation, the initial state of each of the three qubits.
}
\subexer{
Write down Alice's measurement operators in the computational basis and in both Dirac and matrix notation.
}
\subexer{
Convince yourself that the probabilities of the four possible outcomes are independent of the initial state $A_1$.
}
}

\exer{
\emph{Teleporting the state \ket{0}.}
Let's work through an example of teleportation to reinforce our understanding of the protocol.
Rather than teleporting a general pure state \ket{\psi}, we wish to teleport the state \ket{0}.
\subexer{
Write down the initial state of all three qubits in the computational basis.
}
\subexer{
Rewrite the initial state, such that Alice's qubits $A_1$ and $A_2$ are written in the Bell basis, while Bob's qubit is in the computational basis.
}
\subexer{
How many possible outcomes of Alice's Bell-basis measurement are there?
What are their corresponding probabilities?
}
\subexer{
How many classical bits of information does Alice need to communicate to Bob to inform him of the measurement outcome?
}
\subexer{
What correction operations does Bob need to apply to ensure he receives the state \ket{0} after the teleportation protocol is finished?
}
}

\exer{
\emph{Teleporting a mixed state.}
So far we have only looked at the case when Alice wishes to teleport a pure state.
Let's teleport a mixed state, for example a state that is diagonal in the computational basis,
\begin{equation}
    \rho = a \ket{0}\bra{0} + b \ket{1}\bra{1}.
\end{equation}
\subexer{
Write down the initial state of all three qubits in the computational basis.
}
\subexer{
Compute the probabilities of Alice's measurement outcomes on the two qubits.
Did you expect your answer?
}
\subexer{
Find the state of Bob's qubit when Alice obtains \ket{\Phi^+} as the outcome of her measurement.
}
\subexer{
Find the state of Bob's qubit when Alice obtains \ket{\Phi^-} as the outcome of her measurement.
}
\subexer{
What are the correction operations that Bob needs to apply in order to successfully receive the state $\rho$, regardless of Alice's measurement outcome?
}
}

\exer{
\emph{Different initial entangled states.}
Up until now, we have consider the initial entangled state between Alice and Bob to be \ket{\Phi^+}.
Let's see what happens when we use the other Bell states.
\subexer{
Go through the teleportation protocol in order to find out Bob's correction operations in the following cases.
}
\subexer{
When the shared state is \ket{\Phi^-}.
}
\subexer{
When the shared state is \ket{\Psi^+}.
}
\subexer{
When the shared state is \ket{\Psi^-}.
}
\subexer{
Can you see a pattern?
}
}

\end{exercises}

\chapter{BB84: Single-photon QKD}
\label{sec:9_bb84}

This chapter introduces single-photon quantum key distribution (QKD), specifically the first single-photon QKD protocol known as BB84, named for Charles Bennett and Giles Brassard, who developed it in 1984. First, to place the use of QKD in context, we provide a three-phase framework for secure communication. Second, we discuss the types of cryptosystems and the distinction between public key and symmetric key systems. We then present the basic idea of the protocol and the eavesdropper detection mechanism before introducing some of the early QKD testbed networks.

\section{Three phases of cryptographically secure communication}
\label{sec:crypto-phases}

Secure communication using encryption proceeds roughly in the following three phases: first, the parties \textbf{\emph{authenticate}} each other, meaning that they prove they really are who they say they are and not somebody else.
Second, they select or \textbf{\emph{generate a key}} that they will use for encoding their messages. 
Third, they \textbf{\emph{encode}} their messages and encrypt their data, then send them to the other party where they will be decrypted -- the actual bulk data transfer phase of the conversation.

Authentication between Alice and Bob usually begins in the following way.
Alice begins by saying ``I'm Alice'' to Bob.
Bob replies with ``Okay, you're Alice, I'm Bob''.
The problem that both Alice and Bob face now is that anybody could have sent those messages.

Generally, the procedure two parties use to authenticate each other depends on one of three things\footnote{This applies in the real world as well; trying to get into a private club or a 1920s U.S. speakeasy might involve personal recognition, a physical token such as an actual key, or a password you have been given by someone else.}:
\begin{itemize}
    \item something you \textbf{\emph{are}},
    \item something you \textbf{\emph{have}},
    \item something you \textbf{\emph{know}}.
\end{itemize}
In particular the last one is quite important, because you know that the party that you are trying to communicate knows something, for example a secret pre-shared between the two parties, or generally and flexibly the private key corresponding to a particular public key that can be used as an authentication mechanism.

After authentication, Alice and Bob have to generate a \textbf{\emph{cryptographic key}}\index{cryptographic key} (or just ``key'').
There are many different ways to exchange keys, such as Diffie-Hellman key exchange, which we will not go into here.
The key will be used for encrypting the message.
One problem with the key is that once it is generated, it cannot be used forever.
It has to be changed at regular intervals in order to ensure that the encrypted data remains secure.

Once the key is generated, Alice encrypts her message with the key.
She sends it to Bob, where he will use his share of the generated key to decrypt the message and read it. 
Typically, this \textbf{\emph{bulk data encryption}}\index{bulk data enxryption} phase of the conversation uses a \textbf{\emph{symmetric key cryptosystem}}\index{symmetric key cryptosystem}, where Alice and Bob use the same key to encrypt and decrypt the messages.
Common forms include the now-outdated Data Encryption Standard (DES), the stronger (and still in use) 3-DES, and the modern Advanced Encryption Standard (AES)\index{Advanced Encryption Standard (AES)}. Less common is one-time pad (OTP)\index{one-time pad (OTP)}.
Later we will describe the one-time pad, which is also known as the \textbf{\emph{Vernam cipher}}\index{Vernam cipher}.

Quantum key distribution, or QKD, is used in the second phase of this process. 
Before we go into QKD, let's look briefly at how keys are established today when you use the Internet.

\section{Key agreement and use}
\label{sec:key_agreement_use}

We have two parties, Alice and Bob, who are trying to communicate over a public channel.
``Public'' means that anybody has access to this channel, so any message that they transmit can be heard and intercepted by any other party.
This channel is therefore not secure.
And yet, Alice and Bob need to somehow authenticate each other, and prepare the key to be used for the bulk data encryption.

The simplest method is via a \textbf{\emph{pre-shared key (PSK)}}\index{pre-shared key (PSK)}.
Alice and Bob meet somewhere, face to face, long before they want to communicate. 
At this meeting, they create a shared secret between them, known to no one else. 
This key can then be used later to authenticate each other, and to help generate the keys used for the bulk encryption. 
Of course, for this method to work as a general mechanism for any two parties, every person must meet every other person. 
That requires $O(N^2)$ meetings for $N$ people, which is clearly not practical.

To compensate for this shortcoming, instead we use mechanisms that do not require this pairwise prior agreement.
One method is known as \textbf{\emph{public key cryptography}}\index{public key cryptography}.
Alice generates two keys, one is known as the ``public key'' and the other one is known as the ``private key''. 
They are a mathematically related pair.
The public key is used to encrypt the message, but it cannot be used to decrypt the message. 
The private key is needed to decrypt the message.
For this reason, public key cryptography is sometimes called \textbf{\emph{asymmetric encryption}}\index{asymmetric encryption}.
Bob's public key is published in some fashion that allows Alice both to find Bob's key when she needs it, and trust that it is really Bob's key.
Alice uses that to encrypt her message, and then sends it to Bob.
Bob then uses the private key (which he has kept secret) to decrypt Alice's message and read it.
The most prominent form of public key cryptography is known as RSA, designed by Rivest, Shamir and Adelman.

This process works, but it has some disadvantages.
It is slow and computationally expensive.
While it is theoretically possible for Alice to encrypt her actual message to Bob using public-key cryptography, which would combine the functions of authentication and encryption while eliminating the need for a separate key, it is not practical. 
AES, on the other hand, is easier to compute.
Therefore, we usually use \textbf{\emph{three separate functions}} as described in the last section, with RSA for authentication, Diffie-Hellman for session key generation, and AES\index{Advanced Encryption Standard (AES)} for bulk data encryption.

More important than the performance issues with public key cryptography, the security of both RSA and Diffie-Hellman is based on the idea that they are \textbf{\emph{computationally secure}}\index{computational security}.
Anybody listening to the channel is able to record and store the classical messages exchanged.
In principle, an eavesdropper can break the encryption if they have access to a fast enough computer, either now (unlikely) or in the future (harder to guarantee)~\footnote{This is sometimes called ``harvest now, decrypt later''.}.
Therefore, RSA and Diffie-Hellman are not unconditionally secure~\footnote{Cryptographers are working to replace these mechanisms, in a broad push called \emph{post-quantum cryptography (PQC)}.
As of this writing, the first major phase of this process is nearing completion.}.
A crucial point is that quantum computers can in principle break some computationally secure protocols, notably RSA and Diffie-Hellman, with relative ease~\footnote{The details of this possibility are a very long discussion, but what you read in the popular press often over-simplifies this issue.}.
So, how can we actually establish a secure connection between Alice and Bob?

The bulk encryption mechanism can be AES, introduced in the last section, but the details are beyond the scope of this book.
Alternatively, we can encrypt one bit of message using one bit of the key. 
One way to do this is by using an exclusive-OR operation (XOR).
Its truth table is shown in Tab.~\ref{tab:xor_truth_table}.
\begin{table}[t]
    \setcellgapes{5pt}
    \renewcommand\theadfont{}
    \makegapedcells
    \centering
    \begin{tabular}{ccc}
        \hline
        \textbf{Input bit $a$} & \textbf{Input bit $b$} & \textbf{Output bit $c$} \\
        \hline
        0 & 0 & 0 \\
        0 & 1 & 1 \\
        1 & 0 & 1 \\
        1 & 1 & 0 \\
        \hline
    \end{tabular}
    \caption[XOR operation.]{Truth table for the XOR operation.}
    \label{tab:xor_truth_table}
\end{table}
The output bit $c$ of the XOR operation for two input bits $a$ and $b$ is the following,
\begin{equation}
    c = a \oplus b = \begin{cases}
        0, \quad \text{if } a = b, \\
        1, \quad \text{if } a \neq b.
    \end{cases}
\end{equation}
If $m_i$ is the $i$-th bit of the message, $k_i$ is the $i$-th bit of the key, the $i$-th bit of the ciphertext (encrypted message) is given by $c_i = m_i \oplus k_i$. 
This approach is known as \textbf{\emph{one-time pad}}\index{one-time pad (OTP)}.
This technique cannot be broken, provided that the key was generated in a secure and random fashion and, critically, is used only once.
If Alice has a message of $n$ bits that she would like to send to Bob, she requires a secret key that iss at least $n$ bits long.
Once she uses that key to encrypt her message she cannot use it again. If she has some other thing to say to Bob, they require a completely new and fresh secret key to ensure security.
This makes OTP secure but requires a large number of secret bits.

OTP is actually pretty good in theory, but there is one remaining question if we want to use the one-time pad, and that's how to actually distribute this key.
Perhaps our hypothetical key generator is used only during Alice and Bob's face-to-face meeting.
They could generate enough key bits for a year's worth of communication or more, store them somewhere secure and use when needed.
But, we already know that that is impractical.
Our only alternative is to use the public channels available.
The classical public channel is subject to being recorded, as noted above, but a quantum channel, even if nominally public, offers unique properties that allow us to guarantee that no one is listening in.
We will see how this can be achieved in the next section.

\section{BB84 Protocol}
\label{sec:bb84-protocol}

BB84 is a quantum protocol for generating a string of shared and secret classical bits between two parties.
Alice and Bob can utilize a public quantum channel, as well as their public classical channel.
The key is established through appropriately initialized qubits which are communicated by Alice over the quantum channel, and measured by Bob.
Classical information is exchanged between Alice and Bob over the classical channel.
The protocol has the following four stages:
\begin{itemize}
    \item Qubit preparation,
    \item Quantum communication,
    \item Classical communication,
    \item Classical post-processing.
\end{itemize}
Let's discuss these stages in more depth.

\emph{Qubit preparation.}
Alice begins the protocol by generating two $n$-bit random strings, denoted by $a$ and $b$,
\begin{align}
    a & = a_1 a_2 \ldots a_n, \\
    b & = b_1 b_2 \ldots b_n,
\end{align}
where $a_i$ and $b_i$ are the $i$-th bits of bit strings $a$ and $b$, respectively.
Alice uses the values of individual bits from $a$ and $b$ to prepare her qubits in a very particular way.
The basis for the encoding is determined by the bit string $b$, while a particular state of the qubit is chosen according to the bit string $a$.
Qubit $i$ is encoded in the Pauli $Z$ basis if $b_i=0$, and in the Pauli $X$ basis if $b_i=1$.
If bit $a_i=0$ then the qubit is prepared in the +1 eigenstate of the chosen basis.
If $a_i=1$ then the qubit is prepared in the -1 eigenstate of the chosen basis.
I short,
\begin{equation}
    a_i = \begin{cases}
        0 \rightarrow +1\text{ eigenstate}, \\
        1 \rightarrow -1\text{ eigenstate},
    \end{cases}
    \quad
    b_i = \begin{cases}
        0 \rightarrow Z\text{ basis}, \\
        1 \rightarrow X\text{ basis}.
    \end{cases}
\end{equation}
Writing the state of qubit $i$ as \ket{\psi_{a_ib_i}}, we have the following four possible states that Alice will prepare,
\begin{equation}
    \ket{\psi_{00}} = \ket{0}, \quad \ket{\psi_{10}} = \ket{1}, \quad \ket{\psi_{01}} = \ket{+}, \quad \ket{\psi_{11}} = \ket{-}.
    \label{eq:BB84-four-states}
\end{equation}
The full state of all $n$ qubits that Alice prepares is given by the tensor product of the states of individual qubits,
\begin{equation}
    |\psi\rangle=\bigotimes_{k=1}^n\left|\psi_{a_k b_k}\right\rangle.
\end{equation}

Notice that the four states in Eq.~(\ref{eq:BB84-four-states}) that Alice prepares are not all orthogonal.
For example, if we take the inner product between \ket{\psi_{00}} and \ket{\psi_{01}}, we get,
\begin{equation}
    \braket{\psi_{00}}{\psi_{01}} =\frac{1}{\sqrt{2}} \neq 0.
\end{equation}
In this particular case, the inner product is $1/\sqrt{2}$. It will be the same if we take, for example, 
$\braket{\psi_{10}}{\psi_{11}} = 1/\sqrt{2} \ne 0$.
When the inner product is non-zero, it means that the two states are not perfectly distinguishable.
This observation is a crucial ingredient in the BB84 protocol. 

Why is this non-distinguishability important?
It is not used directly in the generation of the key but it is necessary if Alice and Bob have a suspicion that a third malicious party is trying to eavesdrop on their conversation.

Let's look at an example of the encoding, shown in Tab.~\ref{tab:bb84-example}.
Let's say that Alice generates the two random strings $a = 01101$ and $b=11001$.
She starts encoding.
First, she looks at the first bit in her string $b$, which is $1$, so she knows, ``Now I have to encode in the $X$ basis'', and the state that she prepares is a \ket{+} state because her first bit in the $a$ string is a $0$.
To prepare the second qubit, she looks at the second bit in her string $b$, which is again $1$, so again she knows she has to prepare a state from the $X$ basis, and the state is given by the second bit in string $a$, which is $1$, therefore she prepares the state \ket{-}.
She goes on until she prepares all five qubits.
\begin{table}
    \setcellgapes{3pt}
    \renewcommand\theadfont{}
    \makegapedcells
    \centering
    \begin{tabular}{cccccc}
        \hline
        & \textbf{1} & \textbf{2} & \textbf{3} & \textbf{4} & \textbf{5} \\
        \hline
        \boldmath$a_i$ & 0 & 1 & 1 & 0 & 1 \\
        \boldmath$b_i$ & 1 & 1 & 0 & 0 & 1 \\
        \textbf{Encoding basis} & $X$ & $X$ & $Z$ & $Z$ & $X$ \\
        \textbf{Encoded qubit} & \ket{+} & \ket{-} & \ket{1} & \ket{0} & \ket{-} \\
        \hline
    \end{tabular}
    \caption[BB84 encoding example.]{Alice encodes the qubits she wishes to communicate, choosing a basis using a bit from $b$ and an eigenstate based on the corresponding bit from $a$.}
    \label{tab:bb84-example}
\end{table}

\emph{Quantum communication.}
Alice sends the prepared qubits to Bob over the public quantum channel. Let's consider what Bob knows at this time.
Alice has not shared the secret string $b$ containing information about the preparation basis with him.
All Bob knows is that he is receiving qubits that could be any of the four possible states \ket{0}, \ket{1}, \ket{+} or \ket{-}.
He creates his own random bit string $b'$.
Because he's expecting $n$ qubits, he generates $n$ bits: $\{b'_1, b'_2, \ldots b'_n\}$.
He uses this bit string to pick the basis for measuring each of the qubits he receives.
If $b'_i=0$, he measures qubit $i$ in the Pauli $Z$ basis.
If $b'_i=1$, he measures qubit $i$ in the Pauli $X$ basis.
If the outcome of the $i$-th measurement is $+1$, he assigns $0$ to bit $a'_i$.
If the outcome is $-1$, he assigns $1$ to bit $a'_i$.
This way Bob generates a new random string of bits $a'$.

\emph{Classical communication.}
Next, Alice and Bob share classical information with each other over the public classical channel.
Alice shares her randomly generated bit string $b$, and Bob shares his randomly generated bit string $b'$.
This exchange shares information about the basis in which the qubits were prepared and in which they were measured.

Does this classical communication compromise the security of the protocol?
After all, the strings $b$ and $b'$ are used in generating the secret key.
Sharing of these two random bit strings does not make the protocol insecure, provided the communication occurs \textbf{\emph{after}} the qubits are measured by Bob.

\emph{Classical post-processing.}
If Bob measured in the same basis that Alice used during the preparation of the qubit, they will keep the corresponding bits from $a$ and $a'$.
If they measured in different bases, then they just discard the bits $a_i$ and $a'_i$.
If Alice prepares a qubit in a certain basis and Bob measures it in the same basis, the two possible states are orthogonal, meaning they are perfectly distinguishable by the measurement in that basis.
We denote the bits that they keep as $\bar{a}$ and $\bar{a}'$.
These bits are perfectly correlated, meaning that these two shorter bit strings that they generated are equal.
Finally, Alice and Bob share a secret key that they can use to encrypt their message.

Let's look at an example, again considering a case where $n=5$, as shown in Tab.~\ref{tab:complete-bb84}.
Alice has randomly generated two five-bit strings, $a$ and $b$.
Bob generates his own random $b'$ string.
This bit string is different from Alice's $b$ because he does not know Alice's $b$ at this time.
He measures in the basis given by $b'$.
You can see that in the first position Alice's preparation basis and Bob's measurement basis agree, so their classical bits of the secret key $\bar{a}_1$ and $\bar{a}'_1$ are the same.

On the other hand, Alice prepared the second qubit in the $Z$ basis.
Bob's random bit $b'_2 = 1$, meaning that he measured it in the $X$ basis.
With fifty percent probability he will obtain $+1$, with fifty percent probability he will obtain $-1$.
Therefore Alice and Bob cannot be sure that they are really sharing a correlated bit, therefore they discard it.
For the third qubit, $b_3 = b'_3$ and again Alice prepared in the same basis as Bob measured.
Alice and Bob know with certainty their classical bits are correlated Therefore they keep them as part of the secret key..
In this way, they are keeping a random subset of the bits, leaving a shorter string of key bits, all of which are one hundred percent correlated.
The shared key that they have is given by $\bar{a} = \bar{a'} = \{1, 0, 1\}$, as marked with the check marks in the table.

\begin{table}
\setcellgapes{3pt}
    \renewcommand\theadfont{}
    \makegapedcells
    \centering
    \begin{tabular}{cccccc}
        \hline
        & \textbf{1} & \textbf{2} & \textbf{3} & \textbf{4} & \textbf{5} \\
        \hline
        \boldmath$a_i$ & 0 & 1 & 1 & 0 & 1 \\
        \boldmath$b_i$ & 1 & 1 & 0 & 0 & 1 \\
        \textbf{Alice's basis} & $X$ & $X$ & $Z$ & $Z$ & $X$ \\
        \textbf{Encoded qubit} & \ket{+} & \ket{-} & \ket{1} & \ket{0} & \ket{-} \\
        \boldmath$b'_i$ & 1 & 1 & 1 & 0 & 0 \\
        \textbf{Bob's basis} & $X$ & $X$ & $X$ & $Z$ & $Z$ \\
        \boldmath$a'_i$ & 1 & 0 or 1 & 0 & 0 or 1 & 1 \\
        \textbf{keep or discard} & \textcolor{mygreen}{\checkmark} & \textcolor{myred}{\xmarksmall} & \textcolor{mygreen}{\checkmark} & \textcolor{myred}{\xmarksmall} & \textcolor{mygreen}{\checkmark} \\
        \boldmath$\bar{a}_i,\bar{a}_i'$ & 1 & - & 0 & - & 1 \\
        \hline
    \end{tabular}
    \caption[A complete BB84 example for five bits.]{A complete BB84 example for five bits.  The check marks indicate bits where the selected preparation and measurement bases agree, giving a bit that can be used in the final agreed-upon key.}
    \label{tab:complete-bb84}
\end{table}


So far, we have only considered the ideal scenario where there was no eavesdropper\footnote{In fact, we have not even checked for the presence of an eavesdropper yet.}.
Now let's say that somebody is listening to both the public classical channel and the public quantum channel.
Next, we are going to consider the effect of an eavesdropper, whom we are going to name Eve, and see what effect she has on the protocol and how the protocol can discover the presence of such an eavesdropper.

\section{Eavesdropper detection}
\label{sec:eavesdropper-detection}

In the previous section, we saw how the protocol works under idea conditions.
Let's see what happens when we include the effect of an eavesdropper trying to gain access to the secret key that Alice and Bob are generating generate.
This time, we place Eve in between Alice and Bob.
Eve also has access to the public quantum and classical channels that are used to communicate the qubits and to exchange the classical information.

\begin{figure}[t]
    \centering
    \includegraphics[width=0.8\textwidth]{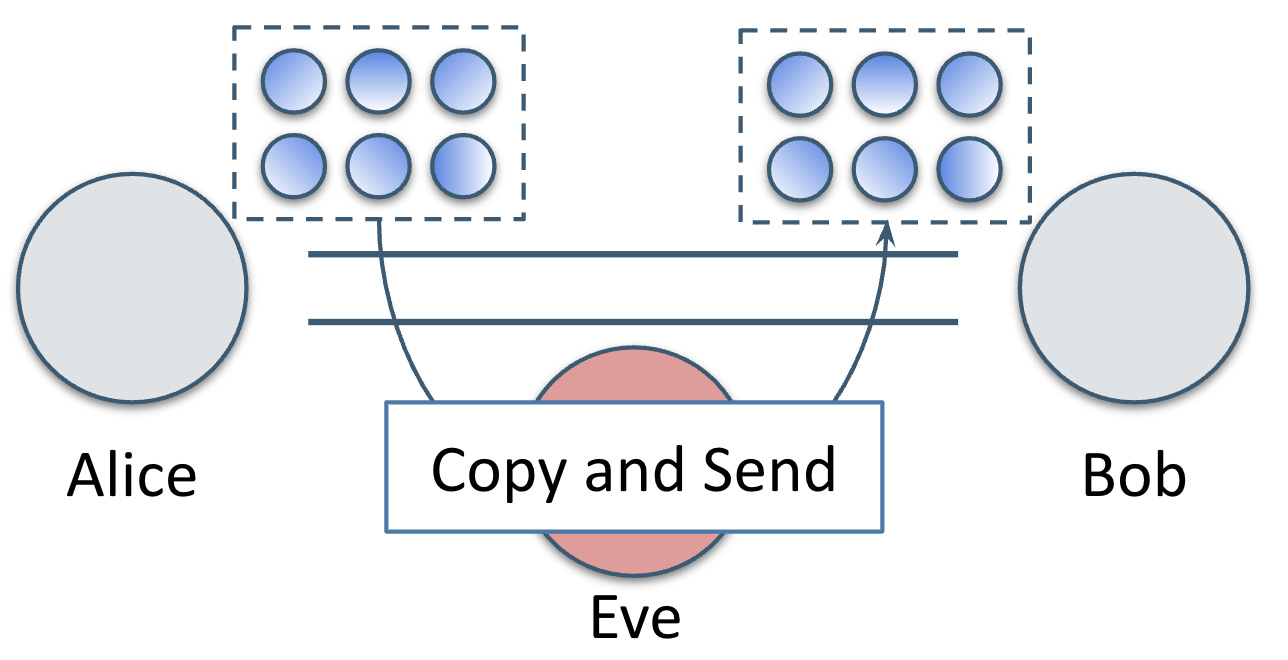}
        \caption[Eve's ideal (but impossible) eavesdropping arrangement.]{Eve's ideal eavesdropping arrangement would be to copy the qubits sent by Alice, sending one copy to Bob and keeping the other copy for herself, but the no-cloning theorem makes that impossible.}
    \label{fig:eve-copy-and-send}
\end{figure}

What can Eve do?
Let's imagine that she can intercept the qubits that Alice is sending over the public quantum channel, copy them and then resend them to Bob, as in Fig.~\ref{fig:eve-copy-and-send}.
If she could do that, she could hold her copy of the qubits, wait until Alice and Bob announce the qubit preparation and measurement bases, then measure her copies.
Doing that, she would discover which qubits are used for the generation of the secret key, and she would also know in which basis to measure them in order to generate a key that is perfectly correlated with the one hared between Alice and Bob.
Luckily, it is impossible for Eve to do that due to the no-cloning theorem\index{no-cloning theorem} we saw in the previous Chapter~\ref{sec:8-3_no-cloning}.

If Eve cannot copy the qubits and hold them, then in order to gain any access to the information that Alice is trying to share with Bob, she has to measure the qubits and forward them to Bob.
What can happen in this case?
Remember, the preparation basis stored in the bit string $b$ is still kept secret by Alice.
That basis has not yet been communicated over a public classical channel to Bob.
Without access to this information, Eve has to pick either the $X$ or $Z$ basis at random for her measurement.
Eve therefore runs the risk of disturbing and altering the state of hte qubits that she intercepted from Alice.

\begin{figure}[t]
    \centering
    \includegraphics[width=0.8\textwidth]{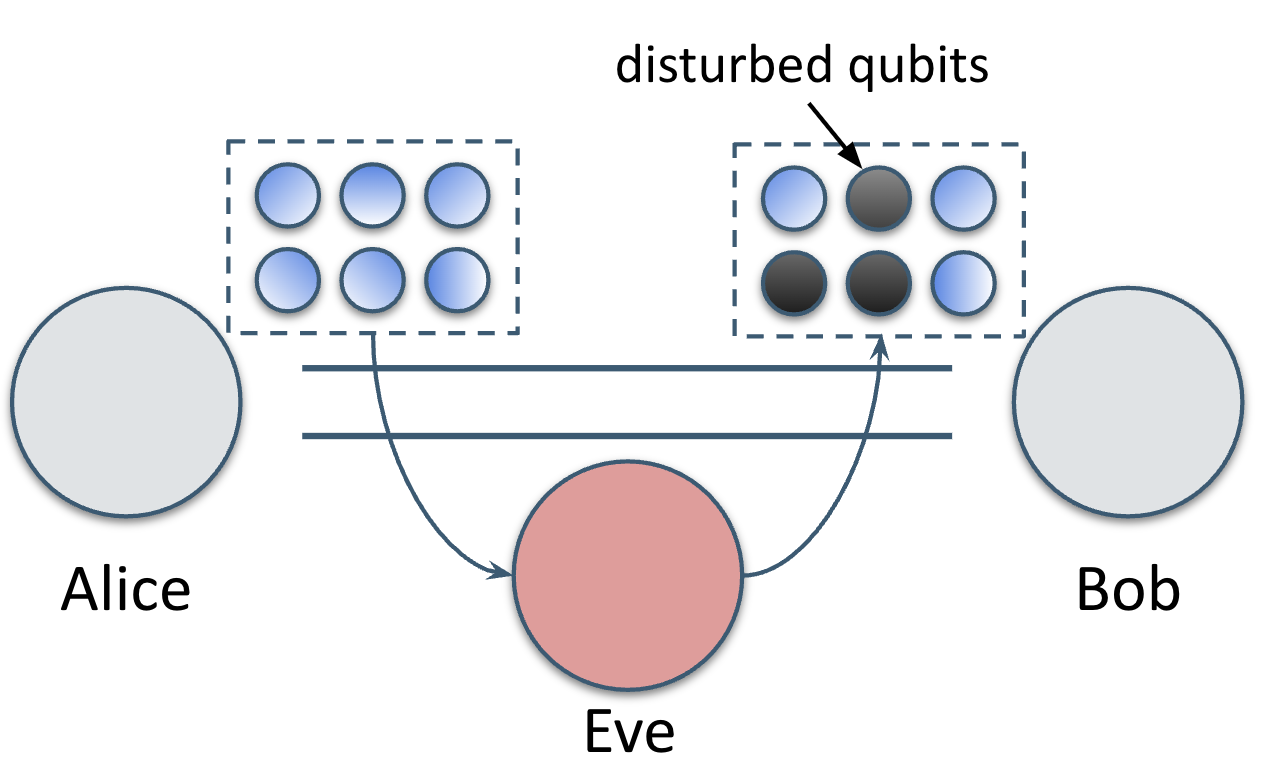}
        \caption[Eve's measurements disturb the qubits.]{If Eve tries to measure  the qubits sent by Alice, then send them on to Bob, the disturbance she introduces will be easily detected.}
    \label{fig:eve-disturbance}
\end{figure}

For example, if Alice prepares the qubit in the Pauli $X$ basis, and Eve also measures in the $X$ basis, there is no disturbance to the state of the qubit.
The qubit is still projected onto the same state that it was prepared in, leaving the state of the qubit unchanged.
However, if Alice's preparation basis is the Pauli $Z$ and Eve's measurement basis is the Pauli $X$, things are different.
The qubit originally prepared by Alice was either in \ket{0} or in \ket{1}.
But by measuring in the $X$ basis, Eve is now projecting the qubit onto either the \ket{+} state or the \ket{-} state.
This disturbs the state of the qubit, forcing it into an eigenstate of the wrong basis.
Similarly, where Alice prepares in the $X$ basis and Eve measures in the $Z$ basis, the state will be projected into a different state than the one that was prepared.
This is the main principle Alice and Bob will use to detect the presence of Eve.

Fig.~\ref{fig:eve-disturbance} shows this scenario.
Alice prepares her qubits, then she starts sending them over the public quantum channel.
Eve intercepts and measures them in a randomly chosen basis, and then she passes on these qubits to Bob.
Sometimes, Eve guesses the correct basis, measuring in the preparation basis and leaves the qubits undisturbed.
But sometimes, she chooses the wrong basis for her measurement.
She disturbs some of the qubits, represented by the black qubits in the figure.
This gives Alice and Bob an opportunity to detect that there is an eavesdropper tampering with the qubits.

To detect Eve, Alice and Bob go through with their BB84 protocol.
They compare their preparation and measurement bases, and keep only those bits where they worked in the same basis, making the new, shorter string $\bar{a}$.

Now comes the new part.
Alice and Bob dedicate a portion of the string $\bar{a}$ to detecting the presence of Eve.
First, let's consider one qubit.
Eve has a $50\%$ chance of measuring in the same basis as Alice's preparation basis for the qubit.
Bob also chooses the same basis as Alice with probability $50\%$. Remember, if Alice prepared in one basis and Bob measured in the same basis, they are expecting this classical measurement outcomes to be the same, giving them a perfectly correlated key. But if Eve measured this qubit in the \textbf{\emph{wrong}} basis and disturbed this qubit, she risks being detected.
Of course, if Bob also selected the wrong basis, then Alice and Bob will discard that bit and Eve's disturbance goes undetected -- she got lucky.
But, if Bob picks the right basis when Eve picks the wrong one, then Alice and Bob will assume the bit to be good but there is a chance that the two classical bits will not coincide. If Alice and Bob compare their values for this bit and they don't coincide, then they know that something went wrong and there's an eavesdropper present trying to gain access to their secret key.  

Let's focus on the string $\bar{a}$, which represents the cases where Alice and Bob picked the same basis. 
From $\bar{a}$, Alice and Bob choose (again, it is important that this selection be random) a subset of the measured bits to use to test for the presence of Eve, reserving the rest for the final, secret key.
The chosen classical bits are disclosed by both Alice and Bob, who compare them and look for discrepancies.
Within this set of bits, the potential detection happens when Eve picked the wrong basis, which happens half of the time, at random.
But even in that case, there is a $50\%$ chance that Eve again gets lucky and Bob's subsequent measurement projected the qubit \textbf{\emph{back}} into the original state that Alice prepared.
In that case, when Alice and Bob compare their results, they will find the same values and Eve slips through undetected.
In fact, Alice and Bob only detect Eve with a probability of $1/4$ using a single bit.

The strength of the protocol comes through repetition of this probabilistic test.
Let's say that Alice and Bob choose to use $n$ of the bits for this detection procedure.
Eve has a probability of $3/4$ of passing each individual test, but put them all together and Alice and Bob have an excellent shot at detecting Eve.
For $n$ test bits, the eavesdropper detection probability is
\begin{equation}
P(n)=1-\left(\frac{3}{4}\right)^n.
\end{equation}

\begin{figure}[t]
    \centering
    \includegraphics[width=0.7\textwidth]{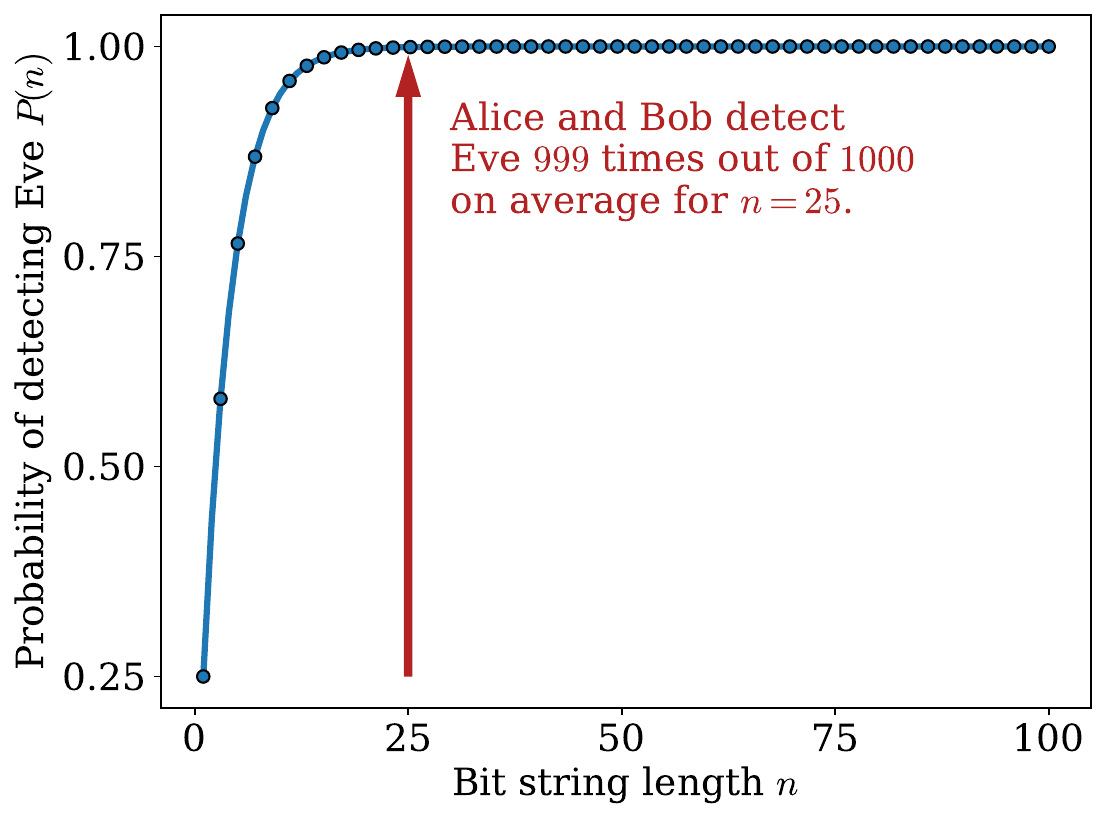}
        \caption[Probability of detecting the presence of Eve.]{The probability of Alice and Bob detecting the presence of Eve reaches $0.999$ with only $25$ samples.}
    \label{fig:catching-eve}
\end{figure}

In Fig.~\ref{fig:catching-eve}, on the horizontal axis, we plot $n$, the length of the bit string that we have dedicated to detecting the eavesdropper.
On the vertical axis, we have the probability of detection $P(n)$.
When we dedicate only a few bits to detect the eavesdropper, $n$ is very small and the probability that Eve is detected is relatively small.
But very quickly, this probability shoots up to near certainty.
With nearly a 100\% probability, Alice and Bob can detect the presence of an eavesdropper.
Even for $n = 25$, dedicating a mere 25 bits to detecting Eve, the chance is only one in a thousand that Eve will not be detected.
With very high probability, Alice and Bob know that somebody's eavesdropping onto their channel, in which case they say to each other, ``We know that we are not sharing a secure secret key, therefore we choose not to continue with the rest of our communication'', preventing Eve from gaining access to the key bits that Alice and Bob had intended to use to encrypt the sensitive messages they had planned to exchange.

From our discussion so far, it would appear that Alice and Bob can distribute a secret key fairly easily using the BB84 protocol.
In particular, any malicious eavesdropper trying to gain information about the secret key seems to be readily discovered using a relatively small number of qubits.
Things are more complicated in real life.
We have discussed the ideal protocol in absence of noise.
Noise can also change the state of the qubits transmitted by Alice.
Qubits affected by noise can flip to their orthogonal state, they can be rotated to a new basis, or more generally, they can become a mixture of pure states.
This means that even in the absence of malicious Eve, there is a substantial probability that some of the qubits reserved for eavesdropper detection will be disturbed and projected onto the wrong state by Bob.
This will lead to Alice and Bob to the erraneous conclusion that someone is trying to eavesdrop on their communication.

In order to avoid this scenario, Alice and Bob must be ready to accept some deviation from the ideal protocol, and make peace with the fact that some level disturbance will be always present.
If the channel is nearly ideal with low levels of noise, the expected amount disturbance is also low.
Noisy channels will result in more disturbance.
The question of what amount of disturbance Alice and Bob are willing to tolerate is crucial.
If they set the acceptable level too high, malicious Eve will have a good chance to go undetected and gain some information about the secret key.
If the acceptable level is set too low, Alice and Bob will have a high chance of unneccessarily rejecting the secret key, leading to a waste of network resources.
Picking the middle ground is a challenging task that goes beyond the scope of this book.


\section{Existing QKD network testbeds}
\label{sec:existing-qkd-networks}


In the previous sections, we saw how Alice and Bob can use quantum mechanics to discover an eavesdropper while generating a secret random key.
The exciting thing about QKD using single photons is that it is not just a theoretical concept that exists only on paper.
It is not even a one-off experimental demonstration under ideal conditions.
QKD networks have been built and tested in the real world.
In this section, we will learn about some of the QKD network testbeds, their capabilities and scale.

The very first network that was built was the DARPA QKD network in 2004.
The network was built in the state of Massachusetts in the USA.
It consisted of ten nodes, and a number of different physical links.
Communication over free space and communication over a fiber were both incorporated, and optical switching allowed different nodes to establish connections.
This setup was unlike any of the previous experimental implementations, where secure quantum communication in the form of a single photon-based QKD was done only on a point-to-point basis over a single link.
This was the first network experiment demonstrating the viability of quantum key distribution in a more realistic setting.
The work was led by Chip Elliott of BBN, and included a version of the IPsec Internet protocol security suite adapted for QKD.
IPsec usually uses the canonical set of RSA or pre-shared secret for authentication, Diffie-Hellman for key establishment, and 3-DES or AES\index{Advanced Encryption Standard (AES)} for bulk data encryption.  In the DARPA QKD network, the Diffie-Hellman process was replaced with keys generated by QKD.

The next experimental testbed, built in Europe around the year 2008, was known as the SECOQC QKD network.
SECOQC stands for ``Secure Communication based on Quantum Cryptography''.
It was launched in Vienna, and it comprised six nodes and eight links. This project developed a layered architecture for the communication services.

Another notable QKD network testbed was built in Tokyo around 2010.
The University of Tokyo Hongo campus and a facility in Otemachi, a place in central Tokyo where many telecommunication providers converge, are separated by only a few kilometers.
The network reached all the way to the National Institute of Information and Communications Technology (NICT) in Kogane, in the western Tokyo suburbs.
The Tokyo QKD network demonstrated for the first time a quantum-secured video conference.
This was another step toward showing that quantum networks are viable in real-world communications.

As of this writing in early 2023, by far the most extensive QKD network in the world is in China.
As of late 2020, the network covered much of the geographic span of China, with 2,000 kilometers of fiber reaching from Shanghai to Beijing and a satellite link that reaches almost to the border with Kazakhstan in Xinjiang.
The eastern fiber network includes 700 individual links, and the network serves 150 users.

\newpage
\begin{exercises}

\exer{
\emph{One-time pad.}
Alice would like to transmit her plaintext message $m$ to Bob, encoding it using the one-time pad with a secret key $k$ that she shares with Bob.
The ciphertext is given by
\begin{equation}
    c = m \oplus k.
\end{equation}
\subexer{
If the plaintext $m$ contains $n$ bits, what is the minimum length of the key?
}
\subexer{
Show that in order to obtain the plaintext $m$, Bob needs to apply the one-time pad to the received ciphertext.
}
\subexer{
Let's say that Alice's plaintext and shared key are
\begin{align}
    m & = 0101010111, \\
    k & = 1010100111,
\end{align}
respectively.
What is the ciphertext?
}
\subexer{
Check that applying the key to the ciphertext decodes the message.
}
}

\exer{
\emph{Two-time pad.}
We mentioned that the one-time pad is secure provided that it is used only once.
Let's consider the scenario where Alice communicates two messages $m_1$ and $m_2$, but she is careless and uses the same key $k$ to encode them before transmission.
Since the same key is used twice, it is called two-time pad.
\subexer{
Write down the two ciphertexts corresponding to Alice's messages. 
}
\subexer{
Eve intercepts both transmissions.
What information about Alice's messages $m_1$ and $m_2$ can she obtain and how? 
}
}

\exer{
\emph{Indistinguishability of quantum states.}
In Sec.~\ref{sec:bb84-protocol}, we asserted that \ket{0} and \ket{+} are not orthogonal, and made statements about measurement probabilities without working through the mathematics.
\subexer{
Prove that the states are not orthogonal.
}
\subexer{
Find the probability of the $\pm 1$ outcomes when measuring both of those states in the $Z$ basis, using calculations similar to those in Sec.~\ref{sec:measurement}.
}
\subexer{
Find the probability of the $\pm 1$ outcomes when measuring both of those states in the $X$ basis, using calculations similar to those in Sec.~\ref{sec:measurement}.
}
}

\exer{
\emph{Biased random number generator.}
Consider the scenario where Eve secretly tampers with Alice's random number generator and changes its output to be biased,
\begin{equation}
    P(0) = \frac{1}{4}, \quad\text{and } \quad P(1) = \frac{3}{4}.
\end{equation}
This random number generator is used by Alice to generate both of her random bit strings $a$ and $b$, as well as Eve's choice of measurement basis.
Bob's random number generator is not compromised.
\subexer{
What are Alice's probabilities of preparing each of the basic encoding state,
\begin{equation}
    P(\ket{0}), \quad P(\ket{1}), \quad P(\ket{+}), \quad P(\ket{-})?
\end{equation}
}
\subexer{
What is the probability that Eve disturbs the intercepted state by measuring it in the wrong basis?
}
\subexer{
What is the probability that Bob projects a disturbed state onto the wrong state?
}
\subexer{
What is the probability that Alice and Bob detect Eve in a single round?
How does this chance compare to the case of ideal random number generator?
}
}

\end{exercises}

\chapter{E91: Entanglement-based QKD}
\label{sec:10_E91}

Continuing with our study of quantum key distribution, in this Chapter we discuss an entanglement-based protocol, known as E91.
This protocol was introduced by Artur Ekert in 1991.
Unlike BB84, the E91 protocol relies on entanglement shared between Alice and Bob to establish a correlated secret key.
We will see that this difference offers Alice and Bob a very powerful tool when it comes to verifying the security of their key.

\section{Introduction}
\label{sec:E91-introduction}

In the previous chapter, we learned about the BB84 single photon-based quantum key distribution protocol.
However, that is not the end of the story for QKD.
In fact, we are more concerned with entanglement-based services than single-photon states in this book.
Before we get into the \textbf{\emph{how}} of entanglement-based QKD, let's look at one reason \textbf{\emph{why}}, beginning with a little review of BB84.

In BB84, Alice and Bob use a public quantum channel to establish a secret key.
Alice prepares qubits in four different states, chosen at random from $\{\ket{0}, \ket{1}, \ket{+}, \ket{-}\}$.
She transmits these states to Bob, who randomly measures them in either the $X$ or $Z$ basis.
After the measurement is finished, Alice and Bob exchange information about the preparation and measurement bases.
If the bases coincide, they keep the results for those measurements, forming the basis for their secret key.
If they dedicate a portion of the key material to eavesdropper detection, they can determine if anyone is listening in on the state, thanks to the non-orthogonality of the original encoded qubit states.

Let's consider how Eve's knowledge of the preparation bases affects the security of the protocol.
In Chapter~\ref{sec:9_bb84}, we saw that if Eve has no knowledge of Alice's preparation bases, the probability that her eavesdropping is detecting is $1/4$.
In the exercises following that Chapter, we saw that having some knowledge of the preparation bases gives Eve a better chance of staying undetected.

\begin{figure}[t]
    \centering
    \includegraphics[width=0.8\textwidth]{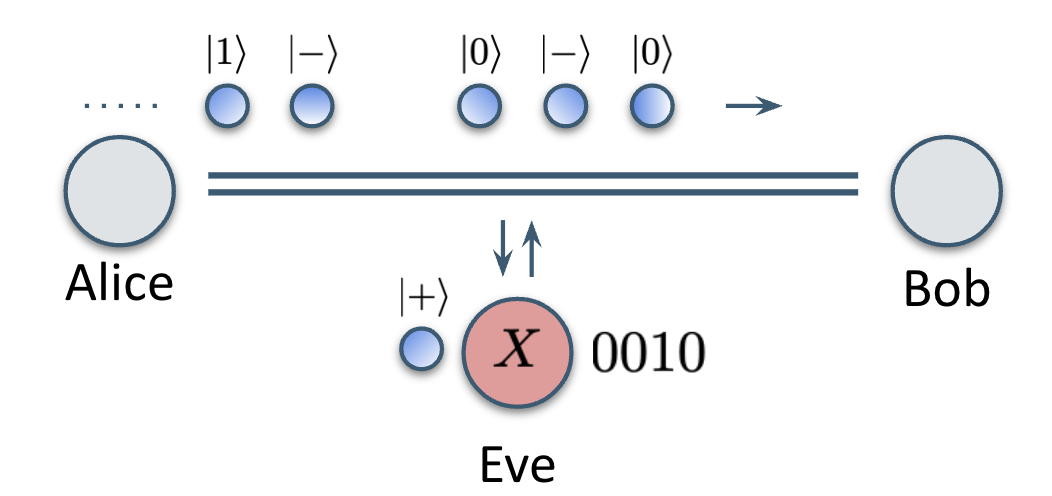}
    \caption[Successful eavesdropping on BB84.]{If Eve can learn the qubit preparation bases for the BB84 photons, she can measure the qubits without disturbing them, remaining undetected as she forwards the qubits to Bob.}
    \label{fig:eve-bb84}
\end{figure}

Let's consider the limit of this situation, where Eve has perfect knowledge of the preparation bases, as shown in Fig.~\ref{fig:eve-bb84}.
She intercepts the first qubit.
Because she knows that this qubit was prepared in the $Z$ basis, she measures it in the $Z$ basis and obtains the corresponding classical bit, and then she sends the qubit on to Bob\footnote{In many implementations, actually, she would create an identical photon that she sends on to Bob.}.
She intercepts the second qubit, and again because she knows the information about Alice's preparation basis, she measures in the appropriate basis, for example the $X$ basis.
In this case, if the state is \ket{-}, she obtains a classical bit one. She repeats this procedure for every qubit.
Although she is measuring these qubits, she is not disturbing them at all because she always measures in the same bases in which they were prepared.
In this way, she can build up a secret key that's perfectly correlated with the key that Alice and Bob end up sharing sharing.
This sneaky measurement attack by Eve is a big problem, causing the whole procedure of BB84 to fail.
Even though Alice and Bob can try to detect Eve as they would in the normal protocol, she has not disturbed any of the qubits.
Therefore, they will never detect her presence.

\begin{figure}[H]
    \centering
    \includegraphics[width=0.8\textwidth]{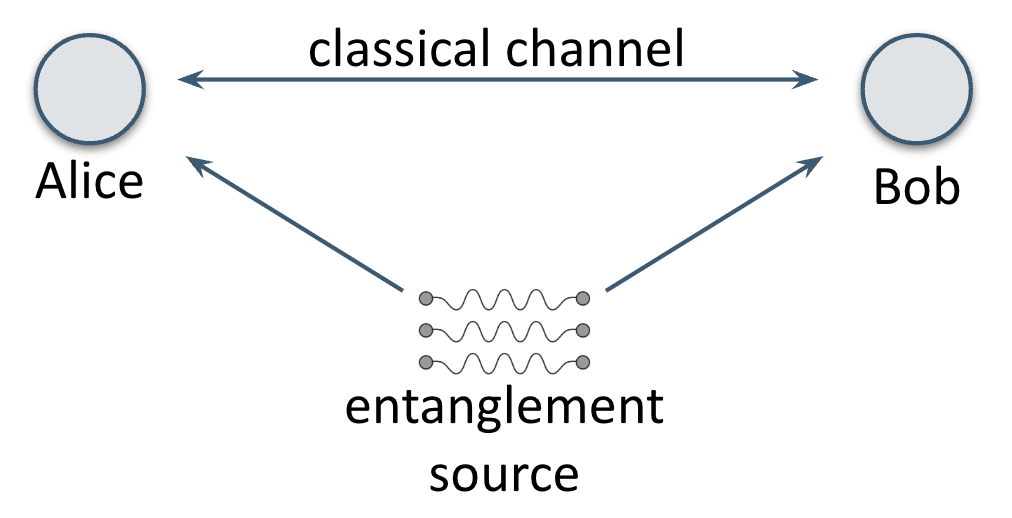}
        \caption[E91 setup.]{In E91, a source of Bell pairs distributes them to Alice and Bob.}
    \label{fig:e91-setup}
\end{figure}

\begin{figure}[H]
    \centering
    \includegraphics[width=0.8\textwidth]{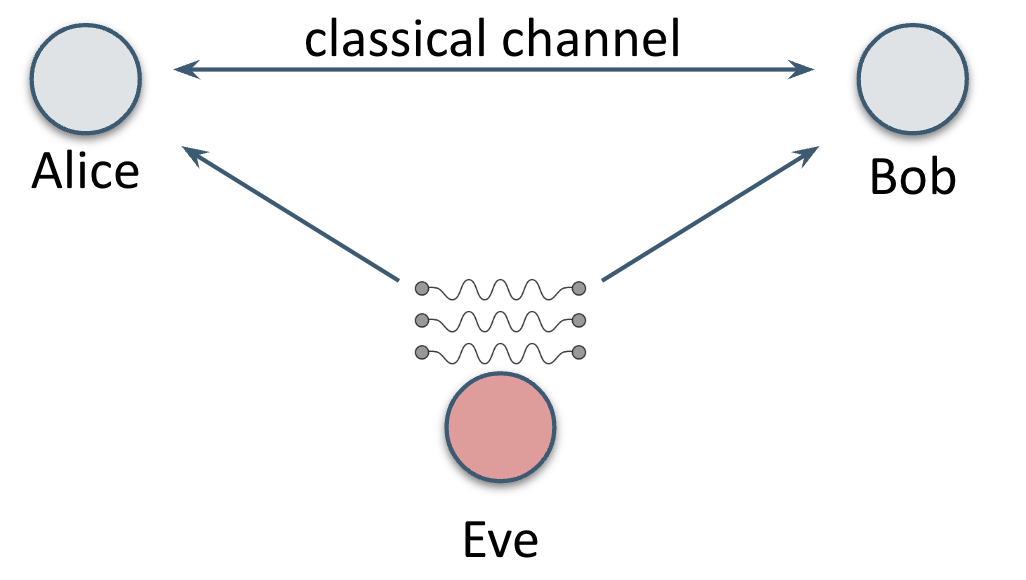}
    \caption[E91 with Eve as Bell pair source.]{Even if Eve controls the Bell pair source, Alice and Bob can maintain secure operation.}
    \label{fig:eve-e91}
\end{figure}

In this chapter, we will show that the E91 protocol is more robust against these kind of attacks.
The protocol relies on pre-shared entanglement between Alice and Bob.
We will assume that Alice and Bob can communicate over a classical channel, and also that there is some source of entangled states, as in Fig.~\ref{fig:e91-setup}.
This source generates multiple copies of an entangled state and distributes the qubits to Alice and to Bob.
We will see that in this protocol, even if Eve controls the source of the qubits, as in Fig.~\ref{fig:eve-e91}, the protocol still remains secure in the sense that Alice and Bob can easily detect an eavesdropping Eve.

\section{Basic ingredients}
\label{sec:E91-basic-ingredients}

There are two basic ingredients to entanglement-based QKD protocols.
The first ingredient is the procedure for establishing a secret key.
For that purpose, we will use an entangled state of two qubits.
Let's consider the case where Alice and Bob are sharing the following Bell pair,
\begin{align}
    |\Psi^{+}\rangle = \frac{1}{\sqrt{2}}(\ket{01}+\ket{10}).
\end{align}
If we measure these qubits in the same basis, the outcomes will be
correlated or anti-correlated, depending on which basis is used to measure them, and the probability of these outcomes is uniformly random.

Let's look at an example to demonstrate how this measurement works.
Let's say that both Alice and Bob measure in the $X$ basis.
We can compute the probabilities of all four possible outcomes.
\begin{align}
\begin{aligned}
    \operatorname{Prob}\left\{\ket{++}_{A B}\right\} & = \frac{1}{2}, \quad  \operatorname{Prob}\left\{\ket{+-}_{A B}\right\} = 0, \\
    \operatorname{Prob}\left\{\ket{--}_{A B}\right\} & = \frac{1}{2}, \quad \operatorname{Prob}\left\{\ket{-+}_{A B}\right\} = 0.
\end{aligned}
\end{align}
The probability that both Alice and Bob obtain a correlated result of \ket{++} is given by a half.
The probability that they get a \ket{--} outcome is also a half,
leaving the other wiht no probability of being observed.

In this way, when Alice measures state \ket{+}, Bob always measures state \ket{+}.
When Alice measures state \ket{-}, Bob always measures state \ket{-}.
The classical bits that are the outcomes of their measurements will always be either 00 with probability a half, or 11 with the same probability.
In this way, they can establish a secret, random correlated key.

What if they measure in the $Z$ basis? The scenario is very similar, although now the results are anti-correlated,
\begin{align}
\begin{aligned}
    \operatorname{Prob}\left\{\ket{00}_{A B}\right\} & = 0, \quad \operatorname{Prob}\left\{\ket{01}_{A B}\right\} = \frac{1}{2}, \\ 
    \operatorname{Prob}\left\{\ket{11}_{A B}\right\} & = 0, \quad \operatorname{Prob}\left\{\ket{10}_{A B}\right\} = \frac{1}{2}.
\end{aligned}
\end{align}
The probability of correlated outcomes is zero this time.
When Alice and Bob both measure in the $Z$ basis, the outcomes will never be $|00\rangle_{AB}$ or $|11\rangle_{AB}$.
The outcomes are always anti-correlated, meaning that when Alice's outcome is $|0\rangle_A$, Bob will measure $|1\rangle_B$.
Similarly, when Alice measures $|1\rangle_A$, Bob's outcome will be $|0\rangle_B$.
An important thing to note is that both possibilities are equally probable, with $50\%$ probability they share $|01\rangle_{AB}$, and with the same probability they share $|10\rangle_{AB}$.
The corresponding classical keys are anti-correlated as well.
All that Bob has to do is flip his bits in order to obtain a random, correlated key, which can then be used to encrypt data for communication.

The second ingredient of en entanglement-based QKD protocol is to verify that they have an entangled state.
Why do they need to do this verification?
The first reason, as we just saw, is that entangled states can be used to generate a correlated random key, so it is crucial to confirm that we really have entanglement before we try to use it.
There is a very important second reason: entanglement can be used for security as well, in a manner that goes beyond what BB84 can achieve. Maximally entangled states are guaranteed to be secure due to \textbf{\emph{monogamy of entanglement}}\index{monogamy of entanglement}.

Monogamy of entanglement is a fundamental property of quantum states,
and it constrains how correlated multiple qubits can be.
In particular, if Alice and Bob share a maximally entangled state,
then we are guaranteed that they cannot share any correlations with a third party, such as Eve.
This makes monogamy of entanglement an important tool for QKD security.
If Alice and Bob can demonstrate and verify that they have a maximally entangled state, they are automatically demonstrating that whatever
key they establish is secure and Eve does not have any information about their secret key.

In general, there is a trade-off: if Alice and Bob share a non-maximally entangled state, they can still share some correlations with Eve.
The stronger the entanglement that they share, the less correlated they are with Eve, until the point where they
are maximally entangled and therefore they share no correlations with Eve.
\textbf{\emph{Stronger entanglement between Alice and Bob implies more a secure key between Alice and Bob.}}

How can Alice and Bob verify that they share a maximally entangled key?
This is done via the use of the \textbf{\emph{CHSH inequality}}\index{CHSH inequality}, which we hinted at back in Sec.~\ref{sec:chsh-game} when we introduced the CHSH game.

Let's start by considering four classical random variables, denoted $A$,
$\bar{A}$, $B$, and $\bar{B}$, each of which can have values $+1$ or $-1$.
Let's say that we form the function
\begin{equation}
    A(B+\bar{B})+\bar{A}(B-\bar{B})=\pm 2
\end{equation}
You can easily convince yourself that for any combination of $+1$ or $-1$, the maximum value that you can get for this expression is $+2$, and the minimum value that you can get is $-2$.

Now imagine that we are constantly generating these random variables.
We are interested in the average value of this function,
\begin{equation}
    |\langle A(B+\bar{B})\rangle+\langle\bar{A}(B-\bar{B})\rangle| \leq 2.
\end{equation}
The angular brackets $\langle\cdot\rangle$ denote the average value of the expression.
The two extremes values are $+2$ and $-2$.
The expectation value will therefore be somewhere in between, depending on the details of the probability distributions for these random classical variables.
We expand these expectation values, then take the absolute value of the whole sum, which is constrained to be less than or equal to two.
We get the following expression which we are going to denote by the symbol $\mathcal{S}$,
\begin{equation}
    \mathcal{S}=|\langle A B\rangle+\langle A \bar{B}\rangle+\langle\bar{A} B\rangle-\langle\bar{A} \bar{B}\rangle| \leq 2.
    \label{eq:chsh-inequality}
\end{equation}
This inequality is known as the CHSH inequality.
Any set of classical random variables $A$, $\bar{A}$, $B$, and $\bar{B}$, have to satisfy this constraint, even if $A$ and $B$ are classically correlated.

What happens in the quantum case?
We can consider $A$, $\bar{A}$, $B$, and $\bar{B}$ to be the measurement outcomes when state \ket{\psi} is measured in a certain basis.
Just to remind you, the expectation value of an observable where Alice measures observable $A$ and Bob measures observable $B$ is given by the expression
\begin{equation}
    \langle A B\rangle=\langle\psi|A \otimes B| \psi\rangle.
\end{equation}
Given that the measurement outcomes are still $\pm1$, we might expect that the CHSH inequality in Eq.~(\ref{eq:chsh-inequality}) applies to the quantum case too.

Amazingly, for some quantum states, we can violate the CHSH inequality.
By ``violating'', we mean that we can obtain a value $\mathcal{S}$ that is larger than 2.

In particular, in an experiment where we measure and compute these four expectation values and then we sum them up in this manner, if we obtain a CHSH expression which is less than two,
then we can say \textbf{\emph{maybe}} the states are classically correlated.
But, if we measure a CHSH expression which is larger than two, then we can, say that \textbf{\emph{definitely}} these states are entangled.
In quantum mechanics, the CHSH expression can go all the way up to a value of $2\sqrt{2}$, which happens for maximally entangled states.

\begin{figure}[H]
    \centering
    \includegraphics[width=0.8\textwidth]{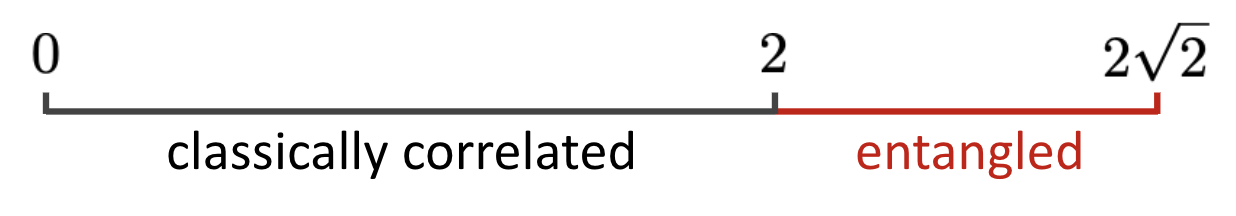}
    \caption{Possible CHSH values.}
    \label{fig:chsh-values}
\end{figure}

Let's consider a particular example.
Take one of the Bell pairs, \ket{\Psi^+}.
For the measurement settings, we consider the following: $A$ is the Pauli $Z$ basis, $\bar{A}$ is the $X$ observable, while $B$ and $\bar{B}$ are given by combinations of $Z$ and $X$, giving us rotated measurement bases $B=\frac{1}{\sqrt{2}}(Z-X)$ and $\bar{B}=\frac{1}{\sqrt{2}}(Z+X)$.

We can go through the algebra of computing the expectation values.
We find that for a maximally entangled state, we obtain the CHSH expression of $\mathcal{S}=2\sqrt{2}$.
Keep in mind that this is a statistical measure; we have to produce and measure a lot of copies of the state \ket{\psi} to estimate each of the four averages in Eq.~\ref{eq:chsh-inequality}.
The four measurement bases and four possible outcomes mean that the calculation requires sixteen values.
An example set of values is shown in Tab.~\ref{tab:chsh-calculation}.  Each entry $P_{ij}$ is the probability that the measurement basis on the left produces the outcomes $i$ at Alice and $j$ at Bob, normalized from counts so that the four $P_{ij}$ values in each line sum to 1.
For this set of values,
\begin{equation}
\begin{aligned}
    \mathcal{S} &= |\langle A B\rangle+\langle A \bar{B}\rangle+\langle\bar{A} B\rangle-\langle\bar{A} \bar{B}\rangle| \\
    &= | -0.71 -0.71 - 0.65 - 0.65 | \\
    &= 2.72 \geq 2,
\end{aligned}
\label{eq:chsh-example}
\end{equation}
giving us a strong CHSH inequality violation.

\begin{table}[t]
    \setcellgapes{3pt}
    \renewcommand\theadfont{}
    \makegapedcells
    \centering
    \begin{tabular}{cccccc}
        \hline
        & \boldmath$P_{++}$ & \boldmath$P_{+-}$ & \boldmath$P_{-+}$ & \boldmath$P_{--}$ & \boldmath$\langle MN \rangle$ \\
        \hline
        \boldmath$AB$ & 0.04 & 0.26 & 0.60 & 0.10 & -0.71 \\
        \boldmath$A\bar{B}$ & 0.04 & 0.26 & 0.60 & 0.10 & -0.71 \\
        \boldmath$\bar{A}B$ & 0.16 & 0.34 & 0.48 & 0.02 & -0.65 \\
        \boldmath$\bar{A}\bar{B}$ & 0.16 & 0.34 & 0.48 & 0.02 & 0.65 \\
        \hline
    \end{tabular}
    \caption[Calculation of the $\mathcal{S}$ value.]{Calculation of the CHSH $\mathcal{S}$ value. $P_{++}$ is the probability of getting the $+1$ eigenstate for both measurements when measuring in the basis in the left column, etc. $\langle MN \rangle$ is the expectation value calculated using $\langle MN \rangle = \bra{\psi}M\otimes N\ket{B} = P_{++} + P_{--} - P_{+-} - P_{-+}$, where $M$ is either $A$ or $\bar{A}$, and $N$ is $B$ or $\bar{B}$,  giving the bases in the left column.}
    \label{tab:chsh-calculation}
\end{table}

This calculation gives us a way of verifying entangled states, and particularly verifying maximally entangled states, a very important condition for the entanglement-based QKD protocol.
If we can demonstrate that we violate the CHSH inequality maximally, with a value of $S = 2\sqrt{2}$, then we can certify that
the state Alice and Bob share is a maximally entangled state.
If the states we are making are maximally entangled, then the principle of monogamy of entanglement\index{monogamy of entanglement} tells us that they are \textbf{\emph{not}} correlated with Eve, and therefore we can guarantee the security of Alice and Bob's secret key.

\section{Protocol}
\label{sec:E91-protocol}

We have described the two basic ingredients of the E91 protocol.
Let's put them together.
The setting is the following: Alice and Bob can communicate over a classical channel, and they share multiple copies of a maximally entangled state \ket{\Psi^+}.
These copies can be generated by Eve herself.

\begin{figure}[H]
    \centering
    \includegraphics[width=0.7\textwidth]{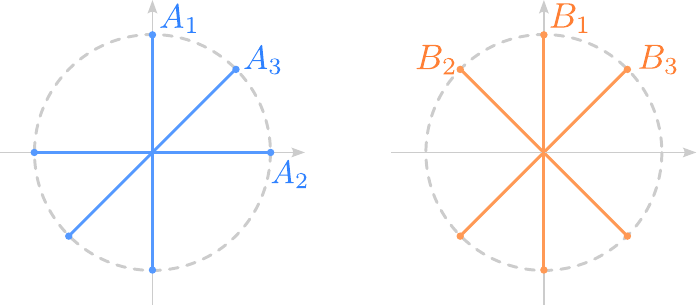}
    \caption[E91 measurement bases.]{E91 basis choices for Alice and Bob. The circle is the $XZ$ plane of the Bloch sphere.}
    \label{fig:e91-bases}
\end{figure}

Alice and Bob randomly choose a measurement basis in which they measure their qubits.
Alice chooses from three measurement bases, as shown in Fig.~\ref{fig:e91-bases}.
The circle represents the $XZ$ plane of a Bloch sphere.
Alice's measurement setting or measurement basis $A_1$ corresponds to measurement in the $Z$ basis.
If she chooses this basis, she projects the state either onto \ket{0} or onto \ket{1}.
She can also measure in the $A_2$ basis, which corresponds to $X$ basis, given by the horizontal direction.
She can also measure in a rotated basis $A_3=\frac{1}{\sqrt{2}}(Z+X)$, which is a linear combination of $Z$ and $X$.
Bob can also measure in the $Z$ basis, given by $B_1$, or in the rotated basis $B_2 = \frac{1}{\sqrt{2}}(Z-X)$, or in the basis $B_3=\frac{1}{\sqrt{2}}(Z+X)$.

Why do we have three different measurements for Alice and three different measurements for Bob rather than two, like we had in the previous protocol, BB84?

Notice that some of these measurement bases overlap.
If both Alice and Bob measure the entangled state in the same basis, they can use the classical outcomes to generate and establish a classical, correlated random key.
This follows similar logic to BB84.
Data from the basis choices $(A_1, B_1)$ or $(A_3,B_3)$ can be used for key generation.
On the other hand, we need some rotated bases in order to compute the CHSH expression $\mathcal{S}$, and see if it violates the classical CHSH inequality.
Violation of the inequality establish that Alice and Bob are really sharing an entangled state.
For this calculation, we use the basis choices $(A_1,B_3)$, $(A_1,B_2)$, $(A_2,B_2)$, $(A_2,B_3)$.

\begin{figure}[H]
    \centering
    \includegraphics[width=0.8\textwidth]{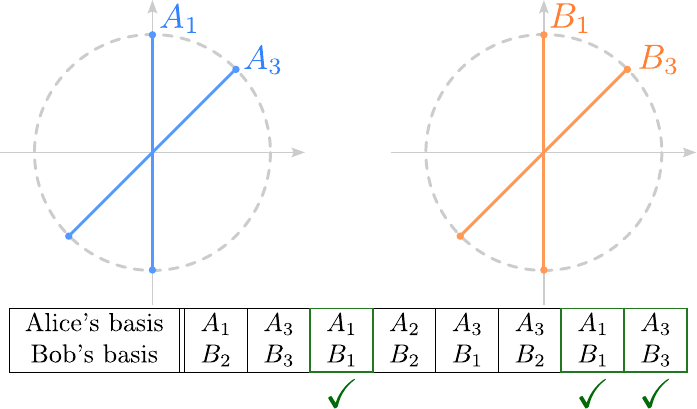}
    \caption[E91 example of a series of basis choices for measurements.]{E91 example of a series of basis choices for measurements. The $A_1$ and $A_3$ bases for Alice and $B_1$ and $B_3$ bases for Bob can be used to generate key bits when Alice and Bob select the same basis, as shown with the green check marks.}
    \label{fig:e91-example}
\end{figure}

In order to establish the key, Alice measures either in $A_1$ or $A_3$, and Bob measures in $B_1$ or $B_3$. They randomly measure their multiple copies of entangled states, and then they exchange information about the basis of their measurements. 
They exchange the information about these bases and look at the places where their measurement basis choices coincide, as shown in Fig.~\ref{fig:e91-example}.
Each partner then keeps the measurement outcome as a bit in their shared secret.

\begin{figure}[t]
    \centering
    \includegraphics[width=0.7\textwidth]{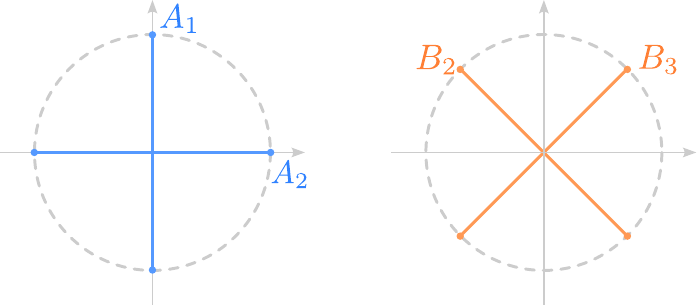}
    \caption[E91 - CHSH bases.]{When the basis choices don't coincide the measurement outcomes can be still used for the CHSH test.}
    \label{fig:10-3_e91_example_CHSH}
\end{figure}

Those two cases where Alice and Bob chose the same measurement basis take care of generating the key.
In some other cases, they will not measure in bases that coincide.
They don't discard these results, but instead  use them to compute the CHSH expression and check for a violation of the classical bound on the CHSH inequality.
In particular, they look for scenarios where they chose $(A_1,B_2), (A_1,B_3), (A_2,B_2),$ or $(A_2,B_3)$ as shown in Fig.~\ref{fig:10-3_e91_example_CHSH}.
Using these measurement outcomes, Alice and Bob then calculate the CHSH correlation function $\mathcal{S}$,
\begin{equation}
    \mathcal{S} = |\langle A_1B_2\rangle + \langle A_1B_3\rangle + \langle A_2B_2\rangle - \langle A_2B_3\rangle|.
    \label{eq:CHSH-inequality-basis}
\end{equation}

This way, they don't need to discard information as was done in BB84. They get to use the information to calculate either the secret correlated key or the CHSH violation.
If they obtain a CHSH correlation function such that $\mathcal{S} \leq 2$, they say ``okay, we cannot conclude whether or not we have an entangled state, but it's safer to just abort.''
If they have $\mathcal{S} > 2$, then they conclude, ``yes, we are sharing an entangled state, therefore we can proceed with the protocol.''
Remember, we said that monogamy of entanglement ensures that if they have an entangled state, then Eve cannot be not strongly correlated with either of them.
In particular, if Alice and Bob have a maximally entangled state, then Eve is not correlated with them at all.

So far, we have considered the case where everything was ideal, with no noise.
But what happens in real life, where noise is always present? How does noise affect the E91 protocol?
In real life, the CHSH value will not equal exactly $2\sqrt{2}$, as we saw in Tab.~\ref{tab:chsh-calculation}.
Moreover, Alice and Bob will not be able to generate a perfectly correlated key, meaning that either noise or the tinkering of Eve will introduce some inconsistencies into the key.

Alice and Bob have to decide on the acceptable security risk even if the keys are not perfectly correlated.
They have to agree, ``Okay, if the correlation is not one hundred percent, but it's very close, we can still use this to do something useful.''
If they agree to this in principle, then they have to engage in two more protocols.
One is called  \textbf{\emph{information reconciliation}}\index{information reconciliation}, which takes the initial secret key that is not perfectly correlated and produces a more correlated key.
It increases the correlation between the secret bit strings obtained by Alice and Bob.
As cryptosystems are generally designed so that a single bit difference in the keys changes half of the bits in the encrypted message~\footnote{A characteristic sometimes called the \textbf{\emph{avalanche effect}} or \textbf{\emph{strict avalanche criterion}} by cryptographers.}, they require that exactly the same key be used at both ends.
Alice and Bob can also perform something known as \textbf{\emph{privacy amplification}}\index{privacy amplification}, where they take their generated secret key and produce a shorter, more secure key.
Privacy amplification is a procedure that attempts to eliminate any possible correlations with Eve.

Having cover the basics of both single-photon and entanglement-based QKD, can we say which protocol is better more secure?
BB84 is based on the indistinguishability of single photons prepared in non-orthogonal bases to discover an eavesdropper, while E91 uses quantum correlations between two qubits to bound the information that is potentially compromised.
Let's compare the two protocols from the point when the secret key is generated.

In the case of BB84, candidate bits for the key are generated when Alice generates her random string $a$ right at the beginning of the protocol.
Remember, she generates two random $n$-bit strings.
String $b$ encodes the information about the basis of preparation, and string $a$ encodes the bits themselves.
If Alice chooses the $Z$ basis, the state she prepares is a \ket{0} or a \ket{1}, whereas if she chooses the $X$ basis, she prepares a \ket{+} or a \ket{-}.
The secret key (or an extended form of it before photons are lost and their corresponding bits are discarded) exists right from the beginning, before any communication between Bob and Alice takes place.

In BB84, a clever Eve can find a way to obtain some information about this secret bit string.
In particular, we can consider a very paranoid scenario where the random number generator (RNG) that Alice is using to generate a random bit string was built by Eve herself.
Eve constructed the RNG in such a way that any information about the random bit string that the device produces gets passed on to Eve iwthout Alice's knowledge.
We saw at the beginning of this chapter that this weakness poses a huge security risk for the BB84 protocol.

In contrast, in the E91 protocol, the secret key is really generated after the entangled pairs of qubits are measured.
The key is not produced when the entangled qubit pairs are created, nor when they arrive at Alice and Bob, but only after Alice and Bob measure them in their random bases.
Even if Eve controls the generation and distribution of entangled pairs, she cannot defeat E91.
\emph{However}, if Eve controls or can predict the output of the RNGs that Alice and Bob use for basis selection, she can defeat E91.
Coupled with the ability to measure the entangled states in the quantum channel before Alice and Bob receive them, she can measure only the key pairs and recover the key in parallel with Alice and Bob, undetected. This process is harder for Eve, so perhaps we can say that E91 is more secure than BB84.

Let's conclude this chapter by talking about some entanglement-based QKD experiments.
We saw in the discussion of BB84 that there were network testbeds for single photon QKD networks.
The development of entanglement-based QKD is not as advanced, but it exists at the level of establishing a secret key over a single link.
One such experiment was performed over free space, meaning that the entangled photons traveled through through air.
This experiment was done over a distance of 144 kilometers between La Palma and Tenerife, two islands in the Canary Islands.
Photon pairs were produced by the spontaneous parametric down-conversion (SPDC)\index{spontaneous parametric down-conversion (SPDC)} process which we saw in Sec.~\ref{sec:4-4_spdc}.
The qubits were encoded in the polarization of the photons.
The obtained CHSH value, $\mathcal{S} =2.508$, was quite a substantial amount above the maximum classical value of 2, giving a strong CHSH violation.

A different, more recent experiment was done over a distance of hundreds of kilometers, but in a lab over optical fiber wound around a spool.  Two distances were tested.
The first distance was 311 kilometers over standard fiber, and the second distance was 404 kilometers over an ultra-low loss fiber.
In this experiment, the obtained bit rate for the secret key was of the order of $10^{-3}$ bits per second for the shorter distance, or $10^{-4}$ bits per second for the longer distance. These bit rates don't actually include the information reconciliation and privacy amplification parts of a full protocol.
If we wish to use this scheme in real life, we would have to add these functions, further lowering the bit rate.

Another fantastic experiment was performed with satellites, where the satellite was use to distribute entangled pairs between two ground stations 1,120 kilometers apart.
Remember, we said that the light travels in a straight line, so by using a satellite we can overcome the complication of curved earth's surface (and reduce the atmospheric disturbance as well) to establish a quantum key over much longer distances.
The measured CHSH value was $\mathcal{S} = 2.56$, again showing a strong CHSH violation.
The obtained bit rate was 0.12 bits per second.
One could ask the question, what if we actually used a single fiber to connect these two ground stations?
Well, the paper that reported these results estimated that the fiber would have been around eleven orders of magnitude less efficient than using the satellites.
At the beginning of this chapter, we discussed the security-related reason for preferring E91 to BB84; arguably, an even bigger reason is simply the loss in fiber, which we will address in Ch.~\ref{sec:11_long-distance} and beyond.

Thus concludes our discussion of quantum key distribution protocols as a key use of quantum effects in communication systems and this block of lessons.
In the final two blocks, we will investigate the construction of quantum repeaters and networks.

\newpage
\begin{exercises}
\exer{
\emph{More measurement bases in E91.}
In E91, Alice and Bob each have three choices of measurement basis, which they are expected to select randomly.  We described how six different pairs of choices are used to generate the secret key and to calculate the CHSH inequality to check for the presence of an eavesdropper.
\subexer{
Sharp-eyed readers may have recognized that, in the explanation, we neglected three choices of basis.  Which ones did we not mention, and why?  What should be done with the data collected when those bases are chosen?
}
\subexer{Taking into account the answer to the prior question, what fraction of total measurements are dedicated to key generation, and what fraction to CHSH testing?}
}

\exer{
\emph{Optimality of CHSH violation.}
On the Bloch sphere, the four basis choice angles for CHSH are 45\degree apart.
In this exercise we will explore why we measure at these angles and not some other ones.
\subexer{
Let's fix Alice's measurements to be the same like in Eq.~(\ref{eq:CHSH-inequality-basis}).
But Bob's measurement basis can be different. It is at some angle $\theta$ as shown in the figure below.
Write down the two observables, $B_1$ and $B_2$, that Bob measures in terms of the angle $\theta$.
}
\subexer{
Write down the four expectation values $\langle A_1B_1\rangle$, $\langle A_1B_2\rangle$, $\langle A_2B_1\rangle$, and $\langle A_2B_2\rangle$ in terms of the angle $\theta$.
}
\subexer{
Construct the CHSH value and find the angle $\theta$ that maximizes it.
}
}

\exer{
\emph{E91 with other Bell pairs.}
We described E91 using \ket{\Psi^+} Bell pairs.
Let's explore what happens when we use one of the other Bell pairs,
\begin{equation}
    |\Phi^+\rangle = \frac{1}{\sqrt{2}} (\ket{00} + \ket{11}).
\end{equation}
\subexer{
Compute the CHSH value $\mathcal{S}$ given in Eq.~(\ref{eq:CHSH-inequality-basis}). Did you expect this value?
}
\subexer{
Write a new CHSH expression tailored to the new Bell pair $|\Phi^+\rangle$. Does it recover the expected violation of $\mathcal{S}=2\sqrt{2}$?
}
\subexer{
What is maximum value of this new CHSH value if Alice and Bob only classical states?
}
\subexer{
Write down a CHSH inequality for the state $|\Psi^-\rangle$.
}
}

\exer{
\emph{CHSH with noisy states.}
Typically, the distributed entangled states will not be pure.
Let's consider an imperfect source of entangled states, which produces the desired state \ket{\Psi^+} with probability $\operatorname{Pr}(\ket{\Psi^+})=1-p_{\epsilon}$.
With equal probability the source produces one of the other Bell pairs, $\operatorname{Pr}(\ket{\Phi^+})=\operatorname{Pr}(\ket{\Phi^-})=\operatorname{Pr}(\ket{\Psi^-})=p_{\epsilon}$.
\subexer{
Write down the density matrix $\rho$ of the two-qubit state that the source generates in Dirac notation.
}
\subexer{
Compute the CHSH value $\mathcal{S}$ for the desired output state \ket{\Psi^+} in terms of the error parameter $p_{\epsilon}$.
}
\subexer{
Compute the fidelity $F$ of the mixed state $\rho$.
}
\subexer{
Plot the CHSH value $\mathcal{S}$ and fidelity $F$ as functions of the error parameter $p_{\epsilon}$.
}
\subexer{
At what value of the error parameter $p_{\epsilon}$ can we no longer guarantee that the state is entangled? What is the fidelity of such a state?
}
}

\end{exercises}

\newpage
\section*{Quiz}
\addcontentsline{toc}{section}{Quiz}

A quiz is available in the online learning system.


\section*{Further reading for chapters 8-10}
  \addcontentsline{toc}{section}{Further reading for chapters 8-10}

Anton Zeilinger, Alain Aspect and John Clauser were awarded the 2022 Nobel Prize in Physics for their experimental demonstrations of the existence of entanglement, with teleportation being referenced indirectly.

A good, popular science treatment of the history of this area is \emph{The Age of Entanglement}, by Louisa Gilder~\cite{gilder08:_age_of_entanglement}.

{\bf Chapter 8}

This chapter introduced one the most fundamental protocols of quantum communication, teleportation. “Mike \& Ike” Chapter 1 goes through the mathematics of the protocol. The original paper is a great read and we highly recommend it:

Charles H. Bennett, Gilles Brassard, Claude Crépeau, Richard Jozsa, Asher Peres, William K. Wootters, Teleporting an unknown quantum state via a dual classical and Einstein-Podolski-Rosen Channels, \emph{Physical Review Letters} 70, 1895 (1993)~\cite{bennett:teleportation}.

{\bf Chapter 9}

An enlightening discussion of the BB84 protocol can be found in Chapter 12 of “Mike \& Ike” along with some exercises that will deepen your understanding of the protocol.
The original 1984 paper can be found here, along with modern commentary:

Charles H. Bennett, Gilles Brassard, Quantum cryptography: Public key distribution and coin tossing, \emph{Theoretical Computer Science} 560, 7 (2014)~\cite{bennett:bb84}.

For a further discussion of classical cryptography aimed at those working in quantum computing, see the forthcoming paper "What Every Quantum Researcher and Engineer Should Know about Classical Cryptography", by Van Meter and Aono.  Portions of this paper are available on Van Meter's blog.

The Chinese QKD network is described in ~\cite{chen2021integrated}.

{\bf Chapter 10}

Entanglement-based QKD was first introduced in:

Artur K. Ekert, Quantum cryptography based on Bell’s theorem, \emph{Physical Review Letters} 67, 661 (1991)~\cite{ekert1991qcb}.

This paper is unfortunately behind a paywall but you should be able to access it through your university’s library system.
A brief discussion of entanglement-based QKD can be also found in Chapter 12 of “Mike \& Ike”.

The exact security statistics necessary for effective QKD protocols, whether BB84 or E91, are a complex matter.  Information reconciliation and privacy amplification are major research topics in their own right.
Fantastic introduction to all these topics as well as other parts of quantum cryptography can be found in the following new book:

Thomas Vidick, and Stephanie Wehner, \emph{Introduction to Quantum Cryptography} (2023)~\cite{vidick_wehner_2023}.

\part{Fundamentals of Quantum Repeaters}

\begin{partintro}
\partintrotitle{Introduction to the fourth lesson block}
So far we have learned how to describe quantum states, we touched on the physical processes that are used in producing and manipulating these, and we have seen examples of fundamental applications of distributed quantum states.
However, we have not addressed the question of what it takes to distribute these fragile quantum states.
Now, it is time to reach for distance.
First, we discuss some of the key ideas behind and requirements for long distance communication, in particular we will focus on the challenges, both classical and quantum, that long distance presents.
Achieving this long distance communication in the quantum realm requires the use \emph{quantum repeaters}, the central technology in this book. After introducing the concepts of repeaters, this lesson block concludes with a look at the technological components used as building blocks of quantum networks.
\end{partintro}

\chapter{Long-distance communication}
\label{sec:11_long-distance}

This chapter covers the challenges that we face when trying to communicate over long distances.
We focus on the main sources of noise and error that affect classical communication, and discuss ways of mitigating their deleterious effects.
Finally, we consider how these sources of error affect quantum communication and discover that classical mitigation techniques are not suitable for use in quantum networking.

\section{Introduction}
\label{sec:ld-intro}

Let's go back to the year 1852 and consider some of the historical background of long-distance communication.
\begin{table}
    \setcellgapes{3pt}
    \renewcommand\theadfont{}
    \makegapedcells
    \centering
    
    \begin{tabular}{cc}
    \hline
    \textbf{From London to} & \textbf{Days} \\
    \hline
    \textbf{New York City} & 12 \\
    \textbf{Bombay} & 33 \\
    \textbf{Singapore} & 45 \\
    \textbf{Sydney} & 73 \\
    \hline
    \end{tabular}
    \caption[Communication before the telegraph.]{Duration a letter took to deliver from London the destination city. Prior to the invention and deployment of the telegraph, long-distance communication was very wslow.}
    \label{tab:london-letter}
\end{table}
Just before the widespread deployment of the telegraph, communication was extremely slow. To give you some idea, a letter sent from London to New York City took around twelve days to arrive, as shown in Tab.~\ref{tab:london-letter}. Delivering the letter to Sydney took staggering seventy three days. 
A faster method was needed. 

Long-distance communication became considerably faster with the invention of the electric telegraph.
The first long-distance demonstration of the power of electric telegraphy was performed by Samuel Morse and Alfred Vail in 1844.
The telegram was sent from the Capitol in Washington, D.C. to Baltimore, a distance of 44 kilometers.
The virtually instantaneous communication over this formidable distance made it clear that in order to achieve fast-speed communication at global scale, submarine cables carrying electric signals were be needed.
In 1850, the first submarine telegraph cable was laid across the English Channel, connecting England and France.
This demonstrated the feasibility of telegraphic communication using submarine cables, and sparked serious interest from wealthy industrialists in funding the laying of transoceanic telegraph cables.
For the next fifteen years various telegraph companies made numerous attempts at laying a functional transatlantic cable.
In 1858, one such attempt was successful enough for Queen Victoria of the United Kingdom and President Buchanan of the United States to exchange a message. 
However, it was not until 1866 that the first truly successful transatlantic cable was laid and operated.
Despite the enormous challenges, by 1871, all of the continents except Antarctica were connected by telegraphic cables.
Between 1902 and 1906, the first transpacific cables were laid, connecting mainland US with Hawaii, Guam, later the Philippines, and finally Japan~\footnote{rdv finds it an interesting historical anecdote that one of Japan's first undersea cables came ashore in the seaside town of Kamakura, where he lives, in 1931.  It connected to Midway, Hawaii, then the continental U.S. It was in use for around a dozen years.}.

The telegraphic cables were eventually phased out and were replaced by telephone cables.
The first transatlantic telecommunications cable, named TAT-1, was laid in 1956.
This coaxial telephone cable was initially able to handle 36 concurrent telephone conversations, which was later upgraded to 72 ocncurrent conversations. 
Compare that with the state two decades later, when the last of the coaxial copper telephone cables, TAT-7, was in use.
It could handle a staggering 4000 telephone conversations concurrently.
The first transatlantic fiber optic cable, the TAT-8, was laid in 1988 and increased the bandwidth to the equivalent of forty thousand telephone conversations.

All of these cables, both copper coaxial cables as in TAT-1 through TAT-7 and fiber optic cables from TAT-8 to the present day, carry more than one conversation over a single physical connection.
The earliest telephone wires, in the nineteenth century, supported only a single conversation, because they used \textbf{\emph{baseband communication}}\index{baseband communication}.
In baseband communication, the signal to be sent is sent on the wire unadorned.
If you want to send a 3kHz sine wave, then the wire carries your 3kHz sine wave.
This requires one physical connection for every concurrent conversation, which obviously incurs an enormous cost.
It would not be practical to have one pair of copper wires for each phone conversation crossing the Atlantic.
Instead, \textbf{\emph{signal modulation}}\index{signal modulation} is used.
The original data signal is used to modulate, or modify, a \textbf{\emph{carrier signal}}\index{carrier signal}.
For example, if your data signal is just binary data, the simplest form of modulation might be \textbf{\emph{on-off keying}}\index{on-off keying}, where the carrier is simply turned on and off, like blinking a light.
For both analog and digital data signals, many modulation schemes have been developed, but this topic is beyond the scope of this book.
See the ``Further Reading'' at the end of the chapter block for more.

Generally speaking, the carrier signal must be a higher frequency than the baseband signal you are modulating.
By using separate carrier frequencies, we can have a copper cable or optical fiber carry more than one conversation at the same time. These frequencies are divided into \textbf{\emph{channels}}\index{channel (frequency or wavelength)}~\footnote{In this book, usually when we use the term ``channel'', we are referring to the physical fiber or connection, instead of this frequency-based concept. Sometimes, we are referring to an abstraction that carries a qubit through space or time, usually in the context of error analysis and correction, as in ``bit flip channel''.}.
This is a form of \textbf{\emph{frequency division multiplexing}}, or FDM\index{frequency division multiplexing (FDM)}, the same scheme used for radio stations and television channels.
When applied to fiber optics, this technique is usually referred to as \textbf{\emph{wavelength division multiplexing}} (WDM)\index{wavelength division multiplexing (WDM)}, instead of FDM.
We will see other multiplexing schemes for sharing physical resources in the next chapter.

In this chapter, we are going to focus on the main considerations when designing these cables, particularly bandwidth and noise.

\textbf{\emph{Bandwidth}}\index{bandwidth} tells us how much information a cable or fiber can carry.
The bandwidth depends on both the physical properties of the fiber itself, as well as how clever we are when it comes to encoding this information before it gets transmitted.
With modern techniques such as \textbf{\emph{dense wavelength division multiplexing}} (DWDM)\index{dense wavelength division multiplexing (DWDM)}, we are able to reach some staggering speeds.
For example, in a standard cable over the distance of 6,600 kilometers, speeds of around 65 terabits per second can be reached.
In a more specialized cable and over shorter distances, this can be increased to over 150 terabits per second.
Over very short distances, we achieved speeds of one petabit per second\footnote{For your reference, a terabit is $10^{12}$ bits, while a petabit is $10^{15}$ bits.}.

No system is perfect, and what we put in is not exactly what we are going to get out at the end of the cable.
We must consider the main sources of loss in optical fibers.
We will consider the following five losses: mode dispersion, absorption, scattering, bending of the fiber, and coupling.
The combined result of all of these sources of loss is that our signal becomes \textbf{\emph{attenuated}}\index{attenuation}, meaning that the output signal will have less power than the input signal.
It is very important to know how much the signal becomes attenuated and how we can combat this attenuation.

Another important factor that affects how much the optical signal is attenuated is the wavelength of the light used for encoding.
Different wavelengths have different absorption coefficients.
This naturally makes certain wavelengths more suitable for long-distance communication.
The three main wavelengths used for fiber optic communication are near 850, 1300, and 1550 nanometers, which all lie in the infrared spectrum.
The wavelength of light is important in quantum communication for other reasons as well.
Not all sources of light can produce single photons of the desired wavelength.
Detector efficiencies depend on the wavelength of the photons, and quantum memories can couple to light only in certain wavelength regions. When multiple connections (whether classical or quantum) share the fiber, they are assigned separate wavelengths.
If the assigned wavelengths are too far apart, then fewer channels can be carried; if they are too close together, \textbf{\emph{filtering}}\index{filtering}, or selecting out, the desired channel is hard and inter-channel interference becomes a problem. 
All these factors introduce complications which must be accounted for when designing and implementing quantum networks.

First, we will talk about the dispersion in optical fibers, then we will consider the other sources of attenuation. Next, we will move on to overcoming these losses, and finally, we will consider the challenges these sources of loss present in the context of quantum communication.

\section{Mode dispersion}
\label{sec:11-2_mode_dispersion}
\begin{figure}
    \centering
    \includegraphics[width=\textwidth]{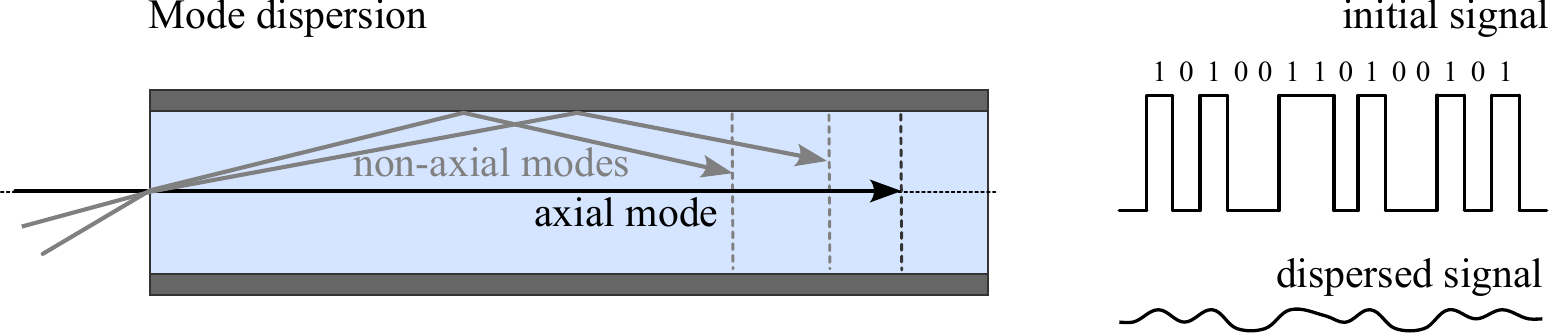}
    \caption[Mode dispersion.]{Different modes traverse the length of the fiber in different times, leading to mode dispersion. Initially sharp signals become difficult to decode.}
    \label{fig:11-2_mode_dispersion}
\end{figure}
\textbf{\emph{Mode dispersion}}\index{mode dispersion} is the first source of signal degradation in the fiber that we are going to consider.
Let's examine the propagation of different modes in a multimode fiber\index{multimode fiber}, as in Fig.~\ref{fig:11-2_mode_dispersion}.
A multimode fiber can contain many modes, all traveling along different paths.
The shortest route is the \textbf{\emph{axial path}}\index{axial path}, which travels directly down the fiber.
The light can also take non-axial paths where it undergoes total internal reflection\index{total internal reflection}.
The total length travelled by non-axial modes is longer than for the axial mode, resulting in a longer propagation time for those modes.

It is important to quantify the time difference between the different modes.
From Fig.~\ref{fig:11-2_mode_dispersion}, it is intuitive that depending on the angle with which the light is coupled to the fiber, the path that it takes will be different.
The fastest mode, travels directly down the center of the fiber.
This is the axial mode.
The slowest mode is the one that is incident on the cladding just at the critical angle, meaning it just barely gets internally reflected.

The initial digitized signal may look something like the clean modulated square wave in the upper right of Fig.~\ref{fig:11-2_mode_dispersion}. It is easy to read out the value of the individual bits.
But as the different modes propagate down the fiber, the whole packet spreads and disperses.
After some distance, it resembles the noisy signal at the bottom right of the Fig.~\ref{fig:11-2_mode_dispersion}.
This means that the readability of the output signal worsens, making it easier to make a mistake during decoding of the signal.

\begin{figure}
    \centering
    \includegraphics[width=0.8\textwidth]{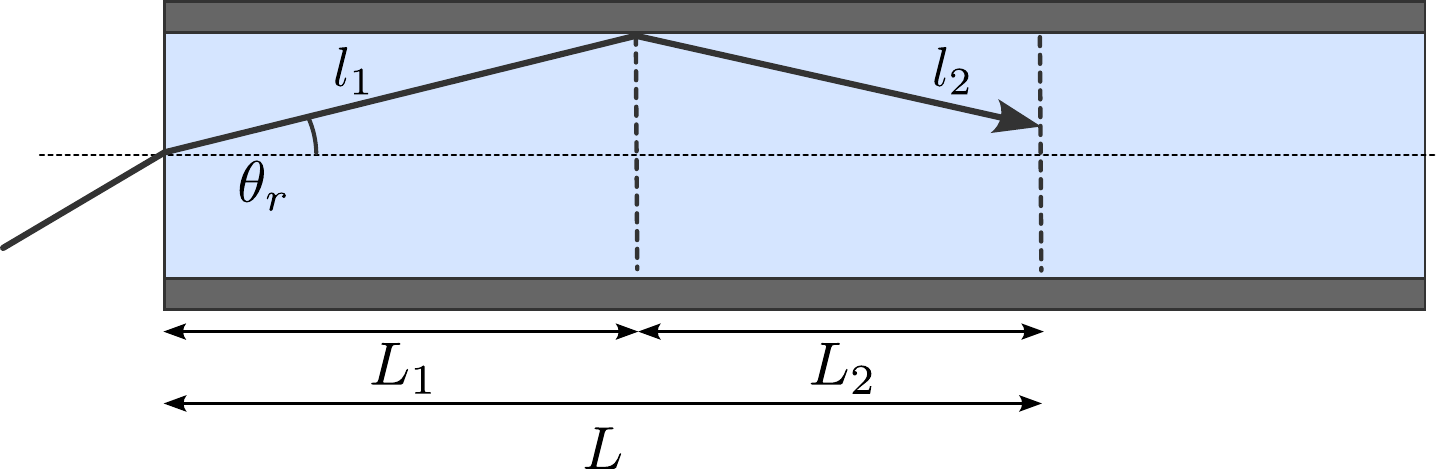}
    \caption[Time delay due to mode dispersion.]{Time delay due to mode dispersion.}
    \label{fig:11-2_dispersion_delay}
\end{figure}

How quickly this dispersion occurs can be quantified by the time between the fastest and slowest modes.
We are going to consider a horizontal length $L$ that the axial mode traverses.
The time that the axial mode takes to cover distance $L$ is the following,
\begin{equation}
    t_{\min } =\frac{L}{v_f}.
\end{equation}
The speed of of light in the fiber, $v_f$, depends on the refractive index of the material, $n_f$, that the fiber is made of, $v_f = c / n_f$.
Therefore the minimum time can be written as
\begin{equation}
    t_{\min } =\frac{Ln_f}{c}.
\end{equation}

Before considering the time $t_{\max}$ that the slowest mode takes to cover the horizontal distance $L$, we consider a general non-axial mode as shown in Fig.~\ref{fig:11-2_dispersion_delay}.
The total distance $l$ that this mode travels is larger than $L$.
The distance traveled before the internal reflection is $l_1$, while the distance traveled after the internal reflection is $l_2$.
Naturally, we have $l = l_1 + l_2$.

Let's start by computing $l_1$,
\begin{equation}
    l_1=\frac{L_1}{\cos \theta_r},
\end{equation}
where $\theta_r$ is the angle of refraction at the interface between the core of the fiber and the fiber's surroundings.
A similar expression is true for $l_2$, namely $L_2 / \cos \theta_r$.
Therefore, the total distance that the light travels is given in terms of the axial distance $L$ and the angle of refraction $\theta_r$,
\begin{equation}
    l=\frac{L_1}{\cos \theta_r}+\frac{L_2}{\cos \theta_r}=\frac{L}{\cos \theta_r}.
\end{equation}
The time needed for this non-axial mode to cover the horizontal distance $L$ is
\begin{equation}
    t = \frac{l}{v_f} = \frac{L}{v_f \cos \theta_r} = \frac{L n_f}{c \cos \theta_r}.
    \label{eq:11-2_time_nonaxial}
\end{equation}

We are now in position to compute $t_{\max}$.
The slowest mode is the one that has to cover the longest distance $l$.
In order for this to happen, the angle at which this mode is incident at the cladding is maximized while still being completely reflected.
The mode needs to be incident at the cladding at the critical angle $\theta_c$, which in turn fixes the maximum angle of refraction such that
\begin{equation}
    \cos\theta_r = \sin\theta_c = n_c / n_f,
    \label{eq:11-2_cos_theta_r}
\end{equation}
where the last equality comes from Eq.~(\ref{eq:7-4_crit_angle}).
Substituting for $\cos\theta_r$ in Eq.~(\ref{eq:11-2_cos_theta_r}), we obtain the time it takes for the slowest mode to traverse the horizontal distance $L$,
\begin{equation}
    t_{\max }=\frac{L n_f^2}{c n_c}.
\end{equation}

It is now easy to put the expressions for $t_{\mathrm{max}}$ and $t_{\mathrm{min}}$ together, and obtain the time delay between the fastest and slowest modes, 
\begin{equation}
    \Delta t = t_{\max} - t_{\min} = \frac{L n_f}{c}\left(\frac{n_f}{n_c}-1\right).
    \label{eq:11-2_time_delay}
\end{equation}

Let's consider an example to gain some intuition about the time delays introduced by mode dispersion and their effect on the encoded message.
Consider a fiber with refractive index $n_f = 1.500$, and cladding with refractive index $n_c = 1.489$.
Using Eq.~(\ref{eq:11-2_time_delay}), we obtain the following time delay per kilometer,
\begin{equation}
    \frac{\Delta t}{L}=37 \mathrm{~ns} / \mathrm{km}.
\end{equation}
This does not seem much.
Let's see the consequences of this small delay.
The speed of light in this fiber is given by $c$ divided by the refractive index of this fiber, which is $v_f=c / n_f=2 \times 10^8 \mathrm{~m} / \mathrm{s}$.
This time delay between the fastest and slowest modes means that our pulse spreads over a distance as it travels through the fiber. We can quantify this dispersion and obtain the value $7.4$ meters per kilometer -- our slowest signal is about three-quarters of one percent slower than our fastest. Every kilometer that our signal travels, it becomes more and more spread by this distance of $7.4$ meters.
Because our wave packets are becoming more spread out, we are losing the sharpness of our signal and they become more difficult to decode.
In order to be able to read the output signal clearly, we demand that the pulses that are coming out of our fiber be separated by twice the value of the spread.
This means that in order for the dispersed signal that is coming out of the fiber to be intelligible, we require that the input pulses also be separated by at least $2\times 7.4 = 14.8$ meters even when connecting over a distance of only one kilometer.
This inadvertently places a limit on how fast we can transmit information because the pulses have to be separated by a certain amount.
Dispersion has a direct consequence on the maximum frequency of the input signal.

\section{Attenuation}
\label{sec:11-3_attenuation}

First, let's look at \textbf{\emph{absorption}}\index{absorption}.
Absorption occurs due to the interaction of light with the material of the fiber.
There are two types of absorption: \textbf{\emph{intrinsic absorption}}\index{intrinsic absorption} and \textbf{\emph{extrinsic absorption}}\index{extrinsic absorption}.
Intrinsic absorption is responsible for attenuating the signal due to electronic and vibrational resonances of the molecules in the fiber.
For silica molecules, electronic resonances occur in the ultraviolet region, and vibrational resonances in the infrared region.
These resonances cause the light to transfer energy to the molecules in the fiber resulting in a power loss in the signal.
This is an intrinsic property of the fiber and we cannot really affect it; it's something that we just have to live with and accept.
Luckily, it's not a very significant source of error, especially when we compare it to the other type of absorption.

Extrinsic absorption is due to impurities introduced during the manufacturing process.
On one hand, this is a much more significant source of absorption, but on the other we can control it by perfecting the manufacturing process.
Generally, the manufacturers of fiber optic cables try to keep the the amount of impurities below one percent.

\begin{figure}[t]
    \centering
    \includegraphics[width=\textwidth]{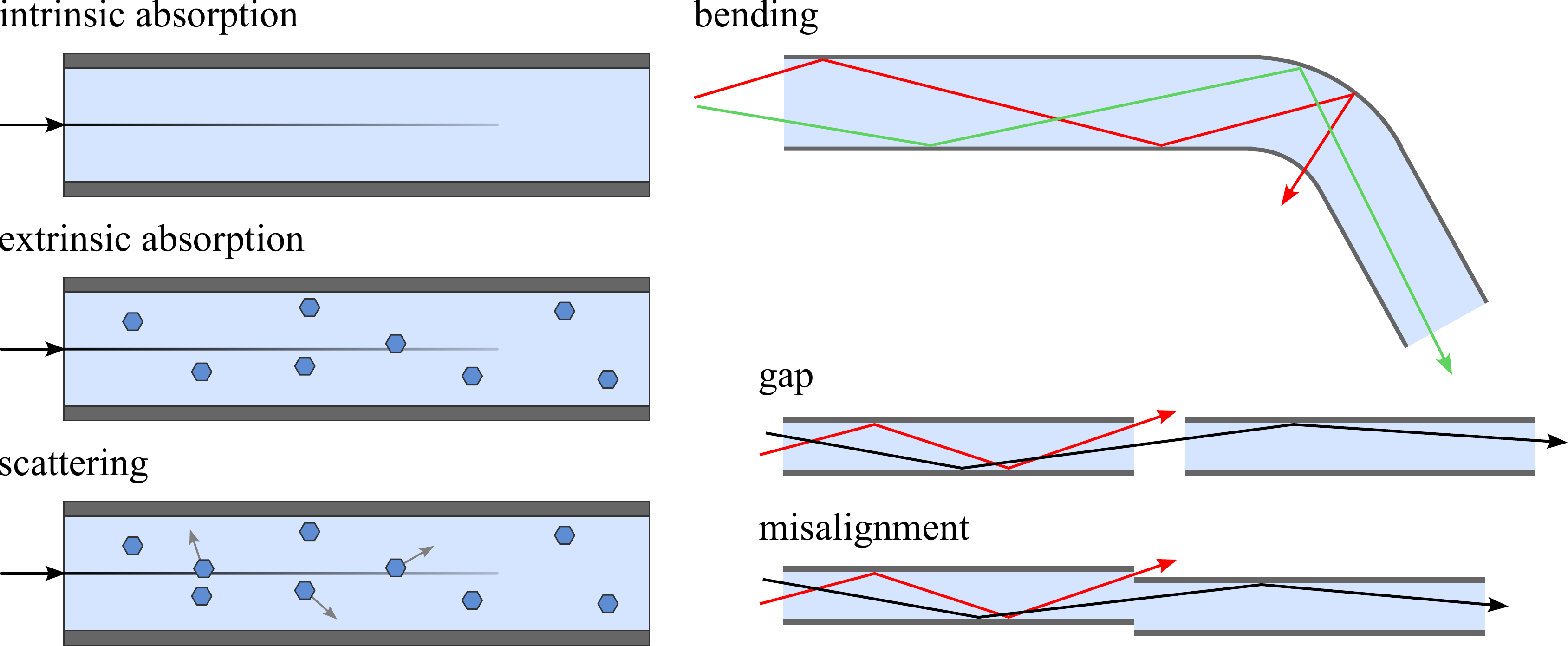}
    \caption[Attenuation losses.]{Various types of attenuation losses.}
    \label{fig:11-3_attenuation}
\end{figure}
In this section, we will consider the rest of the sources of losses that we listed in Sec.~\ref{sec:ld-intro}.
They are all pictured in Fig.~\ref{fig:11-3_attenuation}.

Another source of attenuation is \textbf{\emph{scattering}}\index{scattering}, which can be due to small fluctuations of the density in the fiber material which are introduced during manufacturing process.
This results in a non-uniform refractive index in the material causing scattering of light.
Other sources of scattering include molecules absorbing the light and re-emitting it in a random direction.

The final two sources of attenuation are \textbf{\emph{bending}}\index{bending} and \textbf{\emph{coupling}}\index{coupling}.
Losses due to bending occur when the fiber is bent at a sharp enough angle that the light leaks from it.
Remember, we said that the angle of incidence is crucial for total internal reflection to take place, and therefore for the signal to propagate down the fiber.
Typically, the manufacturers specify a minimum bending radius, usually around 10-20 times the diameter of the fiber.
Inevitably, we will have to join two fibers together.
If the coupling between the fibers introduces a gap, or if the fibers are not aligned correctly, some of the light will leak leading to attenuation of the signal.

Now that we have covered the main sources of loss, let's discuss how to quantify the attenuation in the fiber.
The degree of attenuation is expressed by a unit called the \textbf{\emph{decibel}}\index{decibel}.
Decibel is the ratio of the power that is received at the end of the fiber, $P_o$, given the initial power of the signal, $P_i$.
The number of decibels is defined as the following expression,
\begin{equation}
    \# \text { of } \mathrm{dB}=-10 \log _{10} \frac{P_o}{P_i}
    \label{eq:decibels}
\end{equation}
Remember, the power out $P_o$ is smaller than power in $P_i$.
This ratio is smaller than one, meaning the logarithm is negative.
Applying a minus sign in front of the whole expression lets us talk in terms of positive numbers of decibels.
For convenience, we multiply by ten (``deci'' means ten) and talk about decibels.
Why take the logarithm?
Often, the powers in and out are separated by orders of magnitude.
Taking the logarithm produces a nicer scale for the number of decibels.
Moreover, since every kilometer of fiber multiplies the power loss, using a logarithmic scale makes computations of total loss easier.

\begin{table}
    \setcellgapes{3pt}
    \renewcommand\theadfont{}
    \makegapedcells
    \centering
    \begin{tabular}{cc}
        \hline
        \boldmath$P_o/P_i$ & \textbf{dB} \\
        \hline $1/10$ & 10 \\
        $1/100$ & 20 \\
        $1/1000$ & 30 \\
        \hline
    \end{tabular}
    \caption[Decibels example.]{Examples of calculating loss in decibels (dB).}
    \label{tab:decibels}
\end{table}

We can see few examples what various levels of decibels mean in Tab.~\ref{tab:decibels}.
Ratio of power out to power in as $1/10$, meaning that ninety percent of the power is lost, corresponds to 10 decibels (10dB), as you can convince yourself by substituting for $P_0/P_i$ into Eq.~\ref{eq:decibels}.
Ratio is $1/100$ corresponds to 20 decibels, and ratio of $1/1000$ corresponds to 30 decibels.
You can see that when the ratio of the power out over power in shrinks by an order of magnitude, the loss in decibels increases by ten decibels.

The quality of an optical fiber is often characterized by the \textbf{\emph{attenuation parameter}}\index{attenuation parameter} $\alpha$.
Its units are decibels per kilometer (dB/km).
Dividing Eq.~(\ref{eq:decibels}) by the length of the fiber $L$, and after quick rearrangement obtain hte following,
\begin{equation}
    \alpha = -\frac{10}{L} \log _{10} \frac{P_o}{P_i}, \quad \rightarrow \quad \frac{P_o}{P_i} = 10^{-\alpha L / 10}.
    \label{eq:attenuation_parameter}
\end{equation}
The ratio $P_o/P_i$ does not only represent how weaker the output signal is compared to the initial signal, but also the \textbf{\emph{probability of a single photon travelling the length of the fiber without being lost}}.

We can compute the attenuation parameter for the optical fiber that manages to transmit just $1\%$ of the initial power over a distance of one kilometer.
This was state-of-the-art around 1970.
Using Eq.~(\ref{eq:attenuation_parameter}), the attenuation parameter is $\alpha=20$ dB/km.
The transmission probability rose to around $96\%$ over a kilometer two decades later, which corresponds to an attenuation level of 0.18 dB/km. Three decades later, attenuation levels remain around this value, though attenuation parameters as low as 0.12 dB/km have been demonstrated.

\begin{figure}
    \centering
    \includegraphics[width=0.6\textwidth]{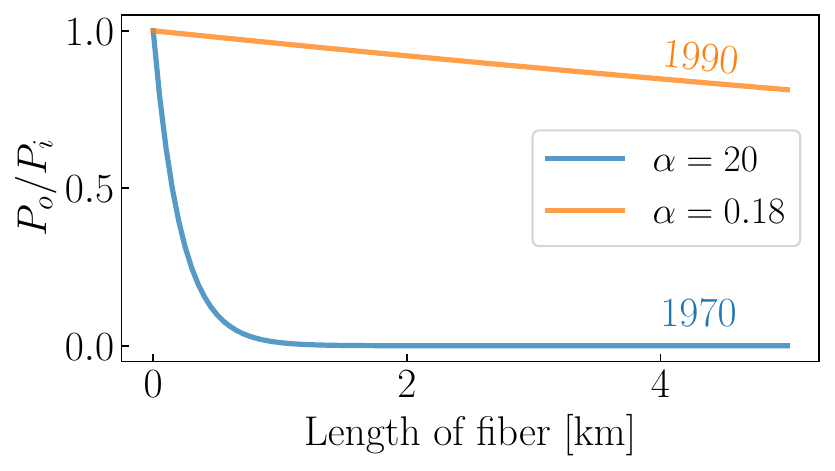}
    \caption[Power loss in fiber.]{Loss in the fiber with distance has improved substantially over the last few decades.}
    \label{fig:11-3_attenuation_dB}
\end{figure}
Figure~\ref{fig:11-3_attenuation_dB} shows the loss versus distance for these two values of attenuation loss.
On the horizontal axis, we plot the length of the fiber through which our signal is traveling, and on the vertical axis we have the ratio of the power out over power in.
If the length of the fiber is zero, then of course the ratio is one.
The signal has not traveled anywhere.
As it travels through the fiber, the blue line corresponds to the attenuation levels that were achieved in 1970, with $\alpha=20$ dB/km, while the orange line is the attenuation parameter achieved in 1990 with $\alpha=0.18$ dB/km.
We can see how quickly the ratio approaches zero for the high attenuation parameter, whereas for the very low attenuation parameter, it decreases much more slowly.

\section{Overcoming losses}
\label{sec:11-4_overcoming_losses}

Let's begin with tackling mode dispersion as it is one of the most straightforward sources of error to overcome.
We discussed how a signal propagating in a multi-mode fiber deteriorates due to the difference in lengths of the various paths that the modes take.
Decreasing the cross-section of the fiber has the effect of limiting the number of modes that the fiber can support.
\textbf{\emph{Single-mode fibers}}\index{single-mode fiber} are narrow enough that they only support the axial mode.
Having only a single mode propagating in the fiber eliminates mode dispersion completely since there is nothing that can get dispersed.

We mentioned that absorption and scattering can be reduced by improving the manufacturing process.
Bending loss is reduced by keeping in mind manufacturer's advice about the maximum bending radius.
Coupling errors can be eliminated, at least partially, by ensuring that the fibers are aligned properly and there is no gap between them.

No matter how many precautions we take when manufacturing and using the fiber, there will always be attenuation resulting in deterioration of the signal.
In Fig.~\ref{fig:11-3_attenuation_dB}, even for a fiber with low attenuation parameter of $\alpha=0.18$ dB/km, the signal becomes attenuated.
The power decreases to 80\% over a short distance of 5 kilometers.
After 20 kilometers, the power drops to 44\%; after 50 kilometers to a mere 13\%, and at a hundred kilometers it is virtually zero.
Hundred kilometers is not such a long distance.
We require systems that can transmit signals over thousands of kilometers.

\begin{figure}[t]
    \centering
    \includegraphics[width=0.6\textwidth]{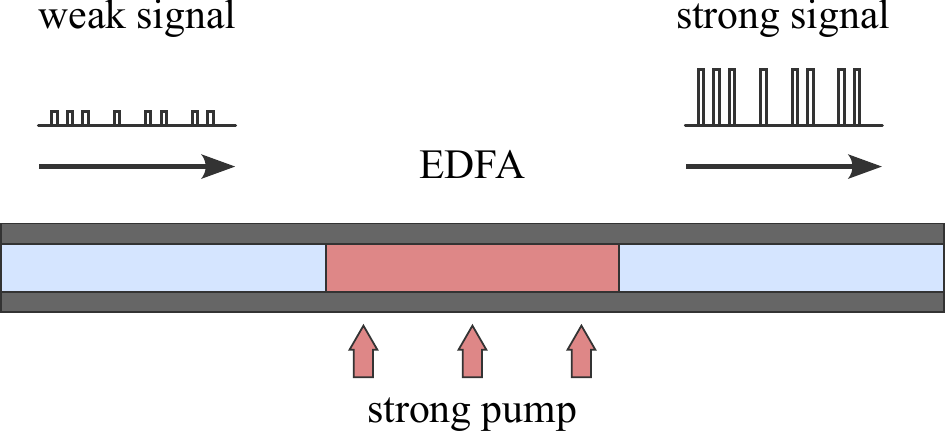}
    \caption[Erbium-doped fiber amplifier (EDFA).]{Basic working principle of an erbium-doped fiber amplifier.}
    \label{fig:11-4_edfa}
\end{figure}

Long-distance transmission is achieved with the help of \textbf{\emph{repeaters}}\index{repeater (classical)}~\footnote{Just to remind you, we will encounter this term "repeater" in the next chapter as well, where we will be talking about quantum repeaters, but they work on a very different basis than classical repeaters. Here, we are talking only about classical repeaters.}, which are devices that are used to boost the signal strength.
There are many different kinds of repeaters based on different physical principles; as an example, we introduce the \textbf{\emph{erbium-doped fiber amplifiers}}\index{amplifier}, or EDFA\index{erbium-doped fiber amplifier (EDFA)} for short.
Fig.~\ref{fig:11-4_edfa} illustrates the basic idea behind EDFA.
Erbium atoms are introduced into a portion of the fiber.
The erbium atoms are excited into their higher energy states using a strong pump, creating population inversion.
As the weak signal passes through this segment of the fiber, it stimulates emission from the erbium atoms.
We know that in the process of stimulated emission, the photons that are emitted are of the same kind as the incoming signal photons.
They have the same phase and travel in the same direction.
Basically, an EDFA is using the principle that is behind lasing to boost the weak signal.
The EDFA amplifies the signal, allowing it to travel further before it needs amplification again.
Commercially available amplifiers are capable of producing gains of up to 40 dB, and are fairly compact with the active region (red in Fig.~\ref{fig:11-4_edfa}) being only few centimeters long.

There is one problem with this approach that we have to keep in mind.
EDFAs not only amplify the signal, they also amplify any noise as well.
The excited erbium atoms amplify the signal when they undergo stimulated emission caused by the photons in the signal.
On occasion, these excited atoms emit photons via spontaneous emission, as we discussed in Section~\ref{sec:5-2_coherent_vs_incoherent}.
Photons emitted this way travel in a random direction.
It is not a big issue when they are emitted in the opposite direction from the incoming signal, as the propagating signal does not get polluted by these spontaneously emitted photons.
On the other hand, when the spontaneously emitted photons travel in the same direction as the signal, they contribute to the noise and are responsible for deterioration of the \textbf{\emph{signal-to-noise ratio (SNR)}}\index{signal-to-noise ratio (SNR)}.
Furthermore, these noisy photons may interact with other excited erbium atoms, stimulating their emission and inadvertently producing more noisy photons.

\section{Quantum challenges}
\label{sec:quantum_challenges}

So far, we have considered classical signals comprised of large number of photons.
In quantum communication, the signal is at the level of individual photons.
Photon loss in the fiber therefore becomes a major problem.
Losing a photon in from a classical signal slightly decreases the power of the signal.
Loss of a photon in quantum communication results in losing the entire signal itself!

Can we combat photon loss the way it is done in classical communication?
Namely, can we create backup copies of the single photons at regular intervals and resend those?
We can, but only if we limit ourselves to communication with orthogonal states.
For example, if we only wish to communicate qubits \ket{0} and \ket{1}, creating backup copies presents no issues and can be done.
This, however, would limit us to classical communication only.
All the magic of quantum communication lies in non-orthogonal states and entangled states.
We saw that creating copies of such states is not allowed by the no-cloning theorem in Sec.~\ref{sec:8-3_no-cloning}.

Since copy and resend strategy is fundamentally not allowed in quantum communication, it would seem we have to resort to send the single photons and hope for the best that they arrive at the destination.
Let's do a quick calculation that will demonstrate that this also is not a viable strategy.
The probability that we transmit the photon through a fiber with attenuation parameter $\alpha$ over distance $L$ is given by the following,
\begin{equation}
    P_{\text {success }}=10^{-\alpha L / 10}.
\end{equation}
Let's plug in some numbers to give us some intuition of what this probability is.
Consider a long fiber of 1000 km (which in the context of global communication is not that long), and assume a best-case scenario where the fiber has ultra-low attenuation of a mere $0.1$ dB/km.
The probability that a single photon gets successfully transmitted is $10^{-10}$.
This is indeed a very small number, but to gain some intuition of how small, consider the case of a source that produces one photon every second.
Every second a single photon gets sent down the fiber.
How long do we need to wait in order for the photon to get successfully transmitted?
On average, we expect to wait 317 years!
We can see that even with ultra-low loss fibers over moderate distances, sending a single photon down the fiber and hoping for the best is not very practical.

Loss is only one source of error that we have to contend with in long distance quantum communication.
Other sources include unitary errors such as Pauli errors, namely $X$ errors that randomly flip the state of our photons, or $Z$ errors where we introduce a phase to the photons.
There are several types of non-unitary errors such as relaxation that can affect the state of qubits.
We do not have to deal with most of these errors in classical communication.
The situation looks dire for quantum communication.
However, we will see in the next Chapter that quantum problems require quantum solutions, and indeed there are clever ways that address these issues.

\newpage
\begin{exercises}

\exer{
\emph{Pulse broadening due to mode dispersion.}
We would like to send a burst of light through a fiber of length $L$.
The light rays enter the fiber simultaneously but at different angles $\theta$, given by some distribution $d(\theta)$.
\subexer{
Draw the shape of the input signal.
}
\subexer{
What is the shape of the output signal if all light rays are incident onto the fiber core along the fiber's axis? What is the delay in the output signal?
}
\subexer{
What is the shape of the output signal if all light rays are incident a the maximum acceptance angle $\theta_{\text{max}}$? What is the delay in the output signal?
}
\subexer{
What is the shape of the output signal if the distribution of the incident angles is
\begin{equation}
    d(\theta) = \frac{\delta(0)+\delta(\theta-\theta_{\text{max}})}{2},
\end{equation}
that is half of the light rays are incident along the fiber's axis, and half are incident at the maximum acceptance angle? What is the delay in the output signal?
}
\subexer{
Write some code that lets the user input an arbitrary distribution of angles $d(\theta)$, refractive indices of the core the core and cladding, and computes the shape of the output pulse.
}
}

\exer{
\emph{Direct transmission of classical signals.}
Consider a signal of some initial power $P_i$.
After travelling in the fiber for length $L$, the power decreases to $P_o$.
\subexer{
How much of the initial signal remains if the signal undergoes a loss of 1 dB, 2 dB, 5 dB, 10 dB, and 50 dB?
}
\subexer{
Plot the fraction of the remaining power of the signal, $P_o/P_i$, as a function of the number of dB.
}
\subexer{
For each of the cases in subexercise \textbf{a)}, compute the attenuation parameter $\alpha$, if the length of the fiber is $L=10$ km.
}
}

\exer{
\emph{Direct transmission of single photons.}
Consider transmitting single photons through the fiber of length $L$.
\subexer{
Compute the probability of the photon being transmitted if it suffers loss of 1 dB, 2 dB, 5 dB, 10 dB, and 50 dB?
}
\subexer{
Check the claim that we would need to wait 317 years on average to receive a photon sent through a 1000 km long fiber with attenuation parameter 0.1 dB/km, given the repetition rate is 1 Hz.
}
\subexer{
How long would we expect to wait if the attenuation rate was 0.2 dB/km, but the repetition rate increased to 1 GHz?
}
}

\exer{
\emph{Amplification of classical signals.}
We have seen how Eq.~(\ref{eq:decibels}) describes decrease in power of the classical signal as it propagates through a fiber.
But decibels can describe power gain, that is amplification, as well.
\subexer{
Can you think of a way of altering Eq.~(\ref{eq:decibels}) such that it describes power gain in dB?
}
\subexer{
We suggested that amplifiers are used every 50km or so.  See if you can find more detailed information about the fiber loss and the spacing of amplifiers in real-world deployments and the amount of gain an EDFA can generate.
}
}

\end{exercises}

\chapter{Quantum repeaters}
\label{sec:12_quantum_repeaters}

\section{The need for repeaters}
\label{sec:12-1_need_for_repeaters}

\begin{figure}[t]
    \centering
    \includegraphics[width=0.85\textwidth]{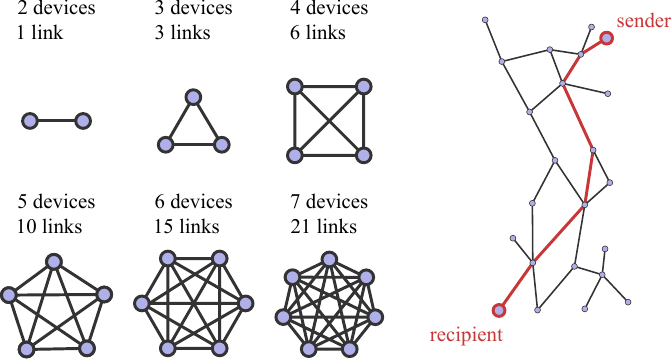}
    \caption[All-to-all coupling between network nodes.]{The left panel shows how many links are required to have all-to-all coupling. The right panel shows a more realistic network topology and the path along which a qubit needs to be teleported from sender to a receiver.}
    \label{fig:12-1_all_to_all}
\end{figure}

In the previous chapter, we saw that there is one big problem when trying to communicate over long distances, and that's photon loss in fibers.
The farther we are trying to communicate, the more likely we are to lose the photon.
There is also another problem: the number of devices that are connected to the network. Currently (2023) there are 8 billion people, and an estimated 31 billion Internet of Things devices.
Despite this staggering number, all of these devices can communicate with each other.
How is this achieved?
One way is to establish a direct connection between each device that is present in the network, as shown in the left panel of Fig.~\ref{fig:12-1_all_to_all}.
Two devices are connected by a single link.
Three devices require three links, four devices require six links and so on.
For $N$ devices, $N (N - 1) / 2$ links are needed in order to have all-to-all coupling.

Real networks do not use all-to-all coupling.
In all-to-all coupling, each individual connection is a link requiring dedicated hardware that enables communication between two devices in the network.
Adding a single device to the network would also require us to add links to every existing device.
Such an approach is clearly not practical even for relatively small networks, let alone worldwide.
The right panel of Fig.~\ref{fig:12-1_all_to_all} shows a much more realistic network, where an arbitrary sender can still transmit a message to the intended receiver, even in the absence of a direct link between the two.

In quantum networks, information encoded in qubits is not transmitted directly.
Rather, we use entangled pairs of qubits to teleport the state of the qubit, as discussed in Chapter~\ref{sec:8_teleportation}.
Each link connecting two devices shares a Bell pair that is used to pass the state of the qubit carrying the quantum message from one device to the next.
We can go back to the right panel of Fig.~\ref{fig:12-1_all_to_all}, and this time think of it as a quantum network, where the quantum sender wishes to pass the state of a qubit to the recipient.
The state of the sender's qubit can be teleported hop-by-hop along the red path until it reaches the recipient.
A problem with this approach is that the operations required by the teleportation protocol as well as the memories used to store Bell pairs are not perfect.
This decreases the fidelity of the teleported qubit.
Repeating the teleportation in this hop-by-hop approach degrades the fidelity of the entangled state, resulting in a garbled quantum state being teleported to the recipient.
One way around this problem is to use the link-level Bell pairs to create a direct entangled connection between the sender and the recipient so that the quantum state can be teleported in one hop.
The quantum nodes tasked with achieving this splicing of link-level entanglement are called \emph{\textbf{quantum repeaters}}\index{quantum repeater}, and they are indispensable in the design of long-distance quantum networks.

We will address four requirements in this chapter.
First, we will show how to establish entanglement between neighboring nodes of a quantum network.
This is known as \textbf{\emph{link-level entanglement}}\index{link-level entanglement}.
Next, we will discuss how the link-level entanglement can be used to create a long-distance entangled connection between end nodes using \textbf{\emph{entanglement swapping}}\index{entanglement swapping}.
After this, we will deal with the issues presented by the adverse effects of noise.
Finally, we will look at management of networks, routing, multiplexing and resource management.

\section{Making link-level entanglement}
\label{sec:12-2_making_link_level_rantanglement}

\begin{figure}[t]
    \centering
    \includegraphics[width=0.8\textwidth]{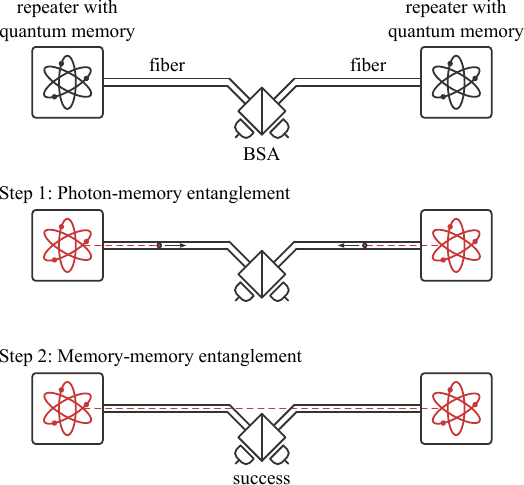}
    \caption[MIM link architecture.]{Memory-interfere-memory (MIM) link architecture.}
    \label{fig:12-2_MIM}
\end{figure}

In this section, we consider the task of creating link-level entanglement between two neighboring repeaters of a quantum network.
One method for entangling two quantum memories and creating link-level entanglement, known as the \emph{\textbf{memory-interfere-memory}} (MIM)\index{memory-interfere-memory (MIM)} link architecture, is pictured in Fig.~\ref{fig:12-2_MIM}.
Another name for this architecture is ``meet-in-the-middle''.
We tend to avoid using this term to reduce the chance of confusion with man-in-the-middle, which is a type of security attack.
Each repeater node is equipped with a quantum memory, and is coupled to an optical fiber.
The fibers lead to a \emph{\textbf{Bell state analyzer}} (BSA)\index{Bell state analyzer (BSA)}, an optical device which is capable of measuring two incoming photons in the Bell basis.
We will discuss an implementation of the BSA in Section~\ref{sec:bell-state-measurements-II}.

The protocol to generate entanglement between the quantum memories is the following.
Each quantum memory emits a photon that is entangled with it, represented by the red dashed line in Step 1 of Fig.~\ref{fig:12-2_MIM}.
These photons are captured and coupled to a fiber, which guides them to the BSA.
At the BSA they are measured in the Bell basis and destroyed in the process.
The success probability of the Bell state measurement depends on the implementation of the BSA.
Straightforward implementation using only linear optics elements results in success probability of at most $50\%$.
This is assuming ideal couplers (to collect photons emitted by the memories), fibers, and detectors at the BSA.
Once the measurement is successful, the entanglement between the memory-photon pairs is transferred to be between the memories as seen in Step 2.

A crucial point about this scheme, and other link-level architectures, is that the photons arriving at the BSA to be measured in the Bell basis must be \textit{\textbf{indistinguishable}}\index{indistinguishable photons}.
This is achieved by arranging for the photons to arrive at the BSA simultaneously, and for the them to have identical spectral properties. 
Even relatively small delays in their arrival times reduce the indistinguishability of the photons, resulting in a diminishing success probability of the Bell state measurement.

\begin{figure}[t]
    \centering
    \includegraphics[width=0.85\textwidth]{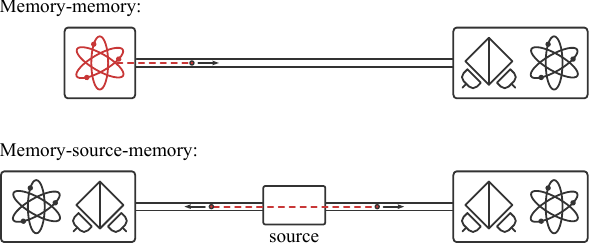}
    \caption[MM and MSM architectures.]{Memory-memory and Memory-source-memory link-level architectures.}
    \label{fig:12-2_MM_MSM}
\end{figure}

Figure~\ref{fig:12-2_MM_MSM} shows two more link-level architectures.
The first one is \textit{\textbf{memory-memory}} (MM)\index{MM link}, where the BSA device is included in one of the repeater nodes.
Entanglement is established in a fashion similar to the MIM architecture, but this time the photon generated by the quantum memory on the left travels the whole length of the fiber to the right node.
There, it interferes with a photon emitted by the right quantum memory.
This architecture sometimes is called ``sender-receiver ''architecture.
The bottom panel of Fig.~\ref{fig:12-2_MM_MSM} shows the \textit{\textbf{memory-source-architecture}} (MSM)\index{MSM link} architecture.
A source of entangled photon pairs is located between the quantum repeaters.
These photons are sent to the quantum repeaters where they interfere with photons emitted locally by the quantum memories. This architecture sometimes is called ``mid-point source'' architecture.

\section{Reaching for distance: Entanglement swapping}
\label{sec:12-3_reaching_for_distance}

\begin{figure}[t]
    \centering
    \includegraphics[width=0.8\textwidth]{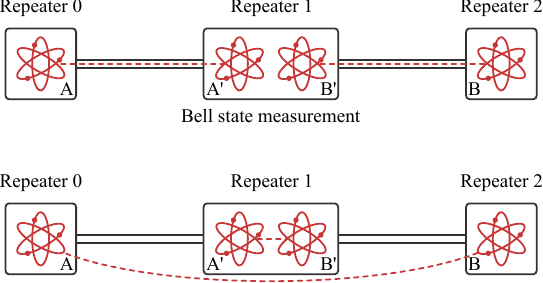}
    \caption[Entanglement swapping.]{Entanglement swapping at Repeater 1 creates entanglement between Repeaters 0 and 2.}
    \label{fig:12-3_entanglement_swapping}
\end{figure}

In the previous section, we saw how to establish link-level entanglement.
We will now extend it to establishing long-distance entanglement spanning multiple hops.
Consider three network nodes represented by Repeater 0, Repeater 1, and Repeater 2, as pictured in Fig.~\ref{fig:12-3_entanglement_swapping}.
Repeaters 0 and 2 each have one qubit, while Repeater 1 has two.
We assume that link-level entanglement has been established between Repeater 0 and one of the qubits at Repeater 1, and the other qubit at Repeater 1 and Repeater 2.
The goal is to establish entanglement between Repeater 0 and Repeater 2.

This is achieved by measuring the two qubits at Repeater 1 in the Bell basis.
This measurement projects them onto one of the four Bell pairs.
At the same, the effect of this measurement is to also project the qubits at Repeater 0 and Repeater 2 onto a Bell pair, creating entanglement between these two distant repeaters.
This procedure is known as \emph{\textbf{entanglement swapping}}\index{entanglement swapping}, and it is an indispensable tool in creating long-distance entanglement in quantum networks.
This entanglement is between two repeaters which are not directly connected.

Let's now get a little technical and show more rigorously that Repeater 0 and Repeater 2 are indeed entangled.
We label the qubit at Repeater 0 $A$, and the qubit with which it is entangled at Repeater 1 $A^{\prime}$.
The other qubit at Repeater 1 is denoted by $B^{\prime}$, and its entangled partner at Repeater 2 by $B$.
The state of the $AA^{\prime}$ qubits is one of the Bell pairs, let's say $|\Phi^+\rangle$.
The same goes for the pair of qubits $BB^{\prime}$.
The total state of the four qubits can therefore be written as
\begin{align}
    \begin{aligned}
        \ket{\psi}_{AA^{\prime}B^{\prime}B} & = \ket{\Phi^{+}}_{A A^{\prime}}\ket{\Phi^{+}}_{B^{\prime} B} \\
        & = \frac{1}{2} ( |0000\rangle+|0011\rangle+|1100\rangle+|1111\rangle ).
        \label{eq:12-3_computational_basis}
    \end{aligned}
\end{align}

In the next step of our calculation, we will use a little trick.
We rewrite the state of the qubits at Repeater 1, $A^{\prime}B^{\prime}$, as superpositions of the Bell states,
\begin{align}
    |00\rangle & = \left( |\Phi^{+}\rangle + |\Phi^{-}\rangle\right) / \sqrt{2}, \\
    |01\rangle & = \left( |\Psi^{+}\rangle + |\Psi^{-}\rangle\right) / \sqrt{2}, \\
    |10\rangle & = \left( |\Psi^{+}\rangle - |\Psi^{-}\rangle\right) / \sqrt{2}, \\
    |11\rangle & = \left(|\Phi^{+}\rangle - |\Phi^{-}\rangle\right) / \sqrt{2}.
\end{align}
We can now substitute these identities in Eq.~\ref{eq:12-3_computational_basis}, and rewrite the total state of the four qubits as follows,
\begin{align}
    |\psi\rangle_{AA^{\prime}B^{\prime}B} = & \frac{1}{2} \Biggl[ |0\rangle \underbrace{\frac{|\Phi^+\rangle + |\Phi^-\rangle}{\sqrt{2}}}_{=\ket{00}} |0\rangle + |0\rangle \underbrace{\frac{|\Psi^+\rangle + |\Psi^-\rangle}{\sqrt{2}}}_{=\ket{01}} |1\rangle \nonumber\\
    & + |1\rangle \underbrace{\frac{|\Psi^+\rangle - |\Psi^-\rangle}{\sqrt{2}}}_{=\ket{10}} |0\rangle + |1\rangle \underbrace{\frac{|\Phi^+\rangle - |\Phi^-\rangle}{\sqrt{2}}}_{=\ket{11}} |1\rangle \Biggr].
\end{align}
We can group the qubits that are going to be measured on the left, and the qubits that we are not going to measure on the right,
\begin{align}
    |\psi\rangle_{A^{\prime}B^{\prime}AB} = & \frac{1}{2} \left[ \frac{|\Phi^+\rangle + |\Phi^-\rangle}{\sqrt{2}}|0\rangle|0\rangle + \frac{|\Psi^+\rangle + |\Psi^-\rangle}{\sqrt{2}}|0\rangle|1\rangle \right. \nonumber\\
    & + \left. \frac{|\Psi^+\rangle - |\Psi^-\rangle}{\sqrt{2}}|1\rangle|0\rangle + \frac{|\Phi^+\rangle - |\Phi^-\rangle}{\sqrt{2}}|1\rangle|1\rangle \right].
\end{align}
We have not really done anything apart from rearranging the qubits.
Finally, we collect all terms with the same Bell pair on qubits $A^{\prime}B^{\prime}$,
\begin{align}
    |\psi\rangle_{A^{\prime}B^{\prime}AB} = & \frac{1}{2} \left[ |\Phi^+\rangle \frac{|0\rangle|0\rangle + |1\rangle|1\rangle}{\sqrt{2}} + |\Psi^+\rangle \frac{|0\rangle|1\rangle + |1\rangle|0\rangle}{\sqrt{2}} \right. \nonumber\\
    & + \left. |\Psi^-\rangle \frac{|0\rangle|1\rangle - |1\rangle|0\rangle}{\sqrt{2}} + |\Phi^-\rangle \frac{|0\rangle|0\rangle - |1\rangle|1\rangle}{\sqrt{2}} \right].
    \label{eq:12-3_almost_final}
\end{align}
Looking at the state of qubits $AB$ in Eq.~(\ref{eq:12-3_almost_final}), we recognize that the qubits are in fact entangled.
We can make this even more explicit,
\begin{align}
    |\psi\rangle_{A^{\prime}B^{\prime}AB} & = \frac{1}{2} \left[ |\Phi^+\rangle |\Phi^+\rangle + |\Psi^+\rangle |\Psi^+\rangle + |\Psi^-\rangle |\Psi^-\rangle + |\Phi^-\rangle |\Phi^-\rangle \right].
    \label{eq:12-3_final}
\end{align}

We can see that if the Bell-state measurement at Repeater 1 results in the outcome $|\Phi^+\rangle_{A^{\prime}B^{\prime}}$, then the state of the qubits at Repeaters 0 and 2 is $|\Phi^+\rangle_{AB}$.
Similarly if the measurement outcome at Repeater 1 is $|\Psi^+\rangle_{A^{\prime}B^{\prime}}$, then Repeaters 0 and 2 share the state $|\Psi^+\rangle_{AB}$, and so on for the other measurement outcomes.

We see that the outcome of the Bell-state measurement is not deterministic.
Each possibility can occur with equal probability of $1/4$.
Performing only the measurement is not enough to distribute entanglement between Repeaters 0 and 2.
Remember, that the measurement takes place far away from Repeaters 0 and 2.
The outcome of the measurement must be communicated to them via a \emph{\textbf{classical message}}.
Only after receiving this message from Repeater 1 will they share a pure Bell pair.
Repeater 1 does not need to send the outcome of the measurement to both Repeaters 0 and 2.
It is enough to send it just to one of them, but both of the Repeaters need to be notified that the procedure has been carried out successfully.
This introduces a time delay that limits how fast the process of establishing entanglement can progress.

Maybe you have noticed that entanglement swapping is very similar to creation of link-level entanglement described in Section~\ref{sec:12-2_making_link_level_rantanglement}.
Mathematically the two procedures are very similar.
At the  link level, memory-photon entanglement is swapped by measuring the photons at the Bell state analyzer, creating memory-memory entanglement.
Whereas in end-to-end entanglement, we are only swapping entanglement between between pairs of memories.
Physically, this is quite a big difference: using linear optics, the maximum success probability of performing entanglement swapping on photons is limited to $50\%$, but entanglement swapping between memories can be done \textbf{\emph{deterministically}}, provided that we have good experimental techniques and we can limit the effects of noise.

An alternative way of understanding entanglement swapping is through teleportation.
By measuring its two qubits in the Bell basis, Repeater 1 is really teleporting the state of qubit $A^{\prime}$ to qubit $B$.
Since qubit $A^{\prime}$ is entangled with the qubit at Repeater 0, this entanglement is also transferred to Repeater 2 along with the state of qubit $A^{\prime}$.

\section{Detecting errors: purification}
\label{sec:12-4_purification}

In this section, we will address what happens when we also include errors in our considerations.
We will learn how to  handle these errors and create a state between Repeater 0 and Repeater 2 of acceptable fidelity.
Our desired state that we want to share between Repeater 0 and 2 is given by the maximally entangled state $\ket{\Phi^+}$.
In the density matrix form, we write it as the outer product,
\begin{align}
    \rho_{AB} = \ket{\Phi^{+}}\bra{\Phi^{+}}.
\end{align}
In reality, there will always be some noise affecting the system.
The state $\rho_{AB}$ will be a mixture of the desired pure state $\ket{\Phi^+}$, and some other unwanted noisy term $\rho_{\text{noise}}$.
With probability given by the fidelity $F$, we will have the desired state $\ket{\Phi^+}$, and with probability $1-F$, we will have some noisy state $\rho_{\text{noise}}$,
\begin{align}
    \rho_{AB} = F |\Phi^{+}\rangle\langle\Phi^{+}|+(1-F)\rho_{\text{noise}}.
\end{align}

Rather than considering the general case of how noise affects our maximally entangled state, we will consider the specific example of a bit-flip channel, which we saw in Sec.~\ref{sec:3-3_density_matrices}.
This channel leaves the state unaffected with probability $F$, otherwise it applies the Pauli $X$ operator to one of the qubits,
\begin{align}
    \rho_{AB} & = F|\Phi^{+}\rangle\langle\Phi^{+}|+(1-F) X_A| \Phi^{+}\rangle\langle\Phi^{+}| X_A \nonumber\\
    & = F \left|\Phi^{+}\right\rangle\left\langle\Phi^{+}|+(1-F)| \Psi^{+}\right\rangle\left\langle\Psi^{+}\right|.
    \label{eq:12-4_bitflip_mixed_state}
\end{align}
We have applied the bit-flip channel to qubit $A$, but it does not matter whether we apply it to qubit $A$ or $B$.
We can easily check that applying the Pauli $X$ operator on qubit $B$ yields the same expression for the state $\rho_{AB}$.

Keep in mind that this is just one possible source of error out of many.
One way of dealing with errors is to use \textbf{\emph{quantum error correction}}\index{quantum error correction (QEC)}.
QEC can detect and correct errors, but usually comes with a large overhead.
Here, we will look at a less ambitious procedure of simply detecting errors, known as \textit{\textbf{purification}}\index{purification}.

Purification is a test for the state $\rho_{AB}$ that checks whether the state is affected by an error.
The test is not perfect and sometimes succeeds even when the state has undergone an error.
If the probability of this ``false positive'' is low enough, then the overall fidelity of the state increases.
It is this sense that we say that the state has been purified.

There are two important things that we need to keep in mind when designing the purification procedure.
First, measurements can be used to reveal information about the state, but they are also very intrusive and destroy the entanglement that we are trying to preserve.
Second, the two qubits of the entangled state are spatially separated and held in separated repeaters.
This distance can be of the order of tens of kilometers for a link, or much longer over a network.

We can get around the first issue of destructive measurement by using another Bell pair to test whether the original pair has undergone the unwanted Pauli $X$ flip.
This is done by entangling the second pair with the original one, and then measuring the second pair in order to extract useful information about the first pair without destroying it.
The second issue of the qubits being at distant locations can be overcome by simple classical communication.

\begin{figure}[t]
    \centering
    \includegraphics[width=0.7\textwidth]{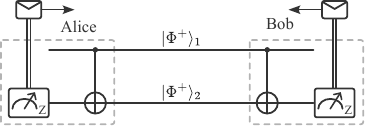}
    \caption[Entanglement purification for Pauli X errors.]{Entanglement purification for detecting Pauli $X$ errors. In this diagram, each line represents a Bell pair, rather than a single qubit, shared between Alice ande Bob.}
    \label{fig:12-4_purification}
\end{figure}
Let's introduce the detailed protocol for entanglement purification for the Pauli $X$ error, pictured in Fig.~\ref{fig:12-4_purification}.
The horizontal wires represent Bell pairs shared between Alice and Bob.
Ideally, these states are $\ket{\Phi^+}_1$ and  $\ket{\Phi^+}_2$, but due to noise they will be mixed states $\rho_1$ and $\rho_2$.
Alice applies a CNOT gate on the qubits in her possession.
Her qubit from the first pair acts as the control and her qubit from the second pair is the target.
Bob also performs a CNOT gate on his two qubits with the same control/target choice.
Alice and Bob then measure the qubits of the second pair in the Pauli $Z$ basis and send the measurement outcomes to each other using classical messages.
They keep the first pair if the measurement outcomes are the same, meaning they both measured the state \ket{0} or they both measured the state \ket{1}.
We say that the measurement outcomes are \textbf{\emph{correlated}}.
Otherwise they discard the first pair, as different measurement outcomes signal that an error has been detected.

Why this particular sequence of steps works will become more clear when we write them out step-by-step.
We start by considering the ideal case where Alice and Bob share two copies of \ket{\Phi^+}.
We label the two qubits that are sent to Alice as $A_1$ and $A_2$, while Bob's qubits are $B_1$ and $B_2$.
Expanding the initial state and applying the CNOT gates,
\begin{align}
    |\Phi^{+}\rangle_{1} |\Phi^{+}\rangle_{2} & = (|00\rangle + |11\rangle)_{A_1B_1} (|00\rangle + |11\rangle)_{A_2B_2} \nonumber\\
    & = |00\rangle_{A_1B_1} |00\rangle + |00\rangle |11\rangle + |11\rangle |00\rangle +|11\rangle |11\rangle \nonumber\\
    & \stackrel{\mathrm{CNOT}_{A}}{\longrightarrow} |00\rangle|00\rangle+|00\rangle|11\rangle+|11\rangle|10\rangle+|11\rangle|01\rangle \\
    & \stackrel{\mathrm{CNOT}_{B}}{\longrightarrow} |00\rangle|00\rangle+|00\rangle|11\rangle+|11\rangle|00\rangle+|11\rangle|11\rangle \nonumber \\
    & = |\Phi^{+}\rangle_{1} |\Phi^{+}\rangle_{2} \nonumber.
    \label{eq:12-4_purification_ideal}
\end{align}
In the above, we have omitted the normalization constants for clarity.
We see that the CNOT gates have left our initial state unaffected.
Measuring the second pair in the Pauli $Z$ basis will give us the state \ket{00} with probability $1/2$, or the state \ket{11} with probability of $1/2$.
The outcomes will always be correlated and therefore Alice and Bob will always keep the first pair.
As well as they should since the first pair is not affected by an error.

Let's see what happens when the first pair is affected by a Pauli $X$ error.
Repeating the same calculation as in Eq.~(\ref{eq:12-4_purification_ideal}), we get
\begin{align}
    |\Psi^{+}\rangle_{1} |\Phi^{+}\rangle_{2} & = (|01\rangle + |10\rangle)_{A_1B_1} (|00\rangle + |11\rangle)_{A_2B_2} \nonumber\\
    & = |01\rangle |00\rangle + |01\rangle |11\rangle + |10\rangle |00\rangle +|10\rangle |11\rangle \nonumber\\
    & \stackrel{\mathrm{CNOT}_{A}}{\longrightarrow} |01\rangle|00\rangle+|01\rangle|11\rangle + |10\rangle|10\rangle+|10\rangle|01\rangle \\
    & \stackrel{\mathrm{CNOT}_{B}}{\longrightarrow} |01\rangle|01\rangle+|01\rangle|10\rangle+|10\rangle|10\rangle+|10\rangle|01\rangle \nonumber \\
    & = |\Psi^{+}\rangle_{1} |\Psi^{+}\rangle_{2} \nonumber.
    \label{eq:12-4_purification_error}
\end{align}
This time the error from the first pair has propagated onto the second pair.
Measuring the second pair now yields either \ket{01} or \ket{10}, both with equal probability of $1/2$.
Since the measurement outcomes are anti-correlated, Alice and Bob discard the first pair.

Now we are in a position where we can consider both initially distributed pairs to be in the mixed state given by Eq.~(\ref{eq:12-4_bitflip_mixed_state}),
\begin{equation}
    \rho_1 \otimes \rho_2 = \left[ F |\Phi^+\rangle \langle\Phi^+| + (1 - F) |\Psi^+\rangle \langle\Psi^+| \right]_1 \otimes \left[ F |\Phi^+\rangle \langle\Phi^+| + (1 - F) |\Psi^+\rangle \langle\Psi^+| \right]_2.
\end{equation}
We have already covered the situation where both Bell pairs are unaffected by noise, which occurs with probability of $F^2$, and where the first pair is bit-flipped, occurring with probability of $(1-F)F$.
The remaining two possibilities of only the second pair being bit-flipped and both pairs being bit-flipped can be calculated in exactly same way.
We summarize all 4 possibilities in Tab.~\ref{tab:12-4_purification_X_error}.
\begin{table}[t]
    \setcellgapes{3pt}
    \renewcommand\theadfont{}
    \makegapedcells
    \centering
    \begin{tabular}{cccccc}
        \hline
        \textbf{Pair 1} & \textbf{Pair 2} & \textbf{Probability} & \textbf{Measurement result} & \textbf{Action} & \textbf{Result} \\
        \hline
        $\left|\Phi^{+}\right\rangle$ & $\left|\Phi^{+}\right\rangle$ & $F^{2}$ & 00 or 11 & \textcolor{mygreen}{Keep} & $\left|\Phi^{+}\right\rangle$ \\
        $\left|\Phi^{+}\right\rangle$ & $\left|\Psi^{+}\right\rangle$ & $F(1-F)$ & 01 or 10 & \textcolor{myred}{Discard} & N/A \\
        $\left|\Psi^{+}\right\rangle$ & $\left|\Phi^{+}\right\rangle$ & $(1-F) F$ & 01 or 10 & \textcolor{myred}{Discard} & N/A \\
        $\left|\Psi^{+}\right\rangle$ & $\left|\Psi^{+}\right\rangle$ & $(1-F)^{2}$ & 00 or 11 & \textcolor{mygreen}{Keep} & $\left|\Psi^{+}\right\rangle$ \\
        \hline
    \end{tabular}
    \caption[X error purification.]{Purification for a Pauli $X$ error.}
    \label{tab:12-4_purification_X_error}
\end{table}

We can make an interesting observation.
When both initial pairs are affected by Pauli $X$ errors, the purification procedure still tells Alice and Bob to keep the first pair because the measurement outcomes on the second pair are correlated.
This occurs with probability $(1-F)^2$ and is the ``false positive'' that we mentioned earlier in this section.

We can see from Tab.~\ref{tab:12-4_purification_X_error} that the total success probability of this purification procedure is
\begin{equation}
    \text{Pr} \{\text{success}\} = F^2 + (1 - F)^2.
\end{equation}
The state that Alice and Bob share after a successful purification is
\begin{equation}
    \rho' = F' |\Phi^+\rangle\langle\Phi^+| + (1 - F') |\Psi^+\rangle\langle\Psi^+|,
\end{equation}
where the new fidelity $F'$ is related to the initial one via
\begin{align}
    F^{\prime}=\frac{F^{2}}{F^{2}+(1-F)^{2}}.
\end{align}
The expression for the new post-purification fidelity is very interesting.
When $F > 0.5$, the fidelity of the purified state increases, $F' > F$.
The name for the process, purification, is now more clear.
We start with a noisy state and end up with a new state of higher fidelity, a more pure state.

\section{Making a network}
\label{sec:making-a-network}

We have considered two important ingredients that are needed to make a network.
We looked at how to create link-level entanglement between neighboring nodes, and how to extend it to end-to-end entanglement with entanglement swapping.
There are a few missing parts however.
The main two are routing routing, and multiplexing.

At the beginning of this chapter, we saw in Fig.~\ref{fig:12-1_all_to_all} that the job of a quantum network is to establish end-to-end entanglement between distant nodes, which is then consumed by applications.
The network needs a way of picking a suitable path along which to execute entanglement swapping that leads to this end-to-end connection.
The second issue is that the figure shows only one connection request being considered.
Of course, a network will most likely be used by more than one application at a time.
Therefore, the network should be able to satisfy multiple simultaneous connection requests between distinct end nodes.

\begin{figure}[t]
    \centering
    \includegraphics[width=0.8\textwidth]{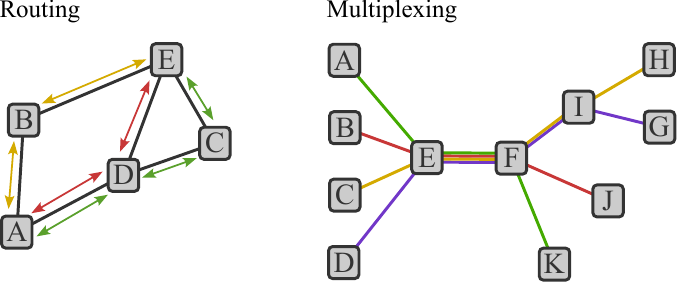}
    \caption[Routing and multiplexing.]{Picking the best path along which to entanglement swap is the job of routing. The quantum network also needs to decide how to allocate its scarce resources in order to satisfy multiple simultaneous connection requests.}
    \label{fig:12-5_routing_multiplexing}
\end{figure}

How the network picks an appropriate path where it performs entanglement swapping is determined by the \emph{\textbf{routing algorithm}}\index{routing algorithm} using information about the topology and condition of the network.
The left panel of Fig.~\ref{fig:12-5_routing_multiplexing} shows a simple case of a quantum network with five nodes.
Let's say that nodes A and E would like to establish an entangled connection.
There are three possible paths that the network can choose.
It can swap entanglement at node B (option 1), or at node D (option 2), or at nodes C and D (option 3).
Each option requires the establishment of different patterns of link-level entanglement.
Option 1 requires link-level entanglement between A-B and B-E before the entanglement swapping can be executed.
Similarly, option 3 requires link-level entanglement on the A-D, D-C, and C-E links.

In order to evaluate which path is best, we need to look at the cost for establishing the corresponding link-level entanglement, and then how to combine these costs in order to establish the cost for the entire path.
One possible way to evaluate the link-level entanglement is \emph{\textbf{seconds per Bell pair at a threshold fidelity}}.
We discussed how establishing a link-level entanglement involves photon emission from quantum memories, and subsequent measurement of these photons in the Bell basis.
These processes are all probabilistic, and so is loss of the photons in fiber.
Depending on the properties of the fiber as well as the quantum hardware at each node, the average time required to successfully establish link-level Bell pair will vary.

How do we combine these costs for each individual link in order to know the full cost of establishing end-to-end entanglement over a path?
One possibility is to use \emph{\textbf{Dijkstra's shortest path first algorithm}}\index{Dijkstra's shortest path first}\label{dijkstra}.
We can sum the link-level costs for an individual path in order to estimate the total cost.
We have seen that sometimes we need to purify Bell pairs, which reduces our effective throughput by at least a factor of two.
In order to balance the trade off between the rate of making raw Bell pairs and fidelity and make the comparison between two links fair, we demand that the link-level Bell pairs be of certain threshold fidelity, achieved using purification if necessary.
The exact value of this threshold is dictated by the needs of the application that is requesting the Bell pairs.

The next issue that networks have to deal with is \emph{\textbf{multiplexing}}\index{multiplexing}.
The network will have to satisfy multiple simultaneous connection requests for end-to-end entanglement.
A good multiplexing scheme helps the network decide how to allocate its resources to satisfy these requests in a timely, reliable and fair manner.
The right panel of Fig.~\ref{fig:12-5_routing_multiplexing} shows four simultaneous requests for entanglement between nodes A-K, B-J, C-H, and D-G.
All of these requests need to use link-level entanglement between E-F, creating \textbf{\emph{contention}}\index{contention} on this link.
Another link with contention is F-I, although only two connections need to use that link.
The network must know how to handle such requests, where a single link needs to be shared between multiple users or connections.

In this chapter, we have seen the four requirements for building a quantum network.
We started with the basic building block of establishing link-level entanglement between neighboring nodes of a network and how to handle photon losses.
Next, we discussed how to use entanglement swapping to splice the link-level entanglement into long-distance end-to-end connections between non-neighboring nodes.
Finally, we briefly discussed routing and multiplexing in quantum networks.

\newpage
\begin{exercises}
\exer{
\emph{Fully connected networks.}
How many connections are needed for a fully connected network covering all of the IoT devices in the global classical Internet?
}

\exer{
\emph{Entanglement swapping with projectors.}
We demonstrated that the procedure for entanglement swapping between qubits $AB_1$ and $B_2C$ works by rewriting the state of qubit $B_1$ and $B_2$ in the bell basis.
Let's go through the derivation again but using the formalism of projectors. (If you need to get a refresher on projectors, take a look at the Exercises in Chapter \ref{sec:3_pure_mixed}.)
\subexer{
The initial state of the two entangled pairs is
\begin{equation}
    \ket{\psi} = \ket{\Phi^+}_{AB_1} \ket{\Phi^+}_{B_2C}.
\end{equation}
Write down the set of four projectors corresponding to the four possible measurement outcomes at Node $B$.
}
\subexer{
Compute the probabilities of the four measurement outcomes using the these projectors.
}
\subexer{
Write down the post-measurement states of qubits $AC$ corresponding to the four measurement outcomes using the projectors.
}
}

\exer{
\emph{Entanglement swapping with noisy states.}
Consider the case where the link-level entangled state between Nodes $A$ and $B$ is not the pure Bell pair \ket{\Phi^+}, but is affected by the bit flip channel.
The Bell pair shared between Nodes $B$ and $C$ remains noiseless.
\subexer{
If the fidelity of the Bell pair between Nodes is $A$ and $B$ if $F_{AB}$, write down the total initial state.
}
\subexer{
Write down the probabilities of the four outcomes of the measurements performed by Node $B$.
}
\subexer{
What are the corresponding post-measurement states of qubits $A$ and $C$?
}
\subexer{
How do the fidelities of the post-measurement states of qubits $A$ and $B$ depend on the fidelity $F_{AB_1}$ of the initial link-level entanglement? 
}
\subexer{
Consider the case where the both entangled links are affected by the bit flip channel, and have fidelities $F_{AB_1}$ and $F_{B_2C}$.
Compute the fidelity of the pair $AC$ obtained after entanglement swapping as a function of these initial fidelities.
}
}

\exer{
\emph{Entanglement swapping and MIM link architecture.}
Recall that in the MIM link architecture, quantum memories are made to emit photons, which are then measured at the BSA, creating entanglement between the quantum memories.
This architecture relies on entanglement swapping performed on photonic and matter qubits.
Fig.~(TODO) shows a simplified atomic structure of the a quantum memory emitting a photon.
If the memory transitions from the excited state \ket{e} to the ground state \ket{a_1}, it emits a horizontally polarized photon \ket{H}\footnote{To be precise, the memory emits a circularly polarized photon which is then converted to a linearly polarized photon.}.
If the memory transitions to ground state \ket{a_2}, it emits a vertically polarized photon \ket{V}.
\subexer{
Write down the state of the memory-photon pair, given that the memory is equally likely to transition to states \ket{a_1} and \ket{a_2}.
}
\subexer{
Write down the total state of both memory-photon pairs.
Keep good track of the physical systems with subscripts.
}
\subexer{
The two photons travel to the BSA where they get measured in the Bell basis.
Write down the projectors corresponding to this measurement.
}
\subexer{
What are the post-measurement states of the quantum memories?
}
\noindent The situation is actually more complicated than this.
Bell-state measurement on photonic qubits cannot be implemented deterministically.
Linear optics elements cannot distinguish all four photonic Bell pairs.
We will learn why this is the case in our next book.
}

\exer{
\emph{Purification and Pauli $Z$ errors.}
We described the purification procedure on Bell pairs affected by the bit flip channel.
Let's look at what happens when the Bell pairs undergo a Pauli $Z$ error.
\subexer{
Assume that only Alice's qubit $A_1$ is affected by the Pauli $Z$.
Write down the total state of both Bell pairs.
}
\subexer{
Apply the purification procedure.
Is this the result you expected?
}
\subexer{
If both Bell pairs are mixed states of fidelity $F$, what is the fidelity of the Bell pair after successful purification procedure?
}
}

\end{exercises}

\chapter{Physical Layer Components}
\label{sec:physical-layer-components}

Now that we have a basic knowledge of the working principles behind a quantum repeater, it is time to delve a little deeper.
In this chapter, we will see how individual physical components of a quantum repeater work.

\section{Introduction}
\label{sec:phys_layer_intro}

\begin{figure}[t]
    \centering
    \includegraphics[width=0.7\textwidth]{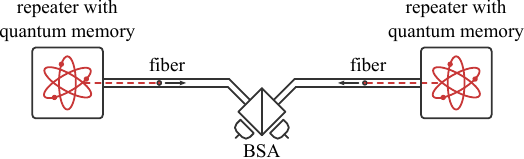}
    \caption[MIM link hardware.]{The MIM link. The nodes at each end are repeaters with memories, shown by the atom symbols.  The node in the middle is the Bell-state analyzer (BSA). The red circles represent photons in flight, entangled with the memories.}
    \label{fig:13-MIM-link}
\end{figure}

Fig.~\ref{fig:13-MIM-link} shows the basic MIM architecture for a quantum repeater link.
What are the individual physical components of a quantum repeater system?
We require a physical system that can be used as a suitable \textbf{\emph{quantum memory}}\index{quantum memory}, represented by the red atom symbols in the figure.
Ideally, we would like to interact with this system optically, meaning it is possible to use laser pulses to manipulate the state of the quantum memory, and make it emit photons.
The difference between classical memories and quantum memories is quite substantial.
Classical memories store 0s and 1s (classical bits), whereas quantum memories have to be able to store not only \ket{0} or \ket{1}, but also entangled state. 
This state could be a pure superposition, or it could be part of a distributed entangled state.
In fact, the latter case will be the usual scenario in our considerations.

The photons emitted from the quantum memories must be collected into an optical fiber.
We have covered optical fibers in Chapters~\ref{sec:7_waveguides}\index{optical fiber} and \ref{sec:11_long-distance} already.
Finally, we need to think how to implement the BSA in the middle.
We know what the function of a BSA is, but we have not discussed the physical components needed that can implement it.

The first half of this chapter is dedicated to Bell-state measurements.
We will discuss two different types of these measurements.
The first type is used in entanglement swapping between two quantum memories.
This Bell-state measurement is used to extend link-level entanglement to longer distances as seen in Chapter~\ref{sec:12_quantum_repeaters}.
The second type is photonic Bell-state measurement.
We will see that these two types of measurement have very different physical implementation, with different consequences, even though they both implement the same function.

In the latter half of this chapter, we will talk about quantum memories.
We will begin with consideration of what a good quantum memory should be like before moving to candidate systems.
We say ``candidate'' because at the moment, there is no leading physical system that is considered to be the best quantum memory. All of the existing candidate systems have their own advantages and drawbacks.

\section{Bell State Measurements I}
\label{sec:bell-state-measurements-I}

In this Section, we focus on the Bell state analyzer from Fig.~\ref{fig:13-MIM-link}, particularly how one can perform measurements in the Bell basis.
We have seen how to describe this measurement mathematically in Chapter~\ref{sec:4_entanglement}.
However, this tells us little about a real-world implementation.
The first step is to break down the Bell state measurement into more elementary operations in order to understand how we can implement it using a real physical system.

\begin{figure}[t]
    \centering
    \includegraphics[width=0.6\textwidth]{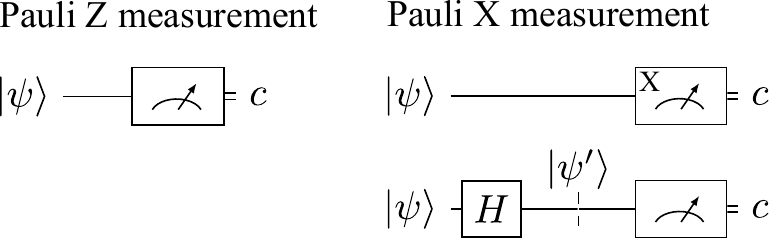}
    \caption[Changing the basis of measurements.]{Changing the basis of a measurement can be achieved by applying a suitable unitary to the state and then measuring in the Pauli $Z$ basis.}
    \label{fig:13-2_measurementPauli}
\end{figure}

Let's step back a little and consider something simpler first, like measuring a single qubit in the Pauli $Z$ basis, depicted in the left panel of Fig.~\ref{fig:13-2_measurementPauli}.
The meter icon represents a measurement in the Pauli Z basis which outputs a single classical bit $c$.
The value of this classical bit can be either $+1$ or $-1$, which is normally written as $c\in\{+1,-1\}$.
For an arbitrary initial state $|\psi\rangle = \alpha |0\rangle + \beta |1\rangle$, the probability that the measurement outcome is $c=+1$ is given by $\text{Pr}(+1)=|\langle0|\psi\rangle|^2=|\alpha|^2$.
The probability that the measurement outcome is $c=-1$ is given by $\text{Pr}(-1)=|\langle1|\psi\rangle|^2=|\beta|^2$.

We can also measure the qubit in the Pauli $X$ basis, as shown in the top part of the right panel in Fig.~\ref{fig:13-2_measurementPauli}.
The measurement icon includes an X to remind us that we are measuring in the Pauli $X$ basis.
The probability that we obtain $c=+1$ is $\text{Pr}(+1)=|\langle+|\psi\rangle|^2=|\alpha+\beta|^2/2$, and the probability of the outcome $c=-1$ is $\text{Pr}(-1)=|\langle-|\psi\rangle|^2=|\alpha-\beta|^2/2$.

But what if we cannot perform the Pauli $X$ measurement directly?
Sometimes, in an experiment, it is straightforward to implement a Pauli $Z$ measurement, but much less clear how to perform a measurement in a different basis.
We can still measure in the desired basis by rotating the state with an appropriate unitary and then performing a measurement in the Pauli $Z$ basis.
In the case of a Pauli $X$ measurement, the appropriate unitary is a Hadamard gate $H$, as shown in the lower part of the right panel in Fig.~\ref{fig:13-2_measurementPauli}.
The state immediately before the measurement in the Pauli $Z$ basis is
\begin{align}
    |\psi'\rangle & = H|\psi\rangle \nonumber\\
    & = \alpha |+\rangle + \beta |-\rangle \nonumber\\
    & = \frac{\alpha + \beta}{\sqrt{2}} |0\rangle + \frac{\alpha - \beta}{\sqrt{2}} |1\rangle.
\end{align}
We can immediately see that the probabilities of obtaining measurement outcomes $+1$ or $-1$ are the same as the probabilities of measuring in the Pauli $X$ basis directly.

How do we know that the unitary that is needed is the Hadamard $H$?
The trick lies in realizing that the probability of obtaining the $c=+1$ outcome via a direct Pauli $X$ measurement must be the same as first rotating the state with some unitary operation $U$ and then measuring in Pauli $Z$ basis.
Written more formally, we require that
\begin{equation}
    |\langle+|\psi\rangle|^2 = |\langle0|U^{\dagger}|\psi\rangle|^2.
    \label{eq:13-2_prob_condition}
\end{equation}
Eq.~(\ref{eq:13-2_prob_condition}) is satisfied when $\langle+|=\langle0|U^{\dagger}$, or $|+\rangle=U|0\rangle$ if you prefer to think in terms of kets rather then bras.
The same unitary $U$ must also satisfy
\begin{equation}
    |\langle-|\psi\rangle|^2 = |\langle1|U^{\dagger}|\psi\rangle|^2,
\end{equation}
that is, the probabilities of obtaining the outcome $c=-1$ via direct measurement in Pauli $X$ basis and via rotating the state before measuring in the Pauli $Z$ basis must be the same.
The unitary operation $U$ that achieves this transformation is the Hadamard $H$, as we saw in Sec.~\ref{sec:2-2_unitary_operations}.

\begin{figure}
    \centering
    \includegraphics[width=0.9\textwidth]{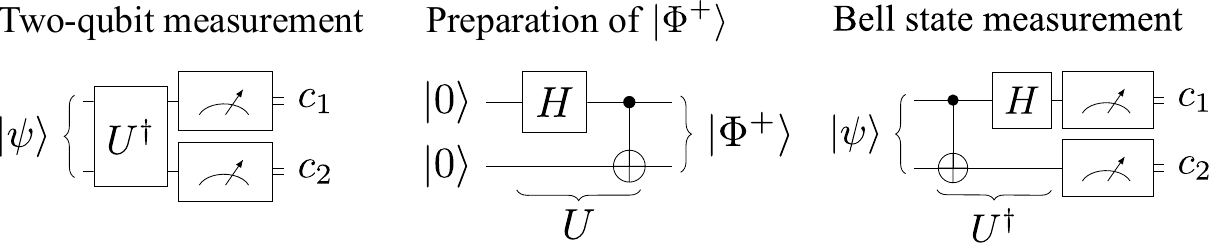}
    \caption[Bell state measurement via Pauli Z measurements.]{Performing the Bell state measurement can be achieved by the right unitary $U$ followed by two measurements in the Pauli $Z$ basis.}
    \label{fig:13-2_measurementBell}
\end{figure}

Transforming the measurement basis by applying a suitable unitary operation on the state does not stop with single qubits but works for multi-qubit measurements as well.
Let's turn our attention to the Bell state measurement.
What is the two-qubit unitary that we must apply such that subsequent measurements of both qubits in the Pauli $Z$ basis have the same effect as a Bell state measurement?
We can apply the same technique that we used in the case of single-qubit Pauli $X$ measurement and look for a unitary $U$ which satisfies
\begin{equation}
    |\langle\Phi^+|\psi\rangle|^2 = |\langle00|U^{\dagger}|\psi\rangle|^2,
\end{equation}
where $|\psi\rangle$ is now an arbitrary two-qubit state, as depicted in the left panel of Fig.~\ref{fig:13-2_measurementBell}.
The outcomes of the measurements on the first and second qubit are denoted by $c_1$ and $c_2$, respectively.
We see that our desired unitary $U$ must satisfy $\langle\Phi^+| = \langle00|U^{\dagger}$, which can be expressed in the ket notation by taking the adjoint of both sides,
\begin{equation}
    |\Phi^+\rangle = U |00\rangle.
\end{equation}
This means we need to find the unitary $U$ which transforms the input $|00\rangle$ into the Bell pair $|\Phi^+\rangle = (|00\rangle + |11\rangle) / \sqrt{2}$.
This unitary operation is pictured in the middle panel of Fig.~\ref{fig:13-2_measurementBell},
\begin{equation}
    U = CNOT_{12} \cdot (H \otimes I).
    \label{eq:13-2_U}
\end{equation}
Starting from the initial state where both qubits are initialized in the $|0\rangle$, we need to apply the Hadamard operation $H$ to the first qubit, followed by a controlled-NOT gate with the first qubit being the control and the second qubit the target.
All that remains to be done is to take the adjoint of Eq.~(\ref{eq:13-2_U}),
\begin{equation}
    U^{\dagger} = (H \otimes I) \cdot CNOT_{12},
\end{equation}
which swaps the order of the Hadamard and the CNOT gates.
This is the unitary operation that we need to apply to the state \ket{\psi} in order to be able to perform a Bell state measurement using measurements in the Pauli $Z$ basis, as shown in the right panel of Fig.~\ref{fig:13-2_measurementBell}.
This trick is very useful in both quantum computation and quantum communication.

This form of the Bell-state measurement is suited for quantum memories.
The three operations required are the Hadamard gate, CNOT gate, and measurement in Pauli $Z$ basis.
A candidate system for a quantum memory usable in a quantum repeater must be able to implement all three of these operations.

\section{Bell State Measurements II}
\label{sec:bell-state-measurements-II}

In this section, we will turn our attention to Bell-state measurement of photonic qubits.
We will see that even though the abstract function of the BSA is the same as entanglement swapping, the implementation differs substantially from the case of quantum memories discussed in Sec.~\ref{sec:bell-state-measurements-I}.
The implementation scheme depends on the particular encoding chosen.
We will consider the case of encoding the state of a qubit using the photon's polarization.
There are two reasons for this choice.
First, it is intuitive and simple to visualize so it makes a good pedagogical example.
Second, it is one of the most commonly used encodings in real experiments.

In this encoding, a qubit in the state \ket{0} is represented as a single photon that is polarized in the horizontal direction, which we write as \ket{H}.
The other computational basis state \ket{1} is represented as a single photon polarized in the vertical direction, and we write \ket{V}.

\begin{figure}[t]
    \centering
    \includegraphics[width=0.6\textwidth]{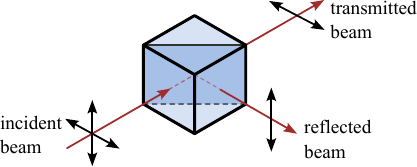}
    \caption[A polarizing beam splitter (PBS).]{The action of a polarizing beam splitter (PBS).  The vertical rays represent vertically polarized light \ket{V}, while the diagonal rays represent horizontally polarized light \ket{H}.  The direction of propagation, red arrows, is normal to the \ket{H}-\ket{V} plane.}
    \label{fig:13-PBS}
\end{figure}

The first question that we should ask is, ``how do we implement a Pauli $Z$ measurement with this encoding?''
We need to distinguish horizontal and vertical polarizations.
This can be done with a piece of crystal called a \emph{\textbf{polarizing beam splitter}} (PBS)\index{polarizing beam splitter (PBS)}, as shown in Fig.~\ref{fig:13-PBS}.
A PBS transmits light of horizontal polarization only, and reflects vertically-polarized light.
An incident beam of arbitrary polarization gets split into two beams by the PBS.
The relative strength of the two beams depends on the polarization of the input light.
The PBS is not creating or changing polarization, but rather sorting the input light into the two categories.
Thinking in terms of computational states, we now have two beams, one for our \ket{0} and one for our \ket{1}.

\begin{figure}[t]
    \centering
    \includegraphics[width=0.7\textwidth]{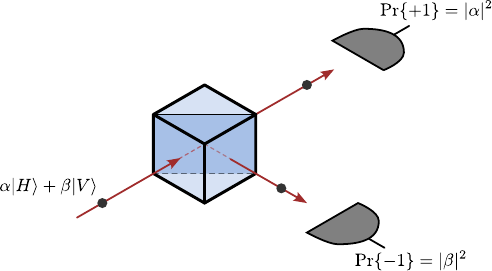}
    \caption[A polarizing beam splitter (PBS) measuring a qubit.]{A PBS measuring an arbitrary photonic qubit in the \{$|H\rangle$, $|V\rangle$\} basis.}
    \label{fig:13-PBS-measure}
\end{figure}

Assume that we have a single photon incident onto the PBS.
We can determine which output path was taken by the photon by placing two detectors into the two possible output paths, as shown in Fig.~\ref{fig:13-PBS-measure}.
If the photon is horizontally polarized, it gets transmitted through the polarizing beam splitter.
It has no chance of being reflected and gets detected by the detector placed in the transmitted path with probability one.
This represents our measurement outcome of $+1$.
On the other hand, if the initial photon is vertically polarized, it always gets reflected and travels down into the bottom detector.
In that case, the probability of the outcome $-1$ is 1, and the probability of the outcome $+1$ is always 0.

What happens if we put in a superposition of the two linear polarizations, as in Fig.~\ref{fig:13-PBS-measure}?
Our input state is given by $\alpha\ket{H} + \beta \ket{V}$.
The photon has a chance to get transmitted, with probability given by $|\alpha|^2$, and it also has a chance to get reflected and travel down into the other detector corresponding to the measurement outcome $-1$ (bottom), with probability of $|\beta|^2$.
This shows how this arrangement implements a measurement in the Pauli $Z$ basis.

\begin{figure}[t]
    \centering
    \includegraphics[width=0.6\textwidth]{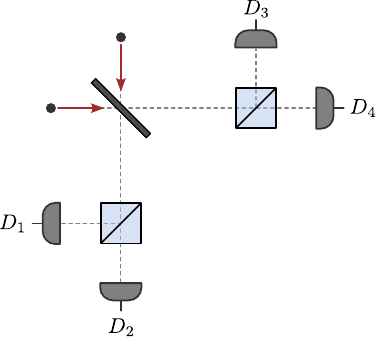}
    \caption[A four-detector Bell-state analyzer (BSA).]{A BSA measuring a pair of photonic qubits, one arriving at the initial beam splitter from above and one from the left.  Different click patterns on the four detectors may confirm projection of the pair into a Bell state, or be ambiguous.}
    \label{fig:13-BSA-clicks}
\end{figure}

We have seen in the previous step that for a Bell-state measurement, we need two measurements in the $Z$ basis and a suitable unitary preceding the measurements.
We have just learned that $Z$ measurements can be implemented with a single PBS and two detectors.
Two $Z$ measurements will therefore require two PBS and four detectors.
The last remaining ingredient that is needed is the unitary transformation changing the basis of the measurements.
This unitary is given by a regular, non-polarizing, beam splitter as pictured in Fig.~\ref{fig:13-BSA-clicks}.

Let's investigate the behavior of the optical arrangement in Fig.~\ref{fig:13-BSA-clicks}.
Assume we have two incoming photons, one coming from the top and one coming from the left, arriving at the regular beam splitter simultaneously.
A complete analysis requires a lot more quantum optics than we have studied so far, so instead of the full derivation, here we will give the result.
Depending on which of the four detectors click, we may learn which Bell-state has been measured.
Let's see what the different patterns are.

If we get a joint detection in detectors $D_1$ and $D_4$, meaning that both detectors click, then we have implemented a successful Bell measurement, and the outcome corresponds to the projection onto the state \ket{\Psi^-}.
If we get a joint detection in $D_2$ and $D_3$, then we can also say that we have implemented a successful Bell measurement, and the result corresponds to the state \ket{\Psi^-}.
A different pattern is a joint detection at the two detectors in the lower branch of our Bell-state analyzer (both $D_1$ and $D_2$ click), from which we can conclude that we have a state \ket{\Psi^+}, corresponding to another successful Bell measurement.
Equally, if both detectors in the right branch of our Bell-state analyzer ($D_3$ and $D_4$) click, then we can also conclude that we have a Bell-state \ket{\Psi^+}.

It is also possible that both photons travel into a single detector. (Detection in any of the four detectors is equally probable.) However, because more than one Bell state can result in this happening, the answer we get is ambiguous: we cannot say that whether we have \ket{\Phi^+} or \ket{\Phi^-}.
This is a problem because we cannot fully implement a Bell-state measurement.
We cannot distinguish all four Bell states, only two of them, \ket{\Psi^+} and \ket{\Psi^-}.
Using the arrangement in Fig.~\ref{fig:13-BSA-clicks}, we can only implement a partial Bell-state measurement.

\begin{table}
    \setcellgapes{3pt}
    \renewcommand\theadfont{}
    \makegapedcells
    \centering
    \begin{tabular}{cccc}
    \hline
        \textbf{Pattern}  & \textbf{Result} & \textbf{Reason} & \textbf{Action} \\
        \hline
        $D_1$ and $D_4$ & \ket{\Psi^-} & & \textcolor{mygreen}{keep} \\
        $D_2$ and $D_3$ & \ket{\Psi^-} & & \textcolor{mygreen}{keep} \\
        $D_1$ and $D_2$ & \ket{\Psi^+} & & \textcolor{mygreen}{keep} \\
        $D_3$ and $D_4$ & \ket{\Psi^+} & & \textcolor{mygreen}{keep} \\
        single click & \ket{\Phi^\pm}  & two photons together/only one arrived & \textcolor{myred}{discard} \\
        no click & N/A & photons lost/detection failure & \textcolor{myred}{discard} \\
        other pattern & N/A & detection error & \textcolor{myred}{discard} \\
        \hline
    \end{tabular}
    \caption[Four-detector BSA click patterns.]{4-detector BSA click patterns and their meaning, based on Fig.~\ref{fig:13-BSA-clicks}.}
    \label{tab:bsa-clicks}
\end{table}

A complete, unambiguous Bell-state measurement cannot always be successfully implemented with linear optics.
In fact, even with 100\% probability of receiving both photons, the maximum probability of a successful Bell measurement is limited to only 50\% at most.
Of course, as we saw when discussing the loss of photons in fiber, the loss of one or both photons is highly probable, increasing the ambiguity of interpreting the result of one click.
These cases (along with the case where something goes wrong in the hardware) are summarized in Tab.~\ref{tab:bsa-clicks}. 
Moreover, we have to take into account that the two photons coming into our Bell-state analyzer have to be synchronized.
If they come in so close together that two detectors click \emph{almost} simultaneously, but not close enough that the photons are truly indistinguishable, we may misinterpret the result.
The Bell-state measurement has failed and we did not establish entanglement between the network nodes, but may not realize it; when we accumulate statistics about the link, this results in lowered fidelity.

\section{Stationary and flying qubits}
\label{sec:statinary-and-flying-qubits}

Let's begin by considering the basic requirements for a good quantum memory, known as \textbf{\emph{DiVincenzo criteria}}\index{DiVincenzo criteria}.
These criteria were introduced in the context of quantum computation, but they also apply in the context of quantum networking, with slightly different emphasis on which ones are important.
The extended list of these criteria, with the first five for quantum computing and the latter two for quantum communication, is as follows,
\begin{enumerate}
    \item A well-defined qubit.
    \item The qubit can be initialized.
    \item Long lifetime.
    \item Universal gate set.
    \item Efficient measurement.
    \item Convert or entangle stationary and flying qubits.
    \item Able to carry flying qubits long distances.
\end{enumerate}

If we want to build a good quantum computer, we need a \textbf{\emph{well-defined qubit}}.
Qubits do not come for free in nature.
Most naturally occurring systems have a complicated energy level structure.
In order to have a well-defined qubit, we must be able to take a system for which we can address two of those energy levels, distinguish them and control them as a pair, without slipping into the other energy levels.
(Sometimes a third level is used as a temporary state to achieve certain effects such as emitting a photon, but the computation is done by restricting actions to the two levels we want to use.)

Second, we need to be able to \textbf{\emph{initialize}} this qubit.
Initialization is important because knowing the initial state of the system is crucial for computation.
Without reliable initialization, we would always get a different answer to out problem.
(This initialization also sometimes involves the temporary use of a third state.)

Third, the qubits need to have \textbf{\emph{long lifetimes}}. This means that the state of qubit, particularly a superposition state, stays stable for a sufficiently long time such that the required operations can be applied to it.
Long lifetimes allow us to carry out longer quantum computations, which we need if we want to solve harder problems.
This criterion is especially important in quantum networking, as we will see below.

Fourth, we must be able to implement a \textbf{\emph{universal set of gates}}.
The physical system implementing a qubit dictates how we can interact with and manipulate the state of the qubit.
Typically this is done with a set of laser pulses in the case of ion qubits sitting in a magnetic trap, or with microwave pulses in the case of superconducting qubits.
If is usually difficult to implement an arbitrary operation with a single pulse.
Luckily, any complicated operation can be broken down and approximated by a series of simpler operations.
Usually, it is enough to be able to implement simple rotations of single qubits and apply entangling operations between two qubits.

Fifth, we need \textbf{\emph{efficient measurements}}.
Just carrying out transformations of the state in a quantum manner is not enough.
We must extract the information from the physical system at the end of the quantum computation.
The measurements process should be accurate with low probability of error, and should be able to be performed in a timely manner.

In the context of quantum communication, we have two more requirements to consider.
The first of these is \textbf{\emph{conversion or entanglement }}between stationary and flying qubits, which we already saw earlier in this chapter.
Stationary qubits are those qubits that are sitting in the quantum network nodes, loaded into the quantum memories. 
Flying qubits are those qubits that are used for entanglement swapping in the BSAs to create link level entanglement between between the quantum memories.
We must be able to entangle photons (the flying qubits) with the stationary qubits inside the memories, but also we must be able to use entanglement swapping to create end-to-end entanglement.

Lastly, we also must be able to \textbf{\emph{transport flying qubits}} over long distances.
The physical system used for flying qubits needs to be robust enough to survive the long journey between the network nodes.
It is for this reason that photons are a good physical system used for communication between distant nodes.

In this Section, we will look at the memory lifetime, and the two communications requirements.

We mentioned already that the ratio of gate speed to memory lifetime is important in quantum computing.
Longer memory lifetimes and shorter gate speeds let us carry out more complex calculations
In the context of quantum communications, we often store qubits for long periods of time without acting on them, as we await messages from partners in the network.
Therefore, what is more important is not the gate speed itself, but the ratio of memory lifetime to communication time, 
provided that the gate time is short compared to the \textbf{\emph{round trip time (RTT)}}\index{rount trip time (RTT)}. 

Let's consider how we establish link level entanglement using the MIM architecture.
The quantum memories emit photons entangled with their respective memory.
These photons are guided by the optical fiber to the BSA, where they get get measured in the Bell basis.
The BSA then communicates the outcome of the measurement back to the  network nodes via a classical message.
The total time needed for the emitted photon to reach the BSA and the classical message to reach the node holding the quantum memory is the round trip time~\footnote{RTT for a link can be node to BSA and back again in some contexts, or all the way memory node to memory node and back again in others. It should be clear from the context which we mean.}.
If the memory lifetime is shorter than the RTT, then we cannot reliably establish the link-level entanglement.
Even if we successfully perform the Bell-state measurements on the photon pairs at the BSA, by the time the return messages are received, our memories will have decohered and are not useful anymore.

Table~\ref{tab:rtt} shows some typical RTTs for various length of fiber.
The speed of light in a fiber is approximately $0.2$ meters per nanosecond, so if our nodes are one kilometer apart, one round trip from one node to the other and back takes 10 microseconds. For a hundred kilometers, it increases to one millisecond, and for ten thousand kilometers it goes all the way up to 100 milliseconds (0.1 seconds) per round trip time.
The values \textbf{\emph{five nanoseconds per meter one way}} and \textbf{\emph{10 microseconds per km round trip}} are easy metrics to remember.

\begin{table}
    \setcellgapes{3pt}
    \renewcommand\theadfont{}
    \makegapedcells
    \centering
    \begin{tabular}{cc}
    \hline
    \textbf{distance (km)}  & \textbf{RTT in fiber} \\
    \hline
    1     & $10\mu$sec \\
    10    & $100\mu$sec \\
    100   & $1$msec \\
    1,000 & $10$msec \\
    10,000 & $100$msec \\
    \hline
    \end{tabular}
    \caption{Round trip times in optical fiber.}
    \label{tab:rtt}
\end{table}

What processes degrade the quantum memories?
The two main processes are \textbf{\emph{energy relaxation}}\index{energy relaxation} and \textbf{\emph{dephasing}}\index{dephasing}.
They are characterized by two different time scales, referred to as the $T_1$ time\index{$T_1$ time} scale and the $T_2$ time\index{$T_2$ time} scale, respectively.

Let's consider the energy relaxation time, given by $T_1$, first.
The two states of the physical system encoding the qubit often have different energies.
All physical systems have the tendency to try and seek the lowest energy state.
So a system initialized in the excited state will not stay there indefinitely, eventually it will lose energy and transition to a state of lower energy.
Typically, the lower-energy state is chosen to encode a \ket{0}, while the higher-energy state encodes a \ket{1}.
Of course that choice is by convention, not an immutable fact of physics.
The relaxation time $T_1$ measures at what time scales the qubit relaxes and undergoes the transition $\ket{1}\rightarrow\ket{0}$.
It tells the probability that a qubit initialized in \ket{1} is still in that state after time $t$,
\begin{equation}
    \operatorname{Prob}(\ket{1})=e^{-t / T_1}.
\end{equation}
The probability that after $T_1$ seconds we still find our state in \ket{1} is given by $\frac{1}{e}$.

The dephasing time $T_2$ tells us about the time scale of a different decoherence process, namely the loss of phase coherence in the qubit.
Unlike the relaxation process, the qubit does not necessarily lose energy, but its phase between the states \ket{0} and \ket{1} becomes uncertain over time.
The $T_2$ time tells us how quickly superpositions are washed out and the qubit decoheres and loses its wonderful quantum properties.
If we start in an equal superposition of \ket{0} and \ket{1} (the \ket{+} state), the dephasing process spreads the Bloch vector corresponding to \ket{+} around the equator of the Bloch sphere.
The fully dephased state of the qubit is then an equal mixture,
\begin{equation}
    \rho = \frac{1}{2} ( \ketbra{+}{+}+\ketbra{-}{-} ) = \frac{1}{2} ( \ketbra{0}{0}+\ketbra{1}{1} ) = \frac{I}{2}.
\end{equation}

In Sec.~\ref{sec:3-3_density_matrices}, we saw the crucial difference between complete mixtures and equal superpositions. Here, if we prepare the state in the pure state, after some time $t$ we will have the mixed state,
\begin{equation}
    \rho = P |+\rangle\langle+| + (1-P)\frac{I}{2} \quad \text{, where }  P=e^{-t / T_2}.
\end{equation}
With probability $P$, the state of the qubit will still be the initial \ket{+}, and with probability $1-P$, it will have decohered into a completely mixed state.

Both of these processes, the relaxation process and the dephasing process, are \textbf{\emph{Poisson processes}}\index{Poisson process}.
It is a little bit ironic since we are talking about memories, but these processes are also called \textbf{\emph{memoryless decay processes}}\index{memoryless decay process}, meaning that their past history does not matter, only their current state.

Now that we have talked about the lifetimes of memories, why they are important, and given some characteristic time scales for long-distance communication, let's address the question of how atoms can be entangled with photons at the physical level.

\begin{figure}[t]
    \centering
    \includegraphics[width=0.7\textwidth]{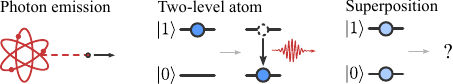}
    \caption[First idea for memory.]{Our first idea for a memory qubit is to use two energy levels of a system, such as ground and excited states of an atom.}
    \label{fig:13-memory-first}
\end{figure}

First we should consider what states of the atom represent the qubit basis states \ket{0} and \ket{1} in the quantum memory?
Natural choice, though not the only one, for the state \ket{0} is to use the ground state of the atom, denoted by \ket{\textrm{g}}.
The other computational basis state of the qubit, \ket{1}, is then represented by one of the excited states of the atom, denoted by \ket{\textrm{e}}.
Fig.~\ref{fig:13-memory-first} shows a two-level atom where $\ket{0}\equiv\ket{\textrm{g}}$ and $\ket{1}\equiv\ket{\textrm{e}}$.

Next, what states of the photon can we use to represent the basis states of the flying qubit?
One possibility is to use the presence or absence of a photon as a qubit, giving the representation $\ket{0}\equiv\ket{\textrm{no photon}},\ket{1}\equiv\ket{\textrm{photon}}$.
We can see in the middle panel of Fig.~\ref{fig:13-memory-first} that if the atom decays from the excited state into its ground state, it emits a photon.
The total state of the quantum memory and the flying photon undergoes the following transition,
\begin{equation}
    \ket{\textrm{e}}\ket{\textrm{no photon}} \rightarrow \ket{\textrm{g}}\ket{\textrm{photon}}.
    \label{eq:memory-photon-emit}
\end{equation}
The summary of the representations for the quantum meory and the flying qubits is shown in Tab.~\ref{tab:physical_representation}.
\begin{table}
    \setcellgapes{3pt}
    \renewcommand\theadfont{}
    \makegapedcells
    \centering
    \begin{tabular}{ccc}
    \hline
    & \textbf{quantum memory} & \textbf{flying qubit} \\
    \hline
    \textbf{physical system} & two-level atoms & emitted photons \\
    \boldmath\ket{0} & \ket{\textrm{g}} & \ket{\textrm{no photon}} \\
    \boldmath\ket{1} & \ket{\textrm{e}} & \ket{\textrm{photon}}\\
    \hline
    \end{tabular}
    \caption{Physical representation for the quantum memory and the flying qubits.}
    \label{tab:physical_representation}
\end{table}

We may now ask whether the emitted photon is entangled with the memory.
We saw in Eq.~\ref{eq:memory-photon-emit} that if the atom is prepared in the excited state \ket{\textrm{e}}, the memory-photon total state is not entangled.
We may try a different approach of initializing the atom in a superposition state $(\ket{\textrm{g}}+\ket{\textrm{e}})/\sqrt{2}$.
Let's write out the total memory-photon state step-by-step,
\begin{align}
    \frac{1}{\sqrt{2}}\left( \ket{\textrm{g}}+\ket{\textrm{e}} \right) \ket{\textrm{no photon}} & = \frac{1}{\sqrt{2}} \left( \ket{\textrm{g}}\ket{\textrm{no photon}} + \ket{\textrm{e}}\ket{\textrm{no photon}} \right) \nonumber\\
    & \rightarrow \frac{1}{\sqrt{2}} \left( \ket{\textrm{g}}\ket{\textrm{no photon}} + \ket{\textrm{g}}\ket{\textrm{photon}} \right) \nonumber\\
    & = \frac{1}{\sqrt{2}} \ket{\textrm{g}} \left( \ket{\textrm{no photon}} + \ket{\textrm{photon}} \right)
    \label{eq:memory-photon-superpostion}
\end{align}
We see that the atom has a 50\% probability to be found in the ground state initially, where it cannot emit any energy.
In that case, our photonic qubit will be in the \textbf{\emph{no photon}} state.
The atom also has a 50\% probability of being in the excited state, from where it can emit a photon.
The final state of the photon is a superposition of both being and not being emitted.
The initial superposition of the atom has been transferred to the photon.
However, we can clearly see from Eq.~(\ref{eq:memory-photon-superpostion}) that the memory is entangled with the flying photon.

\begin{figure}[t]
    \centering
    \includegraphics[width=0.3\textwidth]{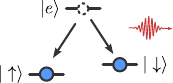}
    \caption[Our second idea for memory.]{Our second idea for a memory qubit is to use the ground-state spin of an atom. This allows us to use a polarized photon as a flying qubit robust against loss.}
    \label{fig:13-memory-second-idea}
\end{figure}

This is a naive picture that demonstrates some of the basic principles of how stationary and flying qubits interact.
However, it suffers from a few important shortcomings.
The first one being the robustness of the encoding for the quantum memory.
Using the excited state \ket{\textrm{e}} to represent one of the computational basis states makes the memory susceptible to energy relaxation, characterized by the $T_1$ time.
Depending on the rate of energy relaxation and round trip times required, the state stored in the quantum memory might decohere and become useless.
Second, our choice of encoding for the flying qubit is not suitable due to the attenuation of light in fiber. 
If we are waiting for a message at the end of the fiber and we do not receive a photon, we cannot be sure if the original message was really \ket{\textrm{no photon}} (\ket{0}), or if the initial message was \ket{\textrm{photon}} (\ket{1}) and the photon just got lost along the way.
To fix these two problems, we will now introduce a slightly more complex version of the atom for the quantum memory and a less ambiguous encoding for the flying qubit.

Figure~\ref{fig:13-memory-second-idea} shows the new atomic structure for quantum memory.
The ground state is now degenerate~\footnote{Meaning ``having the same energy''; no moral weakness implied.}, and spanned by two orthogonal states.
The two ground states are distinguished by their spin.
One is spin\index{spin} up \ket{\uparrow}, and the other is spin down \ket{\downarrow}.
Suitable encoding for the flying qubit is the linear polarization of the photon.
State \ket{0} is represented by vertical polarization \ket{V}, and \ket{1} is represented by horizontal polarization \ket{H} as we saw at the beginning of this chapter.
This new representation is summarized in Tab.~\ref{tab:physical_representation_updated}.
\begin{table}
    \setcellgapes{3pt}
    \renewcommand\theadfont{}
    \makegapedcells
    \centering
    \begin{tabular}{ccc}
    \hline
    & \textbf{quantum memory} & \textbf{flying qubit} \\
    \hline
    \textbf{physical system} & ground state spin & linear polarization \\
    \boldmath\ket{0} & \ket{\uparrow} & \ket{V} \\
    \boldmath\ket{1} & \ket{\downarrow} & \ket{H}\\
    \hline
    \end{tabular}
    \caption{New physical representation for the quantum memory and the flying qubits.}
    \label{tab:physical_representation_updated}
\end{table}

Let's see how this representation leads to an entangled memory-photon pair.
We start by preparing the atom in the excited state \ket{\textrm{e}}.
The atom has a 50-50 probability of decaying either to the spin up \ket{\uparrow} or the spin down \ket{\downarrow} ground state.
If the atom decays into the \ket{\uparrow} state, the emitted photon will be polarized in the \ket{V} state. 
If the atom decays into the spin \ket{\downarrow} state, the photon will be polarized in the \ket{H} state.
Since we do not know the polarization of the emitted photon until we measure it, the total memory-photon state is an equal superposition of the two possibilities.
We can write this transformation as
\begin{equation}
    \ket{\textrm{e}} \rightarrow \frac{1}{\sqrt{2}} \left( \ket{\uparrow}\ket{V}+\ket{\downarrow}\ket{H} \right).
\end{equation}
This is clearly an entangled state of the quantum memory and the flying qubit required by most of the link-level architectures we studied in Sec.~\ref{sec:12-2_making_link_level_rantanglement}.

\begin{figure}[t]
    \centering
    \includegraphics[width=0.7\textwidth]{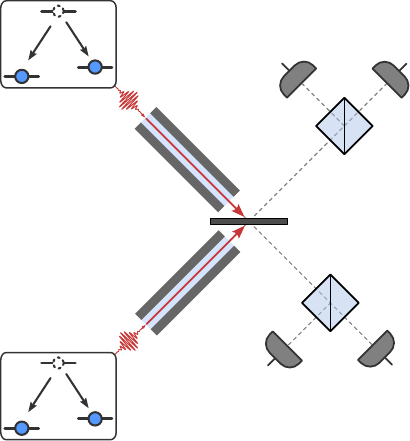}
    \caption[Physical implementation of the MIM link.]{A different view of an MIM link. On the left are the energy level diagrams for the qubits, which decay into a superposition of two states while emitting photons entangled with the memories. On the right is a large view of the Bell state analyzer with three beam splitters and four photon detectors.}
    \label{fig:13-MIM-energy}
\end{figure}

Fig.~\ref{fig:13-MIM-energy} brings this all back to a concrete representation of our Bell-state analyzer.
So far, we have discussed the BSA only abstractly but now we have a much better idea how to make it work in practice.
In the middle are the two single-mode fibers.
On the left are the two quantum memories at our two repeater nodes, separated by some distance, represented by their energy level diagrams.
Each memory is prepared initially in the excited state \ket{\textrm{e}}, which decays into one of its ground states, either spin up \ket{\uparrow} or spin down \ket{\downarrow}.
We do not know which state it decays into, leaving us with a flying photon that is entangled with its respective memory.
These flying photons travel through the single mode fibers, then hit the central beam splitter, where (if all goes well) they interfere and we perform a Bell-state measurement using the setup we discussed in Sec.~\ref{sec:bell-state-measurements-II}.
In this way, we can establish link-level entanglement between the atomic memories sitting at the ends of the link.
This concludes our discussion of the physical implementation of the MIM link architecture.
In fact out discussion applies to the MM link architecture as well, we just need to place the BSA inside one of the nodes.
Addition of a source of entangled photon pairs, such as the one we discussed in Sec~\ref{sec:4-4_spdc} would cover also the physical implementation of the MSM link architecture.

\newpage
\begin{exercises}
\exer{
\emph{Bell-state measurement revisited.}
Let's explore the quantum circuit for a Bell-state measurement in Fig.~\ref{fig:13-2_measurementBell} a bit more detail.
\subexer{
Consider the qubits to be initialized in \ket{11} state.
What are the probabilities of the four measurement outcomes?
}
\subexer{
How do the probabilities change when the input state is \ket{\Phi^+}?
}
\subexer{
What input state should be used if the desired measurement outcome is $c_1=1$ and $c_2=1$?
}
}

\exer{
\emph{Measurement in the graph-state basis.}
We have seen how to describe Bell-state measurements using the language of quantum circuits in Sec.~\ref{sec:bell-state-measurements-I}.
In this exercise, we will explore a two-qubit measurement in a different but related basis.
\subexer{
Two-qubit graph state is an entangled state that is related to the Bell states,
\begin{equation}
    \ket{G_{00}} = I \otimes H \ket{\Phi^+}.
    \label{eq:graph-state-G00}
\end{equation}
Write down the graph state \ket{G_{00}} in Dirac notation.
}
\subexer{
Just like there are four Bell states, there are four orthogonal graph states,
\begin{equation}
    \ket{G_{mn}} = Z^m \otimes Z^n \ket{G_{00}}.
\end{equation}
Write down all four of these states in Dirac notation, and verify they are orthogonal.
}
\subexer{
Draw the quantum circuit that prepares the state \ket{G_{00}}, given the input qubits are both initialized in the \ket{0} state.
}
\subexer{
Draw the quantum circuit representation for a measurement in the graph state basis.
}
}

\exer{
\emph{Round trip time and decohering memories.}
Consider the MIM link architecture with memories that are affected by depolarizing noise with $T_1=100\mu$s.
\subexer{
Assume that the state of the memory-photon immediately after the photon emission is
\begin{equation}
    \ket{\Psi}_{\textrm{mem-phot}} = \frac{1}{\sqrt{2}} \left( \ket{\uparrow}\ket{V} + \ket{\downarrow}\ket{H} \right).
\end{equation}
What is the fidelity of this state after time $t$ that the photon has spent in flight? 
}
\subexer{
Assume that the distance of the BSA to either of the memories is $d=25$ km, and that the BSA succeeds without introducing any noise.
What is the fidelity of the Bell pair shared between the quantum memories immediately after the BSA completes?
}
\subexer{
The BSA communicates the success of entanglement swapping to the nodes.
What is the fidelity of the Bell pair after the nodes receive the classical BSA message.
}
\subexer{
Consider using a better quantum memory with $T_1=10$ ms.
What is the fidelity of the memory-memory Bell pair after receiving the BSA message?
}
}

\end{exercises}

\newpage
\section*{Quiz}
  \addcontentsline{toc}{section}{Quiz}
A quiz for this block of chapters can be found in the online system.


\section*{Further reading chapters 11-13}
  \addcontentsline{toc}{section}{Further reading chapters 11-13}

{\bf Chapter 11}

Our discussion of mode dispersion closely followed Section 5.6 of Hecht’s textbook and we encourage you to read it for the extra details that can be found in the book.

A qualitative review of classical amplifiers (with just the right amount of technical detail) can be found here:

Emmanuel Desurvire, The Golden Age of Optical Fiber Amplifiers, \emph{Physics Today} 47, 20 (1994)~\cite{desurvire1994golden}.

Unfortunately, this article is behind a paywall so you will have to use your university’s online system to access it.

{\bf Chapter 12}

Quantum repeaters are the "bread and butter" of quantum networks. The names Wolfgang D\"ur and Hans Briegel cannot be mentioned often enough in the history of repeaters. Those interested in the paper that introduced the idea of a quantum repeater (and are not scared off by maths) might have a look here:

Hans J. Briegel, Wolfgang Dür, Juan I. Cirac, Peter Zoller, Quantum repeaters: The role of imperfect local operations in quantum communication, \emph{Physical Review Letters} 81, 5932, (1998)~\cite{briegel98:_quant_repeater}.

The paper is behind a paywall and needs to be accessed through your university’s library online services.

One place to learn more is Prof. Van Meter's previous book,\\
Rodney Van Meter, \emph{Quantum Networking}, Wiley-ISTE, 2014~\cite{van-meter14:_quantum_networking}.

{\bf Chapter 13}

A great popular article about the physical layer components of quantum networks can be found here:
Dan Hurley, The quantum internet will blow your mind. Here’s what it will look like, \emph{Discover Magazine}, 2020~\cite{hurley2020quantum}.

Another fantastic review of physical layer components can be found here:

Nicolas Sangouard, Christoph Simon, Hugues de Riedmatten, Nicolas Gisin, Quantum repeaters based on atomic ensembles and linear optics, \emph{Review of Modern Physics} 83, 33, 2011~\cite{sangouard2011quantum}.

Again, the published version is behind a paywall. Be warned though, this paper starts with an excellent introduction but the technical details ramp up quickly after that and rely on good grasp of quantum optics. So if you get lost after the introduction, don’t worry. You can come back to those parts later.

\part{Quantum Repeater Systems}
\begin{partintro}
\partintrotitle{Introduction to the fifth lesson block}
In the next two Chapters, we reach the culmination of our work: learning how to integrate the services provided by large-scale quantum repeater networks into complete classical sensing, computating and communication platforms, and how we can build and operate a true Quantum Internet out of a large set of independent, autonomous, and often heterogeneous individual networks.
\end{partintro}

\chapter{Entanglement Revisited}

In this chapter, we will return to entanglement and discuss it in the context of quantum repeater networks.
We begin with revisiting bipartite entanglement in section~\ref{sec:14-1_bipartite}.
We focus on how to determine the quality of the entanglement shared between distant nodes of a network and why it matters.
In section~\ref{sec:14-2_multipartite}, we will introduce multipartite entanglement shared between more than two nodes of a quantum network.
We will discuss how it differs from bipartite entanglement before moving on to examples of multipartite entangled states.
In the remainder of the chapter, we will shift our focus to applications of quantum repeater networks.
In section \ref{sec:14-3_clock_sync}, we discuss clock synchronization, and distributed blind quantum computation in section ~\ref{sec:14-4_distributed_bqc}.

\section{Bipartite entanglement}
\label{sec:14-1_bipartite}

\begin{figure}[t]
    \centering
    \includegraphics[width=\textwidth]{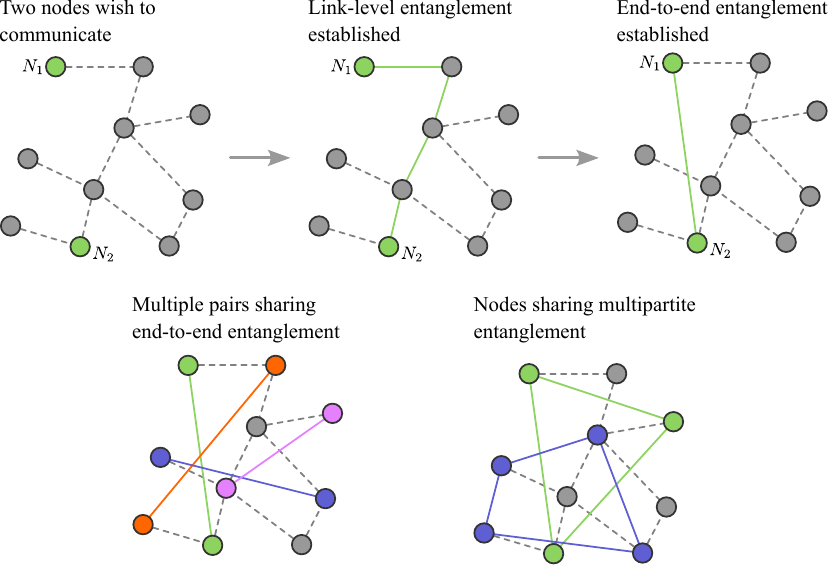}
    \caption[Bipartite and multipartite entanglement.]{Bipartite and multipartite entanglement distribution in quantum networks.}
    \label{fig:14-1_bipartite_multipartite}
\end{figure}

One of the main jobs of a quantum network is to distribute entanglement.
Let's start by considering a concrete example.
Figure~\ref{fig:14-1_bipartite_multipartite} shows a quantum network.
The circles represent quantum nodes while the dashed lines represent physical links connecting neighboring quantum nodes.
Nodes $N_1$ and $N_2$ wish to establish shared entanglement in order to engage in quantum communication.
We saw in the previous chapter that in order to satisfy this request, the quantum network first creates link-level entanglement between neighboring nodes along the path connecting the nodes $N_1$ and $N_2$.
Entanglement swapping is then used to create direct end-to-end bipartite entanglement between these nodes.
This example is rather constrained in the sense that it only shows two of the nodes trying to engage in quantum communication.
A well-designed quantum network should be able to accommodate multiple simultaneous requests for end-to-end entangled connections between multiple pairs of nodes, as shown in Fig.~\ref{fig:14-1_bipartite_multipartite}. 

\begin{figure}[t]
    \centering
    \includegraphics[width=0.4\textwidth]{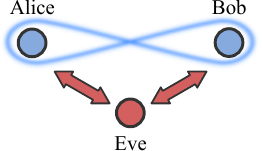}
    \caption[Eavesdropper Eve.]{Eavesdropper Eve tampering with the distributed entangled state between Alice and Bob in order to gain information about the shared secret key.}
    \label{fig:14-1_QKD}
\end{figure}

The quality of the distributed entanglement, whether bipartite or multipartite, matters greatly.
We saw that in entanglement-based QKD, the quality of the entanglement that is shared between Alice and Bob directly impacts the key that they are trying to establish.
If the entangled state shared by Alice and Bob is not perfect then the obtained key will also be only partially correlated.
This has two important consequences.
First, it affects how well Alice and Bob can communicate.
More crucially, the quality of the entangled state affects the \emph{\textbf{security of the key}} that they can establish.
The entangled state may be imperfect due to natural noise introduced by the physical channels during the distribution process.
But it could also be a result of an eavesdropper Eve, trying to listen in on the secret communication between Alice and Bob as shown in Fig.~\ref{fig:14-1_QKD}.
If Alice and Bob want to maintain security of the information that they are wishing to communicate, they must assume that any imperfections in the entangled state have been introduced by Eve.
So the quality of the entangled state places an upper bound on the security of the key that can be established between Alice and Bob in the sense that \emph{\textbf{better quality of the entangled state implies stronger security}}.

We discussed in Section~\ref{sec:3-5_fidelity} that one way of quantifying the quality of an entangled state is the fidelity.
Given an ideal state \ket{\psi} that the quantum network is trying to distribute, the fidelity of the actual mixed state $\rho$ shared between Alice and Bob is given by the expectation value of $\rho$ with respect to \ket{\psi},
\begin{equation}
    F \left( \rho, \ket{\psi} \right) = \bra{\psi}\rho\ket{\psi}.
\end{equation}
It is usually the case in quantum networks that the ideal state is one of the Bell pairs.
If the ideal state is \ket{\Phi^+}, then the fidelity is $F (\rho, \ket{\Phi^+}) = \bra{\Phi^+}\rho\ket{\Phi^+}$.
For noiseless quantum channels and operations, perfect quantum memories and in the absence of an eavesdropper, the distributed state will be the ideal state, $\rho = \ket{\Phi^+}\bra{\Phi^+}$.
The resulting fidelity is
\begin{equation}
    F \left( \rho, \ket{\Phi^+} \right) = \langle\Phi^+\ket{\Phi^+} \langle\Phi^+\ket{\Phi^+} = 1.
\end{equation}
On the other hand, the distributed state may have some error, for example a Pauli $Z$ error on Bob's qubit.
This transforms the entangled state into an orthogonal Bell pair, $I \otimes Z \ket{\Phi^+} = \ket{\Phi^-}$, resulting in fidelity of
\begin{equation}
    F \left( \rho, \ket{\Phi^+} \right) = \langle\Phi^+\ket{\Phi^-} \langle\Phi^-\ket{\Phi^+} = 0.
    \label{eq:14-1_fidelity_0}
\end{equation}
Typically, the distributed state will be mixed, and the fidelity will take values between $ 0 \leq F(\rho, \ket{\Phi^+}) \leq 1$.

Note that the lowest possible value of the fidelity does not imply that the distributed state is useless -- far from it.
As we saw in Eq.~(\ref{eq:14-1_fidelity_0}), the fidelity vanishes when the distributed state is orthogonal to the ideal state \ket{\Phi^+}.
This means that the distributed state is also maximally entangled and can be used for quantum communication after an appropriate transformation.
For the case discussed in Eq.~(\ref{eq:14-1_fidelity_0}), the distributed state can be transformed into the ideal one by either Alice or Bob applying a local Pauli $Z$ operation,
\begin{align}
    \text{Alice applies } Z: \quad & Z \otimes I \ket{\Phi^-} = Z \otimes Z \ket{\Phi^+} = \ket{\Phi^+}, \\
    \text{Bob applies } Z: \quad & I \otimes Z \ket{\Phi^-} = I \otimes Z^2 \ket{\Phi^+} = \ket{\Phi^+}.
\end{align}
We used the fact that applying the Pauli $Z$ operation twice cancels its effect, $Z^2 = I$.

We saw that both extremes of the fidelity, $F(\rho,\ket{\psi}) = 1$ and $F(\rho,\ket{\psi}) = 0$, suggest that the distributed state is useful for quantum communication.
Does this mean that distributed states with intermediate values of fidelity are useful too?
Certainly not.
Let's consider a fully decohered state of two qubits given by the maximally mixed state,
\begin{equation}
    \rho = \frac{1}{4} \left( \ket{00}\bra{00} + \ket{01}\bra{01} + \ket{10}\bra{10} + \ket{11}\bra{11} \right).
\end{equation}
The fidelity of this state with respect to the ideal Bell pair \ket{\Phi^+} is
\begin{align}
    F \left( \rho, \ket{\Phi^+} \right) & = \frac{1}{4} \bra{\Phi^+} \left( \ket{00}\bra{00} + \ket{01}\bra{01} + \ket{10}\bra{10} + \ket{11}\bra{11} \right) \ket{\Phi^+} \nonumber\\
    & = \frac{1}{4} \left( \frac{1}{2} + \frac{1}{2} \right) = \frac{1}{4}.
\end{align}
Fidelity value of $F=0.25$ therefore signifies the distributed state is ``completely useless''.

Fidelity is not the only useful metric when it comes to quantifying the quality of the distributed state.
The other method relies on the CHSH inequality which we saw plays a crucial role in the E-91 QKD protocol in Chapter~\ref{sec:10_E91}.
Named after its inventors Clauser, Horne, Shimony, and Holt, the CHSH inequality is a test of the quantumness of the correlations shared between Alice and Bob.
We saw that to test that Alice and Bob share the \ket{\Psi^+} Bell pair, they have to repeatedly measure in the following bases,
\begin{align}
    \text{Alice's measurement bases:}& & A_1 & = Z, & A_2 & = X \\
    \text{Bob's measurement bases:}& & B_1 & = \frac{Z - X}{\sqrt{2}}, & B_2 & = \frac{Z + X}{\sqrt{2}}.
\end{align}
Using the outcomes of the measurements, they can then compute the CHSH correlation,
\begin{equation}
    \mathcal{S} = \langle A_1 \otimes B_1\rangle + \langle A_1 \otimes B_2\rangle + \langle A_2 \otimes B_1\rangle - \langle A_2 \otimes B_2\rangle.
    \label{eq:14-1_CHSH_Psi_plus}
\end{equation}
For the ideal state, the expectation values can be computed according to $\langle A_i B_j \rangle = \bra{\Psi^+} A_i \otimes B_j \ket{\Psi^+}$, leading to the CHSH correlation of
\begin{equation}
    \left| \mathcal{S} \right| = 2\sqrt{2}.
\end{equation}
For the distributed state given by a mixed state $\rho$, the expectation values can be calculated via the more general expression $\langle A_i \otimes B_j \rangle = \text{Tr} \{ A_i \otimes B_j \rho \}$.
The distributed state is entangled provided that
\begin{equation}
    \left| \mathcal{S} \right| > 2.
\end{equation}
The CHSH inequality acts as a witness of entanglement, with $|\mathcal{S}| = 2$ being the critical threshold.
When Alice and Bob measure $|\mathcal{S}| > 2$, they know they share an entangled state.
On the other hand, when $|\mathcal{S}| < 2$, they cannot conclude anything. The distributed state could be entangled but it could be separable as well.
CHSH inequality violation is also directly linked to the security of the key that Alice and Bob can establish.
\textit{\textbf{Stronger violation of the CHSH inequality leads to more secure secret key.}}

What if the ideal state that Alice and Bob are trying to share is some other Bell pair, let's say \ket{\Phi^+}?
Computing the CHSH violation in Eq.~(\ref{eq:14-1_CHSH_Psi_plus}) but computing the expectation values with respect to the new ideal state \ket{\Phi^+}, we get
\begin{align}
    \langle A_1 \otimes B_1 \rangle & = \frac{1}{\sqrt{2}} & \langle A_1 \otimes B_2 \rangle & = \frac{1}{\sqrt{2}} \\
    \langle A_2 \otimes B_1 \rangle & = - \frac{1}{\sqrt{2}} & \langle A_2 \otimes B_2 \rangle & = \frac{1}{\sqrt{2}}.
\end{align}
This leads to a vanishing CHSH correlation, $\mathcal{S} = 0$.
This is because the expression for the correlation in Eq.~(\ref{eq:14-1_CHSH_Psi_plus}) is not designed to witness entanglement when the ideal state is \ket{\Phi^+}.
We have encountered a similar situation when we computed the fidelity between two orthogonal Bell pairs.
We can use a similar expression to Eq.~(\ref{eq:14-1_CHSH_Psi_plus}) that witnesses entanglement when the ideal state is \ket{\Phi^+}.
All we need to do is exchange the plus and minus signs in front of the last two terms,
\begin{equation}
    \mathcal{S} = \langle A_1 \otimes B_1\rangle + \langle A_1 \otimes B_2\rangle - \langle A_2 \otimes B_1\rangle + \langle A_2 \otimes B_2\rangle.
    \label{eq:14-1_CHSH_Phi_plus}
\end{equation}
Eq.~(\ref{eq:14-1_CHSH_Phi_plus}) yields $|\mathcal{S}| = 2\sqrt{2}$ when the expectation values are computed with respect to \ket{\Phi^+}.

We have seen that both the fidelity as well as the CHSH inequality offer suitable tools to quantify the quality of the distributed bipartite state.
We have to remember to take care when computing both quantities.
In order to obtain meaningful results, we need to use the correct ideal state in the case of fidelity, and the appropriate form of the correlation in the case of the CHSH inequality.

\section{Multipartite entanglement}
\label{sec:14-2_multipartite}

We have seen how quantum networks can establish bipartite entanglement between two of its nodes.
We may wish to distribute entanglement shared between more than two nodes as shown in Fig.~\ref{fig:14-1_bipartite_multipartite}.
Sometimes three clients may wish to engage in a quantum communication protocol that might require all of them to share a tripartite entangled state.
Any entangled state of more than two subsystems is called an \textit{\textbf{entangled multipartite state}}\index{multipartite entanglement}.
It is the job a quantum network to distribute multipartite entangled states between any disjoint subsets of nodes that ask for them.

We begin by considering entangled states of three qubits.
We saw that in the case of two qubits, there are four possible computational basis states, \ket{00}, \ket{01}, \ket{10}, and \ket{11}.
How many computational basis states are there in this case?
Writing all of them out,
\begin{align}
    \ket{000}, \; \ket{001}, \; \ket{010}, \; \ket{011}, \; \ket{100}, \; \ket{101}, \; \ket{110}, \; \ket{111},
\end{align}
we see there are eight different basis states.
Any general three-qubit state can be expanded as a superposition of these computational basis states,
\begin{align}
    \ket{\psi} & = c_0\ket{000} + c_1\ket{001} + c_2\ket{010} + c_3\ket{011} \nonumber\\
    & + c_4\ket{100} + c_5\ket{101} + c_6\ket{110} + c_7\ket{111}.
\end{align}
This superposition is often written in a compact form as
\begin{equation}
    \ket{\psi} = \sum_{i=0}^7 c_i \ket{i},
\end{equation}
where the notation means that the ``$i$'' in \ket{i} is the binary number corresponding to the index in the coefficients $c_i$.
Depending on the values of the coefficients, we obtain different states, some of which will be entangled.

Let's look at a few examples of important entangled multipartite states.
The first state is one of the simplest but also one of the most important.
It is an equal superposition of \ket{000} and \ket{111}, and is known as the \emph{\textbf{GHZ state}}\index{GHZ state}, named after Greenberger, Horne, and Zeilinger,
\begin{equation}
    \ket{GHZ} = \frac{1}{\sqrt{2}} \left( \ket{000} + \ket{111} \right).
\end{equation}
Let's see some of the properties of the GHZ state.
A good start is to measure one of the qubits and see what the state of the remaining qubits is.
Consider measuring the first qubit in the Pauli $Z$ basis.
The projectors corresponding to the +1 and -1 outcomes are
\begin{align}
    \Pi^Z_+ & = \ket{0}\bra{0} \otimes I \otimes I, \\
    \Pi^Z_- & = \ket{1}\bra{1} \otimes I \otimes I,
\end{align}
respectively.
The probability of the measurement outcome being +1 is
\begin{align}
    \text{Prob} \{+1\} & = \text{Tr} \{ \Pi^Z_+ \ket{GHZ}\bra{GHZ} \} \nonumber\\
    & = \bra{GHZ} \left( \ket{0}\bra{0} \otimes I \otimes I \right) \ket{GHZ} \nonumber\\
    & = \frac{1}{2} \left( \bra{000} + \bra{111} \right) \left( \ket{0}\bra{0} \otimes I \otimes I \right) \left( \ket{000} + \ket{111} \right) \nonumber\\
     & = \frac{1}{2} \bra{000} \left( \ket{0}\bra{0} \otimes I \otimes I \right) \ket{000} \nonumber\\
     & = \frac{1}{2}.
\end{align}
Similarly, the probability of the measurement outcome being -1 is
\begin{equation}
    \text{Prob}\{-1\} = \text{Tr}\{\Pi^Z_-\ket{GHZ}\bra{GHZ}\} = \frac{1}{2}.
\end{equation}
If the measurement outcome is +1, the post-measurement state is
\begin{align}
    \frac{1}{\sqrt{\text{Prob}\{+1\}}} \Pi^Z_+ \ket{GHZ} & = \sqrt{2} \left( \ket{0}\bra{0} \otimes I \otimes I \right) \frac{1}{\sqrt{2}} \left( \ket{000} + \ket{111} \right) \nonumber\\
    & = \left( \ket{0}\bra{0} \otimes I \otimes I \right) \ket{000} \nonumber\\
    & = \ket{000}.
\end{align}
We see that the post-measurement state is correlated, all qubits are in the same state, but it is no longer entangled.
Measurement of a single qubit in the Pauli $Z$ basis is enough to destroy the entanglement shared between all three qubits.
This is also the case if the measurement outcome is -1,
\begin{equation}
    \frac{1}{\sqrt{\text{Prob}\{-1\}}} \Pi^Z_- \ket{GHZ} = \ket{111}.
\end{equation}

The next example of an entangled state of three qubits is called the \emph{\textbf{W state}}\index{W state}.
In the computational basis, it is a superposition of three terms,
\begin{equation}
    \ket{W} = \frac{1}{\sqrt{3}} \left( \ket{001} + \ket{010} + \ket{100} \right).
\end{equation}
Measuring the first qubit in the computational basis, the probabilities of the two possible outcomes are
\begin{align}
    \text{Prob}\{+1\} & = \text{Tr} \{ \Pi^Z_+ \ket{W}\bra{W} \} = \frac{2}{3}, \\
    \text{Prob}\{-1\} & = \text{Tr} \{ \Pi^Z_- \ket{W}\bra{W} \} = \frac{1}{3}.
\end{align}
This time the probabilities of the two outcomes are different from each other, unlike in the case of the GHZ state.
What is more interesting, though, are the post-measurement states corresponding to the outcomes of the measurement,
\begin{align}
    \text{outcome +1:} & \quad \ket{0}\ket{\Psi^+}, \label{eq:14-2_W_post_meas_0} \\
    \text{outcome -1:} & \quad \ket{1}\ket{00},
\end{align}
We see that when the outcome of the measurement is +1, the state of the remaining two qubits is a maximally entangled Bell pair.
This is in stark contrast to the GHZ state, where a single measurement destroys all entanglement regardless of the outcome.
This points at a fundamental difference between the GHZ state and the W state in terms of their entanglement properties.

This difference goes even deeper.
We can consider what happens to the entanglement if one of the qubits is lost.
Losing a single qubit of a GHZ state will completely destroy the entanglement.
On the other hand, a loss of a single qubit from a W state leaves the remaining two qubits entangled, albeit not maximally.
The W state is more robust to qubit loss than the GHZ state.

\begin{figure}[t]
    \centering
    \includegraphics[width=\textwidth]{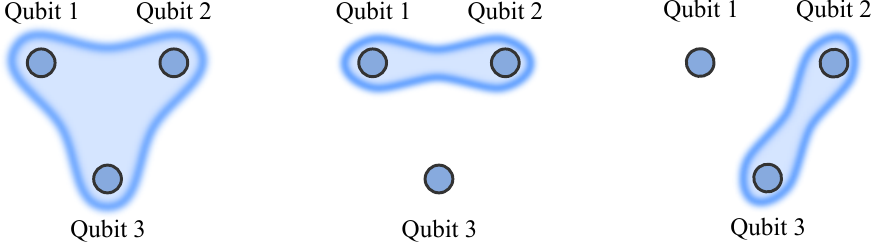}
    \caption[Monogamy of entanglement.]{Monogamy of entanglement restricts how three qubits can be entangled.}
    \label{fig:14-2_monogamy}
\end{figure}

Before moving onto $N$-partite entangled states, we will pause and think how the entanglement is shared between the three qubits.
There is a fundamental tradeoff to how strongly each pair of the qubits can be correlated, known as \emph{\textbf{monogamy of entanglement}}\index{monogamy of entanglement}.
It states that the strength of entanglement between a pair of qubits places an upper bound on the amount of entanglement that the third qubit can share with either of the first two qubits.
The left panel of Figure~\ref{fig:14-2_monogamy} shows a general entangled state of three qubits, where all three qubits are entangled with each other.
Examples corresponding to this situation are the GHZ and W states.
If a pair of qubits is in a maximally entangled state, for example \ket{\Phi+}, then monogamy of entanglement tell us that the third qubit must be in a separable state with the first pair.
This is portrayed by the middle panel in Fig.~\ref{fig:14-2_monogamy}.
Similarly, if qubits 2 and 3 are maximally entangled like in the right panel of Fig.~\ref{fig:14-2_monogamy}, then the first qubit must be completely disentangled with either of them.
An example of this situation is given by the post-measurement state of Eq.~(\ref{eq:14-2_W_post_meas_0}), where we measured the first qubit of a W state in Pauli $Z$ basis and obtained the +1 outcome.

Monogamy of entanglement is one of the most fundamental properties of quantum mechanics.
It is unlike anything that exists in classical physics, and is one of the cornerstones and building blocks which we use in quantum communication.
Monogamy of entanglement is instrumental in guaranteeing security in the E91 QKD protocol.
The more strongly Alice's and Bob's qubits are entangled, the less information an eavesdropper can learn about the secret key.

We can extend our discussion of GHZ and W states to $N$ qubits.
The $N$-qubit GHZ state has the following form,
\begin{equation}
    \ket{GHZ} = \frac{1}{\sqrt{2}} ( \underbrace{\ket{00\ldots0}}_{N\text{ zeroes}} + \underbrace{\ket{11\ldots1}}_{N \text{ ones}} ).
\end{equation}
Notice that the normalization factor has not changed, even though this is now a GHZ state of $N$ qubits.
The W state can be extended to arbitrary number of qubits as well,
\begin{equation}
    \ket{W} = \frac{1}{\sqrt{N}} ( \underbrace{\ket{00\ldots01} + \ket{00\ldots10} + \ldots \ket{10\ldots00}}_{N \text{ terms}} )
\end{equation}
Since the $N$-qubit W state is an equal superposition of $N$ terms, he normalization factor is $1 / \sqrt{N}$.

\begin{figure}[t]
    \centering
    \includegraphics[width=0.6\textwidth]{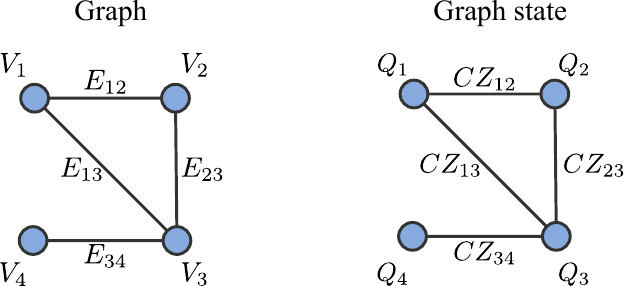}
    \caption[Graph state.]{Graph state of four qubits and its underlying graph.}
    \label{fig:14-2_graph_state}
\end{figure}

GHZ and W states are not the only important examples of multipartite entangled states.
A prominent example of $N$-partite entanglement is the \textbf{\textit{graph state}}\index{graph state}.
This state can be represented visually as a set of vertices $V$ which are connected by a set of edges $E$.
Such an object is known as a \textit{\textbf{graph}}\index{graph}\footnote{Not to be confused with a graph of a function.}.
Figure~\ref{fig:14-2_graph_state} shows an example of a graph with vertex set $V=\{V_1, V_2, V_3, V_4\}$, connected by edges from the edge set $E=\{ E_{12}, E_{23}. E_{13}, E_{34} \}$.
The corresponding graph state is obtained by placing a qubit at each vertex of the graph, initialized in $\ket{+} = (\ket{0} + \ket{1}) / \sqrt{2}$, and applying the \textbf{\emph{control-phase gate}}\index{control-phase gate} $CZ_{ij}$ between qubits whose vertices are connected by and edge $E_{ij}$.
The control-phase gate $CZ$ is a two-qubit gate defined as
\begin{equation}
    CZ_{ij} = \ket{0}\bra{0}_i \otimes I_j + \ket{1}\bra{1}_i \otimes Z_j = \begin{pmatrix}
        1 & 0 & 0 & 0 \\
        0 & 1 & 0 & 0 \\
        0 & 0 & 1 & 0 \\
        0 & 0 & 0 & -1
    \end{pmatrix}.
\end{equation}
\begin{figure}[t]
    \centering
    \includegraphics[width=\textwidth]{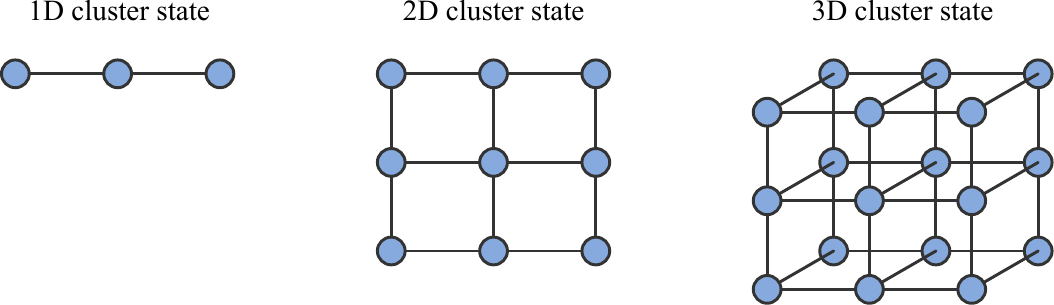}
    \caption[Cluster state.]{Cluster states are examples of graph states with regular structure.}
    \label{fig:14-2_cluster_state}
\end{figure}
The control-phase gate does nothing to the target qubit $j$ if the control qubit $i$ is in the \ket{0} state.
If the control $i$ is in the \ket{1} state, the control-phase gate applies Pauli $Z$ to the target qubit $j$.
Applying this rule we can find the state vector for the graph state \ket{G} in Fig.~\ref{fig:14-2_graph_state},
\begin{align}
    \ket{G} & = CZ_{12} CZ_{23} CZ_{13} CZ_{34} |++++\rangle_{1234} \nonumber\\
    & = \frac{1}{4} ( |0000\rangle + |0001\rangle + |0010\rangle - |0011\rangle \nonumber\\
    & \; \; \; \; + |0100\rangle + |0101\rangle - |0110\rangle + |0111\rangle \\
    & \; \; \; \; + |1000\rangle + |1001\rangle - |1010\rangle + |1011\rangle \\
    & \; \; \; \; - |1100\rangle - |1101\rangle - |1110\rangle + |1111\rangle \\
    & = \frac{1}{2} ( |0+0+\rangle + |0-1-\rangle + |1-0+\rangle - |1+1-\rangle ).
\end{align}
If the underlying graph has a regular structure, as in Fig.~\ref{fig:14-2_cluster_state} the resulting graph state is usually called a \textbf{\emph{cluster state}}\index{cluster states}.

Graph states and cluster states play an important role in quantum computation and quantum communication.
Cluster states are a resource state for a particular computational model known as ``measurement-based quantum computation''.
Graph states are useful in quantum error-correction as well as in many protocols in quantum communication.

\section{Clock synchronization}
\label{sec:14-3_clock_sync}

\begin{figure}[t]
    \centering
    \includegraphics[width=0.8\textwidth]{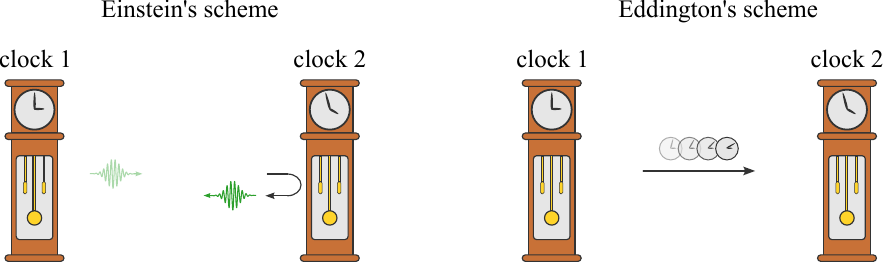}
    \caption[Clock synchronization.]{Clock synchronization via the use of photons or via slow transport of a small clock.}
    \label{fig:14-3_clock_sync}
\end{figure}

So far, we have discussed two applications of quantum networks; quantum teleportation in Chapter ~\ref{sec:8_teleportation} and quantum key distribution in Chapters~\ref{sec:9_bb84} and \ref{sec:10_E91}.
Before concluding this chapter, we will briefly discuss two other applications.
In this Section, we look at \textbf{\emph{clock synchronization}}\index{clock synchronization}.

Establishing a universal time standard is fundamentally important in many areas of modern life.
A telecommunication network requires accurate synchronization in order to deliver reliable operation of its multitude of components.
Synchronization plays an important role in the Global Positioning System (GPS), ensuring accurate position tracking.
Further examples of why clock synchronization is important include financial markets and transportation networks.

There are two main traditional methods of clock synchronization.
\textbf{\emph{Einstein's clock synchronization}}\index{Einstein's clock synchronization} relies on sending light signals between distant clocks.
Figure~\ref{fig:14-3_clock_sync} shows clocks 1 and 2 that we wish to synchronize.
Clock 1 fires a light pulse towards clock 2.
Once this pulse reaches clock 2, it is reflected back via a mirror and clock 2 is set ticking.
By measuring the time of emission as well as time of arrival of the light pulse, clock 1 can estimate when clock 2 started ticking.
This way, clock 1 can adjust its time in order to match that of clock 2.
Another method is know as \textbf{\emph{Eddington's clock synchronization}}\index{Eddington clock synchronization}.
Using this approach, we first locally synchronize clock 1 with another small clock.
We then transport this smaller clock very slowly to the location of clock 2, which subsequently locally synchronizes itself to it.

\begin{figure}[t]
    \centering
    \includegraphics[width=0.8\textwidth]{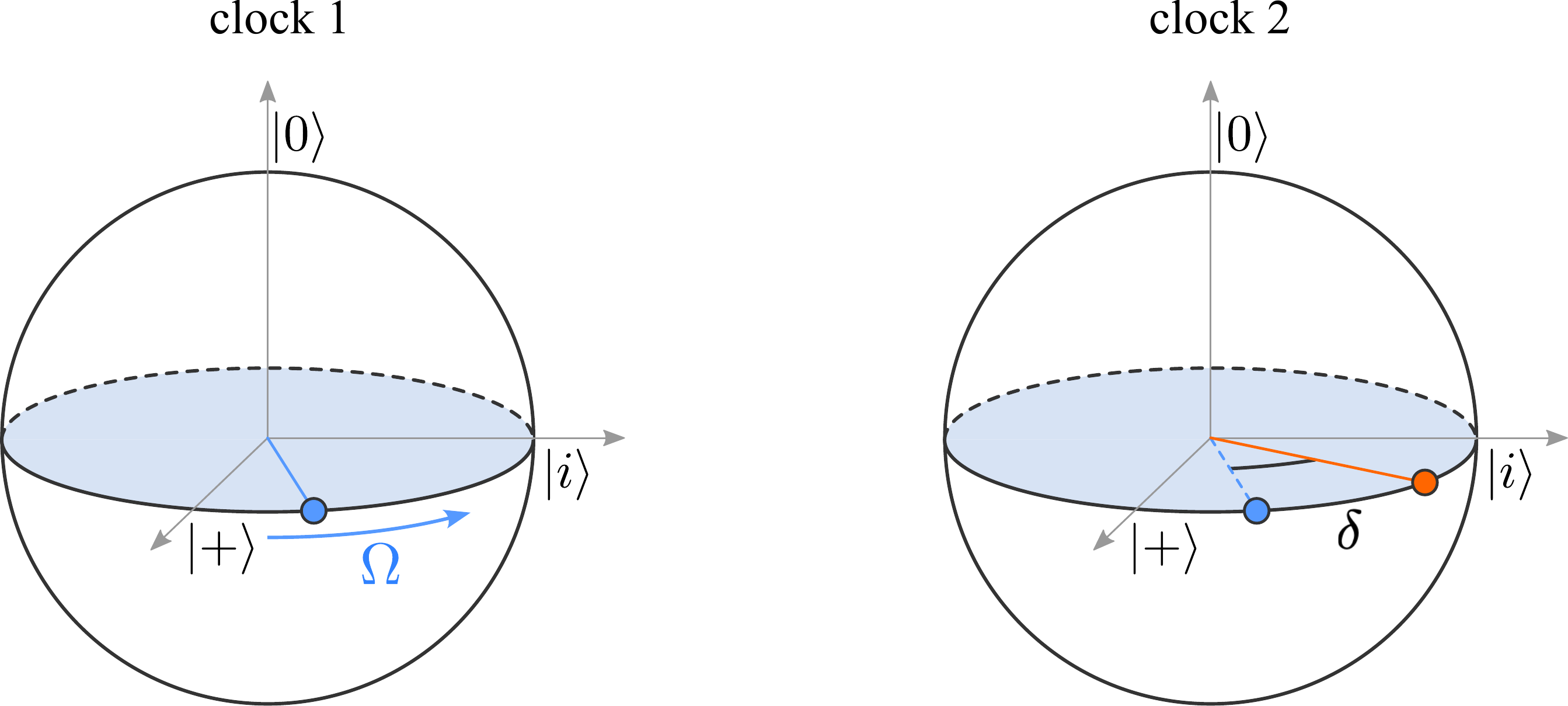}
    \caption[Qubit clocks.]{Two qubits acting as two small clocks.}
    \label{fig:14-3_bloch}
\end{figure}

Qubits can also act as small clocks.
Imagine preparing the state of a qubit in $|+\rangle = (|0\rangle + |1\rangle) / \sqrt{2}$.
By applying a suitable operation we can make the state of the qubit precess in the $x$-$y$ plane of the Bloch sphere at a constant angular velocity $\Omega$ as shown in the left panel of Fig.~\ref{fig:14-3_bloch}.
The state of the qubit at time $t$ will then be given by
\begin{equation}
    |\psi(t)\rangle = \frac{|0\rangle + e^{i\Omega t} |1\rangle}{\sqrt{2}}.
\end{equation}
The dynamics of this qubit has a period of $T = 2\pi / \Omega$, meaning that the state of the qubit will repeat itself after every period, $|\psi(t+T)\rangle = |\psi(t)\rangle$.

We can now consider a second qubit, also precessing at the angular frequency $\Omega$, but initialized at a slightly different state, given by the offset $\delta$, as shown in the right panel of Fig.~\ref{fig:14-3_bloch}.
Eliminating this phase lag is very similar to the traditional problem of clock synchronization.
The schemes of Einstein and Eddington can be applied in this scenario as well.
However, they both have the downside that they require direct transmission of light or matter.
Depending on the scenario, this may be impractical or outright impossible.
Thankfully, we can employ distributed entanglement between the two parties in order to synchronize their qubit clocks.
The precise steps of the protocol that achieves this are beyond the scope of this book but more details can be found in the Further Reading section.

Clock synchronization is another example where entanglement is used as a global resource.
The correlations of a Bell pair can be used to establish a single time frame shared between two distant locations.
Clock synchronization is not just another application of quantum networks but often is a hidden assumption behind many of the applications of quantum technologies, some of which we have discussed already.

\section{Distributed and blind quantum computation}
\label{sec:14-4_distributed_bqc}

\begin{figure}[t]
    \centering
    \includegraphics[width=0.4\textwidth]{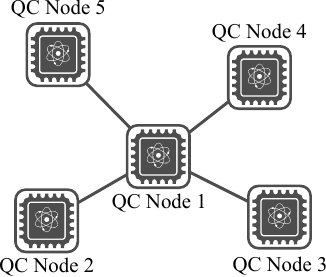}
    \caption[Distributed quantum computation.]{Distributed quantum computation is another application of quantum networks.}
    \label{fig:14-4_distributed_qc}
\end{figure}

Imagine the scenario where a single quantum computer is not capable of executing the computation that we have in mind.
Either the quantum processing unit does not have enough qubits, or they decohere too quickly, or the gates are of insufficiently fidelity.
The situation may seem dire but we need not despair.
We can get around this problem by using \textbf{\emph{distributed quantum computation}}\index{distributed quantum computation}.
Figure~\ref{fig:14-4_distributed_qc} shows five quantum computing nodes connected together via a quantum network.
Each QC node has limited quantum resources in the sense that we mentioned above.
Distributed quantum computation takes the original large problem which cannot be solved by a single QC node, and breaks it into smaller pieces which can be tackled by the QC nodes.
The individual QC nodes share their quantum resources in order to boost the performance or the efficiency of a single QC node.

The primary means of exchanging quantum information between the QC nodes is teleportation using pre-shared Bell pairs.
Another way of achieving distributed quantum computation is to entangle the QC nodes to form a larger distributed multipartite entangled state, which can serve as a substrate for \textbf{\emph{measurement-based quantum computation}}\index{measurement-based quantum computation}.
In such a case, the nodes then need to coordinate their measurements accordingly.
Whatever the means of achieving distributed quantum computation might be, quantum networks play an essential role in connecting the individually weak QC nodes using distributed entangled states.
The main questions of interest in distributed quantum computation are how much entanglement is needed in order to exchange quantum messages between the nodes, and how to efficiently coordinate the computation between the QC nodes.
An important question that needs to be addressed is whether the QC nodes can trust each other, and how much one node is willing to reveal to others about the local computation that it can execute.

The issue of trust brings us to our next application of quantum networks.
Early quantum computers will be sparse due to immense resources needed to run and maintain them.
Users with very limited quantum resources will have to delegate their computation to these quantum servers.
One way is to encode the instructions about the computation into classical bits and send them to the quantum server which then executes the computation.
By doing this, the server will learn the input, the computation itself, as well as the output.
What if the client does not wish to disclose any of this information?
Remarkably, the client can hide all this information by utilizing \textbf{\emph{blind quantum computation}}\index{blind quantum computation} (BQC).
The client only needs very limited quantum resources to engage in BQC.
One of the initial proposals showed that the ability to generate single-qubits is enough for the client to hide the input, the computation, and the output from the server.
The server can only find out an upper limit on the size of the quantum computation that it is being asked to execute.
Different variants of BQC work on the basis that the client can only do single-qubit measurements and it is the server which sends the qubits of a large entangled state one-by-one.
In either case, quantum networks will be needed in order to communicate the information contained in those qubits between the client and the server.

\newpage
\begin{exercises}
\exer{
\emph{CHSH for all Bell pairs.}
Consider Alice's and Bob's measurement bases,
\begin{align}
    A_1 & = Z, & A_2 & = X \\
    B_1 & = \frac{Z - X}{\sqrt{2}}, & B_2 & = \frac{Z + X}{\sqrt{2}}.
\end{align}
We have derived the CHSH correlation function $\mathcal{S}$ for the two of the Bell pairs.
\subexer{
What is the CHSH expression $\mathcal{S}$ that maximally violates the CHSH inequality when applied to the state \ket{\Phi^-}?
}
\subexer{
What is $\mathcal{S}$ that maximally violates the CHSH inequality for the state \ket{\Psi^-}?
}
\subexer{
Consider an initial state \ket{\Phi^+}, under the effect of a bit-flip channel,
\begin{equation}
    \rho = F \ketbra{\Phi^+}{\Phi^+} + (1-F)\ketbra{\Psi^+}{\Psi^+}.
\end{equation}
Compute the appropriate CHSH correlation function $\mathcal{S}$ in terms of the fidelity of the state $F$.
}
\subexer{
At what fidelity $F$ can we not guarantee anymore that the state is entangled?
}
}

\exer{
\emph{Measurements on the $W$ state.}
Let's look at measuring the W state in more detail.
Consider the projectors $\Pi_{\pm 1}^Z$ in Tab.~\ref{tab:3-2_projectors}.  To extend the measurement operators to measuring a subset of qubits, we need to tensor the projection operators with the identity matrix. 
\subexer{
First, consider measuring a single qubit of the W state in the Pauli $Z$ basis.
Calculate the probabilities of finding the $\pm1$ results using
\begin{equation}
\text{Pr} \{\pm 1\} = \langle W | \Pi_{\pm 1}^Z \otimes I \otimes I | W \rangle.
\end{equation}
}
\subexer{Calculate the probabilities of finding the $\pm1$ results when measuring in the Pauli $X$ basis.
}
\subexer{Calculate the probabilities of finding the $\pm1$ results when measuring in the Pauli $Y$ basis.
}
\noindent Now consider measuring two of the three qubits in the W state. You will need four projectors instead of just two.
\subexer{
Calculate the probabilities of finding the $(\pm1,\pm1)$ results when measuring in the $Z$ basis.
}
\subexer{
Calculate the probabilities of finding the $(\pm1,\pm1)$ results when measuring in the $X$ basis.
}
\subexer{
Calculate the probabilities of finding the $(pm1,\pm1)$ results when measuring in the $Y$ basis.
}
}

\exer{
\emph{Measurements on the GHZ state.}
Consider a three-qubit GHZ state.
We saw that measuring its first qubit destroys the entanglement of the state.
\subexer{
Confirm that this is still true when the second or the third qubits are measured.}
\subexer{
Let's consider measuring the first qubit again, but this time in the Pauli $X$ basis.
What are the probabilities of the two possible outcomes.}
\subexer{
Compute the post-measurement state for both measurement outcomes.
Are all of the qubits disentangled?
}
\subexer{
Consider measuring all three qubits of the GHZ state in the Pauli $Z$ basis.
What are the possible outcomes of this measurement on this state?
Can you think of a scenario where this result can be useful?
}
}

\exer{
\emph{Examples of graph states.}
Graph stats are an important class of multipartite entangled states.
Let's consider some basic examples.
\subexer{
Write down the state for a two-qubit graph state \ket{G_2} in the computational basis following the procedure discussed in Sec.~\ref{sec:14-2_multipartite}.
}
\subexer{
Let's rewrite the state \ket{G_2} in new basis, where the second qubit is written in the Pauli $X$ basis.
}
\subexer{
You should be able to see that the state \ket{G_2} is very similar to a Bell pair.
Find a suitable unitary operation $U$ that transforms \ket{G_2} in one of the Bell pairs, for example,
\begin{equation}
    \ket{\Phi^+} = U \ket{G_2}.
\end{equation}
}
\subexer{
Consider the three-vertex graph shown in the left panel of Fig.~\ref{fig:14-2_cluster_state}.
Write down the state of its corresponding graph state \ket{G_3} in the computational basis.
}
\subexer{
Rewrite the state \ket{G_3}, such that the first and the last qubits are written in the Pauli $X$ basis.
}
\subexer{
Does this rewritten form of \ket{G_3} remind you of some three-qubit state that we have discussed?
Find the unitary $U$ that transforms \ket{G_3} into this state.
}
}

\end{exercises}

\chapter{Quantum Internet}

In this chapter, we conclude the book with a discussion of extending from the technologies we have developed throughout the book to a full \textbf{\emph{Quantum Internet}}\index{quantum Internet}, a global network of quantum networks.
We also give some thoughts about developing \textbf{\emph{standards}}\index{standards} for quantum communications, and how understanding quantum communication can lead to a variety of careers in quantum. This chapter is presented as a dialog between the two main authors. It is adapted and updated from the video transcripts and edited for clarity and accuracy, and hence is not a direct transcript.

\section{Networks of networks}

\rrr (laughing) Michal, you look so serious!

So, welcome to chapter fifteen: Quantum Internet.

\mmm Well, sad to say, we've reached the end!

So, what is the Quantum Internet?
Let's begin with the phrase ``network of networks''.
This is a combination of words that you have been hearing quite a lot throughout this book, and here we will give you a slightly more concrete idea of what it really means.

\rrr First off, the (classical) Internet is this really complicated thing. There are a whole bunch of individual networks, and a whole bunch of individual nodes that make up each network. So the global Internet is a ``network of networks''. Let's see a little bit about how that works.

\begin{figure}[t]
    \centering
    \includegraphics[width=1\textwidth]{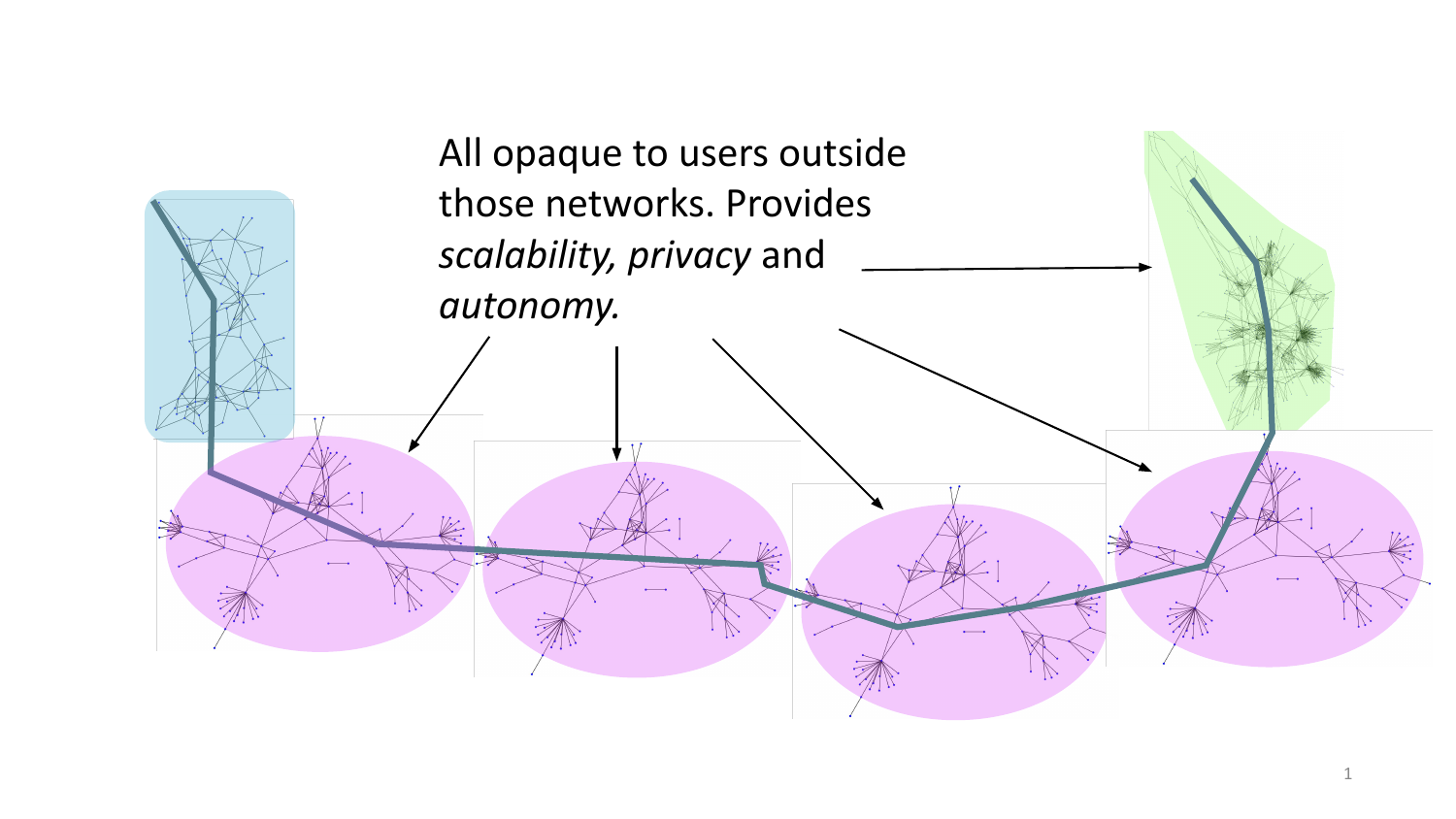}
    \caption[Network of Networks.]{Across an internetwork, a connection passes through many separate, independently managed networks that cooperate with each other for communication purposes. The internal structure of each network is opaque to any node that is not a member of that network; outsiders know only that the network agrees to facilitate communication between neighboring networks.}
    \label{fig:15-1-NofN}
\end{figure}

Assume you want to start in the upper left-hand corner of Fig.~\ref{fig:15-1-NofN}. Your home network is the blue network, and you want to connect to some service that's out on the green network, and you have to get there through the series of pink networks. Michal, how are we going to get there?

\mmm Well, we do not actually know, because we are on the blue network, we just tell the Internet, ``get us there'' and magically it happens!

\rrr We have this path that we have sketched out here on this arbitrary sort of network. The reality is that you are connected to one network and your traffic is going to go from your current location to what is called a \textbf{\emph{gateway}}\index{gateway} to the outside world (also called a \textbf{\emph{border router}}\index{border router}), and from there your traffic will hop across some other networks to get to where it wants to be. But, in order to make the whole system work, the Internet is organized such that you do not know anything about the inside of those other networks. You only know about the blue network.

The other networks are all opaque to the users that are outside of those networks. If you are the operator of one of these networks, this opacity provides you with \textbf{\emph{scalability}}\index{scalability}, \textbf{\emph{privacy}}\index{privacy}, and \textbf{\emph{autonomy}}\index{autonomy}. Scalability is crucial, of course --- otherwise, how are we going to get to billions and billions of nodes on the internetwork?

\mmm True! You can't have every single node on the network know everything about every other node on the entire network, it can not scale that way.

\rrr A second reason for having this information hiding is that the people who own networks want to be able to evolve what is inside the network at their own rate, using their own choices of technology and changing what is going on on the inside. They do not want to share that information outside for business reasons, or for privacy reasons, or for competitive advantage reasons, and this information hiding gives us a certain amount of autonomy. Starting from your network (the blue one on the left), you have to know how to get across your network to that gateway, and then the gateway has to know which network to send things to next. But even that gateway does not have to know anything about the structure inside of that neighboring network.

\mmm Let's have a look where the classical Internet is right now. As of January 2023, there are around seventy-four thousand autonomous systems.

\rrr So what is an autonomous system?

\mmm I was going to ask the same question!

\rrr An autonomous system is one of those networks --- one of the pink ones, or one of the blue ones, or one of the green ones in the figure. Roughly, you can think of an autonomous system as corresponding to an organization, so Keio University would have one network that would be one autonomous system, and the WIDE Project would have a separate network that is one autonomous system~\footnote{The WIDE Project is AS2500. You can learn more about any individual AS at e.g. Hurricane Electric's online information system at \url{https://bgp.he.net/AS2500}.}. The routing protocol then knows which network to go to next --- which one of these autonomous systems is the right ``next hop'' in the AS graph.

\mmm There are around 940,000 IPv4 routing table entries. What does that actually mean? Can you put that into perspective? It sounds like a large number, but what really does it mean?

\rrr It does sound like a large number! But that is not actually the right number to worry about -- that number is how many ``prefixes'', or contiguous blocks of IP addresses, are in use~\footnote{The first video release of this step included an incorrect description of this concept. The description here is more accurate.}.  Many ASes advertise more than one block of addresses.

Instead, we want to know how many \textbf{\emph{neighbors}} each network has in the graph of ASes.  This neighbor relationship is known as an \textbf{\emph{adjacency}}\index{adjacency} or \textbf{\emph{peering}}\index{peering} and is based on a \textbf{\emph{peering agreement}}\index{peering agreement}.
On average, each one of those pink or blue or green networks in Fig.~\ref{fig:15-1-NofN} is connected to about 3.3 other networks. But that average hides a huge range of connections. Of those 74,000 networks, 63,000 are ``stub ASes'' that have only a single neighbor in the AS graph. At the other end of the spectrum, the largest \textbf{\emph{transit network}}\index{transit netwoek} specializing in connecting networks together, rather than serving individual users, has more than 9,500 peers, more than ten percent of all of the networks that make up the Internet.

These network-to-network connections may be direct between two networks (known as \textbf{\emph{private peering}}\index{private peering}), or shared with several networks at places that are called \textbf{\emph{Internet exchange points}}, or IXPs or sometimes just IXes\index{Internet exchange point (IXP)}.

\mmm Right.

\rrr So collectively, each one of those networks shows up in the overall structure for routing across the entire planet, and that entire set of routes is how you figure out how to get from your own network to whatever service you're trying to connect to somewhere on the other side of the planet.

\mmm You know what that reminds me of? 

\rrr What?

\mmm Of the previous animations that we had of the completely connected graph, where we said it does not make sense to just connect all of the nodes to all of the other possible nodes in the network. Same here, you said there are around three to four connections between the autonomous systems. It's the same logic, right?

\rrr Exactly! Actually, if you took that whole 74,000 networks and connected every network to every other network in a \textbf{\emph{complete graph}}\index{complete graph}...how many connections would that be?  What's 74,000 squared divided by two? That would be...two and a half billion adjacencies!  The reality is that they can't all connect to each other, and so the total is about 120,000 adjacencies in the Internet.

Now, the whole evolution of how the Internet got this way, how many networks there are, and the current structure comes from historical business relationships and the like. The quantum Internet will probably evolve in a very different fashion, but this same general idea of having networks that are owned and operated by independent organizations that connect to other networks, that same idea is going to stay.

\mmm I can imagine that.
Developing quantum technology is a very expensive business, so you do not just give everybody access to everything that you have spent so much money in developing and researching.

\rrr Yeah! There are already billions of nodes connected to the Internet (which we talked about earlier in this book) -- computers and mobile phones and IoT devices -- and there is no central node registry, so the true number of these things is not known. If you run a network, you can put more devices on the network (subject to certain constraints) as long as you have the ability to give them addresses and send their traffic out to the broader Internet. In contrast, the phone system has a more centralized authority that gives out phone numbers, and so hypothetically somebody could actually count all of the phone numbers and tell you where they are and you can figure out how many phones there are.  But with the Internet, no one knows how many devices are actually connected to it, because of the autonomy provided by the system allowing each of the individual networks to control what goes on inside of its own network.

\mmm That sounds like a very clever system, it must have taken quite some time to develop this system, no?

\rrr Well, you know, it has been half a century or so since the earliest ARPANET experiments in 1969.  
The modern Internet Protocol suite, known as TCP/IP\index{TCP/IP}, was proposed in 1972-74. IP became the only protocol on the network on January 1, 1983, paving the way for the modern network-of-networks structure. 
The Border Gateway Protocol\index{Border Gateway Protocol (BGP)} is the protocol that exchanges information about the relationships between networks. This modern structure for the Internet was laid out the late 1980s and early 90s. The first document describing BGP was published in 1989~\footnote{\url{https://www.rfc-editor.org/info/rfc1105}}, and the protocol became the central routing mechanism for the Internet in 1994.

\mmm So where do you think the quantum Internet is in relation to the classical Internet?

\rrr We are still decades and decades away from reaching this kind of scale, but we want to apply what we have learned over the last half century in the process of developing the classical internet, and use that knowledge to actually design and develop the quantum Internet as we go.

\mmm Right. Very interesting...

\rrr One other topic that we haven't talked about yet is a big trend in classical information system structures, including both the Internet and computing systems: \textbf{\emph{virtualization}}\index{virtualization}. In the process of virtualization, a computer or even a network might not correspond exactly to a single physical system. You might have a virtual server -- a collection of software and data -- that you think is running on a particular piece of hardware.  However, it might not actually be running \emph{here}, it might be running \emph{over there} instead, somewhere else. There are a lot of mechanisms built into the classical Internet, computer hardware and software to make that virtualization possible, so that services can appear in multiple places and copies of those services can be in different places. All of this allows the services to migrate around based on current conditions or management plans. This virtualization brings a lot of benefits. We are still a long ways away from wanting something like that kind of service virtualization for the quantum Internet, but maybe eventually we will get there.

But a related concept is \textbf{\emph{recursion}}\index{recursion}!

\begin{figure}[t]
    \centering
    \includegraphics[width=0.5\textwidth]{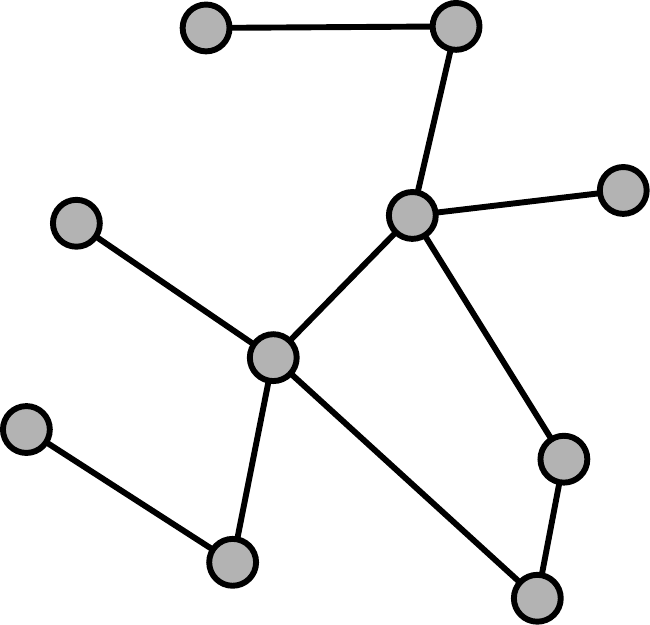}
    \caption[A simple network.]{A simple network.}
    \label{fig:15-2-simple}
\end{figure}

\begin{figure}[t]
    \centering
    \includegraphics[width=.5\textwidth]{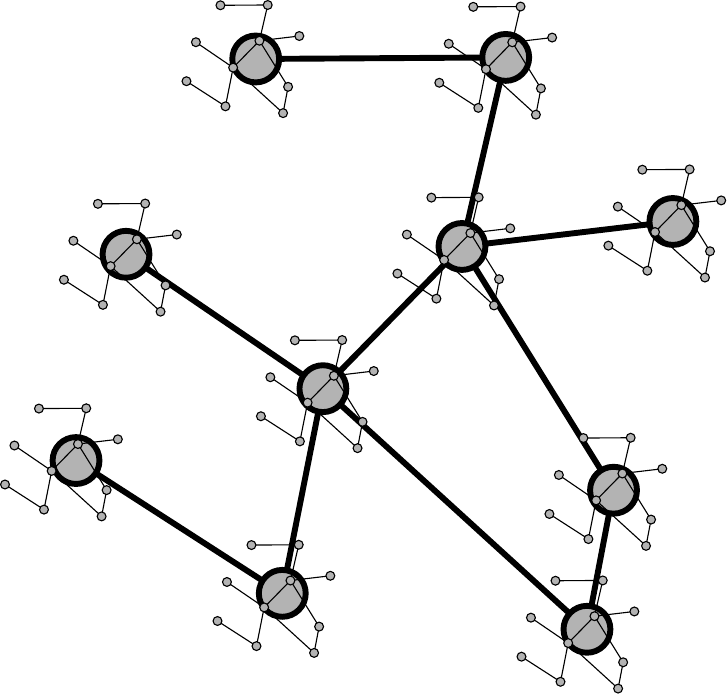}
    \caption[A recursive network.]{A recursive network. In this representation, each ``node'' in the graph is actually an abstraction of an underlying physical network, represented by the smaller nodes and links. In classical Internet terms, the larger nodes and graph could be the graph of autonomous systems. In a  recursive network, two layers are shown here but the recursion could go deeper.}
    \label{fig:15-3-recursive}
\end{figure}

\mmm Oh, I remember recursion from my computing classes.

So we said that the Internet is a network of networks, where every network can be thought of as a single node of some other different layer of a network. Is that what recursion is in the context of networks?

\rrr In Fig.~\ref{fig:15-1-NofN} we had the blue network, and four or five pink networks, and the green network. Each one of those inside has a structure, but we can also talk about the graph that connects that set of networks. You can think about that as the recursion in the system. The system is essentially two layers: the interior gateway protocol and the exterior gateway protocol. The interior protocol is used to manage how you get around inside of your own network. You actually have the choice to use a different protocol or a different algorithm for routing inside your network, as long as you can meet your contractual obligations with others to exchange information with them and to carry traffic back and forth across your network in some reasonable way~\footnote{Commonly used protocols include OSPF and IS-IS, both of which are \textbf{\emph{link state}}\index{link state} protocols where each router learns enough information about the network to run Dijkstra's algorithm (which we saw back on p.~\pageref{dijkstra}) and build a \textbf{\emph{spanning tree}}\index{spanning tree} that it will use to make decisions. Numerous other protocols have been used and continue to be used.}. Today, there is only one choice for the exterior gateway protocol: BGP. BGP is involved in the graph of the relationships between those \textbf{\emph{autonomous systems}}\index{autonomous system}, the Internet-level total architecture between the whole thing. So that is two levels, but the reality is that in most modern local area networks, there's actually another sort of routing system that is built into local area networks as well, so we have already got three layers~\footnote{If you have a network of Ethernet switches, it might in fact be Radja Perlman's spanning tree protocol.}. Conceptually, this process could be repeated indefinitely and we can build what is called a \textbf{\emph{recursive network}}\index{recursive network}.

\mmm In Fig.~\ref{fig:15-2-simple} we see a graph, then in Fig.~\ref{fig:15-3-recursive} see that each of those nodes is really a different network.

\rrr Each one of the individual nodes might be an abstraction, where it is implemented as a network of nodes and links inside. Then inside of that, each one of \emph{those} nodes might, in turn, be implemented as a network inside, and so conceptually at least, it could go to arbitrary depth. Again, that gains us scalability in what we are actually building. It allows us to reuse a lot of the mechanisms that we are designing and building at the level of the global internet; we reuse them when building a wide area network, and then again for an organizational network, and then again for a local area network. Finally, possibly, even inside the individual machines you can actually use similar techniques, and sometimes --- not often, but sometimes --- you will actually have a virtual network inside of your own machine that actually reuses some of those concepts inside, and there's no reason why that process cannot just continue recursively.

\mmm Very cool! So that is the ``network of networks'' concept that underlies today's classical Internet and tomorrow's quantum Internet.  Next, we will see how they work together.

\section{Integration with classical systems}
\label{sec:classical-integration}

\mmm Rod, the title of this section is ``integration with classical systems''. Does that mean that the quantum Internet is not going to replace the classical Internet?

\rrr That is a good question, but yes, the quantum internet will not replace the classical Internet! There will be two networks.

\mmm Right, so how are they going to talk to each other? How are they going to integrate?

\begin{figure}[t]
    \centering
    \includegraphics[width=1\textwidth]{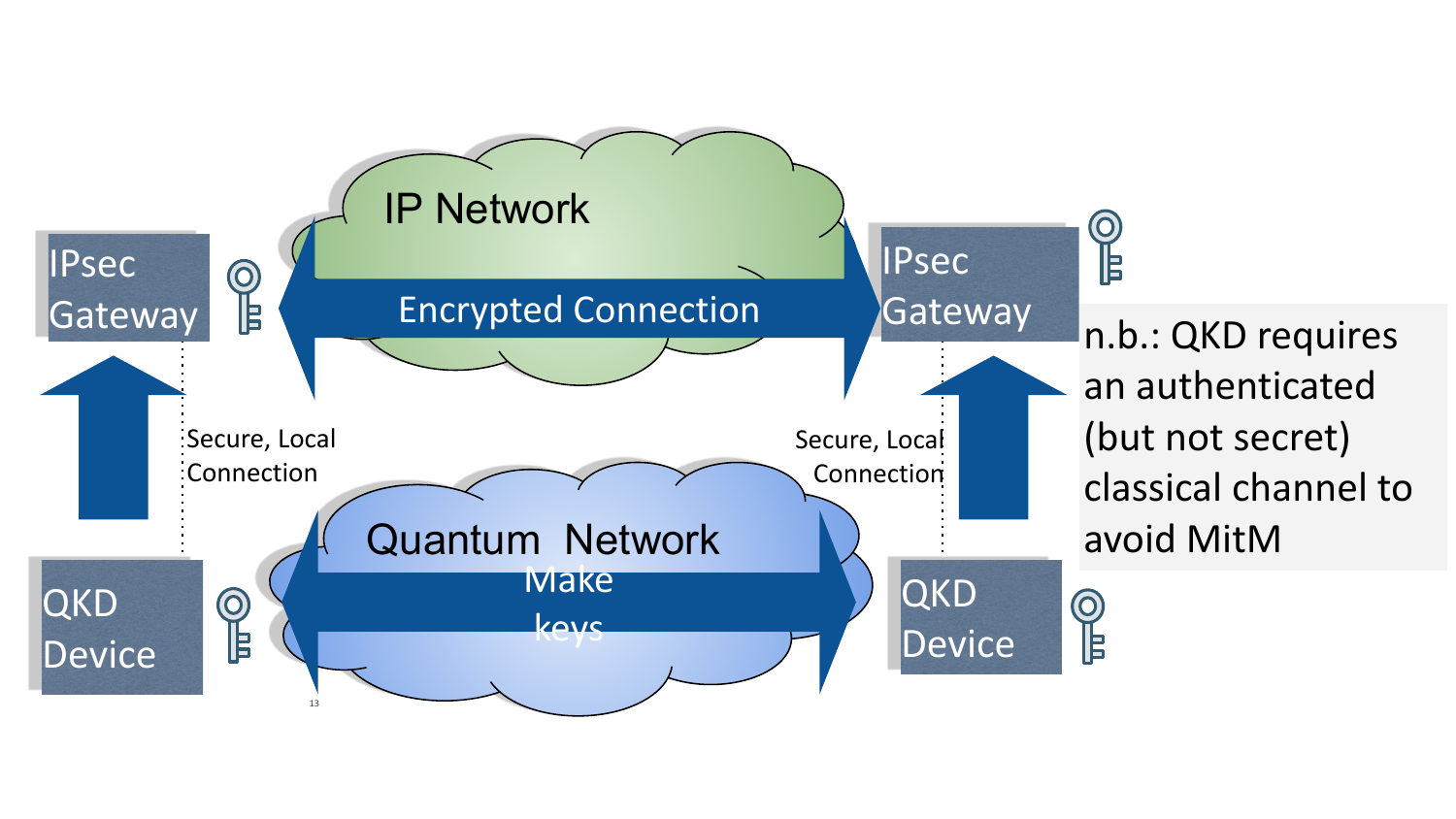}
    \caption[IPsec with QKD.]{The Internet standard encryption protocol known as IPsec can be adapted to use keys generated via QKD rather than Diffie-Hellman key exchange.}
    \label{fig:15-4-ipsec-with-qkd}
\end{figure}

\rrr In Fig.~\ref{fig:15-4-ipsec-with-qkd}, we have two clouds, one labeled ``quantum network'', and one labeled ``IP network''. What does the quantum network do?

\mmm Well, over the course of this module we have seen that the main purpose of a quantum network is to distribute entanglement between parties, whether it be bipartite entanglement or multipartite entanglement, so that it can be used as a resource for communication tasks such as teleportation and QKD (which might be entanglement-based or single photon-based).

\rrr QKD is, in fact, the example that is written in the figure, so what are the QKD devices doing across across the quantum network? Remind us, what's the service they provide?

\mmm Well, they're trying to establish a secret, correlated key between two communicating parties without any eavesdropper knowing what the key is. The key is a secret classical string of bits that we can use to hide our message.

\rrr All right, so it is a string of random bits that are guaranteed to be secret, based on what we have done so far. So what we want to do, once we have them is...

\mmm ...to use them with the classical Internet.

\rrr Yes! So in fact, that is what we are going to do next. The first task is for the quantum network to make those keys, turning quantum states into classical bits, and then the quantum network is going to share those bits with a couple of boxes called IPsec gateways (one at each end), that are connected to an IP network. Have you heard of IPsec\index{IPsec}?

\mmm Not really, no.

\rrr IPsec is one of the standard Internet protocols for encrypting data that you are going to share across the network. One of the other famous protocols is called Transport Layer Security, or TLS\index{Transport Layer Security (TLS)}, which is used for web browsing and for a lot of other things. IPsec is actually slightly older than TLS, and its original design was to connect one network to another network securely. You send data to your nearby IPsec gateway and it encrypts the data and sends it to another IPsec gateway at the other end, where it gets decrypted and sent to your partner over there.

Before the use of QKD, these IPsec gateways worked together to negotiate a key exchange of some sort.

\mmm Back in Sec.~\ref{sec:crypto-phases}, we talked about three phases of an encrypted conversation. The three phases are authentication, key generation and then the bulk data encryption.

\rrr Right. The authentication is often done using \textbf{\emph{public key cryptography}}\index{public key cryptography}. Prior to QKD, and still in most cases today, the key generation was primarily done using a mechanism called \textbf{\emph{Diffie-Hellman key exchange}}\index{Diffie-Hellman key exchange}, which we are going to replace with QKD, as in Fig.~\ref{fig:15-4-ipsec-with-qkd}. Data (random bits) that come out of the QKD network --- or the quantum Internet and the QKD devices attached to it --- will be used to create the keys that are used by the IPsec gateways for encrypting large amounts of data which get sent using an encryption mechanism. Most commonly these days, we use an encryption mechanism called AES, the Advanced Encryption Standard\index{Advanced Encryption Standard (AES)}. Note  that the QKD connection itself actually requires that you also have an authenticated classical channel between the nodes in order to prevent someone standing in the middle of the network and pretending to be Alice in one direction and pretending to be Bob in the other direction, an attack that's called a \textbf{\emph{man-in-the-middle attack}}\index{man-in-the-middle}.

\mmm I see, so let me get this straight. The role of the quantum network really is just one step in a number of steps during a secret communication. So we start classically, we authenticate classically, but then when we require the generation of the key, that is where the quantum part comes in. That is where the quantum magic happens, either through non-fully distinguishable states such as we saw in the BB84 protocol (Ch.~\ref{sec:9_bb84}), or entanglement-based quantum key distribution using the E91 protocol (Ch.~\ref{sec:10_E91}). And whatever the result of that is, it should be a secret, correlated key that is not known to any malicious party. The secret then gets passed back into the classical network, and we proceed classically again.

\rrr Exactly!
And so all of this put together makes for an integrated quantum-and-classical system.
Let's set aside encryption and look at other applications of quantum networking. You are the expert on a couple of these other things that we might want to do with entanglement.

\begin{figure}[t]
    \centering
    \includegraphics[width=1\textwidth]{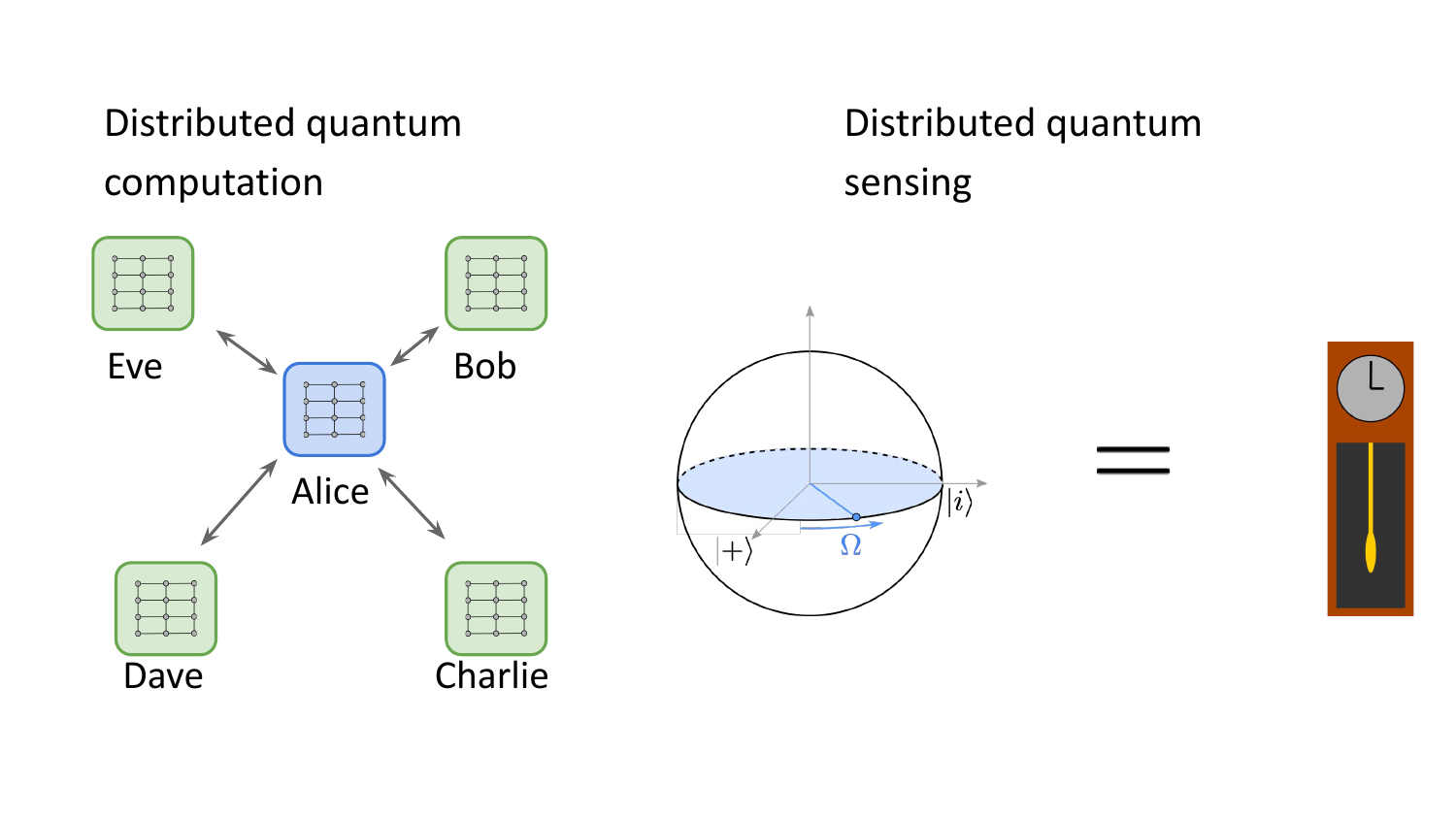}
    \caption[Uses of quantum networking.]{Quantum networks will be used for distributed quantum computation (left) and sensing applications such as clock synchronization (right).}
    \label{fig:15-5-apps}
\end{figure}

\mmm We have talked about distributed quantum computation where Alice, Bob, Eve, Dave and Charlie share some quantum resources, but individually they cannot perform some desired computation because they don't have enough quantum power. They have to figure out how to share their individual quantum power with each other, and to create something that allows them to tackle much larger quantum problems. Fig.~\ref{fig:15-5-apps} illustrates how such a system might be organized. They have to share some quantum resources using the quantum network, but then they have to coordinate with classical messages using the classical network. They basically use the classical control provided by the classical network to coordinate the actions they perform on the quantum part of their systems -- their quantum processors and memories.

\rrr This network of quantum computers that is going to do this distributed computation, is that a complete application? Is that, like, quantum Photoshop or a quantum Oracle database or something? Or...?

\mmm Well, I would say, ``never say never'', but for now the problems that we envision tackling with quantum systems are things that are genuinely quantum, where classical computers can help but are too slow. For example, diagonalizing large matrices, finding new properties of new drugs, new materials, and so on. I do not think that there will be an application of a quantum Photoshop. I do not even know how that would work in reality.

\rrr I do not either! The chemistry one is a really good example, because as I understand it, there will be some classical pre-processing and then we will do a quantum step and then after you get back your quantum answers, then there is going to be some more classical computing to do, is that right?

\mmm That is true. One tool that is used in quantum algorithms for quantum chemistry is known as VQE, or the Variational Quantum Eigensolver, where the quantum part is just one tiny step in a larger framework. It is a very important step that handles a portion of the problem that is particularly slow when you try to solve it using only classical computation and classical devices.

\rrr Then all of that will fit together into an integrated quantum and classical system --- the information system that other people will then use.

\mmm That is right. Since this is an example of a single computation, it might be done by multiple quantum computers inside a single data center, instead of across a wide-area network.

\rrr Okay, tell me about the other application of distributed entanglement that you have in mind.

\mmm Well, the other one is clock synchronization, which we discussed in Sec.~\ref{sec:14-3_clock_sync}. We mentioned that having a global time standard where everybody agrees on the same time is of crucial importance in many, many areas of our modern life. Again, it is not that we want to do something special with quantum time --- the time is classical --- but the way we agree on the same global standard of time comes through using quantum resources such as bipartite entanglement.

\rrr Okay, once you have that, what is it used for?

\mmm The timing signal gets bumped back into the classical network, where important applications might include transportation, such as the global positioning system (GPS). In financial markets even tiny fractions of a second are crucial, they can either gain you millions or lose you millions. So, time is very important and keeping the same time as somebody on the other side of the world is very important as well.

\rrr Right, so all of this will fit together. There is going to be a quantum network and there is going to be some IP network --- the existing classical Internet or what have you --- and these two things have to come together in order to build complete, integrated, hybrid quantum-classical distributed systems.

\mmm And they are all synchronized!
When we were talking about a distributed computation like chemistry, we said that it may be done within a data center. This application is useful in wide-area networks as well.

\rrr That is an important point too.

\mmm Okay, let's shift direction a little bit and see how we can bring these things together.

\rrr All right, let's discuss standardization. If you are a developer, the first thing you are going to do, of course, is to actually build and test some sort of system. Let's say you have a prototype that's up and running. Then what?

\mmm Well then, once that is working, we would like everybody else to use it.

\rrr Yeah, and in communications, that means is that it is nice if my communications device will talk to your communications device.

\mmm Yes! Even though I was the one who made my device, and you were the one who created your device, maybe using completely different means. So how do we standardize these things?

\rrr A standard is a document written and agreed to by a group of people, usually from different organizations~\footnote{The word ``standard'' usually means that it has some official standing, and makers are \emph{required} to follow it, but not all shared specifications have that kind of force; they may just be a temporary, experimental agreement. Here we will not worry about this distinction.}. The information exchange is done in face-to-face meetings, online meetings and extended exchanges on email lists.

For a concrete example, we might think about the physical layer and the network protocol layer. Those two layers might actually be standardized by different organizations. At the physical layer, for example dealing with electrical signals for the Ethernet, a standards organization will define voltage levels and how long a single cable can be~\footnote{e.g., the standard wired electrical Ethernet today is called \emph{Cat 5e}. See \url{https://en.wikipedia.org/wiki/Category_5_cable}.}. One of the things we talked about back in Sec.~\ref{sec:11-2_mode_dispersion} was distortion of signals due to mode dispersion\index{mode dispersion}. We were talking about that in terms of optical fibers, but various types of damage happen with any kind of signal, and so one of the things you have to worry about is how much degradation of the signal is allowed so that you can still guarantee that the end nodes will be able to reconstruct the original message. Engineers come together in a meeting and work out the calculations, usually making some trade-offs between reliability, speed and cost, and decide as a group what is best. Then they write that down in a document called a \textbf{\emph{specification}}\index{specification}. All of that is done for classical signals. What are the equivalents of that for photons?

\mmm Well, we said that in quantum communication, we're interested in exchanging information via single photons, so we have to think about how to get two different devices to talk, particularly at the border between two networks, which will have to exchange photons. How can we ensure that whatever photons I send to the other network will be accepted, and vice versa? I imagine that important factors will include the wavelength of the photons, the shape of the wave packet of the photons, and the time of arrival. When we talked about the BSA in Sec.~\ref{sec:bell-state-measurements-II}, we stressed that in order for interference to work properly, the two photons that are coming in must be indistinguishable. Somehow, we must have a standard procedure that makes sure that whatever photons you are sending me are the same as the photons as I am sending to you. Then when they hit each other and interfere at the BSA, entanglement is created between us, and between the two networks that we are part of.

\rrr The organizations that do standards, like the IETF, the IEEE, ANSI, ITU, will be involved at some point, too. In fact, for QKD, some standards are already being developed; see the ITU-T Y.3800-3999 series~\footnote{\url{https://www.itu.int/ITU-T/recommendations/rec.aspx?id=13990&lang=en}.}

The Internet Research Task Force (IRTF), a sister organization to the Internet Engineering Task Force (IETF), focuses on bringing the Internet community together to develop a shared understanding of problems and solutions that may eventually lead to standards. IRTF work sometimes leads to documents known as \textbf{\emph{Requests for Comments}}\index{Request for Comments}, or RFCs, and the Quantum Internet Research Group published its first quantum repeater network-releated RFC in 2023~\footnote{\url{https://www.rfc-editor.org/info/rfc9340}}.

\section{Putting it all together}
\label{sec:putting-it-all-together}

\mmm Rod, let's put all of these things that we learned in this module (or book) into some context.

\rrr Let's go all the way back to the core idea. There are four things that a repeater has to do, and then once we have  successfully built repeaters, then we can build a network. So the first thing...

\mmm Well, first we have to establish entanglement between two neighboring nodes, so link-level entanglement is item one.

\rrr That involves handling loss of photons!

\mmm Very important.

\rrr Second thing?

\mmm We extend entanglement from neighboring nodes to nodes which are separated by larger distances across multiple hops, so we establish end-to-end entanglement.

\rrr The primary mechanism we used for doing that was entanglement swapping\index{entanglement swapping}. 

\mmm That's right.
Third, we have to think about how to handle state and operation errors. We saw a few examples in our calculations of unitary errors, but there are also other types of errors. The main protocol that we used was purification, which works against both unitary and non-unitary errors.

\rrr All right, and then the last thing?

\mmm Fourth, we have to think more in terms of networking and how to handle things like routing; multiplexing when there is contention for resources; and how to manage all of these things, including security.

\rrr We talked about routing protocols in the multi-level system that is the classical Internet earlier in this chapter, but also back in Chapter~\ref{sec:12_quantum_repeaters} when discussing individual quantum networks or single-level systems.

All right! So all of those are technical requirements for how you go about building boxes and an internet but people who are taking this module (or studying this book) ---

\mmm They're people! 

\rrr Right! \emph{People} are the ones who design and build and operate these networks. This is the first module in what will be an extensive, in-depth sequence on quantum communications and quantum computation.  Our focus is quantum communications, but other groups that are part of this educational effort in Japan are making modules on the other topics, and the set is broad enough that you can choose your specialty. More broadly, as you select from this entire set of modules and complete a focused degree, a minor, or just a class or two for your own interest, many different kinds of jobs are possible leading into various career paths.

\mmm So what are these specializations?

\rrr Well, I am an architect, so I list architecture first. Architecture involves defining the subsystems within a larger system and defining the relationships, so you define block A and block B and the relationship between the two --- the contractual agreement between them, what messages they exchange, what behavior is expected of the different subsystems. All of that allows you to make forward progress by dividing a large problem into a set of smaller problems, and then each of those can be worked on independently to a certain extent.

Someone has to design the protocols! The protocols define the messages that get exchanged on the wire and the behavior: what you do when you get a message, but also what you do when you \emph{don't} get a message. These and other sets of rules you have to follow, including the sort of promises you make as a condition of participating in a particular network communication, for example, all form a protocol.  As we just discussed, the work in protocol design and standardization involves working with many people.

\mmm Hardware is very important. That's closer to my heart as a physicist, but hardware are the physical things that basically allow all of these more abstract things such as architecture and protocols to work. So we need to design the hardware, we need to analyze it, we have to ensure that our quantum boxes or network nodes, including the quantum memories, are distributing entanglement properly. If something goes wrong, what do we do, and how do we test things, and how do we build them?

\rrr Everything we are talking about here is all super critical. You have to have all of these things in order to build a network, but fundamentally if the hardware does not work, we are just toast right off the bat.

You have to have software, too. Some of that is software that is internal to the boxes you are building, and so nobody cares if you change that implementation. What they care about is the external behavior of the box. But software also includes implementation of those protocols that we talked about earlier. And that building hardware is hard, right? Building software, people think it's easy, but ---

\mmm It does not happen automatically.

\rrr It is also the case that software can be iterated more quickly than hardware. You can make small changes to things in software and redeploy software fairly quickly. But! Where software meets protocols, once the protocols are defined, it starts to get to be a lot harder to change. Although protocol software may iterate quickly, because getting that right also takes a lot of work, interoperability is critical and that requires stable interfaces and protocols. You have to have the right balance between hardware and software, both when designing the system and when hiring and assigning people. Although it often seems like it is easier to get software up and running quickly and easily, building a complete and robust system that does everything that is expected and nothing it is not expected to do, that is all hard work, too.

\mmm Right.

\rrr We also have to have operations and management of the networks! So, once all these boxes are built, somebody's job is going to be to order boxes from the companies that provide this stuff and bring them into your building and unbox them and set them up and turn them on and run them and make sure that they connect to the local systems and that everything works and then keep track of what is working, what is not. All of that falls under the category of operations and maintenance. That is an important career path in the classical Internet as well.

\mmm The next one is education and community, exactly what we are doing now through this online course and book! We are educating people ---

\rrr ...and we are participating in the community! You and I are both researchers and educators, but this applies not only at the university level where we are right now, but also includes high school and post-secondary education. It can involve educating the public as well as educating people within a formal school context. All of that is vital, so some people studying this module could wind up as high school teachers. That is a perfectly fine career path for going forward, because I am certain that there are going to be quantum classes in high schools in the future.

Another aspect of community is \textbf{\emph{ethics}}\index{ethics}. There is a lot of discussion these days about the ethics of artificial intelligence, or AI. There is just beginning to be a conversation about the ethics of of quantum as well. In addition to the moral issues, we must consider how all this quantum technology is going to fit into society and our legal system, because in particular, quantum networks and quantum computers both have a significant impact on cryptography, which is considered to be a sensitive issue.

\mmm It seems like a very disruptive technology.  Anything quantum, whether it is communication or computation...

\rrr Businesses and governments both care about this and how it is going to affect their their own operations as well as their own societies.
Finally, we come to theory! That would be you, wouldn't it?

\mmm That would be me, yes! Theory is always needed, particularly when it comes to quantum computation and quantum technologies. One quite important field is information theory: the formal mathematical theory of how to process information, how to distribute information, how to communicate information. These are all included under the umbrella of information theory, which can be either quantum or classical, but quantum is the more fun one -- and often the more surprising one.

The design of new algorithms, of course, is the domain of theory! We always want to know how to do things better or how to do things which we have not thought of, and quantum is the perfect playground for that. There is truly a large opportunity for creativity: let it run free, and see what amazing new algorithms and applications we can come up with.

\rrr It is a fantastic and exciting time to be in quantum, whether quantum networking or quantum communication. I have been doing this since 2003. Prior to that I was doing primarily classical systems, and you know, every year quantum gets to be more and more exciting, and more and more real.

\mmm Particularly with the recent developments by IBM and Google and other new startup companies like IonQ and PsiQuantum, it is a truly mesmerizing time to be in quantum, whether it's computation, communication or sensing!

\rrr And now is a great time for you all to be ``quantum natives'' and to come join all of us in the process of this.

So, thank you all for participating in this module (or reading this book), and we will see you again in the next module (or book)!

\mmm See you, bye!

And a \textbf{\emph{huge}} thanks to all of the people that helped us create and edit both the online course and this textbook. You will find their names in the front of the book.

\newpage
\section*{Quiz}
  \addcontentsline{toc}{section}{Quiz}

A quiz covering this block of lessons is available in the online system.


\section*{Further reading chapters 14-15}
  \addcontentsline{toc}{section}{Further reading chapters 14-15}

{\bf Chapter 14}

For those interested in how a GHZ state can be produced with linear optics, we recommend:

Dik Bouwmeester, Jian-Wei Pan, Matthew Daniell, Harald Weinfurter, Anton Zeilinger, Observation of three-photon Greenberger-Horne-Zeilinger entanglement, \emph{Physical Review Letters} 82, 1345 (1999)~\cite{bouwmeester:ghz}.
A freely accessible preprint version can be found on the arXiv~\footnote{\url{https://arxiv.org/abs/quant-ph/9810035}}.

Extension of the Bell inequality to three particles and its experimental violation is reported here:

Jian-Wei Pan, Dik Bouwmeester, Matthew Daniell, Harald Weinfurter, Anton Zeilinger, Experimental test of quantum nonlocality in three-photon Greenberger-Horne-Zeilinger entanglement, \emph{Nature} 403, 515 (2000)~\cite{pan2000experimental}.

{\bf Chapter 15}

The notion that multiple layers of topology can be combined, that an internetwork is a network of networks, extends back to the early 1970s or before; see Vint Cerf, Yogen Dalal and Carl Sunshine, RFC 675 (1974)~\cite{RFC0675}.  For a modern perspective on designing internetworks, we recommend Dave Clark's book, \emph{Designing an Internet}~\cite{clark2018designing}.

For more on the current state of routing in the Internet, as of this writing, Geoff Huston writes a detailed blog entry each year analyzing trends, published on the APNIC website~\footnote{e.g., the January 2023 entry is at \url{https://blog.apnic.net/2023/01/06/bgp-in-2022-the-routing-table/}}. Up-to-the-moment information can be found at Hurricane Electric's BGP Toolkit~\footnote{\url{https://bgp.he.net/}}.

The idea of unifying many layers into a single, indefinitely recursive structure was proposed by Touch \emph{et al.}~\cite{touch2006recursive}, and adapted to quantum networking as described by Van Meter, Horsman and Touch~\cite{van-meter11:rqrn}.

For discussion of the physical layer of the Quantum Internet we recommend:

H. Jeff Kimble, The quantum internet, \emph{Nature} 453, 1023 (2008)~\cite{kimble08:_quant_internet}.

A freely accessible preprint version can be found on the arXiv~\footnote{\url{https://arxiv.org/abs/0806.4195}}.

Kimble’s paper focuses on the hardware and does not discuss any of the networking aspects. For that we recommend the book you are currently reading, or for more advanced material either the third course in this Q-Leap Communications sequence, simply titled ``Quantum Internet'', or Van Meter’s monograph:

Rodney Van Meter, \emph{Quantum Networking}, Wiley-ISTE, 2014~\cite{van-meter14:_quantum_networking}.

\endmatter

\nocite{*}
\bibliographystyle{mit-chicago}
\bibliography{main}

\printindex
\onecolumn

\end{document}